%% file: thesis.tex
\documentclass[a4paper,twoside,11pt,BCOR20mm,DIV14,headsepline,
chapterprefix,appendixprefix,cleardoubleempty]{scrbook}
\usepackage[dvips]{graphicx}
\usepackage{amssymb,amsmath}
\usepackage{bbm}
\usepackage{dsfont}
\usepackage{tuetitle}
\usepackage[sort&compress,square,comma,numbers]{natbib}
\usepackage[romanian,ngerman,french,UKenglish]{babel}
\usepackage{enumerate}
\usepackage{colortbl}
\definecolor{darkblue}{rgb}{0,0,0.5}
\usepackage[colorlinks,linkcolor=darkblue,citecolor=darkblue,
urlcolor=darkblue]{hyperref}
\usepackage[automark]{scrpage2} 

\newcommand{\eq}[1]{Eq.~(\ref{#1})}
\newcommand{\fig}[1]{Fig.~{\ref{#1}}}
\newcommand{\be}{\begin{equation}}
\newcommand{\ee}{\end{equation}}
\newcommand{\bea}{\begin{eqnarray}}
\newcommand{\eea}{\end{eqnarray}}
\newcommand{\ST}{Slavnov--Taylor }
\newcommand{\YM}{Yang--Mills }
\newcommand{\DS}{Dyson--Schwinger }
\newcommand{\BS}{Bethe--Salpeter }
\newcommand{\w}{\omega}
\newcommand{\e}{\varepsilon}
\newcommand{\al}{\alpha}
\newcommand{\ba}{\beta}
\newcommand{\ga}{\gamma}
\newcommand{\G}{\Gamma}
\newcommand{\de}{\delta}
\newcommand{\De}{\Delta}
\newcommand{\et}{\eta}
\newcommand{\ta}{\tau}
\newcommand{\si}{\sigma}

\newcommand{\la}{\lambda}

\newcommand{\ka}{\kappa}
\newcommand{\ro}{\rho}
\newcommand{\ha}{\frac{1}{2}}
\newcommand{\pd}{\partial}

\renewcommand{\th}{\theta}
\newcommand{\cd}{{\cal D}}
\newcommand{\cs}{{\cal S}}
\newcommand{\cl}{{\cal L}}
\renewcommand{\div}{\vec{\nabla}}
\newcommand{\s}[2]{{#1}\!\cdot\!{#2}}
\newcommand{\ov}[1]{\overline{#1}}
\newcommand{\dk}[1]
{\,\,\,\raisebox{-0.4ex}{\large $\bar{}$}\!\!d\,{#1}\,}

\newcommand{\ev}[1]{<\!\!{#1}\!\!>}
\newcommand{\kslash}{k\hspace{-2mm}\slash}

\newcommand{\dx}[1]{d^4{#1}\,}
\renewcommand{\imath}{\mathrm{i}}

\setkomafont{pageheadfoot}{\fontsize{10pt}{10pt}\scshape\sffamily}
\setkomafont{pagenumber}{\normalfont\scshape\sffamily}
\addtokomafont{caption}{\small} 
\setkomafont{captionlabel}{\sffamily\bfseries}
\deffootnote[1em]{1.1em}{0.3em}{\textsuperscript{\thefootnotemark}}
\allowdisplaybreaks[1]

\begin{document}

\include{title}
\include{dedication}

\include{thanks}

\include{zusammenfassung}

\include{abstract}
\pagestyle{empty}
\clearpage
\pagestyle{scrheadings}
\renewcommand*{\chapterpagestyle}{empty}  
\renewcommand*{\chaptermarkformat}
{\chapapp~\thechapter\autodot\enskip}
\renewcommand{\sectionmark}[1]{\markright{\thesection\ #1}}
\setcounter{page}{1} 
\include{qcd}

\include{ds}
\include{perturbative}
\include{st}

\include{hq}
\include{nGreen}
\include{conclusions}

\include{app.not}
\include{app.standard}

\include{app.check.noncov2pct}

\include{app.noncov3pct}

\include{app.energy.vertex}

\addcontentsline{toc}{chapter}{Bibliography}
\bibliographystyle {utphys2} 
\bibliography{$HOME/bibliography/biblio}

\end{document}

%% file: title.tex
\thispagestyle{empty}

\author{Carina~Popovici}
\birthplace{
\foreignlanguage{romanian}
{Turnu-M\u{a}gurele} (Rum\"anien)}
\dateoforalexamination{09.02.2011}
\examinationyear{2011}
\title{Quark sector of Coulomb gauge Quantum Chromodynamics}
\dean{Prof.~Dr.~Wolfgang Rosenstiel}
\experts{Prof.~Dr.~Hugo Reinhardt \and Prof.~Dr.~Werner~Vogelsang
}
\maketitle

%% file: dedication.tex
\thispagestyle{empty}
\vspace*{8cm}
\begin{center}
{\it To my niece Carina-Elena}
\end{center}
\clearpage{\pagestyle{empty}\cleardoublepage}

%% file: thanks.tex
\chapter*{Acknowledgements}
\pagestyle{empty}
\renewcommand*{\chapterpagestyle}{empty}

First of all, I want to express my gratitude to my supervisor Prof.\
Hugo\ Reinhardt for all his encouragement, help and support during my
PhD.

A big thank you must also go to Dr.\ Peter\ Watson, for his systematic
guidance and great effort he put into training me through all stages
of this thesis. I have profited not only from his constructive ideas
and sharp judgement, but also from his intellectual honesty in
handling all kinds of problems and challenges coming on the way. It
has been an honor and a privilege to be his first PhD student.

For the collaboration during my Master studies and for patiently
leading my fist steps in T\"ubingen, I am grateful to Prof.\
Karl-Wilhelm\ Schmid and Prof.\ Alexandra\ Petrovici.

I also acknowledge the staff in the Institute for Theoretical Physics,
in particular Ingrid\ Estiry and Sabrina\ Metzler, for their great
help and efforts to minimize bureaucracy.

The chats in the celebrated coffee room with Dr.\ Markus\ Quandt, Dr.\
Giuseppe\ Burgio, Davide\ Campagnari, Markus\ Leder, Markus\ Pak,
Tanja\ Branz, Tina\ Oexl, Dr.\ Wolfgang\ Schleifenbaum, my officemate
Dr.\ Wolfgang Lutz and all the colleagues, past and present, may not
always have been about physics but they were always appreciated and
fun. Thanks also to Martina\ Blank for the ``out of the box''
discussions and for sharing her ideas with me.  Thanks to Gabriela\
Vi{\c{s}}{\u a}nescu for creating, every now and then, a very special
Romanian atmosphere in T\"ubingen.

I would also like to acknowledge Prof.\ Michael\ Pennington for giving
me the chance to spend a wonderful time at IPPP in Durham.  It is a
pleasure to recall the numerous and interesting discussions and quite
generally, the nice atmosphere at work. In particular, I would like to
mention Dr.\ J\"org\ J\"ackel, Dr.\ Daniel\ Ma$\hat{\i}$tre, and Dr.\
Emiliano\ Molinaro. A warm thank to Daniel for his very precious
friendship and for taking the responsibility to entertain me.

Thanks also to Prof. Orlando Oliveira for giving me the opportunity to
start work in a new area, for his trust and support over the last
months of this PhD.

For both the generous financial support and the many interesting
events I could take part in, I am indebted to the \emph{Deutscher
Akademischer Austausch Dienst} (DAAD).

My family --- my parents, my sister and my beloved little niece ---
are my greatest source of strength and inspiration.  I wish to take
this opportunity and express my deepest gratitude for their love,
care, encouragement and unlimited support.

Last, but definitely not least, a very-very big-big thank you to
Torsten, for teaching me Quantum Field Theory in the seminar back in
2005, for explaining me over and over ``that thing with the Gribov
copies'', for coping with all my crises, and ultimately for being
there with me through the whole thing.

\clearpage

%% file: zusammenfassung.tex
\pagestyle{empty}
\renewcommand*{\chapterpagestyle}{empty}
\chapter*{Zusammenfassung}

Quantenchromodynamik [QCD] wird heute als die Theorie betrachtet,
welche die Starke Wechselwirkung zwischen den fundamentalen
Konstituenten der Hadronen, den Quarks und Gluonen, korrekt beschreibt
\cite{Peskin:1995ev,
Itzykson:1980rh,Muta:1998vi,Ryder:1985wq,Pokorski}.  Bei gro{\ss}en
Energien (kleinen Abst\"anden) verschwindet die Kopplungskonstante --
ein Ph\"anomen, das gemeinhin als \emph {asymptotische Freiheit}
bezeichnet wird. In dem entsprechenden Impulsbereich wurde die
St\"orungstheorie angewendet und in tiefinelastischen
Streuexperimenten erfolgreich \"uberpr\"uft. F\"ur mittlere und kleine
Impulse hingegen wird die Kopplungskonstante so gro{\ss}, dass die
St\"orungstheorie nicht mehr angewendet werden kann. Daher m\"ussen
andere Methoden angewendet werden, um das \emph {Confinement} zu
erkl\"aren, also das Ph\"anomen, dass im Experiment nur farblose
Hadronenzust\"ande beobachtet werden.  In dieser Dissertation wird der
Quark-Sektor der QCD in Coulomb-Eichung mit Hilfe der
Dyson-Schwinger-Gleichungen untersucht.  In diesem Rahmen nutzen wir
verschiedene Strategien, um die Eigenschaften der QCD sowohl bei
gro{\ss}en als auch bei kleinen Impulsen zu studieren.

Im ersten Kapitel erinnern wir an einige grunds\"atzliche
Eigenschaften der QCD. Ausgehend von der Lagrangefunktion der QCD
pr\"asentieren wir die Eichfixierung und motivieren unsere Wahl der
Coulomb-Eichung. Verschiedene Aspekte des Confinements werden
diskutiert, insbesondere der Gribov-Zwanziger-Confinement-Mechanismus
und seine Relevanz in Coulomb-Eichung.

Das zweite Kapitel besch\"aftigt sich mit der grunds\"atzlichen
Ableitung der Dyson-Schwingen-Gleichungen. Funktionale Methoden werden
eingef\"uhrt und der Quark-Pro\-pa\-ga\-tor sowie die Quark-Beitr\"age
zur Gluon-Zweipunktfunktion und zur Quark-Gluon-Vertexfunktion werden
formal abgeleitet.

Im dritten Kapitel werden diese Funktionen in 1-Loop-St\"orungstheorie
ausgearbeitet. Um die in den Gleichungen auftretenden
nicht-kovarianten Loop-Integrale in Coulomb-Eichung zu behandeln, wird
eine neue Methode basierend auf Differentialgleichungen und partieller
Integration entwickelt. Physikalische Resultate werden verifiziert, so
z.B. die Gültigkeit der analytischen Fortsetzung zwischen Minkowski-
und Euklidischer Raum-Zeit und die Renormierung der Quarkmasse. Des
weiteren wird der Quark Beitrag zum 1-Loop-Koeffizient der
$\ba$-Funktion berechnet.

Das vierte Kapitel ist der Slavnov-Taylor-Identit\"at der
Quark-Gluon-Vertexfunktion gewidmet.  Insbesondere wird das Auftreten
des sogenannten Quark-Geist-Streukerns untersucht.

In Kapitel~\ref{chap:hq} n\"ahern wir uns dem Confinement-Problem
durch Betrachtung des Grenzfalls schwerer Quarks. In diesem Limes
nutzen wir den (vollst\"andig nicht-st\"orungstheoretischen)
funktionalen Formalismus kombiniert mit einer Entwicklung nach
Potenzen des Inversen der schweren Quarkmasse. Durch Einschr\"ankung
auf die f\"uhrende Ordnung in dieser Massen-Entwicklung leiten wir
eine strenge analytische L\"osung f\"ur den Propagator der schweren
Quarks ab.  Anschlie{\ss}end nutzen wir die Gleichungen f\"ur
gebundene Zust\"ande von Mesonen und Baryonen, um das linear wachsende
Potential abzuleiten, das Quark-Confinement erkl\"art.

Kapitel~\ref{chap:nGreen} behandelt die Vier-Punkt Greenschen
Funktionen der Theorie. Ausgehend vom in Kapitel~\ref{chap:ds}
eingef\"uhrten Funktionalformalismus werden diese Funktionen explizit
abgeleitet und ihre Beziehung zu den Gleichungen gebundener Zust\"ande
aus dem vorangegangenen Kapitel diskutiert.

Kapitel~\ref{chap:concl} enth\"alt die Zusammenfassung und das Fazit.
Es folgen die Anh\"ange in denen unter anderem die in
Kapitel~\ref{chap:g^2} abgeleiteten nicht-kovarianten Integrale
explizit \"uberpr\"uft werden und auch einige in der Arbeit
ben\"otigte Zwei- und Drei-Punkt-Integrale berechnet sind.

\clearpage\pagestyle{empty}\cleardoublepage

%% file: abstract.tex
\pagestyle{empty}
\renewcommand*{\chapterpagestyle}{empty}
\chapter*{Abstract}

Quantum Chromodynamics [QCD] is widely believed to be the correct
theory of strong interactions between the fundamental constituents of
the hadrons, the quarks and gluons \cite{Peskin:1995ev,
Itzykson:1980rh,Muta:1998vi,Ryder:1985wq,Pokorski}.  At high energies
(small distances), the coupling between quarks and gluons tends to
zero, a phenomenon known as \emph {asymptotic freedom}.  In this
momentum region, perturbation theory has been applied and successfully
tested in deep inelastic processes.  At intermediate and low momenta
though, the coupling constant becomes strong enough to invalidate
perturbation theory. Different methods must be employed to investigate
\emph {color confinement}, i.e.  the phenomenon that only colorless
hadronic states are observed in the experiment. This thesis deals with
the quark sector of Coulomb gauge QCD and the method we employ is the
\DS equations.  In this framework, we utilize different strategies to
explore both the large and small momentum properties of QCD.

In the first chapter we review some basic properties of strong QCD.
Starting with the QCD Lagrangian, the gauge fixing procedure is
presented, and our choice of using Coulomb gauge is motivated. Certain
aspects of confinement are discussed, in particular the
Gribov-Zwanziger confinement mechanism and its relevance in Coulomb
gauge.

The second chapter is concerned with the formal derivation of the \DS
equations. Functional methods are introduced and the quark propagator,
along with the quark contribution to the gluon two-point functions and
the quark-gluon vertex functions are formally derived.

In the third chapter, these functions are examined at one-loop
perturbative level. To handle the Coulomb gauge noncovariant loop
integrals entering the equations, a new method based on differential
equations and integration by parts technique is developed.  Physical
results are verified, such as the validity of the analytic
continuation between Minkowski and Euclidian space, and the quark mass
renormalization. The quark contribution to one-loop coefficient of the
$\ba$-function is also calculated.

Chapter~\ref{chap:st} is devoted to the \ST identity for the
quark-gluon vertices. In particular, the appearance of the so-called
quark-ghost scattering kernels is explored.

In Chapter~\ref{chap:hq}, we address the problem of confinement by
restricting ourselves to the heavy quark sector of the theory. In this
limit, we employ the (full nonperturbative) functional formalism,
combined with an expansion in the inverse of the heavy quark
mass. Restricting to the leading order in the mass expansion, we
derive an exact, analytical solution for the heavy quark propagator.
We then consider the bound state equations for mesons and baryons and
use them to derive the linearly rising potential which confines the
quarks.

Chapter~\ref{chap:nGreen} is devoted to the four-point Greens
functions of the theory. Based on the functional formalism introduced
in Chapter~\ref{chap:ds}, these functions are explicitly derived and
their connection to the bound state equations from the previous
chapter is discussed.
 
Chapter~\ref{chap:concl} includes the summary and conclusions. It is
followed by appendices where, among others, the noncovariant integrals
derived in Chapter~\ref{chap:g^2} are checked explicitly, and also
some standard two- and three-point integrals needed in this work are
evaluated.

\clearpage\pagestyle{empty}\cleardoublepage

%% file: qcd.tex
\chapter{QCD in Coulomb gauge}
\label{chap:qcd}

A deep understanding of QCD requires a whole toolbox of theoretical
methods: analytical perturbative methods for weak coupling, numerical
lattice gauge theory, along with the canonical approach
\cite{Abers:1973qs}, and functional methods \cite{Itzykson:1980rh}.
The latter two methods have the advantage that they are not restricted
to weak coupling and can still be treated analytically. The functional
methods most commonly used are the renormalization group equations
(see, for example, Refs.~\cite{Gies:2006wv, Pawlowski:2005xe} for
reviews) and the \DS equations of motion for the Green's functions of
the theory \cite{Dyson:1949ha, Schwinger:1951ex, Schwinger:1951hq}.
In this thesis, we employ the functional equation techniques and
derive the \DS equations.

Since QCD is a non-abelian gauge theory, within the functional
approach considered here it is necessary to fix the gauge. Thus, after
briefly introducing the QCD Lagrangian, we will present the gauge
fixing procedure and the problems related to it, with emphasis on
Coulomb gauge. We will then motivate why among various gauges, Coulomb
gauge has the advantage that it is ``physical'': after converting to
first order formalism, we will show that the number of dynamical
variables reduces to the number of the physical degrees of freedom.
Further, the Gribov-Zwanziger confinement scenario
\cite{Zwanziger:1998ez, Gribov:1977wm} will be introduced and its
relevance for Coulomb gauge will be put forward. The alternative
approaches to QCD in Coulomb gauge will be also reviewed, and in
particular, the results obtained within the Hamiltonian formalism will
be outlined.

Although Coulomb gauge seems to be more efficient in identifying the
physical degrees of freedom, noncovariance introduces severe technical
difficulties\footnote{The noncovariant Feynman integrals in Coulomb
gauge will be examined in Chapter~\ref{chap:g^2}.}  and moreover, the
problems related to renormalization have not yet been solved.  From a
practical point of view, Landau gauge has the advantage of being
covariant and thus many infrared investigations have been undertaken,
however studies are still in progress.  A brief review of the results
obtained in this gauge, also in correspondence to the confinement
mechanism, will be presented at the end of this chapter.
\section{QCD as non-abelian gauge theory}

QCD is a non-abelian gauge theory whose matter constituents, the
quarks, are spin $1/2$ fermions and obey the Dirac
Lagrangian\footnote{Initially we use Minkowski metric defined in the
Appendix~\ref{chap:app.not}. }
\be
{\cal L}_q=\bar q_{\al}(x)\left[\imath \gamma_{\mu}D^{\mu}
-m\right]_{\al\ba}q_{\ba}(x),
\textrm{~~~~} \bar q=q^{\dagger}\ga^0 
\label{eq:dirac}
\ee
where the Dirac $\ga^\mu$ matrices satisfy the Clifford algebra,
$\{\gamma^\mu,\gamma^\nu\}=2g^{\mu\nu}$ and the indices $\al,\ba\dots$
commonly denote the Dirac spinor, flavor and (fundamental) color.  The
quark fields $\bar q, q$ transform in the fundamental representation
of the gauge group $SU(N_c)$\footnote{Some relevant formulas for the
group $SU(N_c)$ are collected in the Appendix~\ref{chap:app.not}.},
with $N_c=3$ realized in QCD. The covariant derivative (in the
fundamental representation) is given by
\be
D_{\mu}=\pd_{\mu}-\imath g A_{\mu},
\ee
where $g$ is the coupling constant of the theory and
$A_{\mu}(x)=A_{\mu}^a(x) T^a$.  The non-abelian gauge field
$A_{\mu}^a(x)$ transforms according to the adjoint representation of
the gauge group.  Given an infinitesimal transform $U
(x)=1-\imath\th^a(x)T^a$ the variation of the gauge and quark fields
is
\bea
\de A_{\mu}^a(x)&=&-\frac{1}{g}\hat{D}_{\mu}^{ac}(x)\th^c(x)\\
\de q_{\al}(x)&=&-\imath T^a\theta^a(x)q_{\al}(x)
\label{eq:inftransf}
\eea
where the covariant derivative in the adjoint representation reads
\be
\hat{D}_{\mu}^{ac}=\de^{ac}\pd_{\mu}+gf^{abc}A_{\mu}^b.
\label{eq:covderadj}
\ee
Defining the field strength tensor 
\be
F_{\mu\nu}
=T^{a}(\pd_{\mu}A_{\nu}^a-\pd_{\nu}A_{\mu}^a
+gf^{abc}A_{\mu}^bA_{\nu}^c)
\label{eq:fieldstr}
\ee
we can construct a kinetic term for the non-abelian gauge field
$A_{\mu}$
\be
{\cal L}_{YM}=-\frac12 \textrm{Tr} F_{\mu\nu}F^{\mu\nu}
=-\frac14 F^a_{\mu\nu}F^{a\mu\nu},
\ee
such that the total Lagrange density $\cl_{QCD}=\cl_{q}+\cl_{YM}$ is
invariant under local gauge transformations.

Let us now consider the functional integral
\be
Z=\int\cd[A \bar q q]\exp{\left\{\imath\cs_{QCD}\right\}},
\label{eq:funcint}
\ee
where $\cd[A \bar q q]$ denotes the functional integration measure for
the \YM and quark fields and the QCD action is given by
\be
{\cs}_{QCD}= \int d^4x \left\{\bar q_{\al}(x)
\left[\imath \gamma_{\mu}D^{\mu}
-m\right]_{\al\ba}q_{\ba}(x)-\frac14
F^a_{\mu\nu}F^{a\mu\nu}\right\}.
\label{eq:sqcd}
\ee
The difficulty with the functional integral \eq{eq:funcint} is that
the measure $\cd[A \bar q q]$ runs over infinitely many gauge
equivalent configurations (field configurations that are connected by
gauge transformations), whereas the the action $\cs_{QCD}$ is gauge
invariant.  Hence it introduces for every gauge orbit\footnote {A
gauge orbit contains all configurations connected by gauge
transformations and will be explicitly defined in the next section.} a
divergent factor (the volume of the gauge group). The way to handle
this problem is to use a method introduced by Faddeev and Popov
\cite{Faddeev:1967fc}. The idea is to single out one representative
from each orbit by imposing a gauge fixing condition to the functional
integral \eq{eq:funcint}.  In this thesis we shall be considering only
Coulomb gauge, but also other choices such as Landau gauge are
possible.  As a consequence, a new set of Grassmann fields, known as
ghosts, are introduced.  The gauge fixing procedure, the ghost fields
and the problems associated with them will be discussed in the next
section.
\section {Gauge fixing and ghosts}
\label{sec:gauge.fixing}

\begin{figure}[t]
\centering\includegraphics[width=0.35\linewidth]{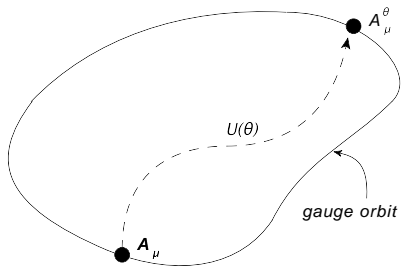}
\caption{\label{fig:gauge.orbit}
Depiction of a gauge orbit containing $ A_{\mu}$ and the gauge
transformed field  $A_{\mu}^{\theta}$}
\end{figure}

We start by defining the {\it gauge orbit} for some configuration
$A_{\mu}$ as the set of all gauge equivalent configurations, i.e. each
point $A_{\mu}^{U}$ on the gauge orbit is obtained by acting upon
$A_\mu$ with the gauge transformation $U$ (see also
\fig{fig:gauge.orbit}):
\be
[A_{\mu}^{orbit}]=:\{ A_{\mu}^{U}= 
 U A_{\mu} U^{\dagger} -\frac{\imath}{g} (\pd_{\mu}
U)U^{\dagger} \mbox{~}| \mbox{~} U\in SU(N_c)\}.
\ee
Then the integral over all the gauge fields $\int{\cal D}A$ can be
written as an integral over a full set of gauge-inequivalent
configurations $\int {\cal D}A^{ref}$, where $A^{ref}$ is a reference
gauge field representative for the orbit (i.e., integral {\it over}
all possible gauge orbits), and an integral {\it around} each gauge
orbit $\int \cd O (A^{ref})$
\be
\int{\cd}A=\int{\cd}A^{ref} \int{\cd}O (A^{ref}).
\ee
Since one integrates over infinitely many equivalent configurations
related by a gauge transformation, the integral around the gauge orbit
$\int \cd O (A^{ref})$ is infinite and must be eliminated.  The
strategy is to extract only one representative gauge-field
configuration out of each orbit by imposing a gauge fixing condition
$\chi[A]=0$.  Each gauge field configuration on the orbit $O
(A^{ref})$ is a gauge transformation of $ A^{ref} $, i.e. one can (up
to a volume factor) rewrite the integral $ \int{\cd}O (A^{ref}) $ as
an integral over the gauge group $ \int{\cd} U$.  The gauge fixing
condition is implemented by inserting the identity
\be
1=\Delta[A]\int {\cd}U \de[\chi(A^U)]
\ee
into the functional integral \eq{eq:funcint}.  Ideally, the gauge
fixing condition $\chi(A)=0$ should be satisfied by only one $A_{\mu}$
of each gauge orbit.~\footnote {But, as will shortly be discussed,
this is in practice impossible due to topological restrictions.}  The
Faddeev-Popov determinant $\Delta[A]= \Delta[A^U] $ accounts for the
functional determinant arising from the argument of the delta
function.  Using the invariance of the action under gauge
transformations, one can rewrite the functional integral of the theory
as
\be
Z=\int\cd U Z_{gf}
\label{eq:funcint1}
\ee
where the gauge fixed amplitude is given by
\be
Z_{gf}=\int\cd[A \bar q q] \Delta [A] \de[\chi(A)]
\exp{\left\{\imath\cs_{QCD}\right\}},
\label{eq:funcint2}
\ee
and the divergent measure $\int {\cal D}U$ has been factorized as an
overall constant which can be absorbed in the normalization. In the
following, we will use the notation $Z$ (instead of $Z_{gf}$) for the
gauge fixed functional integral, since no confusion can arise.

In Coulomb gauge, the (noncovariant) gauge fixing condition is given
by:
\be
\chi[A]= \s{\vec{\nabla}}{\vec{A}^a}=0
\label{eq:gaugefix}
\ee
and hence the temporal and spatial components of the gauge field must
be treated differently.  The gauge fixing condition can be implemented
by rewriting the delta function with the help of a Lagrange multiplier
$\la^a$
\be
\de[\chi(A)]\to \int\cd \la\exp\left[ -\imath \int d^4x
  \textrm{~}\la^a\s{\vec{\nabla}}{\vec{A}^a}\right],
\label{eq:gaugefix1}
\ee
and the Faddeev-Popov determinant
\be
\De [A]=\textrm {Det~}[\s{\vec{\nabla}}{\vec{D}^{ab}}]
\ee
can be written as a functional integral over two new Grassmann valued
fields $c$ and $\bar c$, the so-called ``Faddeev-Popov
ghosts''\footnote {A pedagogic introduction on this topic can be found
in \cite{Ryder:1985wq}.}:
\be
\De [A]\to \int\cd[\bar cc]\exp\left[ -\imath \int d^4x
\textrm{~}\ov{c}^a\s{\vec{\nabla}}{\vec{D}^{ab}}c^b\right].
\label{eq:FPdet}
\ee
In the above, we have introduced the spatial component of the
covariant derivative \eq{eq:covderadj} (in the adjoint representation)
\be
\vec{D}^{ac}=\de^{ac}\vec\nabla-gf^{abc}\vec A^b.
\label{eq:covderadjs}
\ee
Due to their unusual spin-statistics (they obey Fermi-Dirac
statistics, and at the same time are scalar), the ghost fields are
only allowed to appear in closed loops in Feynman diagrams and never
as initial or final states in a physical process. Putting all these
together, we can write for the Coulomb gauge functional integral:
\be
Z=\int\cd[A \bar qq\bar cc\la]\exp{\left\{\imath\cs_{QCD}
+\imath\cs_{FP}\right\}},
\label{eq:ZFP}
\ee
with
\bea
\cs_{QCD}&=&\int d^4x\left\{\bar q_{\al}\left[\imath\gamma^{0}D_{0}
+\imath \vec\gamma\cdot\vec D-m\right]_{\al\ba}q_{\ba} -\frac14
F^a_{\mu\nu}F^{a\mu\nu}\right\},\label{eq:sqcd1}\\
\cs_{FP}&=&\int d^4x\left[-\la^a\s{\vec{\nabla}}{\vec{A}^a}
-\ov{c}^a\s{\vec{\nabla}}{\vec{D}^{ab}}c^b\right].
\label{eq:sqcd2}
\eea
In the QCD action, \eq{eq:sqcd1}, we have separated the covariant
derivative into temporal and spatial components (implicitly in the
fundamental color representation), given by:
\be
D^{0}=\pd^{0}-\imath g T^{c}\si^{c},\;\;\;
\vec D=\vec\nabla+\imath g T^{c}\vec A^{c},
\label{eq:covder}
\ee
where the temporal component of the gauge field $A^{0a}$ has been
renamed to $\si^a$. $\cs_{FP}$ collects the terms originating from the
gauge fixing condition \eq{eq:gaugefix1} and the Faddev-Popov
determinant \eq{eq:FPdet}.
\begin{figure}[t]
\textsf{
\centering
\includegraphics[width=0.75\linewidth]{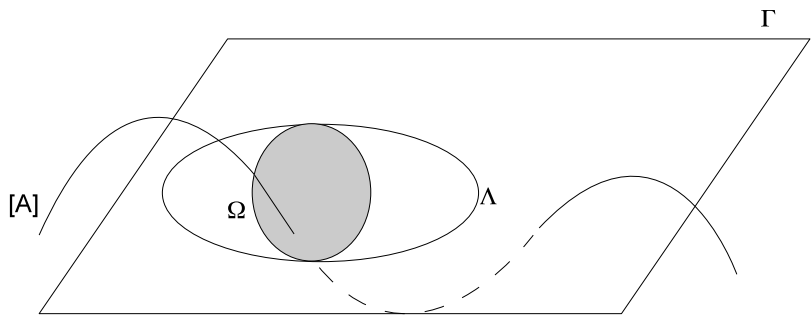}
\caption{\label{fig:gauge.fixing} Illustration of the hyperplane
$\Gamma$ in gauge field configuration space obtained by gauge
fixing. Furthermore, the first Gribov region $\Omega$ and the
fundamental modular region $\Lambda$ are shown. A gauge orbit [$A$]
intersects the hyperplane $\Gamma$ several times thus generating
Gribov copies. The fundamental modular region, by definition, is
intersected only once.}}
\end{figure}

The gauge fixing procedure described above is not yet complete, in the
sense that the simple Faddeev-Popov trick is not sufficient to extract
a single gauge configuration from each gauge orbit.  The reason is the
presence of the so-called {\it Gribov copies} \cite{Gribov:1977wm},
i.e. configurations $A_{\mu}$ connected by a gauge transformation that
produce multiple intersection points of a gauge orbit with the
hyperplane $\G$ generated by the gauge fixing condition (see also
\fig{fig:gauge.fixing}). In order to avoid the Gribov copies, it is
necessary to restrict the hyperplane $\G$ to the so-called {\it Gribov
region} $\Omega$. This is obtained by minimizing the $L^2$-norm of the
vector potential
\be
F_A[U]=\int d^3x \textrm{~tr} \left[A_i^U(x)\right]^2
\ee
 along the gauge orbit \cite{vanBaal:1997gu}. In this region, any {\it
local} minimum of this norm implements the gauge fixing condition (in
this case, Coulomb gauge) and the Faddeev-Popov operator is restricted
to positive eigenvalues:
\be
\Omega\equiv\left\{\vec
A:\s{\div}{\vec{A}}=0;-\s{\div}{\vec{D}}\geq0\right\}.
\ee
Importantly, the Gribov region $\Omega$ contains the trivial
configuration $g \vec A=0$ (i.e., it contains all configurations
relevant for perturbation theory) and therefore the ultraviolet regime
is not influenced by this restriction.  Moreover, the Faddeev-Popov
determinant vanishes on the boundary $\pd \Omega$ of the first {\it
Gribov horizon} \cite{Zwanziger:2003cf}.

In general, $\Omega$ is still not free of Gribov copies\footnote{A
proof of the existence of Gribov copies inside $\Omega$ was given in
\cite{Dell'Antonio:1991xt}.}, thus in principle one has to restrict
further the gauge field configurations to the {\it fundamental modular
region} $\Lambda$ -- the region of {\it global} minima of the norm
defined above:
\be
\Lambda\equiv\left\{\vec A:F_A[\mathbbm{1}] \leq F_A[U]
\textrm{~~}\forall U \right\}.
\ee
This gauge condition is also known as ``minimal Coulomb gauge''
\cite{Zwanziger:1998ez}.  In practise, these configurations are
extremely complicated to identify\footnote{For recent lattice studies
see, for example, \cite{Cucchieri:1997dx, Silva:2004bv,
Alexandrou:2000ja, Bogolubsky:2005wf}.}. However, in the continuum
Zwanziger showed by means of stochastic quantization that Gribov
copies inside the Gribov region do not affect the Green's functions of
the theory \cite{Zwanziger:2003cf}.  Precisely, the dominant
configurations lie on the common boundary of $\Omega$ and $\Lambda$
and hence, in practise, restriction to the Gribov region is
sufficient.  This restriction might eventually generate nontrivial
boundary terms which could influence the derivation of the field
equations of motion (and implicitly the \DS equations), but since by
definition the Faddeev-Popov operator vanishes on the boundary of
$\Omega$, these boundary terms are identically zero.
\section{Approaches to Coulomb gauge QCD}
\subsection{First order functional formalism}
\label{sec:firstorder}

As already mentioned, the gauge of choice in this work is Coulomb
gauge. Although this gauge does not have so many practical advantages
as Landau gauge for example, there are several reasons motivating our
choice: in Coulomb gauge there is a natural picture of confinement,
Gauss law is naturally built in (such that in principle gauge
invariance is fully accounted for) and the total color charge is
conserved and vanishing \cite{Zwanziger:1998ez,Watson:2007fm}.
However, except for the original work of Khriplovich
\cite{Khriplovich:1969aa}, only recently \DS studies in Coulomb gauge
have been undertaken.  The technical barrier stems from the so-called
energy divergence problem -- the unregulated divergences generated by
the ghost loops \cite{Leibbrandt:1996tn,Andrasi:2005xu,Doust:1987yd}.
The way to circumvent this problem is to use the {\it first order
functional formalism}\footnote{However, since we are mainly concerned
with the quark sector of QCD, in the course of our investigations we
will come across few points where restriction to second order
formalism \cite{Watson:2008fb,Watson:2007vc} will be sufficient.}.
Within this formalism these divergences cancel exactly and moreover,
the system reduces automatically to ``would-be-physical'' degrees of
freedom \cite{Zwanziger:1998ez}.  All these aspects will be discussed
in detail in the course of this section.

\subsection*{Motivation and main idea}

The main advantage of working in Coulomb gauge stems from the fact
that in this gauge the Gribov-Zwanziger mechanism of
confinement\footnote{The Gribov-Zwanziger confinement scenario will be
presented in more detail in Section~\ref{sec:landau}.} becomes
particularly transparent.  In this picture, the long range confining
force is provided by the instantaneous Coulomb interaction, which
appears to be enhanced for small three-momenta $\vec q^2\to 0$,
whereas the transverse (colored) gluon is suppressed, reflecting the
absence of the colored states in the spectrum.  In covariant gauges
such as Landau gauge, a quantity that leads to the confinement
potential has not been identified so far.

The conversion to the first order (or phase-space) formalism is
motivated by the fact that within this formalism the famous Coulomb
gauge energy divergence
\cite{Leibbrandt:1996tn,Andrasi:2005xu,Doust:1987yd} can be avoided.
Energy divergence means that the functional integral \eq{eq:ZFP} gives
an ill-defined integration, stemming from the energy-independent ghost
loops\footnote{ The energy independence of the ghost propagator
follows from the fact that the Faddeev-Popov operator involves only
spatial derivatives and spatial components of the gauge fields (see
Refs.~\cite{Watson:2006yq,Watson:2008fb} for a complete derivation and
discussion).} which are integrated over both 3-momentum {\it and}
energy. As an example, consider the following one-loop integral
\cite{Watson:2007mz}:
\be
\int dk_{0}\int d^3\vec k [(\vec k-\vec p)^2 \vec k^2]^{-1}.
\ee
This divergence appears in any number of dimensions and cannot be
regularized by usual dimensional regularization, although when taken
as a full set in the \DS equations, all such loops cancel.  To handle
this type of divergences, Leibbrandt introduced a modified form of the
dimensional regularization, the so-called split dimensional
regularization \cite{Leibbrandt:1996tn}, in which two complex
parameters $\w$ and $\si$ are introduced, $d^3\vec q\to d^{2\w}\vec q
\textrm{~and~} d q_0\to d^{2\si} q_0$ with the limits $\w\to 3/2$ and
$\si\to 1/2$ to be taken after all the integrations have been
completed.  An alternative approach is the so-called negative
integration method (NDIM) \cite{Suzuki:2000ki, Suzuki:1999cua}, where
a ``Feynman-like'' integral is solved, i.e. a loop integral in
negative $D$-dimensional space with propagators raised to positive
powers in the numerator. With these two methods the Coulomb gauge
integrals were studied up to one and two-loop level perturbatively and
results for the divergent part for several of them have been achieved.
These divergences do in principle cancel order by order in
perturbation theory (tested up to two-loops \cite{Heinrich:1999ak}),
but this cancellation is difficult to isolate.

Furthermore, within the first order formalism we are able to cancel
the unphysical ghost fields, i.e. the Faddeev-Popov term.  This means
that we reduce the functional integral \eq{eq:ZFP} to ``physical''
degrees of freedom, the transverse gluon and transverse $\vec\pi$
fields, which in classical mechanics would be the configuration
variables and their momentum conjugates (see below).

We keep the term ``physical'' into quotations marks because it is
realized that the true physical objects are the color singlet states,
their observables being the mass spectrum and the decay widths.

The presentation of the first order formalism, together with the
reduction to the ``physical'' degrees of freedom, follows
\cite{Watson:2006yq} and is discussed in some detail.  We start by
expressing the field strength tensor \eq{eq:fieldstr} in terms of the
chromo-electric and -magnetic fields (recall that the temporal
component $A^{0a}$ has been renamed to $\si^a$)
\be
\vec{E}^a=-\pd^0\vec{A}^a-\vec{\nabla}\si^a+gf^{abc}\vec{A}^b\si^c,
\;\;\;\;
B_i^a=\epsilon_{ijk}\left[\nabla_jA_k^a-\ha gf^{abc}A_j^bA_k^c\right],
\ee
such that the Yang--Mills action can be split into chromoelectric and
-magnetic terms
\be
\cs_{YM}= \int
d^4x\left[\ha\s{\vec{E}^a}{\vec{E}^a}-\ha\s{\vec{B}^a}{\vec{B}^a}\right].
\ee
Next, we consider the chromoelectric term in the action. We linearize
this term and hence convert to the first order formalism by
introducing an auxiliary field $\vec{\pi}$ via the identity
\cite{Zwanziger:1998ez}
\be
\exp{\left\{\imath\int d^4x\ha\s{\vec{E}^a}{\vec{E}^a}\right\}}
=\int\cd\vec{\pi}\exp{\left\{\imath\int d^4x\left[-\ha
\s{\vec{\pi}^a}{\vec{\pi}^a}-\s{\vec{\pi}^a}{\vec{E}^a}\right]
\right\}}.
\label{eq:piintro}
\ee
Classically, the $\vec\pi$ field is interpreted as the momentum
conjugate to $\vec A$.  The $\vec{\pi}$ field is then split into
transverse and longitudinal components using the identity
\bea
\mathrm{const}&=&\int\cd\phi\de\left(\s{\vec{\nabla}}{\vec{\pi}}
+\nabla^2\phi\right)\nonumber\\
&=&\int\cd\left\{\phi,\ta\right\}\exp{\left\{-\imath\int 
d^4x\ta^a\left(\s{\vec{\nabla}}{\vec{\pi}^a}+\nabla^2\phi^a\right)
\right\}}.
\label{eq:constpi}
\eea
After making the change of variables
$\vec{\pi}\rightarrow\vec{\pi}-\vec{\nabla}\phi$ and collecting
together all the parts that contain $\vec{\pi}$, we can write our full
functional integral as ($\Phi$ denotes the collection of all fields):
\be
Z=\int\cd\Phi\exp{\left\{\imath\cs_{q}+\imath\cs_{YM}
+\imath\cs_{FP}\right\}}=
\int\cd\Phi\exp{\left\{\imath\cs_{q}+\imath\cs_{\pi}
+\imath\cs_B+\imath\cs_{FP}\right\}}
\label{eq:func}
\ee
with
\bea
\cs_{q}&=&\int d^4x\textrm{~}\bar q_{\al}\left[\imath\gamma^{0}D_{0}
+\imath \vec\gamma \cdot\vec D-m\right]_{\al\ba}q_{\ba}, 
\nonumber\\
\cs_{\pi}&=&\int d^4x\left[-\ta^a\s{\vec{\nabla}}{\vec{\pi}^a}
-\ha\s{(\vec{\pi}^a-\div\phi^a)}{(\vec{\pi}^a-\div\phi^a)}
+\s{(\vec{\pi}^a-\div\phi^a)}{\left(\pd^0\vec{A}^a+\vec{D}^{ab}\si^b
\right)}\right]\!\!,\nonumber\\
\cs_B&=&\int d^4x\left[-\ha\s{\vec{B}^a}{\vec{B}^a}\right],
\nonumber\\
\cs_{FP}&=&\int d^4x\left[-\la^a\s{\vec{\nabla}}{\vec{A}^a}
-\ov{c}^a\s{\vec{\nabla}}{\vec{D}^{ab}}c^b\right].
\label{eq:act}
\eea

Having derived the functional integral and the full QCD action in the
first order formalism, we are now in the position to show that the
ghost loops indeed cancel and the system reduces to the ``physical''
degrees of freedom. In the next section, we will demonstrate that the
above QCD action simplifies to an expression where only the transverse
$\vec A$ and $\vec \pi$ fields appear.

\subsection*{Reduction to ``physical'' degrees of freedom}

Given the functional integral, \eq{eq:func}, and \YM part of the
action (the quark field is not considered in this discussion),
\eq{eq:act}, we start by rewriting the Lagrange multiplier terms as
$\de$-function constraints, which automatically eliminates the
$\s{\div}{\vec{A}}$ and $\s{\div}{\vec{\pi}}$ terms in the action. In
addition, we also rewrite the ghost terms as the original
Faddeev-Popov determinant.  This has clearly the drawback that the
local formulation and the BRST invariance of the theory\footnote{The
standard BRST and \emph {time-dependent} Gauss--BRST transformations
used in this work will be introduced in Chapter~\ref{chap:st}. }  are
no longer manifest.  The functional integral now reads
\be
Z=\int\cd\Phi\mathrm{Det}
\left[-\s{\vec{\nabla}}{\vec{D}}\de^4(x-y)\right]
\de\left(\s{\div}{\vec{A}}\right)\de
\left(\s{\div}{\vec{\pi}}\right)
\exp{\left\{\imath\cs\right\}}
\ee
with 
\be
\cs=\int d^4x\left[-\ha\s{\vec{B}^a}{\vec{B}^a}
-\ha\s{\vec{\pi}^a}{\vec{\pi}^a}
+\ha\phi^a\nabla^2\phi^a
+\s{\vec{\pi}^a}{\pd^0\vec{A}^a}
+\si^a\left(\s{\div}{\vec{D}^{ab}}\phi^b
+g\hat{\ro}^a\right)\right],
\ee
where we have defined the effective color-charge of the gluons
$\hat{\ro}^a=f^{ade}\s{\vec{A}^d}{\vec{\pi}^e}$.  Next, we use the
fact that the action \eq{eq:act} has become linear in $\si$ (after
introducing the field $\vec\pi$), and write the integral over $\si$ as
a $\de$-function constraint. This enforces the chromo-dynamical
equivalent of Gauss' law giving
\be
Z=\int\cd\Phi\mathrm{Det}
\left[-\s{\vec{\nabla}}{\vec{D}}\de^4(x-y)\right]
\de\left(\s{\div}{\vec{A}}\right)\de
\left(\s{\div}{\vec{\pi}}\right)
\de\left(-\s{\div}{\vec{D}^{ab}}\phi^b-g\hat{\ro}^a\right)
\exp{\left\{\imath\cs\right\}}
\label{eq:ZdetFP}
\ee
with
\be
\cs=\int d^4x\left[-\ha\s{\vec{B}^a}{\vec{B}^a}
-\ha\s{\vec{\pi}^a}{\vec{\pi}^a}+\ha\phi^a\nabla^2\phi^a
+\s{\vec{\pi}^a}{\pd^0\vec{A}^a}\right].
\ee
The implementation of the Gauss' law is very important because this
essentially ensures the gauge invariance of the system.  Defining the
inverse Faddeev-Popov operator $M$:
\be
\left[-\s{\div}{\vec{D}^{ab}}\right]M^{bc}=\de^{ac},
\label{eq:invfp}
\ee
we can factorize the Gau{\ss} law $\de$-function constraint as
\be
\de\left(-\s{\div}{\vec{D}^{ab}}\phi^b-g\hat{\ro}^a\right)
=\mathrm{Det}\left[-\s{\vec{\nabla}}{\vec{D}}\de^4(x-y)\right]^{-1}
\de\left(\phi^a-M^{ab}g\hat{\ro}^b\right).
\ee
The inverse functional determinant cancels the original Faddeev-Popov
determinant from \eq{eq:ZdetFP}, leaving us with
\be
Z=\int\cd\Phi\de\left(\s{\div}{\vec{A}}\right)
\de\left(\s{\div}{\vec{\pi}}\right)\de
\left(\phi^a-M^{ab}g\hat{\ro}^b\right)
\exp{\left\{\imath\cs\right\}}.
\ee
We now use the $\de$-function constraint to eliminate the
$\phi$-field.  Since the inverse Faddeev-Popov operator $M$ is
Hermitian, we can reorder the operators in the action to give us
\be
Z=\int\cd\Phi\de\left(\s{\div}{\vec{A}}\right)
\de\left(\s{\div}{\vec{\pi}}\right)\exp{\left\{\imath\cs\right\}}
\ee
with
\be
\cs=\int d^4x\left[-\ha\s{\vec{B}^a}{\vec{B}^a}
-\ha\s{\vec{\pi}^a}{\vec{\pi}^a}
-\ha g\hat{\ro}^bM^{ba}(-\nabla^2)M^{ac}g\hat{\ro}^c
+\s{\vec{\pi}^a}{\pd^0\vec{A}^a}\right].
\label{eq:actionApi}
\ee
As promised, the action \eq{eq:actionApi} contains only transverse
$\vec{A}$ and $\vec{\pi}$ fields.  All other fields, especially the
unphysical ghosts, have been formally eliminated.  However, the
appearance of the functional $\de$-functions and the inverse
Faddeev-Popov operator $M$ have led to a non-local formalism.  It is
not known how to do practical calculations within this formulation,
but the non-local nature of the above result certainly serves as a
guide to the local formulation.

Before we close this section, few more remarks are in order. Quite
generally, one can argue that first order formalism in Coulomb gauge
is better suited to describe physical phenomena then other gauges such
as Landau gauge. Indeed, the natural decomposition of degrees of
freedom, both physical and unphysical, inherent to the first order
formalism, automatically leads to the cancellations of the unphysical
components.  The subtlety is then to identify how these cancellations
arise, and also it is very important to ensure that approximation
schemes employed respect such cancellations.  For example, the
unphysical ghost loop of the gluon polarization should be cancelled in
the \DS equations.  Given that the ghost propagator is energy
independent, the temporal component of the gluon propagator must
itself have a part that is independent of energy
\cite{Cucchieri:2000hv} in order to cancel this divergence.  Later on
in Chapter~\ref{chap:hq} devoted to heavy quarks, this observation
will be used to show that only color singlet quark-antiquark bound
states are physically allowed, regardless of the specific form of the
energy independent part of the temporal gluon propagator.

On the other hand, one unpleasant feature of this approach is the
large number of fields, but as it turns out this does not have serious
implications for the \DS equations \cite{Watson:2006yq,
Popovici:2008ty}. More important though, the issue of
renormalisability remains unclear, in the sense that a complete proof
of full multiplicative renormalization is still missing and one does
not have a Ward identity in the usual sense \cite{Zwanziger:1998ez}. A
brief overview of the attempts to renormalize Coulomb gauge will be
given at the end of this chapter.

\subsection{Alternative methods}

Currently, the most popular continuum formalism to QCD in Coulomb
gauge is the Hamiltonian formalism \cite{Feuchter:2004mk,
Reinhardt:2004mm, Szczepaniak:2001rg, Reinhardt:2009ks,
Schleifenbaum:2006bq, Epple:2006hv, Epple:2007ut}.\footnote{In
Ref.~\cite{Reinhardt:2009ks}, the results obtained within the
Hamiltonian approach to \YM theory are reviewed.}  In this approach,
one starts by imposing Weyl gauge $A_0^a(x)=0$, and subsequently, the
Coulomb gauge is fixed with the help of the Faddeev-Popov method,
whereas the Gauss law is imposed as constraint.  The \YM
Schr{\"o}\-din\-ger equation is then solved using the variational
principle for the vacuum state, with a Gaussian ansatz for the wave
functional.  Then the vacuum energy is minimized and this leads to a
coupled system of non-linear \DS equations for the gluon energy, the
ghost and the Coulomb form factor and for the curvature in
configuration space \cite{Feuchter:2004mk}.  These have been solved
analytically in the infrared and numerically in the whole momentum
regime \cite{Schleifenbaum:2006bq}.  Similar to Landau gauge (see
\cite{Fischer:2006ub} for a recent review), it has been found that in
Coulomb gauge there are two different infrared powerlaw exponents for
the gluon and the ghost propagator \cite{Epple:2006hv,
Epple:2007ut}. The favored solution is the most singular -- with the
ghost propagator dressing function diverging as $1/|\vec k|$ -- which
generates a linearly rising heavy quark potential at large distances
\cite{Epple:2006hv}.\footnote{In the next section, the connection to
Landau gauge as well as the infrared ghost dominance will be discussed
in more detail.}  Also the gluon energy is divergent in the infrared,
reflecting the absence of the gluons in the physical spectrum at low
energies, which is again a signal of confinement.  Recently, the
static potential between infinitely heavy color sources has been also
studied with a method based on the Dyson equation for the Wilson
loop. In \cite{Pak:2009em}, the authors considered the temporal Wilson
loop with the instantaneous part of the gluon propagator and the
spatial Wilson loop with the static gluon propagator, and solved the
corresponding Dyson equation in Coulomb gauge.  In both cases a
linearly rising potential has been found.

A second approach to Coulomb gauge is the lattice QCD (see, for
example, \cite{Cucchieri:2006hi} for a review of the results obtained
on the lattice). The first lattice calculations have concentrated on
the infrared behavior of the ghost and gluon propagators.  In
particular, in Ref. \cite{Cucchieri:2000hv} it has been shown that the
static transverse gluon propagator is suppressed in the IR limit,
while the time-time component is enhanced. Moreover, in the
infinite-volume limit, it has been found that the transverse gluon is
well described by the Gribov's formula \cite{Gribov:1977wm} but
unfortunately this study was not conclusive in the ultraviolet
\cite{Cucchieri:2000hv, Cucchieri:2000kw}.  Recently, in
\cite{Burgio:2008jr} the residual temporal gauge has been fixed and
the renormalization of the full gluon propagator has been studied.  It
has been found that the static propagator is renormalizable only in
the limit of continuous time, i.e. the lattice Hamiltonian
formulation, and the resulting static propagator satisfies the
Gribov's formula at all momenta.  For the ghost propagator, lattice
results have been reported in \cite{Langfeld:2004qs, Quandt:2008zj},
and more recently in \cite{Burgio:2010wv}.  An infrared divergence
stronger than $1/\vec k^2$ has been found, in agreement with the
horizon condition necessary in the Zwanziger confinement criterion.
Very recently, lattice studies of the Coulomb gauge quark propagator
have been also undertaken and preliminary results have been presented
\cite{Burgio:2010wv}. The ambiguities due to the Gribov copies have
been analyzed and it was found that their influence on the quark
propagator is small. Moreover, the residual gauge has been fixed, and
it appears that the impact of the residual gauge fixing is only
reflected in the time-dependence of the propagator.
\section{Connection to Landau gauge and aspects of confinement}
\label{sec:landau}

Due to its covariance, Landau gauge $\pd_{\mu} A_{\mu}^a=0$ has been
for a long time the preferred gauge for non-perturbative \DS
studies. It preserves Lorenz invariance and it has a distinct property
which makes it very attractive for practical approximations, namely
that the the ghost-gluon vertex remains bare in the
infrared\footnote{Or at least, from a semiperturbative analysis of the
\DS equation for the ghost-gluon vertex (with nonperturbative ghost
and gluon propagators and a bare ghost-gluon vertex), it follows that
deviations from the tree-level vertex are very small
\cite{Schleifenbaum:2004id}.}.

The first non-perturbative calculations in Landau gauge date back to
the late seventies, when the infrared gluon propagator was first
studied by Mandelstam and Bar-Gadda \cite{Mandelstam:1979xd,
BarGadda:1979cz}.  In this calculation, the ghost loop and the
four-gluon vertex, which do not contribute at first level in
perturbation theory, have been neglected and only the (non-abelian)
triple gluon vertex has been considered.  It was found that the gluon
propagator is infrared divergent and moreover, assuming a single gluon
exchange, a linearly rising potential between heavy quarks has been
derived. Today this picture is known as {\it infrared slavery}.  In
Landau gauge this picture has been rather misleading, in the sense
that only very late it has been shown that in this gauge it is not the
gluons, but the ghosts that drive the infrared properties of the
theory\footnote{We refer to the original works
\cite{vonSmekal:1997is,vonSmekal:1997vx}, and for a review, see
Ref.~\cite{Fischer:2006ub}.}.  More precisely, the gluon propagator
vanishes at zero momentum, whereas the ghosts provide for a long range
correlation.  The gluon and ghost propagators are characterized by the
so-called infrared exponents, which are related by a scaling relation,
hence the name {\it scaling solution}.  We briefly mention that there
exists a second solution, the {\it decoupling solution}, which
possesses quite different characteristics: the ghost propagator is not
infrared enhanced but remains bare in the infrared and the gluon
propagator becomes finite instead of going to zero
\cite{Fischer:2008uz,Dudal:2007cw}.  The connection between the two
solutions (scaling and decoupling), different only in the deep
infrared, still remains to be understood.

Let us now analyze the consequences of restricting the configuration
space to a compact region (Gribov or fundamental modular region), for
the infrared properties of the theory.  As discussed in the beginning
of this chapter (Sec.~\ref{sec:gauge.fixing}), the ultraviolet regime
is not affected by this restriction (when the coupling becomes small,
all relevant configurations lie in the vicinity of $g A=0$), whereas
the infrared can be in principle governed by any domain within the
Gribov region.  Zwanziger argued that only the behavior of the gauge
field on the Gribov horizon (i.e., the boundary of the Gribov region)
is important for the infrared properties \cite{Zwanziger:1995cv}.
Pictorially, the situation is similar to a compact sphere of radius
$r$ in high $N$ dimensions, where the probability distribution is
concentrated on the boundary due to the volume measure $r^{N-1} dr$.
In Coulomb gauge, the Coulomb potential between two external color
charges\footnote {This potential is renormalization group invariant
and is an upper bound for the gauge invariant potential from the
Wilson loop -- a statement known as ``no confinement without Coulomb
confinement'' \cite{Zwanziger:2002sh}.}  has been derived
\cite{Epple:2006hv}, and it is related to
\be
\ev{M (-\nabla^2) M}^{ab} (\vec x,\vec y),
\label{eq:coulpot}
\ee 
where $M$ is the {\it inverse} Faddeev-Popov operator, defined in
\eq{eq:invfp}.  Since on the Gribov horizon the Faddeev-Popov operator
develops a zero eigenvalue, the Coulomb energy \eq{eq:coulpot} becomes
very large in the vicinity of the Gribov horizon, leading directly to
an asymptotically linear potential. This mechanism is known as the
\emph {Gribov-Zwanziger scenario of confinement}.  One can also
inspect the relation between confinement and the restriction to the
Gribov region by examining the infrared behavior of the ghost
propagator.  This propagator is given by to the expectation value of
the inverse Faddeev-Popov operator and therefore is strongly
divergent, due to the vanishing eigenvalues of the Faddeev-Popov
operator on the Gribov horizon -- this is known as the {\it horizon
condition} \cite{Gribov:1977wm}. Hence, the ghost propagator becomes
infrared enhanced from the effects of the Gribov horizon.

As outlined above, in Landau gauge the infrared ghost dominance has
been established, but in Coulomb gauge, the Gribov-Zwanziger scenario
has a somewhat different realization, depending on the specific
formalism.  As discussed in the previous section, in the Hamiltonian
formalism the ghost propagator is infrared enhanced, whereas in the
functional formalism, apart from the ghost and the transversal spatial
gluon propagator, there is a third propagator, the temporal gluon
propagator (in the canonical formalism, this would correspond to the
non-abelian color Coulomb potential, \eq{eq:coulpot}). Assuming that
this propagator is largely independent of energy and diverges like
$1/\vec k^4$ in the infrared (as indicated by the lattice results
\cite{Quandt:2008zj}) one directly obtains a linearly rising potential
for color singlet quark-antiquark states, leading back to the old
infrared slavery picture\footnote {We postpone the detailed
presentation of this mechanism to Chapter~\ref{chap:hq}, concerning
the heavy quarks.}.  We also mention that very recently, a relation
between the ghost and temporal gluon propagators has been found
\cite{Watson:2010cn}. Based on the Gribov-Zwanziger scenario, this
result, together with the \ST identities presented in
Ref.~\cite{Watson:2008fb}, provide an important element towards
connecting the infrared slavery with the ghost dominance picture.

\section{Issue of renormalizability}

As already mentioned, the renormalizability of Coulomb gauge in the
continuum has not been proven yet. In the following, we briefly review
the efforts made in this direction. Among the various attempts, the
most sophisticated approach has been pursued in Ref.
\cite{Baulieu:1998kx}.  There, the authors define the so-called
``interpolating gauge''
\be
-a\pd_0A_0+\nabla\cdot\vec A=0
\ee
and recover Coulomb gauge in the limit $a\to 0$.  A linear shift in
the field variables is performed in order to exhibit a symmetry
(called $r$-symmetry) between the Fermi and Bose unphysical degrees of
freedom. Individual closed Fermi-ghost loops and closed unphysical
Bose loops diverge like $1/\sqrt a$, but they cancel in pairs at every
order in perturbation theory by virtue of the $r$-symmetry. Thus in
the Coulomb gauge limit the correlation functions are finite, and this
remains true also for the renormalized correlation functions. However,
there have been also identified one-loop graphs that vanish like
$\sqrt a$, which do not exist in the formal Coulomb gauge (i.e. for
$a=0$). These graphs can not be neglected since they give a finite
contribution at two-loop order, when inserted into the graphs that
diverge like $1/\sqrt a$. One possibility is that these graphs are
merely gauge artifacts and decouple from the expectation values of all
gauge-invariant quantities such as Wilson loop, but up to know this
has not been explicitly shown.

Renormalization of Coulomb gauge QCD has been studied also within the
Lagrangian, second order formalism. In Ref. \cite{Niegawa:2006ey}, a
proof of algebraic renormalizability of the theory has been given with
the help of the Zinn-Justin equation. Through diagrammatic analysis
the authors have shown that in the strict Coulomb gauge $g^2 D_{00}$
($D_{00}$ is the time-time component of the gluon propagator) is
invariant under renormalization, in accordance with a similar result
obtained by Zwanziger \cite{Zwanziger:1998ez}. In a covariant gauge,
no component of the gluon field has this property.

%% file: ds.tex
\chapter{\DS equations}
\label{chap:ds}

\DS equations are the equations of motions in quantum field theory
(analogous to the classical Euler-Lagrange equations) and they relate
the various Green's functions of the theory. They are very powerful
tools to treat nonperturbative phenomena, such as confinement and
chiral symmetry breaking, whereas in the weak coupling regime the
perturbative series is recovered.  The most convenient way to derive
\DS equations (which will also be employed in this work) is to use the
functional method, i.e. to derive these equations directly from the
invariance of the generating functional under the variation of the
field \cite{Schwinger:1951hq, Schwinger:1951ex}.  An alternative
method is the Dyson resummation \cite{Dyson:1949ha}, which reorganizes
perturbative corrections into subdiagrams.

\DS equations built an infinite tower of coupled non-linear integral
equations, providing a complete description of the theory.  Thus, in
order to solve the theory it would be in principle necessary to solve
the whole set of equations. However, this is in practice impossible
and hence the main question is how to truncate the system, i.e. how to
find a way to reduce these equations to a smaller subset of simpler
equations which can be solved. The only systematic truncation relies
on perturbation theory, otherwise one is forced to make an ansatz for
the unknown higher order Green's functions. Importantly, the ansatz
must respect the symmetry properties of the theory, i.e. it must obey
the Ward--Takahashi identities in QED or the \ST identities in QCD.

In this Chapter we present the formal derivation of the \DS equations
of the two- and three-point quark Green's functions (i.e., the quark
gap equation and the quark-gluon vertex functions).  In Chapter
\ref{chap:g^2} we will then present results in the perturbative limit,
and in the second part of this thesis we will derive the four-point
Green's functions and analyze them in the limit of the heavy quark
mass.

\section{Field equations of motion}
\label{chap:feom}

The full generating functional of the theory is constructed from the
functional integral, \eq{eq:func}, by adding the corresponding source
terms.  Explicitly, we have (recall that $\cd\Phi$ denotes the
integration over all fields):
\be
Z[J]=\int\cd\Phi\exp{\left\{
\imath\cs_{q}+ \imath\cs_{YM}+\imath\cs_{FP}+\imath  \cs_{s}\right\}}
\label{eq:Z}
\ee
with the action \eq{eq:act} and the sources defined by
\be
{\cal S}_s=
\int d^4x\left[\ro^a\si^a+\s{\vec{J}^a}{\vec{A}^a}+\ka^a\phi^a
+\s{\vec{K}^a}{\vec{\pi}^a}+
\ov{c}^a\et^a+\ov{\et}^ac^a+\xi^a\la^a+\ov{q}_{\al}\chi_{\al}
+\ov{\chi}_{\al} q_{\al}\right].
\label{eq:source}
\ee

In the derivation of the quark field equation of motion (from which
the \DS equations will be derived) the gauge-fixing term in the
action, $S_{FP}$, and the terms arising from the conversion to the
first order formalism, discussed in the previous Chapter, are
unimportant because the quarks are not connected by a primitive vertex
to any of the corresponding fields, including the ghosts (i.e., there
is no direct coupling term in the quark Lagrange density).  What is
however important later on is that these extra fields will formally
enter the discussion of the Legendre transform (through partial
functional derivatives) which, in principle, gives additional terms
but which will turn to be vanishing at one-loop order perturbatively.

Also, it is important to note that the generating functional,
\eq{eq:Z}, is restricted to the Gribov region.  This restriction might
generate complications due to the presence of the Gribov copies inside
the Gribov region.  However, as discussed in
Section~\ref{sec:firstorder}, the Gribov copies do not influence the
derivation of the \DS equations in the continuum.  Moreover, the
boundary terms that may in principle appear are identically zero due
to the fact that the Faddeev-Popov operator vanishes on the boundary
of the Gribov region.

The quark equation of motion follows from the generating functional
\eq{eq:Z} and from the observation that the integral of a total
derivative vanishes:
\bea
\lefteqn{
\int\cd\Phi\frac{\delta}{\delta\imath \bar q_{x\ga}}
\exp\bigg\{
\imath\cs_{YM}
+\imath \int d^4 x\left[\right.\bar q_{x\al}\left(\imath\ga^{0}D_{0}
+\imath \vec\ga\cdot\vec D-m \right)_{\al\ba}q_{x\ba}
}
\nonumber\\
&&\hspace{2,5cm}+\bar\chi_{x\al}q_{x\al}
+\bar q_{x\al}\chi_{x\al}\left.\right]+\ldots\bigg\}
=0. \hspace{5cm}
\label{eq:gen3}
\eea
In the above, we have inserted the explicit expression for the quark
contribution to the QCD action. Also, we have written the quark
sources explicitly and denoted the rest with dots.  Using the
expression for the components of the covariant derivative,
\eq{eq:covder}, it follows that
\be
\int\cd\Phi
\left\{\left[\imath \gamma^{0}\pd_{0x}
+\imath\vec\gamma\cdot\vec\nabla_{x}
+ gT^{c}\gamma^{0}\si_{x}^{c}
- gT^{c}\vec\gamma\cdot\vec A_{x}^{c}
-m\right]_{\al\ba}q_{x\ba}+\chi_{x\al}\right\}
\exp{\left\{\imath\cs\right\}}=0,
\label{eq:gen5}
\ee
where $\cs$ is the full action plus source terms.  This expression can
be rewritten in terms of derivatives of the generating functional $Z$:
\be
\left[\imath\gamma^{0}\pd_{0x}+\imath
  \vec\gamma\cdot\vec\nabla_{x}-m\right]_{\al\ba}
\frac{\delta Z}{\delta \imath\bar\chi_{x\ba}}
+\left[gT^{c} \gamma^{0}\right] _{\al\ba}\frac{\delta^2 Z}
{\delta \imath\rho_{x}^{c}\delta \imath\bar\chi_{x\ba}}
-\left[gT^{c} \gamma^{k}\right]_{\al\ba}\frac{\delta^2 Z}
{\delta \imath J_{kx}^{c}\delta \imath\bar\chi_{x\ba}}
+\chi_{x\al}Z
\!=\!0
\label{eq:gen6}
\ee
Th equation \eq{eq:gen5} is the starting point for the derivation of
the quark \DS equations. Before we proceed to explicitly derive them,
let us first introduce some notations and briefly review the various
Green's functions of the theory.

In general, the vacuum expectation values of the time-ordered products
of field operators -- the {\it full n-point Green's functions} of the
theory (both connected and disconnected) -- are obtained by functional
differentiation of the generating functional with respect to the
sources:
\be
G_{n}(x_1,\dots, x_n)=\frac{\delta^{n}Z[J]}{\delta \imath
  J(x_1)\dots \delta \imath J(x_n)}.
\ee

However, in practice we work with connected and one-particle
irreducible $n$-point Green's functions.  In order to eliminate the
disconnected vacuum to vacuum diagrams, we use the generating
functional of the connected Green's functions $W$, defined as
\be
Z[J]=e^{W[J]},\label{eq:genfunccon}
\ee
such that the {\it connected Green's functions} are given by
\be
W_{n}(x_1,\dots, x_n)=
\frac{\delta^{n}W[J]}{\delta\imath J(x_1)\dots \delta\imath J(x_n)}.
\ee
We now introduce a bracket notation for the functional derivatives of
$W$ with respect to the sources, such that for a generic
source~$J_{\al}$
\be
\ev{\imath J_{\al}}=\frac{\de W}{\de\imath J_{\al}}.
\label{eq:WJ}
\ee
Explicitly, we have:
\bea
\frac{\delta Z[J]}{\delta \imath\bar\chi_{x\al}}&=&Z[J]
\ev{\imath\bar\chi_{ax}},\\
\frac{\de^2Z[J]}{\de\imath\ro^a_x
\de\imath \ov{\chi}_{x\al}}&=&Z[J]\left[\ev{\imath\ro^a_x\imath
\ov{\chi}_{x\al}}+\ev{\imath\ro^a_x}\ev{\imath\ov{\chi}_{x\al}}
\right].
\label{eq:eqgen7}
\eea

Using the above equations, we convert \eq{eq:gen6} into derivatives of
$W[J]$ and obtain :
\bea
\lefteqn{
\hspace{-1cm}\left[\imath\gamma^{0}\pd_{0x}
+\imath\vec\gamma\cdot\vec\nabla_{x}
-m\right]_{\al\ba}\ev{\imath\bar\chi_{x\ba}}
+g T^{c}\left\{\right.
\gamma^{0}\left[\ev{\imath\rho_{x}^{c}}\ev{\imath\bar\chi_{x\al}}
+\ev{\imath\rho_{x}^{c}\imath\bar\chi_{x\al}}\right]}
\nonumber\\&&
\hspace{4cm}-\gamma^{k}\left[\ev{\imath J_{kx}^{c}}
\ev{\imath\bar\chi_{x\al}}
+\ev{\imath J_{kx}^{c}\imath\bar\chi_{x\al}}\right]\left.\right\}
+\chi_{x\al}
=0. \label{eq:eomW}
\eea
We define the generic classical field (since no confusion can arise,
we use the same notation for the quantum fields which are integrated
over and for the resulting classical fields) to be:
\be
\Phi_{\al}=\frac1Z\int
\cd\Phi\Phi_{\al}\exp{\left\{\imath\cs\right\}}=
\frac1Z\frac{\delta Z}{\delta\imath J_{\alpha}}.
\ee
Explicitly, the classical quark and antiquark fields are given by:
\bea
q_\al(x)=\frac{1}{Z}\int{\cal D}\Phi q_\al(x)\exp{\left\{\imath{
\cal S}\right\}}&=\frac{1}{Z}\frac{\de Z}{\de\imath
\ov{\chi}_\al(x)}&=\ev{\imath\ov{\chi}_\al(x)}\nonumber\\
\ov{q}_\al(x)=\frac{1}{Z}\int{\cal D}\Phi \ov{q}_\al(x)\exp{\left\{
\imath{\cal S}\right\}}&=-\frac{1}{Z}\frac{\de Z}{
\de\imath\chi_\al(x)}&=-\ev{\imath\chi_\al(x)}. 
\label{eq:qbarqclassic}
\eea
Furthermore, we define the effective action (function of the classical
fields) via the Legendre transform of $W[J]$ with respect to the
fields:
\be
\G[\phi,\ov{q},q]=W[J,\ov{\chi},\chi]-\imath J_\al\phi_\al-\imath
\ov{\chi}_\al q_\al-\imath\ov{q}_\al\chi_\al.
\label{eq:legendretr}
\ee
In the above, we have explicitly separated the quark and \YM
components, such that $J_\al$ and $\phi_\al$ denote generic gluonic
(\YM) sources and classical fields, respectively, and we also use the
common convention that summation over all discrete indices and
integration over continuous arguments is implicit.  The generating
functional $\G[\Phi]$ then yields the {\it $n$-point proper} or {\it
one-particle irreducible (1PI) Green's functions}, which are those
Green's functions that are still connected after one internal line is
cut:
\be
\G_{n}(x_1,\dots, x_n)=\frac{\delta^{n}\G[\Phi]}
{\delta\imath \Phi(x_1)\dots \delta\imath \Phi(x_n)}.
\ee

In the following, we introduce a bracket to denote derivatives of $\G$
with respect to fields -- although the notation is similar to the
derivatives of $W$ with respect to the sources, no confusion can arise
since we never mix derivatives with respect to sources and fields.
This gives:
\bea 
\ev{\imath J_{\alpha}}=\frac{\delta W}{\delta \imath J_\al}=
\Phi_{\alpha}\textrm{~~~and~~~}
\ev{\imath\Phi_{\alpha}}=\frac{\delta\G}{\delta\imath\Phi_\al}
=-J_{\alpha}.
\label{eq:JPhi}
\eea
Note that care must be taken to observe the correct minus signs
associated with the quark (Grassmann) fields and sources. Explicitly,
this reads:
\bea
q_\al(x)=\ev{\imath\ov{\chi}_\al(x)},&&\chi_\al(x)=-\ev{\imath
\ov{q}_\al(x)},\nonumber\\
\ov{q}_\al(x)=-\ev{\imath\chi_\al(x)},&&\ov{\chi}_\al(x)=\ev{
\imath q_\al(x)}.
\eea

Having defined the proper (1PI) Green's function, we can now rewrite
the equation of motion, \eq{eq:eomW}, (and from which the \DS
equations will be derived) in terms of proper functions as:
\bea
\ev{\imath \bar q_{x\al}} &=&
-\imath\left[\imath\gamma^{0}\pd_{0 x}
+\imath\vec\gamma\cdot\vec\nabla_{x}
-m \right]_{\al\ba}\imath q_{x\ba}\nonumber\\
&+&g T^{c}\gamma^{0}\left[\si_{x}^{c}q_{x \al}
+\ev{\imath\rho_{x}^{c}\imath\bar\chi_{x \al}}\right]
-g T^{c}\gamma^{k}\left[A_{kx}^{c}q_{x \al}
+\ev{\imath J_{kx}^{c}\imath\bar\chi_{x \al}}\right]. 
\label{eq:qeom}
\eea

In a similar fashion, one can derive the gluon field equations of
motion. They are given by\footnote{The complete derivation, carried on
in the context of the \YM studies, has been presented in
Ref.~\cite{Watson:2006yq} and will not be repeated here.}  (the trace
is over Dirac and fundamental color indices):
\bea
\ev{\imath A_{ix}^a}&=&
-\G_{\bar qqAi\al\ba}^{(0)a} (\imath\bar{q}_{x\al})(\imath q_{x\ba})
-g\mathrm{Tr}
\left\{[T^a\gamma^{i}]_{\al\ba}
\ev{\imath\bar\chi_{x\ba}\imath\chi_{x\al}}\right\}
\nonumber\\
&-&\int\dx{y}\dx{z}\G_{\si AAij}^{(0)cab}(z,x,y)
\left[\ev{\imath J_{jy}^b\imath\ro_z^c}
-\imath A_{jy}^b\imath\si_z^c\right]\nonumber\\
&-&\int\dx{y}\dx{z}\frac{1}{2!}\G_{\si A\si i}^{(0)cab}(z,x,y)
\left[\ev{\imath\ro_y^b\imath\ro_z^c}
-\imath\si_y^b\imath\si_z^c\right]\nonumber\\
&-&\int\dx{y}\dx{z}\frac{1}{2!}\G_{3Aijk}^{(0)abc}(x,y,z)
\left[\ev{\imath J_{jy}^b\imath J_{kz}^c}
-\imath A_{jy}^b\imath A_{kz}^c\right]
+\ldots~,
\label{eq:a_eom}\\
\ev{\imath\si_x^a}&=&
-\G_{\bar qq\si\al\ba}^{(0)a}(\imath\bar{q}_{x\al})(\imath q_{x\ba})
+g\mathrm{Tr}
\left\{[T^a\gamma^{0}]_{\al\ba}\ev{\imath\bar\chi_{x\ba}
\imath\chi_{x\al}}\right\}\nonumber\\
&-&\int\dx{y}\dx{z}\frac{1}{2!}\G_{\si AAjk}^{(0)abc}(x,y,z)
\left[\ev{\imath J_{jy}^b\imath J_{kz}^c}
-\imath A_{jy}^b\imath A_{kz}^c\right]\nonumber\\
&-&\int\dx{y}\dx{z}\G_{\si A\si j}^{(0)abc}(x,y,z)
\left[\ev{\imath J_{jy}^b\imath\ro_z^c}
-\imath A_{jy}^b\imath\si_z^c\right]
+\ldots~
\label{eq:si_eom}
\eea
where $\G_{\bar qqA i\al\ba}^{(0)a}$, $\G_{\bar qq\si\al\ba}^{(0)a}$,
and $\G_{\si AAij}^{(0)cab}, \G_{\si A\si i}^{(0)cab}$,
$\G_{3Aijk}^{(0)abc}$ are the tree-level quark-gluon and triple-gluon
vertices, respectively.  The first two terms in the above gluon
equations of motion, containing the quark-gluon interactions, are
needed for the derivation of the quark contributions to the gluon
proper two-point functions.  The rest of the terms represent the \YM
self-interaction, and are required for the derivation of the \DS
equations for the quark-gluon vertex functions (the dots represent the
terms which are not important for the quark sector of the theory and
have been left aside).  We also mention that in principle one can
derive another two equations of motion, for the $\vec\pi$ and $\phi$
fields, but since the quarks do not directly couple to any of these
fields, the quark field will not give a contribution to the
corresponding proper two-point functions (at least at one
loop-perturbative order).

At this stage, it is useful to introduce multiple functional
derivatives with respect to quark fields and sources, which will be
later on used to derive the \DS equations. Consider the following
partial differentiations, both with respect to sources and fields:
\begin{subequations}
\bea
\frac{\de}{\de\imath \Phi_{\ba}}\ev{X(J)}&=& 
-\imath S[\ga]\ev{\imath \Phi_{\ba}\imath \Phi_{\ga}}\ev{\imath
  J_{\ga} X (J)}\label{eq:mfd10}\\
\frac{\de}{\de\imath J_{\ba}}\ev{Y(\Phi)}&=& 
\imath S[\ga]\ev{\imath J_{\ba}\imath J_{\ga}}\ev{\imath
  \Phi_{\ga} Y(\Phi)}
\label{eq:mfd11}
\eea
\end{subequations}
where $S[\ga]=\pm1$ accounts for the fact that the quark fields are
Grassmann-valued, i.e. a minus sign appears when the index $\ga$
refers to the following combinations
\be
\ev{\dots\imath q_{\ga}}\ev{\imath \bar\chi_{\ga}\dots},\,\,\,
\ev{\dots\imath \chi_{\ga}}\ev{\imath \bar q_{\ga}\dots}.
\ee
For $X(J)=\imath J_{\al}$, we have that 
\be
\pm\imath\frac{\de}{\de\imath \Phi_{\ba}}\ev{\imath J_{\al}} 
= \pm S[\ga]\ev{\imath \Phi_{\ba}\imath \Phi_{\ga}}\ev{\imath
  J_{\ga}\imath J_{\al}}
=\de_{\al\ba}
\label{eq:funcderiv1}
\ee
(the overall sign is negative for $\Phi_{\al}\equiv\bar q_{\al}$).
Taking the functional derivative of this with respect to the source
$\imath J_{\de}$ and using the relation \eq{eq:mfd11}, we find
\bea
\lefteqn{
 \frac{\de}{\de\imath J_{\de}}  S[\ga]
\ev{\imath \Phi_{\ba}\imath \Phi_{\ga}}
\ev{\imath  J_{\ga}\imath J_{\al}}
}\nonumber\\
&&=\imath S[\ga, \ka]
\ev{\imath  J_{\de}\imath J_{\ka}}
\ev{\imath \Phi_{\ka}\Phi_{\ba}\imath \Phi_{\ga}}
\ev{\imath  J_{\ga}\imath J_{\al}}
+\eta_{\de\ba}S[\ga]
\ev{\imath \Phi_{\ba}\imath \Phi_{\ga}}
\ev{\imath  J_{\ga}\imath J_{\de}\imath J_{\al}}=0,\nonumber\\
\label{eq:mfd3}
\eea
where the factor $\eta_{\de\ba}=-1$ if the fields $\de, \ba$
anticommute.  For $X(J)=\imath J_{\de}\imath J_{\al}$, \eq{eq:mfd10}
becomes
\be
\frac{\de}{\de\imath \Phi_{\ba}}\ev{\imath J_{\de}\imath J_{\al}}= 
-\imath S[\ga]\ev{\imath \Phi_{\ba}\imath \Phi_{\ga}}\ev{\imath
  J_{\ga} \imath J_{\de}\imath J_{\al}}
\ee
or, with the help of \eq{eq:mfd3}
\be
\frac{\de}{\de\imath \Phi_{\ba}}\ev{\imath J_{\de}\imath J_{\al}}= 
-\eta_{\de\ba} S[\ga, \ka]
\ev{\imath  J_{\de}\imath J_{\ka}}
\ev{\imath \Phi_{\ka}\Phi_{\ba}\imath \Phi_{\ga}}
\ev{\imath  J_{\ga}\imath J_{\al}}.
\label{eq:mfd4}
\ee
With this notation multiple functional derivatives of arbitrary order
can be efficiently constructed. In particular, this will be used in
the next Section for the derivation of the quark gap equation. Also,
later on in Chapter~\ref{chap:nGreen} the above formula will be
employed in the derivation of the 4-point quark Green's functions.

\section{Quark gap equation}

We start the derivation of the gap equation by taking the functional
derivative of the quark field equation of motion, \eq{eq:qeom}, with
respect to the quark field $\imath q_{w}$, and omit those terms which
will eventually vanish when the sources are set to zero.  We arrive
at:
\bea
\label{eq:qq}
\lefteqn{
\ev{\imath\bar q_{x\al}\imath q_{w\ba}} =
\G^{(0)}_{\bar qq\al\ba}(x) \delta (x-w)
}
\nonumber\\
&&-\int d^4yd^4z\,\de(x-y)\de(x-z)\left[\G_{\bar qq\si \al\ga}^{(0)a}
\frac{\delta}{\delta\imath q_{w\ba}}
\ev{\imath\rho_{y}^{a}\imath\bar\chi_{z\ga}}
+\G_{\bar qq Aj \al\ga}^{(0)a}\frac{\delta}{\delta\imath q_{w\ba}}
\ev{\imath J_{jy}^{a}\imath\bar\chi_{z\ga}}\right]. \nonumber\\
\label{eq:gapeq1}
\eea
where the (configuration space) tree-level quark proper two-point
function $\G^{(0)}_{\bar qq}(x)$ and quark-gluon vertices $\G_{\bar
qq\si}^{(0)a}$, $\G_{\bar qq Aj }^{(0)a}$ are obtained from the quark
equation of motion \eq{eq:qeom}.  In this thesis we will consider the
gap equation both at one-loop perturbative order and in the heavy
quark limit.  Consequently, the explicit form of the corresponding
tree-level quantities will be given separately in
Chapter~\ref{chap:g^2}, which deals with perturbation theory, and in
Chapter~\ref{chap:hq}, where the heavy quark limit is investigated.

 We now use the formula \eq{eq:mfd4} to calculate the functional
derivatives appearing in the bracket.  As already mentioned, simply
because of the presence of the $\vec{\pi}$ and $\phi$ fields arising
in the first order formalism, we must allow for the additional terms
to be generated. For completeness, we keep all these terms for the
moment, bearing in mind that they vanish when we consider the one-loop
order in perturbation theory:
\begin{figure}[t]
\centering\includegraphics[width=0.7\linewidth]{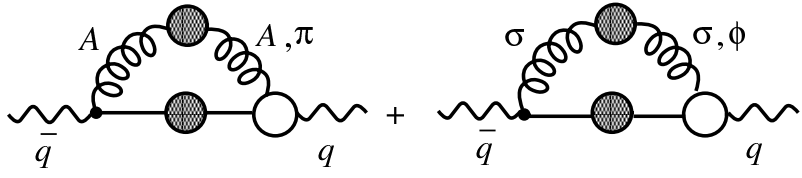}
\caption{\label{fig:gapeq}Full nonperturbative diagram for the quark
self-energy. Filled circles denote dressed propagators and empty
circles denote dressed vertices. Springs denote connected (propagator)
functions, solid lines denote quark propagators and wavy lines denote
the external legs of the proper functions.}
\end{figure}
\bea
\frac{\delta}{\delta\imath q_{w\ba}}
\ev{\imath\rho_{y}^{a}\imath\bar\chi_{z\ga}}&=&
-\int d^4vd^4u
\ev{\imath\bar\chi_{z\ga}\imath\chi_{v\de}}
\ev{\imath\rho_{y}^{a}\imath\rho_{u}^{b}}
\ev{\imath \bar q_{v\de}\imath q_{w\ba}\imath \si_{u}^{b}}
\nonumber\\
&&-\int d^4vd^4u
\ev{\imath\bar\chi_{z\ga}\imath\chi_{v\de}}
\ev{\imath\rho_y^a\imath\kappa_{u}^{b}}
\ev{\imath \bar q_{v\de}\imath q_{w\ba}\imath \phi_{u}^{b}},
\label{eq:rhoqq}\\
\frac{\delta}{\delta\imath q_{w\ba}}
\ev{\imath J_{jy}^{a}\imath\bar\chi_{z\ga}}&=&
-\int d^4vd^4u
\ev{\imath\bar\chi_{z\ga}\imath\chi_{v\de}}
\ev{\imath J_{yj}^{a}\imath J_{uk}^{b}}
\ev{\imath \bar q_{v\de}\imath q_{w\ba}\imath A_{uk}^b}\nonumber\\
&&-\int d^4vd^4u
\ev{\imath\bar\chi_{z\ga}\imath\chi_{v\de}}
\ev{\imath J_{yj}^{a}\imath K_{uk}^{b}}
\ev{\imath \bar q_{v\de}\imath q_{w\ba}\imath \pi_{uk}^{b}}.
\label{eq:jqq}
\eea

At this point, it is useful to introduce our conventions and notations
for the Fourier transform. For a general two-point function (connected
or proper) which obeys translational invariance we have:
\bea
\ev{\imath J_{\al}(y)  \imath J_{\ba}(x)}&=&
\int \dk{k} W_{\al\ba}(k) e^{-\imath k\cdot (y-x)}
\label{eq:fourier1}
\\
\ev{\imath  \Phi_{\al}(y)\imath \Phi_{\ba}(x)}&=& 
\int \dk{k} \G_{\al\ba}(k) e^{-\imath k\cdot (y-x)},
\label{eq:fourier2}
\eea
where $\dk{k}=d^4 k/(2\pi)^{4}$.  The propagator (connected 2-point
function) $W_{\al\ba}(y,x)$ and proper (1PI) two-point function
$\G_{\al\ba}(y,x)$ are related via the Legendre transform. Whereas in
covariant gauges this is simply an inversion, in Coulomb gauge this
may not always be the case. The relation between the connected and
proper two-point functions follows from \eq{eq:funcderiv1}. For the
quark propagator, we find (in momentum space) the standard relation
\be
W_{\al\ga} (k) \G_{\ga\ba} (k) =\de_{\al\ba}.
\ee

Returning to the gap equation, we insert the expressions
\eq{eq:rhoqq}, \eq{eq:jqq} into \eq{eq:qq}, Fourier transform into
momentum space, and we obtain the quark \DS (or gap) equation:
\bea
\lefteqn{\G_{\bar q q \al\ba}(k)=\G_{\bar q q \al\ba}^{(0)}(k)} 
\nonumber\\
&&+\int \dk{\w}
\G_{\bar q q\si \al\ga}^{(0)a}(k,-\w,\w -k)W_{\bar q q \ga\de}(\w)
\G_{\bar q q\si \de\ba}^{b}(\w,-k,k-\w)W_{\si\si}^{ab}(k-\w)
\nonumber\\
&&+\int \dk{\w}
\G_{\bar q q\si \al\ga}^{(0)a}(k,-\w,\w -k)W_{\bar q q \ga\de}(\w)
\G_{\bar q q\phi \de\ba}^{b}(\w,-k,k-\w)W_{\si\phi}^{ab}(k-\w)
\nonumber\\
&&+\int\dk{\w}
\G_{\bar q qAi \al\ga} ^{(0)a}(k,-\w,\w -k)W_{\bar q q \ga\de}(\w)
\G_{\bar qqAj \de\ba}^{b}(\w,-k,k-\w)W_{AAij}^{ab}(k-\w)
\nonumber\\
&&+\int\dk{\w}
\G_{\bar q qAi \al\ga} ^{(0)a}(k,-\w,\w -k)W_{\bar q q \ga\de}(\w)
\G_{\bar qq\pi j \de\ba}^{b}(\w,-k,k-\w)W_{A\pi ij}^{ab}(k-\w).
\hspace{1cm}
\label{eq:gapeqms}
\eea
 The self-energy corrections are presented diagrammatically in
\fig{fig:gapeq}.  We see that the $\vec\pi$ and $\phi$ fields do make
a contribution thanks to the existence of the mixed propagators
$W_{A\pi ij}$ and $W_{\si\phi}$ in the first order formalism.  But, as
emphasized, these contributions will drop out at one-loop order
because of the absence of corresponding tree-level vertices, i.e.,
there exist no direct interaction terms in the action between the
quark fields and the auxiliary fields of the first order
formalism. However, for future studies one has to bear in mind that
additional contributions may arise.

In the second part of this thesis we will also consider the gap
equation in the heavy quark limit. In this case we shall work in the
standard, second order formalism, where the auxiliary fields do not
appear from the first place.

\section{Quark contributions to the gluon propagators}
\label{sec:dsquarkcontr}
\begin{figure}[t]
\centering\includegraphics[width=0.3\linewidth]{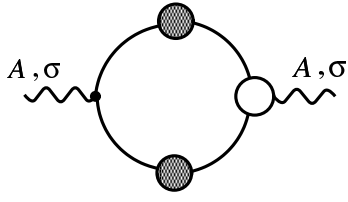}
\caption{\label{fig:Asiqq}One-loop diagram for the quark contributions
to the gluon proper two-point functions.  Filled circles denote
dressed propagators and empty circles denote dressed vertices.  Solid
lines denote quark propagators and wavy lines denote the external legs
of the proper functions.}
\end{figure}

In order to understand the analytic structure of the gluon
propagators, it is necessary to explore the quark contributions to the
gluonic proper two-point functions.\footnote {Apart from the quark
contribution, the gluon propagators contain ghost loops, as well as a
collection of terms generated by the tree-level 4-gluon interactions
(which give rise to tadpole and explicit 2-loop contributions to the
gluon \DS equations). All these will not be considered in this work (a
complete derivation of the \YM terms appearing in the Coulomb gauge
gluon propagators can be found in Ref.~\cite{Watson:2006yq}).}  In
contrast to covariant gauges, the various gluonic degrees of freedom
(temporal and spatial) are being separated into three proper two-point
functions, $\G_{\si\si}$, $\G_{AA}$ and $\G_{\si A}$.  The \DS
equation for the third function, $\G_{\si A}$, can be derived, since
the quark-gluon vertices entering the equation are defined, even
though the function itself does not have a tree level component in the
first order formalism.

Starting with the $\si$ equation of motion \eq{eq:si_eom} and
following the same procedure as for the gap equation we derive the
quark contribution to the proper two-point function $\G_{\si\si}$
(indicated by the index $(q)$) in configuration space:
\be
\ev{\imath\si_x^a\imath\si_w^b}_{(q)}=\!
-\mathrm{Tr}\!\int d^4yd^4zd^4ud^4v
\G_{\bar q q\si\al\ga}^{(0)a}(z,y,x)
\ev{\imath\bar \chi_{y\ga}\imath \chi_{u\ba}}
\ev{\imath \bar q_{u\ba}\imath q_{v\de}\imath\si_{w}^{b}}
\ev{\imath \bar\chi_{v\de}\imath \chi_{z\al}}\!\!\!. 
\label{eq:qqsigma_mom}
\ee
In the above, the trace over Dirac and fundamental color indices is
taken.  Performing the Fourier transform, we get in momentum space:
\be
\G_{\si\si(q)}^{ab}(k)=
-\mathrm{Tr}\int\dk{\w}
\G_{\bar qq\si\al\ga}^{(0)a}(\w-k,-\w,k)W_{\bar qq\ga\ba}(\w)
\G_{\bar qq\si \ba\de}^{b}(\w,k-\w,-k)W_{\bar qq\de\al }(\w-k).
\label{eq:siqdse}
\ee
Similarly we obtain:
\be
\G_{\si Ai(q)}^{ab}(k)=-\mathrm{Tr}\int\dk{\w}
\G_{\bar qq\si\al\ga}^{(0)a}(\w-k,-\w,k)W_{\bar qq\ga\ba}(\w)
\G_{\bar qqAi\ba\de}^{b}(\w,k-\w,-k)W_{\bar qq \de\al}(\w-k),
\label{eq:Aqdse}
\ee
(it is easy to check that $\G_{\si Ai(q)}=\G_{A\si i(q)}$) and
\be
\G_{AAij(q)}^{ab}(k)=-\mathrm{Tr}\int\dk{\w}
\G_{\bar qqAi\al\ga}^{(0)a}(\w-k,-\w,k)W_{\bar qq\ga\ba}(\w)
\G_{\bar qq Aj\ba\de}^{b}(\w,k-\w,-k)W_{\bar qq \de\al}(\w-k). 
\label{eq:siAqdse}
\ee
In Chapter~\ref{chap:g^2} we will consider these loop contributions
(shown collectively in Fig.~\ref{fig:Asiqq}) at one loop perturbative
level, and compare the results with the calculations performed in
covariant gauges. We can already anticipate that since at one loop
perturbative level the quark loop cannot be different from its
covariant analog (the difference is that the spatial and temporal
degrees of freedom are separated into the corresponding proper
two-point functions), the one loop results should equal the covariant
gauge calculations. Later on we will see that this is indeed the case.

\section{Quark-gluon vertex}

\begin{figure}[t]
\centering\includegraphics[width=1.0\linewidth,
height=0.35\textheight]{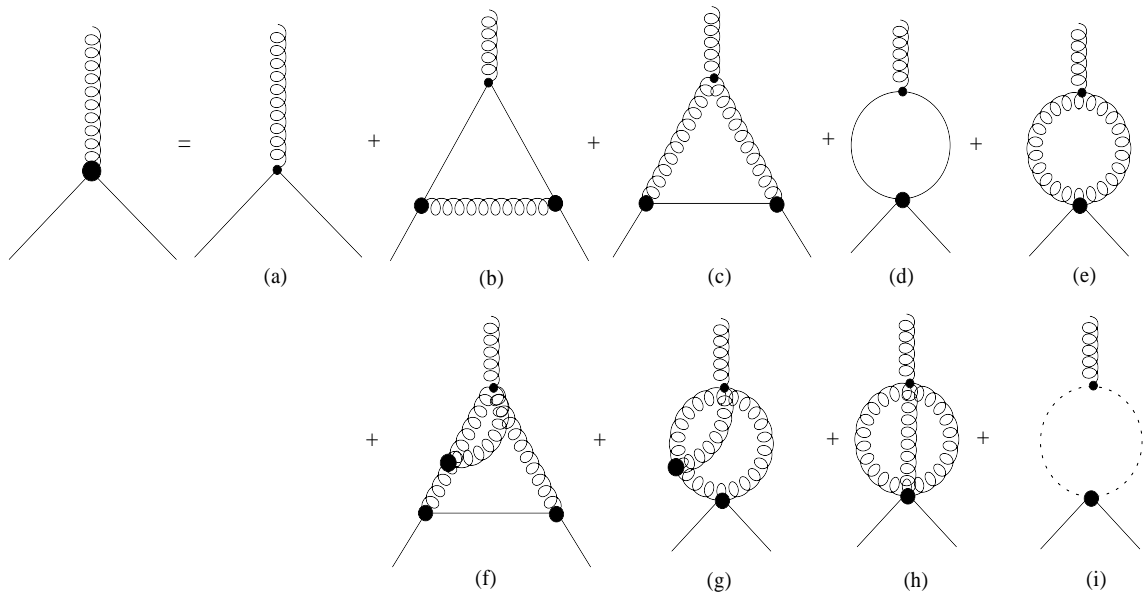}
\caption{\label{fig:quark-gluon-vertex}\DS equation for the
(\emph{spatial} or \emph{temporal}) quark-gluon vertex, written in
terms of proper (1PI) Green's functions. Blobs denote dressed vertices
and all internal propagators are fully dressed. Internal propagators
denoted by springs may be either spatial ($\vec{A}$) or temporal
($\si$) propagators, dashed lines represent the ghost propagator and
solid lines represent the quark propagator. Symmetry factors and signs
have been omitted.}
\end{figure}
\begin{figure}[t]
\centering\includegraphics[width=1.00\linewidth,
 height=0.18\textheight]{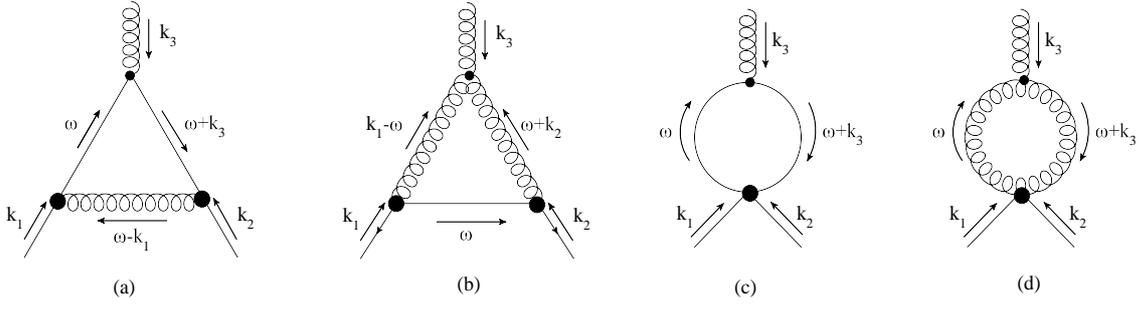}
\caption{\label{fig:mom_rooting}Momentum rooting for the quark-gluon
vertex functions.  Diagrams (a) and (c) represent the QED-like graphs,
and diagrams (b) and (d) are the non-abelian graphs.}
\end{figure}

The quark-gluon vertex, as an important component that relates the \YM
and the quark sector of QCD, has been intensively studied in the last
years, mostly in covariant gauges.  At perturbative level, it has been
analyzed in Ref.~\cite{Davydychev:2000rt}, in arbitrary (covariant)
gauge and dimension.  Also, in Landau gauge, nonperturbative studies
have been carried on (see, for example, \cite{Alkofer:2008tt} and
references therein).

In Coulomb gauge, due to the intrinsic noncovariance, there are $two$
quark-gluon vertices, spatial and temporal.\footnote{In fact, within
the first order formalism one can derive the \DS equations for two
more vertex functions, corresponding to the interaction of the quarks
with the additional fields $\pi$ and $\phi$. However, since there are
no direct interaction terms in the Lagrangian, these terms do not give
a contribution at one loop order in perturbation theory.}  Just as for
the quark contributions to the gluonic two-point functions, the \DS
equations for the spatial and temporal quark-gluon vertices are
obtained by taking the functional derivative of the gluon field
equations of motion \eq{eq:a_eom}, \eq{eq:si_eom} with respect to
$\imath\bar q_{v\de}, \imath q_{w\ga}$ and setting sources to zero.

Defining the Fourier transform for the vertex functions (all momenta
are incoming):
\be
\G(x,y,z)=\int\dk{k_1}\dk{k_2}\dk{k_3}(2\pi)^4\de(k_1+k_2+k_3)e^
{-\imath k_1\cdot x-\imath k_2\cdot y-\imath k_3\cdot z}\G(k_1,k_2,
k_3),
\label{eq:ftvertex}
\ee
\pagebreak
we get for the spatial quark gluon vertex, in terms of proper Green's
functions:
\bea
\lefteqn{\G_{\bar qqAi \al\ba}^d(k_1,k_2,k_3)=-gT^d_{\al\ba}\ga^i }
\nonumber\\
&&-\int\dk{\w}\left\{
\G_{\bar qqAk\al\ga}^c(\w-k_1,k_1,-\w)W_{\bar qq\ga\de}(\w)
\G_{\bar q qAi\de\la}^{(0)d}(k_3,\w,-k_3-\w) 
W_{\bar q q\la\eta}(k_3+\w) \right.\nonumber\\
&&\times\G_{\bar
 qqAj\eta\ba}^b(k_3+\w,k_2,k_1-\w)W_{AAjk}^{bc}(k_1-\w)
 \nonumber\\
&&+
\G_{\bar qq\si\al\ga}^c(\w-k_1,k_1,-\w)W_{\bar qq\ga\de}(\w)
\G_{\bar q qAi\de\la}^{(0)d}(k_3,\w,-k_3-\w) 
W_{\bar qq \la\eta}(k_3+\w) \nonumber\\
&&\times\G_{\bar qq\si\eta\ba}^b(k_3+\w,k_2,k_1-\w)
W_{\si\si}^{bc}(k_1-\w)\nonumber\\
&&+
\G_{\bar qqAk\al\ga}^a(\w-k_1,k_1,-\w)W_{\bar qq\ga\de}(\w)
\G_{\bar qq\si\de\ba}^e(\w,k_2,-k_2-\w) 
W_{\si\si}^{ec}(k_2+\w) \nonumber\\
&&\times G_{\si AAij}^{(0)dbc}(k_2+\w,k_3,k_1-\w)
W_{AAjk}^{ba}(\w-k_1)\nonumber\\
&&+
\G_{\bar qq\si\al\ga}^a(\w-k_1,k_1,-\w)W_{\bar qq\ga\de}(\w)
\G_{\bar qqAk\de\ba}^e(\w,k_2,-k_2-\w)W_{AAjk}^{ec}(k_2+\w) \nonumber\\
&&\times\G_{\si AAij}^{(0)dbc}(k_2+\w,k_3,k_1-\w)W_{\si\si}^{ba}(\w-k_1)
\nonumber\\
&&+
\G_{\bar qqAl\al\ga}^a(\w-k_1,k_1,-\w)W_{\bar qq\ga\de}(\w)
\G_{\bar qqAm\de\ba}^e(\w,k_2,-k_2-\w)W_{AAmk}^{ec}(k_2+\w)
\nonumber\\
&&\times\G_{3Aijk}^{(0)dbc}(k_2+\w,k_3,k_1-\w)W_{AAjl}^{ba}(\w-k_1)
\nonumber\\
&&+
\G_{\bar qq\si\al\ga}^a(\w-k_1,k_1,-\w)W_{\bar qq\ga\de}(\w)
\G_{\bar qq\si\de\ba}^e(\w,k_2,-k_2-\w) W_{\si\si}^{ec}(k_2+\w)
\nonumber\\
&&\times\G_{\si A\si
  i}^{(0)dbc}(k_2+\w,k_3,k_1-\w)W_{\si\si}^{ba}(\w-k_1)
\nonumber\\
&&+
\G_{\bar qq\si\si\al\ba}^{ce}(\w+k_3,k_2,k_1,\w) W_{\si\si}^{ac}(\w)
\G_{\si A\si  i}^{(0)abd}(\w,k_3,-\w-k_3)
W_{\si\si}^{be}(\w+k_3) \nonumber\\
&&+\left.
\G_{\bar qq\bar qq\al\ba\de\ga}^{ce}(\w+k_3,k_1,k_2,-\w)
 W_{\bar qq\de\ga}^{ac}(\w)
\G_{\bar qqAi \la\eta }^{(0)d}(\w,k_3,-\w-k_3)
W_{\bar qq\eta\ga}^{be}(\w+k_3)
\right\}\nonumber\\
&&+\dots. \label{eq:qAvertex_1loop}
\eea
Similarly, the temporal quark gluon vertex is given by:
\bea
\lefteqn{\G_{\bar qq\si\al\ba}^d(k_1,k_2,k_3)=gT^d_{\al\ba}\ga^0}
\nonumber\\
&&-\int\dk{\w}\left\{
\G_{\bar qqAk\al\ga}^c(\w-k_1,k_1,-\w)W_{\bar qq\ga\de}(\w)
\G_{\bar q q\si\de\la}^{(0)d}(k_3,\w,-k_3-\w)
W_{\bar q q\la\eta}(k_3+\w) \right.\nonumber\\
&&\times\G_{\bar qqAj\eta\ba}^b(k_3+\w,k_2,k_1-\w)
W_{AAjk}^{bc}(k_1-\w)\nonumber\\
&&+
\G_{\bar qq\si\al\ga}^c(\w-k_1,k_1,-\w)W_{\bar qq\ga\de}(\w)
\G_{\bar q q\si\de\la}^{(0)d}(k_3,\w,-k_3-\w)
W_{\bar q q\la\eta}(k_3+\w) \nonumber\\
&&\times\G_{\bar qq\si\eta\ba}^b(k_3+\w,k_2,k_1-\w)
W_{\si\si}^{bc}(k_1-\w)\nonumber\\
&&+
\G_{\bar qqAk\al\ga}^a(\w-k_1,k_1,-\w)W_{\bar qq\ga\de}(\w)
\G_{\bar qq\si\de\ba}^e(\w,k_2,-k_2-\w)W_{\si\si}^{ec}(k_2+\w)
 \nonumber\\
&&\times\G_{\si A\si j}^{(0)dbc}(k_2+\w,k_3,k_1-\w)
W_{AAjk}^{ba}(\w-k_1)\nonumber\\
&&+
\G_{\bar qq\si\al\ga}^a(\w-k_1,k_1,-\w)W_{\bar qq\ga\de}(\w)
\G_{\bar qqAk\de\ba}^e(\w,k_2,-k_2-\w)W_{AAjk}^{ec}(k_2+\w)
\nonumber\\
&&\G_{\si A\si j}^{(0)dbc}(k_2+\w,k_3,k_1-\w)W_{\si\si}^{ba}(\w-k_1)
\nonumber\\
&&+
\G_{\bar qqAl\al\ga}^a(\w-k_1,k_1,-\w)W_{\bar qq\ga\de}(\w)
\G_{\bar qqAm\de\ba}^e(\w,k_2,-k_2-\w)W_{AAmk}^{ec}(k_2+\w)
\nonumber\\
&&\times\G_{\si AAjk}^{(0)dbc}(k_2+\w,k_3,k_1-\w)
W_{AAjl}^{ba}(\w-k_1)\nonumber\\
&&+\left.
\G_{\bar qq\bar qq\al\ba\de\ga}^{ce}(\w+k_3,k_1,k_2,-\w) 
W_{\bar qq\de\la}^{ac}(\w)
\G_{\bar qq\si \la\eta }^{(0)d}(\w,k_3,-\w-k_3)
W_{\bar qq\eta\ga}^{be}(\w+k_3)
\right\}+\dots. \nonumber\\
\label{eq:qsivertex_1loop}
\eea
In the expressions \eq{eq:qAvertex_1loop}, \eq{eq:qsivertex_1loop},
the dots represent the interaction of the quarks with the $\pi$ and
$\phi$ fields, which are not regarded here. Further, the (two-loop)
\YM diagrams arising from the 4-gluon vertex are not considered in
this work (see also footnote at the beginning of
Section~\ref{sec:dsquarkcontr}), but for completeness we have included
these two-loop contributions in the graphical representation from
\fig{fig:quark-gluon-vertex}. The momentum rooting for the vertex
functions is shown diagrammatically in \fig{fig:mom_rooting}.

In both equations, the external gluon leg is connected to a bare
vertex in the loop diagrams.  Alternatively, one can start with the
quark equation of motion and take functional derivatives with respect
to the gluon fields. Then in the corresponding \DS equations the
external quark legs are attached to the bare internal vertex.  In the
full theory, both equations should give the same quark gluon vertex,
however in a truncated theory one of these equations is eventually
easier to solve.\footnote{ See also Appendix B of
Ref.~\cite{Alkofer:2008tt} for an extended discussion of this topic.}
In this work we will only use the first ``version'' of the quark-gluon
vertex, based on Eqs.~(\ref{eq:qAvertex_1loop},
\ref{eq:qsivertex_1loop}).

In the next chapter, we will explicitly evaluate the divergences of
the vertex functions at one-loop perturbative level.  Moreover, the
corresponding one-loop expressions, combined with the so-called
quark-ghost kernels, will be then used to show that the \ST identity
for the quark-gluon vertex is satisfied at leading order in
perturbation theory, without explicitly evaluating the integrals.

%% file: perturbative.tex
\chapter{One-loop perturbative results}
\label{chap:g^2}

This chapter is concerned with the limit when the coupling $g$ between
quarks and gluons is small, i.e.  the whole formalism can be expanded
in powers of $g$, giving rise to the perturbative expansion.

We will first derive the Feynman rules for the quark sector of the
theory (basically, these are the lowest perturbative order Green's
functions), and we will establish the general form of the two-point
functions.  Then we will consider the proper two- and three-point
Green's functions derived in the preceding chapter at one-loop order
in perturbation theory, and evaluate the corresponding two-point
dressing functions.  Coming across the problems originating from the
energy-divergence and the noncovariance inherent to Coulomb gauge, we
will derive the required noncovariant massive integrals, using a
technique based on differential equations and integration by parts. We
then consider the renormalization of the quark mass and propagator and
we verify that the corresponding renormalization factors agree with
the results obtained in covariant gauges.  We also evaluate the first
coefficient of the perturbative $\ba$-function, and again we find that
our results agree with the covariant gauge calculations. Moreover, we
will consider the divergent parts of the temporal and spatial
quark-gluon vertices, and shortly discuss the implications for the \ST
identity for the quark-gluon vertices presented in
Chapter~\ref{chap:st}.

\section{Feynman rules }

In this section, we derive the basic Feynman rules and collect all the
tree-level quantities required for our one-loop calculations.  In
addition to tree-level propagators and proper vertices we also derive
the proper two-point functions of the theory.

With the field equation of motion written in the forms \eq{eq:gen5},
\eq{eq:eomW}, we can now derive the Feynman rules for the quark
components of the theory.  We first derive the quark propagator.  From
the quark equation of motion in terms of connected functions,
\eq{eq:eomW}, ignoring interaction terms and functionally
differentiating we get the tree-level propagator in configuration
space
\be
0=\left[\imath\gamma^{0}\pd_{0x}
+\imath\vec\gamma\cdot\vec\nabla_{x}-m\right]
_{\al\ba}\ev{\imath\chi_\ga(z)
\imath\ov{\chi}_\ba(x)}^{(0)}-\imath\de_{\ga\al}\de(z-x).
\ee
For the quark propagator, the Fourier transform \eq{eq:fourier1}
explicitly reads (recall that translational invariance is assumed):
\be
\ev{\imath\ov{\chi}_\ba(z)\imath\chi_\al(x)}= W_{\ov{q}q\ba
\al}(z-x)=\int\dk{k}e^{-\imath k\cdot(z-x)}W_{\ov{q}q\ba\al}(k)
\ee
such that in momentum space, we get
\be
0=\int\dk{k}e^{-\imath k\cdot(z-y)}\left\{
W_{\ov{q}q\ba\al}^{(0)}(k)
\left[\gamma^0 k_0-\gamma^i k_i+m\right]
_{\al\ga}+\imath\de_{\ba\ga}\right\}.
\ee
The solution is:
 \be 
W_{\bar qq\al\ba}^{(0)}(k)=
(-\imath) \frac{\left[\gamma^0 k_0-\gamma^i k_i
+m\right]}{k_0^2-\vec k^2-m^2} \de_{\al\ba}.
\label{eq:treelevelquarkprop}
\ee

In a similar fashion, starting with the quark equation of motion in
terms of proper functions, \eq{eq:qeom} the tree-level quark proper
two-point function is derived. We obtain:
\be
\G^{(0)}_{\bar qq\al\ba}(x)=
\imath\left[\imath\gamma^{0}\pd_{0 x} 
+\imath\vec\gamma\cdot\vec\nabla_{x}-m\right]_{\al\ba}
\ee
or, after Fourier transforming to momentum space: 
\be
\G_{\bar qq \al\ba}^{(0)}(k)=\imath \left[\gamma^0 k_0
-\gamma^i k_i-m\right]\de_{\al\ba}.
\ee
Due to the noncovariance, we have written out explicitly the
components of $\kslash $, but later on (where appropriate), we will
use the usual notation $\kslash=\gamma^0 k_0-\gamma^i k_i$.

 The (spatial and temporal) tree-level gluon propagators needed in
this work have been derived in \cite{Watson:2006yq} and are given by:
\be
W_{A A ij}^{(0)ab}(k)=\de^{ab}
\frac{\imath t_{ij}(\vec k)}{k_{0}^2-\vec{k}^2},
\textrm{~~~}
W_{\si\si}^{(0)ab}(k)=\de^{ab}\frac{\imath }{\vec{k}^2}
\label{eq:treelevelgluonprop}
\ee
where 
\be
t_{ij}(\vec k)=\delta_{ij}-k_ik_j/\vec{k}^2
\label{eq:tij}
\ee
 is the transverse spatial projector.  It is understood that the
denominator factors involving both temporal and spatial components
implicitly carry the Feynman prescription, i.e.,
\be
\frac{1}{(k_0^2-\vec k^2)}\rightarrow\frac{1}{(k_0^2-\vec k^2+\imath 0_+)},
\ee 
such that the analytic continuation to the Euclidean space ($k_0\to
\imath k_4$) can be performed. This will be explicitly verified at
one-loop order in perturbation theory.

There are two tree-level quark-gluon vertices, spatial and temporal,
again obtained by taking the functional derivatives with respect to
quark, antiquark and gluon field:
\bea
\G_{\bar q q\si\al\ba}^{(0)a}&=&\left[gT^{a}\ga^{0}\right]_{\al\ba},
\label{eq:treelevelquarkvertex1}\\
\G_{\bar q qAj\al\ba}^{(0)a}&=&-\left[gT^{a}\ga^{j}\right]_{\al\ba}.
\label{eq:treelevelquarkvertex2}
\eea

Later on, in the evaluation of the vertex functions, we will also need
the tree-level gluonic vertices derived in \cite{Watson:2007vc} (all
momenta are defined as incoming and momentum conservation is assumed):
\bea
\G_{\si AAjk}^{(0)abc}(p_a,p_b,p_c)&=&
\imath gf^{abc}\de_{jk}(p_b^0-p_c^0),\nonumber\\
\G_{\si A\si j}^{(0)abc}(p_a,p_b,p_c)&=&
-\imath gf^{abc}(p_a-p_c)_j,\nonumber\\
\G_{3A ijk}^{(0)abc}(p_a,p_b,p_c)&=&
-\imath gf^{abc}
\left[\de_{ij}(p_a-p_b)_k+\de_{jk}(p_b-p_c)_i
+\de_{ki}(p_c-p_a)_j\right],\nonumber\\
\G_{\ov{c}cA i}^{(0)abc}(p_{\ov{c}},p_c,p_A)&=&-\imath
gf^{abc}p_{\ov{c}i}.
\label{eq:treeleveltriplegluon}
\eea

\section{Two-point functions}

In this section, we introduce the general decompositions of the
two-point functions derived in the preceding chapter, and derive the
one-loop perturbative expressions for the associated dressing
functions. In order to evaluate the noncovariant massive integrals
arising in the one-loop calculations, we employ a technique based on
differential equations and integration by parts. We then give the
perturbative results for the two-point functions.

\subsection{General decomposition}

In order to investigate the \DS equations for the quark gap equation
derived in the previous chapter, in addition to the tree-level forms
given in the previous section, we will also require the general
decompositions for the quark propagator and proper two-point function,
and the relationship between them.  Because we work in a noncovariant
setting, the usual arguments must be modified to separately account
for the temporal and spatial components.  Starting with the quark
propagator, we observe that (recall that $\cd\Phi$ denotes the
integration over all fields)
\be
W_{\ov{q}q \al\ba}(k^0,\vec{k})
\sim \de_{\al\ba}\int\cd\Phi\,\ov{q}q\,\exp{\left\{\imath\cs\right\}}
\ee
such that under both time-reversal and parity transforms, the
propagator will remain unchanged -- the bilinear combination $\ov{q}q$
is scalar.  Since the propagator depends on both $k^0$ and $\vec{k}$,
it has thus \emph{four} components, in distinction to the covariant
case where there are only two (a dressing function multiplying
$\kslash$ and a mass term).  Hence we can write
\be
W_{\bar qq \al\ba}(k)=
\de_{\al\ba}\frac{(-\imath)}{k_0^2-\vec k^2-m^2}
\left\{ k_0\ga^{0}F_t(k)-k_i\ga^{i}F_s(k)+M(k)+k_0k_i\ga^0\ga^iF_d(k)
\right\}
\label{eq:prop_gendecom}
\ee
where all dressing functions are functions of both $k_0^2$ and
$\vec{k}^2$.  At tree-level, one can trivially identify $F_t=F_s=1$,
$F_d=0$ and $M=m$.  The last term with $F_d$ has no covariant
counterpart. The possible appearance of this term will be discussed
below.

For the proper two-point function, the same arguments apply and we
write
\be
\G_{\bar qq \al\ba}(k)=\imath \de_{\al\ba}
\left\{k_0\ga^{0}A_t(k)-k_i\ga^{i}A_s(k)-B_m(k)
+k_0k_i\ga^0\ga^iA_d(k)\right\}
\label{eq:gap_gendecom}
\ee
and we will refer to $A_t, A_s$ and $B_m$ as the temporal, spatial and
massive components, respectively.  Again the last component $A_d$ has
no covariant counterpart. The relationship between the connected and
proper two-point functions is supplied via the Legendre transform
(introduced in Chapter~\ref{chap:ds}) and we have
\be
\G_{\bar qq}(k)W_{\bar qq }(k)=1.
\label{eq:legtr}
\ee

Let us now discuss the possible appearance of the genuinely
noncovariant term corresponding to the dressing function $A_d$ (or
equivalently $F_d$).  At one-loop order in perturbation theory, these
terms are vanishing. This can be deduced from the Dirac structure of
the self-energy loop, stemming from the quark propagator and
tree-level vertices. Namely, the tree-level quark propagator does not
contain a term with $k_0k_i\ga^0\ga^i$ and from the tree-level
vertices we only have either two $\ga^0$ or two $\ga^i$ matrices
together (the gluon propagator is either purely temporal or
spatial). This implies that there is no one-loop contribution that has
the overall structure $\ga^0\ga^i$ and in turn this means that $A_d=0$
at one-loop order perturbatively. Hence, the components $A_d$ and
$F_d$ will only appear (if at all) at two-loop order or beyond. In in
fact, very recently lattice calculations have shown that $F_d=A_d=0$
\cite{Burgio:2010wv}.

Using the definitions \eq{eq:prop_gendecom}, \eq{eq:gap_gendecom},
together with the relation \eq{eq:legtr}, we then get the following
set of relations for the remaining dressing functions:
\bea
F_{t}=\frac{(k_0^2-\vec k^2-m^2)A_t}
{k_0^2A_t^2-\vec k^2A_s^2-B_m^2},\textrm{~~~}
F_{s}=\frac{(k_0^2-\vec k^2-m^2)A_s}
{k_0^2A_t^2-\vec k^2A_s^2-B_m^2},\textrm{~~~}
M=\frac{(k_0^2-\vec k^2-m^2)B_m}{k_0^2A_t^2
-\vec k^2A_s^2-B_m^2}.
\label{eq:b}
\eea
We again emphasize that these relations only hold up to one-loop
perturbatively --- in possible future studies it must be recognized
that the fourth Dirac structure, $\gamma^0\gamma^i$, may enter in a
nontrivial fashion.  In this case, the set of relations \eq{eq:b}
should be replaced with the following set (which includes the
functions $A_d$ and $F_d$):
\bea
F_t&=&\frac{(k_0^2-\vec k^2-m^2)A_t}
{k_0^2A_t^2-\vec k^2A_s^2-B_m^2+k_0^2\vec{k}^2A_d^2},\nonumber\\
F_s&=&\frac{(k_0^2-\vec k^2-m^2)A_s}
{k_0^2A_t^2-\vec k^2A_s^2-B_m^2+k_0^2\vec{k}^2A_d^2},\nonumber\\
M&=&\frac{(k_0^2-\vec k^2-m^2)B_m}
{k_0^2A_t^2-\vec k^2A_s^2-B_m^2+k_0^2\vec{k}^2A_d^2},\nonumber\\
F_d&=&\frac{(k_0^2-\vec k^2-m^2)A_d}
{k_0^2A_t^2-\vec k^2A_s^2-B_m^2+k_0^2\vec{k}^2A_d^2}.
\label{eq:fullwd}
\eea

\subsection{One-loop perturbative expansions}

Let us now consider the one-loop perturbative expansions of the quark
gap equation and the quark contributions to the gluon two-point
functions.  Although so far the formalism has been presented in
4-dimensional Minkowski space, in order to evaluate the resulting loop
integrals we have to convert to Euclidean space. This means that we
make the analytic continuation $k_0\rightarrow \imath k_4$, where
$k_4$ denotes the temporal component of the Euclidean 4-momentum, such
that $k^2=k_4^2+\vec k^2$.  Additionally, to regularize the integrals,
dimensional regularization is employed\footnote{An alternative method
proposed by Leibbrandt is the so-called split dimensional
regularization.  This has been discussed in the introductory chapter
of this thesis.}  with the Euclidean space integration measure
\be
\dk{\w}=\frac{d\w_4 d^d \vec\w}{(2\pi)^{d+1}}
\ee
where $d=3-2\e$ is the spatial dimension.  To preserve the dimension
of the action we must assign a dimension to the coupling through the
replacement
\be
g^2\rightarrow g^2\mu^\e,
\ee
where $\mu$ is the square of a non-vanishing mass scale (which may be
later associated with a renormalization scale).

The perturbative expansion of a generic two-point dressing function is
written as:
\be
\G=\G^{(0)}+g^2\G^{(1)}
\ee 
where the factor $\mu^{\e}$ is included in $\G^{(1)}$ such that the
new coupling and $\G^{(1)}$ are dimensionless.

Let us first consider the (full nonperturbative) gap equation,
\eq{eq:gapeqms} (depicted in \fig{fig:gapeq}). We first insert the
various tree-level vertices and propagators given by
Eqs.~(\ref{eq:treelevelquarkprop}), (\ref{eq:treelevelgluonprop})
(\ref{eq:treelevelquarkvertex1}), and
(\ref{eq:treelevelquarkvertex2}). Then we insert the general
decomposition of the proper two-point functions, \eq{eq:gap_gendecom},
occurring on the left-hand side of the gap equation, take the Dirac
projection, solve the color and tensor algebra and lastly perform the
Wick rotation.  The one-loop temporal, spatial and massive components
of the quark gap equation in Euclidean space read (the Casimir
invariant $C_F=(N_c^2-1)/2N_c$ is listed in the
Appendix~\ref{chap:app.not}):
\bea
A_t(k)&
=&1-g^2\mu^\e C_F\frac{1}{k_4^2}\int\dk{\w}
\left\{\frac{k_4\w_4}{(\w^2+m^2)(\vec k-\vec\w)^2}
-\frac{k_4\w_4(\textrm{d}-1)}{(\w^2+m^2)(k-\w)^2}\right\},
\label{eq:atmassive}\\
A_s(k)&
=&1-g^2\mu^\e C_F\frac{1}{\vec k^2}
\int\dk{\w}\bigg\{
-\frac{2[\vec k\cdot(\vec k-\vec\w)][\vec\w\cdot(\vec k-\vec\w)]}
{(\w^2+m^2)(k-\w)^2(\vec k-\vec\w)^2}
-\frac{\vec k\cdot\vec\w}{(\w^2+m^2)(\vec k-\vec\w)^2}\nonumber\\
&&\hspace{4cm}
+\frac{\vec k\cdot\vec\w(3-\textrm{d})}{(\w^2+m^2)(k-\w)^2}
\bigg\},
\label{eq:asmassive}\\
B_m(k)&=&m+m g^2\mu^\e C_F\int\dk{\w}
\left\{\frac{1}{(\w^2+m^2)(\vec
 k-\vec\w)^2}+\frac{(\textrm{d}-1)}{(\w^2+m^2)(k-\w)^2}\right\}.
\label{eq:bmassive}
\eea
As mentioned earlier, the possible contribution corresponding to the
genuinely noncovariant dressing function $A_d$ does not appear at
one-loop.  To evaluate the integrals occurring in \eq{eq:asmassive},
it is helpful to use the identity:
\be
\s{\vec{k}}{\vec{\w}}=\frac12\left[k^2+\w^2-(k-\w)^2\right]-k_4\w_4,
\label{eq:scalprod}
\ee
which enables us to rewrite $A_s$ as a combination of more
straightforward integrals:
\bea
A_s(k)&=&1-g^2\mu^\e C_F\frac{1}{\vec k^2}
\int\dk{\w}\left\{
-\frac12\frac{(k^2+m^2)^2}{\w^2[(k-\w)^2+m^2]\vec\w^2}
+\frac{2(k^2+m^2)k_4\w_4}{\w^2[(k-\w)^2+m^2]\vec\w^2}
\right.\nonumber\\
&&\left.+\frac{2\e \vec k^2+2k_4^2}{[(k-\w)^2+m^2]\w^2}
+\frac{2\vec k\cdot\vec\w  (1-\e)}{[(k-\w)^2+m^2]\w^2}
-\frac12 \frac{m^2+3 k^2}{[(k-\w)^2+m^2]\vec\w^2}
\right\}.
\label{eq:asmassive1}
\eea

Let us now examine the quark contributions to the various proper
two-point gluon dressing functions given by
Eqs. (\ref{eq:siqdse}-\ref{eq:siAqdse}) (presented in
\fig{fig:Asiqq}). Again, we insert the tree-level factors given by
Eqs.~(\ref{eq:treelevelquarkprop}),(\ref{eq:treelevelgluonprop}),
(\ref{eq:treelevelquarkvertex1}) and (\ref{eq:treelevelquarkvertex2}),
solve the color and tensor algebra and perform a Wick rotation.  The
one-loop integral expressions are (recall that we have $N_f$ flavors
of identical quarks):
\bea
&&\vec k^2\G_{\si\si(q)}^{(1)}(k)=\mu^\e N_{f}2
\int\dk{\w}\frac{\vec\w^2-\w_4^2-\vec\w\cdot\vec k+\w_4k_4+m^2}
{(\w^2+m^2)[(k-\w)^2+m^2]},
\label{eq:Gsisi}\\
&&k_i k_4\G_{\si A(q)}^{(1)}(k)=\mu^\e N_{f}4
\int\dk{\w}\frac{\w_i\w_4-k_i\w_4}{(\w^2+m^2)[(k-\w)^2+m^2]},
\label{eq:GsiA}\\
&&\vec k^2t_{ij}(\vec k)\G_{AA(q)}^{(1)}(k)
+ k_ik_j\bar\G_{AA(q)}^{(1)}(k)\nonumber\\
&&\mbox{~~~}= 2N_f\mu^\e
\int\dk{\w}\frac{2\w_i\w_j-2\w_ik_j+\delta_{ij}
(\w_4k_4+\vec\w\cdot\vec k-\w_4^2-\vec\w^2-m^2)}
{(\w^2+m^2)[(k-\w)^2+m^2]},\hspace{2cm}
\label{eq:GAA}
\eea
where $\G_{AA(q)}^{(1)}$ and $\bar\G_{AA(q)}^{(1)}$ are the
transversal and longitudinal components of the proper two-point
function $\G_{AAij(q)}^{(1)ab}$ (given by \eq{eq:siAqdse}),
respectively\footnote{ The decomposition of the spatial two-point
function $\G_{AAij}$ is explained in detail in
Ref.~\cite{Watson:2006yq} .}.

Having derived the one-loop perturbative expressions for the
propagator dressing functions and the quark contributions to the
gluonic two-point functions, we can now proceed to evaluate the
corresponding loop integrals. There are two categories of integrals
arising: those which can be solved using standard techniques (such as
Schwinger parametrization or Mellin representation --- the details of
these techniques are presented in the Appendices~\ref{chap:app2},
\ref{chap:app3}), and those which require a more complex approach. In
the next section, we will concentrate on this later variety.  Since
this part is rather technical, the reader might skip this and go
directly to Section~\ref{subsec:1loopres}, where the physical results
are presented.

\subsection{Noncovariant massive loop integrals}
\label{chap:app.noncov2pct}

The noncovariant massive loop integrals appearing in the one-loop
expansions from the previous section are studied by using a technique
based on differential equations and integration by parts developed
previously in Ref.~\cite{Watson:2007mz}.  We will consider the two
integrals:
\bea
A_m(k_4^2,\vec{k}^2)&=&
\int\frac{\dk{\w}}{\w^2[(k-\w)^2+m^2]\vec{\w}^2},
\label{eq:adef}\\
A^4_m(k_4^2,\vec{k}^2)&=&
\int\frac{\dk{\w}\,\w_4}{\w^2[(k-\w)^2+m^2]\vec{\w}^2}.
\label{eq:a4def}
\eea

\subsubsection{Derivation of the differential equations}

Let us first write Eqs.~(\ref{eq:adef}) and (\ref{eq:a4def}) in the
general form ($n=0,1$)
\be
I^n(k_4^2,\vec{k}^2)=
\int\frac{\dk{\w}\,\w_4^n}{\w^2[(k-\w)^2+m^2]\vec{\w}^2}.
\label{eq:idef}
\ee
In this derivation $k_4^2$ and $\vec{k}^2$ are treated as variables
whereas the mass, $m$, is treated as a parameter.  The two first
derivatives are:
\bea
k_4\frac{\pd I^n}{\pd k_4}&=&
\int\frac{\dk{\w}\,\w_4^n}{\w^2[(k-\w)^2+m^2]
\vec{\w}^2}\left\{-2\frac{k_4(k_4-\w_4)}{(k-\w)^2+m^2}\right\},
\label{eq:ade0}\\
k_k\frac{\pd I^n}{\pd k_k}&=&
\int\frac{\dk{\w}\,\w_4^n}{\w^2[(k-\w)^2+m^2]
\vec{\w}^2}
\left\{-2
\frac{\s{\vec{k}}{(\vec{k}-\vec{\w})}}{(k-\w)^2+m^2}\right\}.
\label{eq:adevec}
\eea
There are also two integration by parts identities:
\bea
0&=&\int\dk{\w}
\frac{\pd}{\pd\w_4}\frac{\w_4^{n+1}}{\w^2[(k-\w)^2+m^2]\vec{\w}^2}
\nonumber\\
&=&\int\frac{\dk{\w}\,\w_4^n}{\w^2[(k-\w)^2+m^2]\vec{\w}^2}
\left\{n+1-2\frac{\w_4^2}{\w^2}-2\frac{\w_4(\w_4-k_4)}{(k-\w)^2
+m^2}\right\},\label{eq:ibpA}\\
0&=&\int\dk{\w}
\frac{\pd}{\pd\w_i}\frac{\w_i\w_4^n}{\w^2[(k-\w)^2+m^2]\vec{\w}^2}
\nonumber\\
&=&\int\frac{\dk{\w}\,\w_4^n}{\w^2[(k-\w)^2+m^2]\vec{\w}^2}
\left\{d-2-2\frac{\vec{\w}^2}{\w^2}-2
\frac{\s{\vec{\w}}{(\vec{\w}-\vec{k})}}
{(k-\w)^2+m^2}\right\}.
\label{eq:ibpA4}
\eea
Adding these two expressions gives
\be
0=\int\frac{\dk{\w}\,\w_4^n}{\w^2[(k-\w)^2+m^2]\vec{\w}^2}
\left\{d+n-3-2\frac{\w\cdot(\w-k)}{(k-\w)^2+m^2}\right\}.
\label{eq:ibpSum}
\ee
Expanding the numerator factor, we can rewrite \eq{eq:ibpSum} as
\be
0=\int\frac{\dk{\w}\,\w_4^n}{\w^2[(k-\w)^2+m^2]\vec{\w}^2}
\left\{d+n-4+\frac{k^2-\w^2+m^2}{(k-\w)^2+m^2}\right\}.
\label{eq:ibpSum1}
\ee
Combining this with \eq{eq:ibpA} then yields:
\bea
\lefteqn{
\int\frac{\dk{\w}\,\w_4^n}{\w^2[(k-\w)^2+m^2]\vec{\w}^2}
\left\{-2\frac{k_4(k_4-\w_4)}{(k-\w)^2+m^2}\right\}=
}
\nonumber\\
&&\hspace{-0.6cm}
\int\frac{\dk{\w}\,\w_4^n}{\w^2[(k-\w)^2+m^2]\vec{\w}^2}
\left[\frac{2 k_4^2}{k^2+m^2} (n+d-4)-n+1\right]
+\frac{\vec{k}^2+m^2}{k^2+m^2}
\int\frac{\dk{\w} \w_4^n}{[(k-\w)^2+m^2]^2\vec{\w}^2}
\nonumber\\
&&\hspace{-0.6cm}
-2\int\frac{\dk{\w}\w_4^n}{\w^4[(k-\w)^2+m^2]}
-2\int\frac{\dk{\w}\w_4^n}{\w^2[(k-\w)^2+m^2]^2}. 
\eea
This leads to the temporal differential equations for $A_m$ and
$A^4_m$:
\bea
k_4\frac{\pd A_m}
{\pd  k_4}&=&\left[1+2\frac{(d-4)k_4^2}{k^2+m^2}\right]A_m
+2\frac{\vec{k}^2+m^2}{k^2+m^2}\int\frac{\dk{\w}}
{[(k-\w)^2+m^2]^2\vec{\w}^2}\nonumber\\
&-&2\int\frac{\dk{\w}}{\w^4[(k-\w)^2+m^2]}
-2\int\frac{\dk{\w}}{\w^2[(k-\w)^2+m^2]^2},
\label{eq:detemp1}\\
k_4\frac{\pd A^4_m}{\pd k_4}&=&2\frac{(d-3) k_4^2}{k^2+m^2}A_m^4
+2\frac{\vec{k}^2+m^2}{k^2+m^2}
\int\frac{\dk{\w}\,\w_4}{[(k-\w)^2+m^2]^2\vec{\w}^2}
\nonumber\\
&-&2\int\frac{\dk{\w}\,\w_4}{\w^4[(k-\w)^2+m^2]}
-2\int\frac{\dk{\w}\,\w_4}{\w^2[(k-\w)^2+m^2]^2}.
\hspace{3cm}
\label{eq:detemp2}
\eea
In the same manner, we derive the differential equations involving the
spatial components:
\bea
k_i\frac{\pd A_m}{\pd k_i}&=&
\left[2-d+2\frac{(d-4)\vec{k}^2}{k^2+m^2}\right]A_m
-\frac{2\vec{k}^2}{k^2+m^2}
\int\frac{\dk{\w}}{[(k-\w)^2+m^2]^2\vec{\w}^2}\nonumber\\
&+&2\int\frac{\dk{\w}}{\w^4[(k-\w)^2+m^2]}
+2\int\frac{\dk{\w}}{\w^2[(k-\w)^2+m^2]^2},\label{eq:desp1}\\
k_i\frac{\pd A^4_m}{\pd k_i}&=&
\left[2-d+2\frac{(d-3)\vec{k}^2}{k^2+m^2}\right]A^4_m
-\frac{2\vec{k}^2}{k^2+m^2}
\int\frac{\dk{\w}\,\w_4}{[(k-\w)^2+m^2]^2\vec{\w}^2}
\nonumber\\
&+&2\int\frac{\dk{\w}\,\w_4}{\w^4[(k-\w)^2+m^2]}
+2\int\frac{\dk{\w}\,\w_4}{\w^2[(k-\w)^2+m^2]^2}.
\hspace{3cm}
\label{eq:desp2}
\eea
It is in fact possible to write down a mass differential equation, but
in the light of the method presented here this would not bring any new
information.  However, this third differential equation will turn to
be useful in checking our solutions -- the detailed derivation will be
presented in the Appendix~\ref{chap:app3}.

At this point, it is instructive to show how the differential
equations for the massless integrals considered in
Ref.~\cite{Watson:2007mz} are regained.  In the massless limit, there
are potential ambiguities arising in the integrals appearing in
Eqs.~(\ref{eq:detemp1} -\ref{eq:desp2}), because in part, the limits
$m\rightarrow 0$ and $\e\rightarrow 0$ do not interchange.  Let us
start by considering the following integral given by \eq{eq:smi3}
(similar arguments apply to all the integrals appearing in the
differential equations):
\be
I=\int\frac{\dk{\w}}{\vec{\w}^2[(k-\w)^2+m^2]^2}=
\frac{[m^2]^{-1-\e}}{(4\pi)^{2-\e}}\frac{\G(\frac12-\e) 
\G(1+\e)}{\G(3/2-\e)}
{}_{2}F_{1}\left(1,1+\e;3/2-\e;-\frac{\vec{k}^2}{m^2}\right).
\label{eq:stan_ex}
\ee
It is useful here to invert the argument of the hypergeometric with
the help of the formula (see, for instance, Ref.~\cite{abramowitz}):
\bea
{}_{2}F_{1}(a,b;c;t)&=&\frac{\G(c)\G(b-a)}{\G(b)\G(c-a)}(-t)^{-a}
{}_{2}F_{1}\left(a,1-c+a;1-b+a;\frac1t\right)\nonumber\\
&&+\frac{\G(c)\G(a-b)}{\G(a)\G(c-b)}(-t)^{-b}
{}_{2}F_{1}\left(b,1-c+b;1-a+b;\frac1t\right).
\label{eq:hyparginv}
\eea
Then we have:
\bea
I&=&\frac{1}{\vec k^2}\frac{[m^2]^{-\e}\G(\e)}{(4\pi)^{2-\e}}
{}_{2}F_{1}\left(1,\frac 12+\e;1-\e;-\frac{m^2}{\vec{k}^2}\right)
 \nonumber\\
&&+\frac{[\vec k^2]^{-1-\e}}{(4\pi)^{2-\e}}
\frac{\G\left(\frac12-\e\right)\G(-\e)\G(1+\e)}
{\G\left(\frac12-2\e\right)}
{}_{2}F_{1}\left(1+\e,\frac12+2\e;1+\e;-\frac{m^2}{\vec{k}^2}\right).
\label{eq:example_ambiguity}
\eea
In the expression above, the problem of the non-interchangeable limits
is seen explicitly in the first term.  However, when all the integrals
occurring in the various differential equations are put together, such
terms explicitly cancel and only the second term of
\eq{eq:example_ambiguity} (which leads to the correct massless limit)
contributes.

Returning to the differential equations, we evaluate the standard
integrals in terms of $\e$ (see Appendix \ref{chap:app2}) and with the
notation $x=k_4^2$, $y=\vec{k}^2$ we obtain for $A_m$:
\bea
\lefteqn{2x\frac{\pd A_m}{\pd x}=
\left[1-2(1+2\e)\frac{x}{x+y+m^2}\right]A_m} \nonumber\\
&&+2\frac{[m^2]^{-1-\e}}{(4\pi)^{2-\e}}\left\{
\frac{y+m^2}{x+y+m^2}X
{}_{2}F_1\left(1,1+\e;3/2-\e;-\frac{y}{m^2}\right)\right.\nonumber\\
&&- \left.
\frac{\G(-\e)\G(1+\e)}{\G(2-\e)}
{}_{2}F_1\left(2,1+\e;2-\e;-\frac{x+y}{m^2}\right)
-Y{}_{2}F_1\left(1,1+\e;2-\e;-\frac{x+y}{m^2}\right)\right\},
\nonumber\\\label{eq:atde0}\\
\lefteqn{2y\frac{\pd A_m}{\pd y}=
\left[-1+2\e-2(1+2\e)\frac{y}{x+y+m^2}\right]A_m} \nonumber\\
&&-2\frac{[m^2]^{-1-\e}}{(4\pi)^{2-\e}}\left\{
\frac{y}{x+y+m^2}X
{}_{2}F_1\left(1,1+\e;3/2-\e;-\frac{y}{m^2}\right)\right.\nonumber\\
&&-\left. 
\frac{\G(-\e)\G(1+\e)}{\G(2-\e)}
{}_{2}F_1\left(2,1+\e;2-\e;-\frac{x+y}{m^2}\right)
-Y {}_{2}F_1\left(1,1+\e;2-\e;-\frac{x+y}{m^2}\right)\right\}
\nonumber\\
\label{eq:asde0}
\eea
and for the integral $A^4=k_4\ov{A}_m$:
\bea
\lefteqn{2x\frac{\pd \ov{A}_m}{\pd x}=
\left[-1-4\e\frac{x}{x+y+m^2}\right]\ov{A}_m}\nonumber\\
&&+2\frac{[m^2]^{-1-\e}}{(4\pi)^{2-\e}}\left\{
\frac{y+m^2}{x+y+m^2}X
{}_{2}F_1\left(1,1+\e;3/2-\e;-\frac{y}{m^2}\right)\right.\nonumber\\
&&-\left. \frac{Y}{2-\e}
\left[{}_{2}F_1\left(2,1+\e;3-\e;-\frac{x+y}{m^2}\right)+(1-\e)
{}_{2}F_1\left(1,1+\e;3-\e;-\frac{x+y}{m^2}\right)
\right]\right\},\nonumber\\\label{eq:at4de0}\\
\lefteqn{2y\frac{\pd \ov{A}_m}{\pd y}=
\left[-1+2\e-4\e\frac{y}{x+y+m^2}\right]\ov{A}_m} \nonumber\\
&&-2\frac{[m^2]^{-1-\e}}{(4\pi)^{2-\e}}
\left\{\frac{y}{x+y+m^2}X
{}_{2}F_1\left(1,1+\e;3/2-\e;-\frac{y}{m^2}\right)\right.\nonumber\\
&&-\left.\frac{Y}{2-\e} 
\left[{}_{2}F_1\left(2,1+\e;3-\e;-\frac{x+y}{m^2}\right)
+(1-\e) {}_{2}F_1\left(1,1+\e;3-\e;-\frac{x+y}{m^2}\right)
\right]\right\},\nonumber\\
\label{eq:as4de0}
\eea
where
\be
X=\frac{\G(1/2-\e)\G(1+\e)}{\G(3/2-\e)},
\;\;\;\;Y=\frac{\G(1-\e)\G(1+\e)}{\G(2-\e)}.
\ee
\subsubsection{Solving the differential equations}
  
Let us first consider the integral $A_m$.  By the same method as in
Ref.~\cite{Watson:2007mz}, we make the following ansatz:
\be
A_m(x,y)=F_{Am}(x,y)G_{Am}(x,y)
\ee
such that
\bea
2x\frac{\pd F_{Am}}{\pd x}&=&
\left[1-(2+4\e)\frac{x}{x+y+m^2}\right]F_{Am},
\label{eq:agpdxh}\\
2y\frac{\pd F_{Am}}{\pd y}&=&
\left[-1+2\e-(2+4\e)\frac{y}{x+y+m^2}\right]F_{Am},
\label{eq:agpdyh}\\
F_{Am}2x\frac{\pd G_{Am}}{\pd x}
&=&2\frac{[m^2]^{-1-\e}}{(4\pi)^{2-\e}}
\left\{
\frac{y+m^2}{x+y+m^2}X
{}_{2}F_1\left(1,1+\e;3/2-\e;-\frac{y}{m^2}\right)\right.\nonumber\\
&&-
\frac{\G(-\e)\G(1+\e)}{\G(2-\e)}
{}_{2}F_1\left(2,1+\e;2-\e;-\frac{x+y}{m^2}\right).\nonumber\\
&&+\left. Y{}_{2}F_1\left(1,1+\e;2-\e;-\frac{x+y}{m^2}\right)
\right\},\label{eq:agpdx}\\
F_{Am}2y\frac{\pd G_{Am}}{\pd y}
&=&2\frac{[m^2]^{-1-\e}}{(4\pi)^{2-\e}}
\left\{-\frac{y}{x+y+m^2}X
{}_{2}F_1\left(1,1+\e;3/2-\e;-\frac{y}{m^2}\right)\right.\nonumber\\
&&+
\frac{\G(-\e)\G(1+\e)}{\G(2-\e)}
{}_{2}F_1\left(2,1+\e;2-\e;-\frac{x+y}{m^2}\right)
\textrm{~~~~~~~~~~~~~~~~~~~~~~~~~~~~~~} \nonumber\\
&&+\left. Y {}_{2}F_1\left(1,1+\e;2-\e;-\frac{x+y}{m^2}\right)
\right\}.
\label{eq:agpdy}
\eea
By inspection, it is simple to determine the solution for the two
homogeneous equations, \eq{eq:agpdxh} and \eq{eq:agpdyh}:
\be
F_{Am}(x,y)=x^{1/2}y^{-1/2+\e}(x+y+m^2)^{-1-2\e}.
\label{eq:af}
\ee
Since the mass $m$ is treated as a parameter, the (dimensionful)
solution, \eq{eq:af}, may have an integration constant proportional to
$[m^2]^{-1-\e}$. However, returning to the original equations,
\eq{eq:agpdxh} and \eq{eq:agpdyh}, we see that the only consistent
solution is the one for which this constant vanishes.

Let us now make the following ansatz for the function $G_{Am}$, which
will be verified below ($z=x/y$):
\be
G_{Am}(x,y)=G^0_{Am}(x,y)+\tilde{G}_{Am}(z).
\label{eq:ag}
\ee
The component $G^0_{Am}(x,y)$ can be found by adding the differential
equations \eq{eq:agpdx} and \eq{eq:agpdy}, which lead to:
\bea
x\frac{\pd G^0_{Am}}{\pd x}+y\frac{\pd G^0_{Am}}{\pd y}
&=&\frac {1}{(4\pi)^{2-\e}}\left\{\frac{m^2}{\sqrt{x(y+m^2)}}
\ln\left(\frac{\sqrt{1+\frac{m^2}{y}}+1}{\sqrt{1+\frac{m^2}{y}}-1}
\right)
+{\cal O}(\e)\right\}.
\eea
Because the function $G^0_{Am}$ is multiplied by the function $F_{Am}$
(which does not have an $\e$ pole), the term of order ${\cal O}(\e)$
will not contribute.  The solution of this equation is:
\be
G^0_{Am}(x,y)=-\frac{2}{(4\pi)^{2-\e}}\left\{\frac{1}{\sqrt{z}}
\left[\sqrt{1+\frac{m^2}{y}}\ln\left(\frac{\sqrt{1+\frac{m^2}{y}}+1}
{\sqrt{1+\frac{m^2}{y}}-1}\right)-\ln y\right]
+{\cal O}(\e)\right\}+{\cal C}_1.
\label{eq:agcomp0}
\ee

Before we proceed to determine $\tilde G_{Am}$, we justify the ansatz
for $G_{Am}$, given by \eq{eq:ag}. First we observe that:
\be
2x F_{Am}\frac{\pd G_{Am}}{\pd x}=2zF_{Am}\frac{\pd \tilde G_{Am}}
{\pd z}+ 2xF_{Am}\frac{\pd G^0_{Am}}{\pd x}.
\label{eq:agcomp0til}
\ee
Then we subtract the above equation, \eq{eq:agcomp0til}, from
\eq{eq:agpdx}.  This gives:
\bea
 \lefteqn{z\frac{\pd \tilde G_{Am}}{\pd z}=
-x \frac{\pd G^0_{Am}}{\pd x}
+\frac{1}{F_{Am}}\frac{[m^2]^{-1-\e}}{(4\pi)^{2-\e}}
\bigg\{
\frac{y+m^2}{x+y+m^2}X{}_{2}
F_1\left(1,1+\e;3/2-\e;-\frac{y}{m^2}\right)}\nonumber\\
&&+\left. Y{}_{2}F_1\left(1,1+\e;2-\e;-\frac{x+y}{m^2}\right)
-\frac{\G(-\e)\G(1+\e)}{\G(2-\e)}
{}_{2}F_1\left(2,1+\e;2-\e;-\frac{x+y}{m^2}\right)
\right\}. \nonumber\\
\label{eq:tildegz}
\eea
Evaluation to the first order in $\e$ is straightforward and we see
that the right hand side of the above expression is only a function of
the variable $z$.  This allows us to write down a differential
equation for $\tilde G_{Am} (z)$ in the form:
\be
z\frac{\pd \tilde G_{Am}}{\pd z}
=\frac{1}{(4\pi)^{2-\e}}\frac{1}{\sqrt{z}}
\left\{\frac 1\e-\gamma+\ln m^2+\cal O(\e)\right\},
\ee
from which we get immediately
\be
\tilde G_{Am}(z)=-\frac{2}{(4\pi)^{2-\e}}
\frac{1}{\sqrt{z}}\left\{\frac 1\e -\gamma
+\ln m^2+\cal O(\e)\right\}
+{\cal C}_2.
\label{eq:agtilde}
\ee
Returning to the original differential equations (\ref{eq:agpdxh} -
\ref{eq:agpdy}) with the function $G(x,y)= G^0_{Am}(x,y)+\tilde
G_{Am}(z)$, we see that the only consistent solution is the one for
which the overall constant ${\cal C}_1+{\cal C}_2$ vanishes.

We may now put together the solutions Eqs.~(\ref{eq:af}),
(\ref{eq:agcomp0}) and (\ref{eq:agtilde}) and write for the function
$A_{m}$:
\bea
A_m(x,y)
&=&\frac{(x+y+m^2)^{-1-\e}}{(4\pi)^{2-\e}}
\bigg\{-\frac {2}{\e} +2\gamma
+2\ln \left(\frac{x+y+m^2}{m^2}\right) \nonumber\\
&&-2\sqrt{1+\frac{m^2}{y}}
\ln\left(\frac{\sqrt{1
+\frac{m^2}{y}}+1}{\sqrt{1+\frac{m^2}{y}}-1}\right)
+\cal O(\e)\bigg\}.\hspace{4cm}
\label{eq:asol}
\eea
We see that for $m^2=0$ we regain the result from
Ref.~\cite{Watson:2007mz} and that the singularities are located at
$x+y+m^2=0$ (with $m^2, y \ge 0$).  Two more useful checks arise from
the study of the power expansion around $x=0$ and the mass
differential equation (this will be explained in detail in
Appendix~\ref{chap:app3}).

We now proceed in the same fashion to determine the function
$\ov{A}_m(x,y)=F_{\ov{A}m}(x,y)G_{\ov{A}m}(x,y)$.  The resulting
partial differential equations are in this case:
\bea
2x\frac{\pd F_{\ov{A}m}}{\pd x}&=&
\left[-1-4\e\frac{x}{x+y+m^2}\right]F_{\ov{A}m},\\
2y\frac{\pd F_{\ov{A}m}}{\pd y}&=&
\left[-1+2\e-4\e\frac{y}{x+y+m^2}\right]F_{\ov{A}m},\\
F_{\ov{A}m}2x\frac{\pd G_{\ov{A}m}}{\pd x}&=&
2\frac{[m^2]^{-1-\e}}{(4\pi)^{2-\e}}
\left\{\frac{y+m^2}{x+y+m^2}X
{}_{2}F_1\left(1,1+\e;3/2-\e;-\frac{y}{m^2}\right)\right.
\nonumber\\
&&-\frac{Y}{2-\e}
\bigg[{}_{2}F_1\left(2,1+\e;3-\e;-\frac{x+y}{m^2}\right)
 \nonumber\\
&&+\left. (1-\e){}_{2}F_1\left(1,1+\e;3-\e;-\frac{x+y}{m^2}\right)
\bigg]\right\},
\label{eq:a4gpdx}\\
F_{\ov{A}m}2y\frac{\pd G_{\ov{A}m}}{\pd y}&=&
2\frac{[m^2]^{-1-\e}}{(4\pi)^{2-\e}}\left\{
-\frac{y}{x+y+m^2}X
{}_{2}F_1\left(1,1+\e;3/2-\e;-\frac{y}{m^2}\right)\right.\nonumber\\
&&+\frac{Y}{2-\e} 
\bigg[{}_{2}F_1\left(2,1+\e;3-\e;-\frac{x+y}{m^2}\right) \nonumber\\
&&+\left. (1-\e){}_{2}F_1\left(1,1+\e;3-\e;-\frac{x+y}{m^2}\right)
\bigg]\right\}\hspace{4cm},
\label{eq:a4gpdy}
\eea
with $X,Y$ defined previously.  The solution to the first pair is
\be
F_{\ov{A}m}(x,y)=x^{-1/2}y^{-1/2+\e}(x+y+m^2)^{-2\e}.
\label{eq:a4fsol}
\ee
For brevity, in the above expression and also in the derivation of the
function $G_{\ov{A}m}=G^0_{\ov{A}m}+\tilde{G}_{\ov{A}m}$ (the analogue
of $G_{Am}$) we omit the constants of integration -- they vanish as in
the case of the functions $F_{Am}$ and $G_{Am}$.

As before, for $G_{\ov{A}m}(x,y)$ we make the ansatz:
\be
G_{\ov{A}m}(x,y)=G^0_{\ov{A}m}(x,y)+\tilde{G}_{\ov{A}m}(z).
\label{eq:a4g}
\ee
In the limit $\e\rightarrow 0$, the component $G^0_{\ov{A}m}(x,y)$ is
determined from the differential equation:
\be
x\frac{\pd G^0_{\ov{A}m}}{\pd x}+y\frac{\pd G^0_{\ov{A}m}}{\pd y}=
\frac{1}{(4\pi)^{2-\e}}
\left\{\frac{\sqrt{x}}{x+y+m^2}\frac{m^2}{\sqrt{y+m^2}} 
\ln\left(\frac{\sqrt{1+\frac{m^2}{y}}+1}
{\sqrt{1+\frac{m^2}{y}}-1}\right)
+{\cal O}(\e)\right\}. \nonumber\\
\ee
The solution of this equation is:
\bea
\lefteqn{G^0_{\ov{A}m}(x,y)=\frac{1}{(4\pi)^{2-\e}}
\left\{\imath\ln\left(\frac{\sqrt{1+\frac{m^2}{y}}-
\imath\sqrt{z}}{\sqrt{1+\frac{m^2}{y}}+\imath\sqrt{z}}\right)
\ln\left(\frac{\imath\sqrt{z}+1}{\imath\sqrt{z}-1}\right)\right.}
\nonumber\\
&&\left.-\imath \textrm{Li}_2
\left(\frac{1-\imath\sqrt{z}}{1+\imath\sqrt{z}}
\cdot
\frac{\sqrt{1+\frac{m^2}{y}}-\imath\sqrt{z}}{\sqrt{1+\frac{m^2}{y}}
+\imath\sqrt{z}}\right)
+\imath \textrm{Li}_2 
\left(\frac{1+\imath\sqrt{z}}{1-\imath\sqrt{z}}\cdot
\frac{\sqrt{1+\frac{m^2}{y}}-\imath\sqrt{z}}{\sqrt{1+\frac{m^2}{y}}
+\imath\sqrt{z}}
\right)+{\cal O}(\e)\right\},\nonumber\\
\label{eq:a4gcomp0}
\eea
where $\textrm{Li}_2(z)$ is the dilogarithmic function \cite{lewin}:
\be
\textrm{Li}_2(z)=-\int_0^z\frac{\ln(1-t)}{t}dt.
\ee

As before, we check that the ansatz for $G_{\ov{A}m}(x,y)$ given in
\eq{eq:a4g} is correct and derive the differential equation for the
function $\tilde G_{\ov{A}m}$, in the limit $\e\rightarrow 0$:
\be
z \frac{\pd\tilde G_{\ov{A}m}}{\pd z}
=\frac{1}{(4\pi)^{2-\e}}\left\{\frac{\sqrt{z}}{z+1}\left[ \ln
    (1+z)-\ln z-2\ln 2\right]+{\cal O}(\e)\right\}.
\ee
The result we leave for the moment in the form:
\bea
\tilde G_{\ov{A}m}(z)&=&\frac{1}{(4\pi)^{2-\e}}
\left\{\right.-4\ln 2\arctan(\sqrt{z})
+\int_0^z\frac{dt}{\sqrt{t}(1+t)}\ln{(1+t)}\nonumber\\
&-&\int_0^z\frac{dt}{\sqrt{t}(1+t)}\ln{t}+{\cal O}(\e)\left.\right\}.
\label{eq:a4tildegzsol}
\eea
With the solutions, Eqs.~(\ref{eq:a4fsol}), (\ref{eq:a4gcomp0}) and
(\ref{eq:a4tildegzsol}), after some further manipulation we can write
down the following simplified expression for the integral $A_m^4$:
\bea
\lefteqn{A^4_m(x,y)=k_4\frac{(x+y+m^2)^{-1-\e}}{(4\pi)^{2-\e}}
\frac{(1+z+\frac{m^2}{y})}{\sqrt{z}}
\bigg\{
-\int_0^z\frac{dt}{\sqrt{t}(1+t)}\ln{\left(1+t+\frac{m^2}{y}\right)}
}\nonumber\\
&&
+2\ln\left(\frac{\sqrt{1
+\frac{m^2}{y}}+1}{\sqrt{1+\frac{m^2}{y}}-1}\right)
\arctan{\left(\frac{\sqrt{z}}{\sqrt{\frac{m^2}{y}+1}}\right)}
\!+2\ln\left(\frac{m^2}{y}\!\right)\arctan{(\sqrt{z})}
+{\cal O}(\e)\bigg\},\hspace{1cm}
\label{eq:a4sol}
\eea
 with the integral 
\bea \lefteqn{\int_0^z\frac{dt}{\sqrt{t}(1+t)}
\ln{\left(1+t+\frac{m^2}{y}\right)}=\pi\ln{2} -\imath \ln
 \left(\frac{1-\imath\sqrt{z}}{1+\imath\sqrt{z}}\right)2\ln 2}
\nonumber\\
&&+\imath \ln\left(\frac{1-\imath\sqrt{z}}{1+\imath\sqrt{z}}
 \frac{\sqrt{1+\frac{m^2}{y}}-\imath\sqrt{z}}{\sqrt{1+\frac{m^2}{y}}
  +\imath\sqrt{z}}\right)
\ln\left(\sqrt{1+\frac{m^2}{y}}+1\right) \nonumber\\
&&+\imath \ln\left(\frac{1-\imath\sqrt{z}}{1+\imath\sqrt{z}}
\frac{\sqrt{1+\frac{m^2}{y}}+\imath\sqrt{z}}{\sqrt{1+\frac{m^2}{y}}
 -\imath\sqrt{z}}\right)
\ln\left(\sqrt{1+\frac{m^2}{y}}-1\right)\nonumber\\
&&-\imath\ln{(\sqrt{z}-\imath)}\left[\ln{2}+\ln{(1+z+\frac{m^2}{y})}
 -\ln{(1-\imath\sqrt{z})}-\frac12\ln{(\sqrt{z}-\imath)}\right]
\nonumber\\
&&-\imath \textrm{Li}_2{\left(\ha-\frac{\imath}{2}\sqrt{z}\right)}
+\imath \textrm{Li}_2{\left(\ha+\frac{\imath}{2}\sqrt{z}\right)}
-\imath\textrm{Li}_2
\left(\frac{\imath+\sqrt{z}}{-\imath+\sqrt{z}}\right)
+\imath\textrm{Li}_2
\left(\frac{-\imath+\sqrt{z}}{\imath+\sqrt{z}}\right)
\nonumber\\
&&+\imath\ln{(\sqrt{z}+\imath)}\left[\ln{2}+\ln{(1+z+\frac{m^2}{y})}
 -\ln{(1+\imath\sqrt{z})}-\ha\ln{(\sqrt{z}+\imath)}\right]
\nonumber\\
&&+\imath\textrm{Li}_2 \left(\frac{\imath
\sqrt{1+\frac{m^2}{y}}+\sqrt{z}}{-\imath+\sqrt{z}}\right)
-\imath\textrm{Li}_2 \left(\frac{-\imath
\sqrt{1+\frac{m^2}{y}}+\sqrt{z}}{\imath+\sqrt{z}}\right)
\nonumber\\
&&+\imath\textrm{Li}_2 \left(\frac{-\imath
 \sqrt{1+\frac{m^2}{y}}+\sqrt{z}}{-\imath+\sqrt{z}}\right)
+\imath\textrm{Li}_2 \left(\frac{\imath
 \sqrt{1+\frac{m^2}{y}}+\sqrt{z}}{\imath+\sqrt{z}}\right).  \eea

We see that for $m^2=0$ we get the correct limit for the function
$A_m^4$.  We also mention that the singularities are located at
$x+y+m^2=0$ and the apparent singularities at $z=-1$ (i.e., $x+y=0)$
in the expression \eq{eq:a4sol} are canceling out.  This can be easily
seen by making a series expansion of \eq {eq:a4sol} around $z=-1$:
\bea
A^4_m&\stackrel{z\rightarrow-1}{=}&\frac{1}{\sqrt{z}}
\left[\frac{y}{m^2}-\frac{z+1}{2}\left(\frac{y}{m^2}\right)^2
+{\cal O}\left((z+1)^3\right)
\right]\nonumber\\
&&-\frac{\sqrt{1+\frac{m^2}{y}}}{\sqrt{z}(z+1+\frac{m^2}{y})}
\ln\left(\frac{
\sqrt{1+\frac{m^2}{y}}+1}{\sqrt{1+\frac{m^2}{y}}-1}\right)
+\cal O(\e).\hspace{4cm}
\label{eq:a4expz-1}
\eea
Again, the result \eq{eq:a4sol} has been checked by performing an
expansion around $x=0$ and by studying the mass differential equation
(see Appendix~\ref{chap:app3}).

\subsection{Results in the limit  \texorpdfstring{$\e\rightarrow 0$} {epsilon~0}}
\label{subsec:1loopres}

Having derived the noncovariant massive loop integrals in the previous
section, and collecting the results for the standard massive integrals
from Appendix~\ref{chap:app2}, we are now in the position to write
down the perturbative results for the dressing functions under
consideration.  For the temporal, spatial and massive components of
the quark gap equation, Eqs.~(\ref{eq:atmassive}), (\ref{eq:bmassive})
and (\ref{eq:asmassive1}), we find in the limit $\e\rightarrow 0$:
\bea
\lefteqn{
A_t(k)=1
+\frac{C_F g^2}{(4\pi)^{2-\e}}\bigg\{
\frac{1}{\e}-\ga-\ln\frac{m^2}{\mu}}\nonumber\\
&&+1-\frac{m^2}{k^2}
+\left(\frac{m^4}{k^4}-1\right)\ln\left( 1+\frac{k^2}{m^2}\right)
+\cal{O}(\e)
\bigg\},
\label{eq:atmassive_eps}\\
\lefteqn{
A_s(k)=1+\frac{C_Fg^2}{(4\pi)^{2-\e}}
\bigg\{
\frac1\e-\ga-\ln\frac{m^2}{\mu}
}
\nonumber\\
&&+1+8\frac{k^2}{\vec k^2}
+4\frac{m^2}{\vec k^2}-\frac{m^2}{k^2}
+\left(1+\frac{m^2}{k^2}\right)
\left(4\frac{k^2}{\vec k^2}-1+\frac{m^2}{k^2}\right)
\ln \left( 1+\frac{k^2}{m^2}\right)\nonumber\\
&&-\!\left(4\frac{k^2}{\vec k^2}+2\frac{m^2}{\vec k^2}\right)
\sqrt{1+\frac{m^2}{\vec k^2}}
\ln\left(\frac{\sqrt{1+\frac{m^2}{\vec k^2}}+1}
{\sqrt{1+\frac{m^2}{\vec k^2}}-1}\right)
\!-2\frac{k_4^2}{\vec k^4}(k^2+m^2)f_m\left(k_4^2,\vec{k}^2\right)
+\cal{O}(\e)\!\bigg\},\nonumber\\\\
\label{eq:asmassive_eps}
\lefteqn{B_m(k)=m
+m\frac{C_Fg^2}{(4\pi)^{2-\e}}\bigg\{
\frac4\e-4\ga-4\ln\frac{m^2}{\mu}
}\nonumber\\
&&+10-2\sqrt{1+\frac{m^2}{\vec k^2}}
\ln\left(\frac{\sqrt{1+\frac{m^2}{\vec k^2}}+1}
{\sqrt{1+\frac{ m^2}{\vec k^2}}-1}\right)
-2\left(1+\frac{m^2}{k^2}\right)
\ln\left(1+\frac{k^2}{m^2}\right)+\cal{O}(\e)
\bigg\},
\label{eq:bmassive_eps}
\eea
where the function $f_m(x,y)$ is given by ($x= k_4^2, y= \vec k^2,
z=x/y$):
\bea
f_m(x,y)&=&
\frac{2}{\sqrt z}\ln\left(\frac{m^2}{y}\right)\arctan{(\sqrt{z})}
+\frac{2}{\sqrt z}
\ln\left(\frac{\sqrt{1+\frac{m^2}{y}}+1}{\sqrt{1
+\frac{m^2}{y}}-1}\right)
\arctan{\left(\frac{\sqrt{z}}{\sqrt{\frac{m^2}{y}+1}}\right)}
\nonumber\\
&&-\int_0^1\frac{dt}{\sqrt{t}(1+zt)}\ln{\left(1+zt
+\frac{m^2}{y}\right)}.
\eea
The last integral has been rewritten using the identity:
\be
\frac{1}{\sqrt{z}}\int_0^z\frac{dt}{\sqrt{t}(1+t)}
\ln{\left(1+t+\frac{m^2}{y}\right)}=
\int_0^1\frac{dt}{\sqrt{t}(1+zt)}
\ln{\left(1+zt+\frac{m^2}{y}\right)}.
\ee
As a useful check, we can set $m=0$ and show that the results for the
temporal and spatial components are in agreement with the calculation
performed independently using the one-loop massless integrals derived
in Ref.~\cite{Watson:2007mz}.

Another important point is related to the singularity structure of the
above dressing functions. As has been shown in the previous section,
in the noncovariant integrals the singularities appear at
$x+y+m^2=0$. Further, it is easy to see that the standard integrals
have the same singularity structure.  As promised, since the
singularities in both the Euclidean and spacelike Minkowski regions
are absent, we find that the validity of the Wick rotation is
justified. In addition, the results for $A_t$,$A_s$ and $B_m$ can be
compared (allowing for the color factors) and agree with those of
Quantum Electrodynamics \cite{Adkins:1982zk, Malenfant:1987kw}.

Having calculated the dressing functions for the quark proper
two-point Green's function, we are now able to discuss the structure
of the quark propagator.  In \eq{eq:b} we first analyze the
denominator factor.  Let us denote (in Euclidean space):
\be
D(k)=k_4^2A_t^2(k)+\vec k^2A_s^2(k)+B_m^2(k).
\ee
Inserting the expressions from Eqs.~(\ref{eq:atmassive_eps}),
(\ref{eq:asmassive_eps}) and (\ref{eq:bmassive_eps}) into the above
equation, we have:
\bea
D(k)&=&k^2+m^2\left\{1+6\frac{g^2C_F}{(4\pi)^{2-\e}}
\left[\frac1\e-\ga-\ln\frac{m^2}{\mu}+\frac43\right]\right\}
\nonumber\\
&&+(k^2+m^2)\frac{2C_F g^2}{(4\pi)^{2-\e}}
\bigg\{\frac1\e-\ga-\ln\frac{m^2}{\mu}+9
\nonumber\\
&&+
\left(3-\frac{m^2}{k^2}\right)\ln\left(1+\frac{k^2}{m^2}\right)
-4\sqrt{1+\frac{m^2}{\vec k^2}}
\ln\left(\frac{\sqrt{1+\frac{m^2}{\vec k^2}}+1}{\sqrt{1+\frac{m^2}
{\vec k^2}}-1}\right)-2\frac{k_4^2}{\vec k^2}f_m(k_4^2,\vec k^2)
\!\!\bigg\} \!.
\nonumber\\
\label{eq:denfact}
\eea
Defining the renormalized mass, $m_R$, via:
\be
m^2=Z_m^2 m_R^2
\textrm{~~~with~~~} 
Z_m^2=1-6\frac{g^2C_F}{(4\pi)^{2-\e}}
\left\{\frac1\e-\ga-\ln\frac{m^2}{\mu}+\frac43\right\},
\ee
we see that the expression for $D(k)$, \eq{eq:denfact}, then contains
explicitly the overall factor $k^2+m_R^2$. This means that the simple
pole mass of the quark emerges, just as it does in covariant
gauges. For the remaining part, the singularity structure is such that
non-analytic structures do not appear for spacelike or Euclidean
momenta.  Moreover, we see that the renormalization factor, $Z_m$,
which defines the physical perturbative pole mass and hence should be
a gauge invariant quantity, agrees with the result obtained in
covariant gauges \cite{Muta:1998vi}.

Because of the Dirac structure, it is more convenient to write the
quark propagator in Minkowski space, i.e. to analytically continue
$k_4^2\rightarrow-k_0^2$. We have shown that the analytic continuation
of the functions $A_t,A_s, B_m$ (and implicitly, of the function
$D(k)$) back into the Minkowski space is allowed and this enables us
simply to write:
\be
W_{\bar qq \al\ba}(k)
=\imath \de_{\al\ba}\left\{\ga^0k_0A_t(k)
-\ga^ik_iA_s(k)+B_m(k)\right\} D^{-1}(k). 
\ee
Inserting the denominator factor, \eq{eq:denfact}, in the limit
$\e\rightarrow 0$ and replacing the mass with its renormalized
counterpart, the above expression gives:
\be
W_{\bar qq \al\ba}(k)=
-\de_{\al\ba}\frac{\imath}{k_0^2-\vec k^2-m_R^2}\left\{(\kslash+m_R)
\left[1-C_F\frac{g^2}{(4\pi)^{2-\e}}\left(\frac1\e-\ga\right)\right]
+\textrm{~finite~ terms} \right\}.
\ee
We can thus write down for the quark propagator:
\be
W_{\bar qq \al\ba}(k)=(-\imath)\de_{\al\ba}
\frac{\kslash+m_R} {k_0^2-\vec k^2-m_R^2}Z_2
+\textrm{~finite~ terms~}
\label{eq:quarkproponeloop}
\ee
and identify the renormalization constant (omitting the prescription
dependent constants)
\be
Z_2=1-\frac{g^2C_F}{(4\pi)^{2-\e}}\left(\frac1\e-\ga\right).
\ee

Turning to the quark loop contributions to the gluon two-point proper
functions, in evaluating the integral structure of
Eqs.~(\ref{eq:Gsisi}), (\ref{eq:GsiA}) and (\ref{eq:GAA}) we observe
the following relations (in Euclidean space):
\bea
\G_{\si\si(q)}^{(1)}(k)=\G_{\si A(q)}^{(1)}(k)=
-\frac{\vec k^2}{k^2}\G_{AA(q)}^{(1)}(k)
=-\frac{\vec k^2}{k_4^2}\bar\G_{AA,q}^{(1)}(k)
=I(k_4^2,\vec k^2),
\label{eq:AsiST}
\eea
where the integral $I(k_4^2,\vec k^2)$ reads (using the results of
Appendix~\ref{chap:app2}), as $\e\rightarrow 0$:
\bea
I(k_4^2,\vec k^2)&=&\frac{N_f}{(4\pi)^{2-\e}}
\left\{
-\frac23\left[\frac{1}{\e}-\ga-
\ln \frac{k^2}{\mu}\right]-\frac{10}{9}
+\frac23\left(4\frac{m^2}{k^2}+\ln \frac{m^2}{k^2}\right)
\right.\nonumber\\
&&+\left.
\frac23\sqrt{1+\frac{4m^2}{k^2}}\left(1-2\frac{m^2}{k^2}\right)
\ln\left(\frac{\sqrt{1+\frac{4m^2}{k^2}}+1}
{\sqrt{1+\frac{4m^2}{k^2}}-1}\right)
+\cal O(\e)\right\}.
\label{eq:AsiSTint}
\eea
The relations \eq{eq:AsiST} are similar to the \ST identities for the
\YM part of the theory, derived in \cite{Watson:2007vc}.  Also, the
above integral \eq{eq:AsiSTint} agrees with the results obtained in
covariant gauges (see for instance \cite{Muta:1998vi}).  This was in
fact expected, since at one-loop level the quark loop as a whole is
identical with its covariant counterpart --- the only difference is
that the various degrees of freedom (temporal and spatial) are being
separated into the corresponding proper two-point functions, i.e.,
$\G_{AA}$, $\G_{A\si}$ and $\G_{\si\si}$.

Let us now consider the one-loop gluon propagator dressing functions,
in connection to the first coefficient of the perturbative
$\ba$-function.  Analogously to the two-point proper functions, these
are constructed by writing $D=D^{(0)}+g^2D^{(1)}$.  As mentioned
previously, in the first order formalism we have to account for the
presence of the additional $\vec\pi, \phi$ and ghost fields, and
implicitly the corresponding propagators (for example $D_{A\pi}$). We
have already seen that at one-loop the quarks only contribute to three
of the gluon proper two-point functions ($\G_{AA},\G_{A\si}$ and
$\G_{\si\si}$).  But since these gluon two-point functions are related
to the various connected (propagator) two-point functions arising from
the first order formalism, there will be (quark loop) contributions to
many more of these propagators.  The relationship between the
connected and proper gluon two-point functions in the first order
formalism is derived in Ref.~\cite{Watson:2007mz} and will not be
reproduced here.  The full set of quark contributions to these gluonic
type propagators is given by:
\bea
&&D_{AA(q)}^{(1)}(k)=D_{\si\si(q)}^{(1)}(k)=\G_{\si\si(q)}^{(1)}(k)=
I(k_4^2,\vec k^2), \label{eq:letransf}\\
&&D_{A\pi(q)}^{(1)}(k)=-\frac{\vec k^2}{k_4^2}D_{\pi\pi(q)}^{(1)}(k)
=D_{\si\phi(q)}^{(1)}(k)=-D_{\phi\phi(q)}^{(1)}(k)= I(k_4^2,\vec k^2).
\label{eq:letransf1}
\eea
We are now able to identify the first coefficient of the
$\ba$-function. As is well known in Landau gauge, a renormalization
group invariant running coupling can be defined through the following
perturbative combination of gluon and ghost propagator dressing
functions \cite{Fischer:2006ub}:
\be
g^2D_{AA}D_c^{2}\sim g^2\left[1+\frac{g^2}{16\pi^2}\frac{1}{\e}
\left(\frac{11N_c}{3}-\frac{2N_f}{3}\right)\right].
\ee
At one-loop in perturbation theory, the coefficient of the $1/\e$ pole
above is simply minus the first coefficient of the $\ba$-function
($\beta_0=-11N_c/3+2N_f/3$).  By inspecting the relations
\eq{eq:letransf}, containing the quark contribution to the propagator
$D_{AA}$ , and those obtained in Ref.~\cite{Watson:2007mz} for the \YM
part of the propagator $D_{AA}$ and the propagator $D_{c}$, we see
that the same result is achieved in Coulomb gauge.  Moreover, in
Coulomb gauge, a second renormalization group invariant combination of
propagators appears and is given by $g^2 D_{\si\si}$
\cite{Zwanziger:1998ez}.  Again, combining our results
\eq{eq:letransf} and those obtained in Ref.~\cite{Watson:2007mz} we
see that indeed the coefficient of $1/\e$ agrees with this.

\section{Quark-gluon vertices} 

We now evaluate the divergent components of the quark-gluon vertex
functions at one loop perturbative level. Just as for the two-point
functions discussed in the preceding section, we write the the
perturbative expansion of the vertex function as
\be
\G=\G^{(0)}+g^3\G^{(1)}
\ee 
such that the new coupling and $\G^{(1)}$ are dimensionless.
 
 We first consider the temporal part of the quark-gluon vertex,
\eq{eq:qsivertex_1loop}. By the same method as for the two-point
functions, we insert the appropriate tree-level vertices and
propagators from \eq{eq:treelevelquarkprop},
\eq{eq:treelevelgluonprop}, \eq{eq:treeleveltriplegluon},
\eq{eq:treelevelquarkvertex1} and \eq{eq:treelevelquarkvertex2} and
solve the color algebra.  For the one-loop expansion of the temporal
part of the quark-gluon vertex we find, in Minkowski space (the factor
$C_F-N_c/2$ arising from the color algebra is derived in the
Appendix~\ref{chap:app.not}):
\bea
&&\G_{\bar
  qq\si\al\ba}^{(1)d}(k_1,k_2,k_3)=
\left(C_F-\frac{N_c}{2}\right)T^d_{\al\ba}\int\dk{\w}\nonumber\\
&&
\left\{
-\ga^k\frac{\left[\ga^0(k_1-\w)_0-\ga^i(k_1-\w)_i
+m\right]}{(k_1-\w)^2-m^2}
\ga^0\frac{\left[-\ga^0(k_2+\w)_0+\ga^i(k_2+\w)_i
+m\right]}{(k_2+\w)^2-m^2}\ga^j
\frac{t_{jk}(\vec\w)}{\w^2}\right.\nonumber\\
&&\left.
-\ga^0\frac{\left[\ga^0(k_1-\w)_0-\ga^i(k_1-\w)_i
+m\right]}{(k_1-\w)^2-m^2}\ga^0
\frac{\left[-\ga^0(k_2+\w)_0+\ga^i(k_2+\w)_i
+m\right]}{(k_2+\w)^2-m^2}\ga^0
\frac{1}{\vec\w^2}
\right\}\nonumber\\
&&- \frac{1}{2}N_cT^d\int\dk{\w}
\left\{\ga^k\frac{\left[\ga^0(k_1-\w)_0-\ga^i(k_1-\w)_i
+m\right]}{(k_1-\w)^2-m^2}\ga^0
\frac{(2k_3+\w)_j}{(\vec\w+\vec k_3)^3}
\frac{t_{jk}(\vec\w)}{\w^2}\right.\nonumber\\
&&+\ga^0\frac{\left[\ga^0(\w-k_2)_0-\ga^i(\w-k_2)_i
+m\right]}{(k_2-\w)^2-m^2}\ga^k
\frac{(-2k_3-\w)_j}{(\vec\w+\vec k_3)^3}
\frac{t_{jk}(\vec\w)}{\w^2}\nonumber\\
&&\left.+\ga^l\frac{\left[\ga^0(k_1-\w)_0-\ga^i(k_1-\w)_i
+m\right]}{(k_1-\w)^2-m^2}\ga^k
(k_3+2\w)_0\frac{t_{jk}(\vec\w+\vec k_3)}{(\w+k_3)^2}
\frac{t_{jl}(\vec\w)}{\w^2}\right\}
\label{eq:quarksi1loop}
\eea

Since we are only interested in the divergent part of the above
expression, we first make use of the identity \eq{eq:scalprod} to
simplify the non-abelian terms containing two noncovariant denominator
factors, and then we employ a simple power analysis to separate the
convergent and divergent integrals (at leading order in perturbation
theory).  After eliminating the convergent terms and using the
expression \eq{eq:tij} for the transversal projector $t_{ij} (\vec
k)$, we are left with the following divergent part of the temporal
quark-gluon vertex:
\bea
\lefteqn{\G_{\bar qq\si,div}^{(1)d}(k_1,k_2,k_3)=
\left(-\frac{\imath}{2}T^dN_c+\imath T^dC_F\right)}\nonumber\\
&&\int\dk{\w}
\left\{\frac{\ga^k(\ga^0\w_0^2+\ga^i\ga^0\ga^l\w_i\w_l)\ga^j}
{\w^2[(k_1-\w)^2-m^2][(k_2+\w)^2-m^2]}
\left(\de_{jk}-\frac{\w_j\w_k}{\vec\w^2}\right)\right.\nonumber\\
&&+\left.\frac{\ga^0(\ga^0\w_0^2+\ga^i\ga^0\ga^l\w_i\w_l)\ga^0}
{\vec\w^2[(k_1-\w)^2-m^2][(k_2+\w)^2-m^2]}
\right\}\nonumber\\
&&-\imath T^dN_c\int\dk{\w}
\frac{\ga^l\ga^0\ga^k \w_0^2}{[(k_1-\w)^2-m^2](\w+k_3)^2\w^2}
\left[
  \de_{kl}-\frac{\w_k\w_l}{2\vec\w^2}-\frac{(\w+k_3)_k\w_l}
{2(\vec\w+\vec k_3)^2}\right]\hspace{1cm}
\label{eq:quarksi1loopdiv}
\eea

Collecting the results for the divergent factors of the three-point
loop integrals listed in the Appendix~\ref{app:app4}, it is
straightforward to calculate the divergent part of the temporal quark
gluon vertex, in the limit $\e\to 0$. It is given by (in Minkowski
space):
\be
\G_{\bar qq\si,div}^{(1)d}=T^d
\ga^0\frac{C_F}{(4\pi)^2}\frac{1}{\e}+\textrm{finite terms}.
\label{eq:temporaldiv}
\ee

Repeating this calculation for the spatial component of the
quark-gluon vertex, \eq{eq:qAvertex_1loop}, we obtain:
\be
\G_{\bar qqAi,div}^{(1)d}=-T^d
\ga^i\left(C_F+\frac{4}{3}N_c\right)\frac{1}{(4\pi)^2}\frac{1}{\e}
+\textrm{finite terms}.
\label{eq:spatialdiv}
\ee

The results \eq{eq:temporaldiv}, \eq{eq:spatialdiv} will then provide
a useful check for the \ST identity for the quark-gluon vertices. We
will return to this subject at the end of the next chapter, when we
analyze the divergent structure of the \ST identity.

%% file: st.tex
\chapter{\ST identity for the quark-gluon vertices}
\label{chap:st} 

The \ST identities relate Green's functions with a different number of
external legs, analogous to the Ward--Takahashi identities in QED.
They play an important role in connection to nonperturbative \DS
studies, since they provide constraints for the higher $n$-point
functions that enter the \DS equations.

In order to derive these identities, we start with the observation
that the BRST invariance is an unbroken symmetry of the gauge fixed
theory, at the level of the Lagrangian.  The BRST symmetry (and its
variation in Coulomb gauge) will be presented in detail in the first
Section of this chapter.  Next, we will present the formal derivation
of the \ST identities for the quark-gluon vertex functions, based on
the BRST invariance of the QCD Green's functions, and in particular,
we will examine the so-called quark-ghost scattering kernels. We will
then demonstrate that these identities are satisfied at one-loop
perturbative level. The proof is based on the translational invariance
of the loop integrals and it does not require the explicit evaluation
of the loop integrals.

This chapter concerns the \ST for the quark-gluon vertices, but
similar identities can be derived for Green's functions of arbitrary
order. Some of the higher order identities will be explored in the
second part of this thesis, in the limit of heavy quark
mass. Moreover, their implications for the \DS equations will also be
discussed.

\section{Gauss--BRST symmetry}

Regardless of the specific gauge, the gauge fixing term in the QCD
action violates the invariance under the local gauge transformation
\eq{eq:Umatr}.  Except for the axial gauge, the invariance of the full
(gauge-fixed) action can be recovered by performing a local gauge
transformation similar to \eq{eq:Umatr}, but with the following ansatz
\cite{Becchi:1974md,Becchi:1975nq}
\be
\theta^a (x)=c^a (x) \la
\label{eq:BRStheta}
\ee
where both $c$ and $\la$ are Grassman-valued quantities and $\la$ is a
constant. By demanding that the gluon, quark and ghost fields
transform as
\bea
\de A^a&=&\frac{1}{g}\hat{D}^{ac}c^c\de\la,\nonumber\\
\de q_{\al}&=&- \imath \left[T^a\right]_{\al\ba}c^a\de\la q_{\ba} 
\nonumber\\
\de\ov{c}^a&=&\frac{1}{g}\la^a\de\la\nonumber\\
\de c^a&=&-\ha f^{abc}c^bc^c\de\la
\label{eq:BRS0}
\eea
($c$ and $\bar c$ are independent Grassmann fields, and thus we have
independent transformations), we find that the gauge-fixing term
remains also invariant. This new symmetry is called BRST
(Becchi-Rouet-Stora-Tyutin) and is the basis for deriving the
generalized Ward--Takahashi identities (also called \ST identities).
 
While the ``standard'' BRST symmetry is independent of the gauge
considered, in Coulomb gauge the time derivatives are absent and hence
one can further define a {\it time-dependent} --- so-called
Gauss--BRST symmetry \cite{Zwanziger:1998ez}.  As before, the
spacetime-dependent paramether $\theta^a$ is factorized into two
Grassmann components
\be
\theta^a (x)=c^a (x) \de\la_t,
\label{eq:BRS3}
\ee
but in this case $\de\la_t$ (not to be confused with the colored
Lagrange multiplier $\la^a$) is a {\it time dependent} infinitesimal
variation.  Within the standard, second order formalism, the fields
transform as:
\bea
&&\de\si_a=-\frac{1}{g}\partial_0\th^a-f^{abc}
\si^b\th^c,\;\;\;\;
\de\vec{A}^a=\frac{1}{g}\div\th^a-f^{abc}\vec{A}^b\th^c,\nonumber\\
&&\de q_\al=-\imath\th^a\left[T^a\right]_{\al\ba}q_\ba,\;\;\;\;
\de\ov{q}_\al=\imath\th^a\ov{q}_\ba\left[T^a\right]_{\ba\al},
\nonumber\\
&&\de\ov{c}^a=\frac{1}{g}\la^a\de\la_t,\;\;\;\;\de c^a=-\frac{1}{2}
f^{abc}c^bc^c\de\la_t,\;\;\;\;\de\la^a=0.
\eea

The Gauss-BRST transform, \eq{eq:BRS3}, is the starting point for the
derivation of the \ST identities in Coulomb gauge.  These identities,
together with the peculiar features introduced by the time dependent
transformation, will be discussed in the next section.

\section{Formal derivation}

As discussed in the previous section, the full (gauge fixed) QCD
action in the standard, second order formalism, \eq{eq:sqcd1}, is
invariant under a Gauss--BRST transform, \eq{eq:BRS3}.  The Coulomb
gauge \ST identities arise from regarding the Gauss--BRST transform as
a change of integration variables under which the generating
functional \eq{eq:Z} (together with the source terms \eq{eq:source})
is invariant.  Since the Jacobian factor is trivial
\cite{Watson:2006yq} and only the source term varies, we deduce that
\bea
\lefteqn{0=\left.\int{\cal D}\Phi
\frac{\de}{\de\left[\imath\de\la_t\right]
}\exp{\left\{\imath{\cal S}_{QCD}+\imath{\cal S}_{FP}+
\imath{\cal S}_s+\imath\de{\cal S}_s\right\}}\right|_{\de\la_t=0}}
\nonumber\\
&&\hspace{-0.6cm}
=\int{\cal D}\Phi\exp{\left\{\imath{\cal S}_{QCD}+
\imath{\cal S}
_{FP}+\imath{\cal S}_s\right\}}\int d^4x\de(t-x_0)\left\{
-\frac{1}{g}\left(\partial_x^0\ro_x^a\right)c_x^a
+f^{abc}\ro_x^a\si_x^bc_x^c
\right.\nonumber\\
&&\hspace{-0.6cm}-\left.\frac{1}{g}J_{ix}^a\nabla_{ix}c_x^a
+f^{abc}J_{ix}^aA_{ix}^bc_x^c-
\imath\ov{\chi}_{\al x}c_x^aT_{\al\ba}^aq_{\ba x}-\imath c_x^a
\ov{q}_{\ba x}T_{\ba\al}^a\chi_{\al x}+\frac{1}{g}\la_x^a\et_x^a+
\frac{1}{2}f^{abc}\ov{\et}_x^ac_x^bc_x^c
\right\}\!. \nonumber\\
\eea 
Notice the $\de(t-x_0)$ constraint, which arises because the
time-dependent variation $\de\la_t$ and is characteristic to the
Gauss-BRST transform. It leads eventually to a non-trivial energy
injection in the \ST identities which is not present in the covariant
gauge case.

The above identity is best expressed in terms of proper functions.
Using the definitions \eq{eq:JPhi} for the derivatives of $W$ with
respect to the sources and of $\G$ with respect to the classical
fields, we arrive at the identity
\bea
\lefteqn{0=\int d^4x\de(t-x_0)}\nonumber\\
&&
\hspace{-0.5cm}
\times\bigg\{
\frac{1}{g}\left(\partial_x^0\ev{\imath\si_x^a}\right)c_x^a
-f^{abc}\ev{\imath\si_x^a}\left[\ev{\imath\ro_x^b\imath\ov{\et}_
x^c}+\si_x^bc_x^c\right]
-\frac{1}{g}\left[\frac{\nabla_{ix}}{(-\nabla_x^2)}\ev{\imath A_
{ix}^a}\right]\ev{\imath\ov{c}_x^a}
\nonumber\\
&&\hspace{-0.5cm}
-f^{abc}\!\ev{\imath A_{ix}^a}t_{ij}(\vec{x})\left[\ev{\imath J_{jx}
^b\imath\ov{\et}_x^c}+A_{jx}^bc_x^c\right]
-\frac{1}{g}\la_x^a\ev{\ov{c}_x^a}
+\frac{1}{2}f^{abc}\ev{\imath c_x^a}\!\left[\ev{\imath\ov{\et}_x^b
\imath\ov{\et}_x^c}+c_x^bc_x^c\right]
\nonumber\\
&&\hspace{-0.5cm}
+\imath T_{\al\ba}^a\ev{\imath q_{\al x}}\left[\ev{\imath\ov{\chi}
_{\ba x}\imath\ov{\et}_x^a}-c_x^aq_{\ba x}\right]
+\imath T_{\ba\al}^a\left[\ev{\imath\chi_{\ba x}\imath\ov{\et}_x^a}
+c_x^a\ov{q}_{\ba x}\right]\ev{\imath\ov{q}_{\al x}}
\bigg\}.
\eea
Note that functional derivatives involving the Lagrange multiplier
result merely in a trivial identity such that the classical field
$\la_x^a$ can be set to zero \cite{Watson:2008fb}.  Further, to derive
the quark \ST identities, one functional derivative with respect to
$\imath c_z^d$ is needed and then the ghost fields/sources can be set
to zero.  Implementing this then gives
\bea
\lefteqn{0=\int d^4x\de(t-x_0)}
\nonumber\\
&&\hspace{-0.5cm}\times\left\{
-\frac{\imath}{g}\left(\partial_x^0\ev{\imath\si_x^d}\right)
\de(z-x)
-f^{abc}\ev{\imath\si_x^a}\left[\frac{\de}{\de\imath c_z^d}
\ev{\imath\ro_x^b\imath\ov{\et}_x^c}-\imath\si_x^b\de^{dc}
\de(z-x)\right]
\right. \nonumber\\
&&\hspace{-0.5cm}
+\frac{1}{g}\left[\frac{\nabla_{ix}}{(-\nabla_x^2)}\ev{\imath 
A_{ix}^a}\right]\ev{\imath\ov{c}_x^a\imath c_z^d}
-f^{abc}\ev{\imath A_{ix}^a}t_{ij}(\vec{x})\!\left[\frac{\de}{\de
\imath c_z^d}\ev{\imath J_{jx}^b\imath\ov{\et}_x^c}-\imath A_{jx}^b
\de^{dc}\de(z-x) \right]
\nonumber\\
&&\hspace{-0.5cm}
-\imath T_{\al\ba}^a\ev{\imath q_{\al x}}\left[\frac{\de}{\de
\imath c_z^d}\ev{\imath\ov{\chi}_{\ba x}\imath\ov{\et}_x^a}+
\de^{da}\de(z-x)\imath q_{\ba x}\right]
\nonumber\\
&&\hspace{-0.5cm}
\left.
+\imath T_{\ba\al}^a\left[\frac{\de}{\de\imath c_z^d}\ev{\imath
\chi_{\ba x}\imath\ov{\et}_x^a}-\de^{da}\de(z-x)\imath\ov{q}
_{\ba x}\right]\ev{\imath\ov{q}_{\al x}}\right\}.
\label{eq:stid0}
\eea
Two further functional derivatives with respect to $\imath q_{\ga\w}$
and $\imath\ov{q}_{\de v}$ are taken and all remaining fields/sources
set to zero. Now we use the following relation, derived in
Ref.~\cite{Watson:2008fb}
\be
\left.\frac{\de}{\de\imath c_z^d}\ev{\imath\ro_x^b\imath\ov{\et}_
x^c}\right|_{J=0}=\left.t_{ij}(\vec{x})\frac{\de}{\de\imath c_z^d}
\ev{\imath J_{jx}^b\imath\ov{\et}_x^c}\right|_{J=0}=0,
\ee
and obtain the \ST identity for the quark-gluon vertices in
configuration space:
\bea
\lefteqn{0=\int d^4x\de(t-x_0)} \nonumber\\
&&\hspace{-0.5cm}\times\left\{
-\frac{\imath}{g}\left(\partial_x^0\ev{\imath\ov{q}_{\de v}\imath 
q_{\ga\w}\imath\si_x^d}\right)\de(z-x)
+\frac{1}{g}\left[\frac{\nabla_{ix}}{(-\nabla_x^2)}\ev{\imath
\ov{q}_{\de v}\imath q_{\ga\w}\imath A_{ix}^a}\right]\ev{\imath
\ov{c}_x^a\imath c_z^d}
\right.\nonumber\\
&&\hspace{-0.5cm}
+\imath T_{\al\ba}^a\ev{\imath\ov{q}_{\de v}\imath q_{\al x}}
\left[\left.\frac{\de^2}{\de\imath q_{\ga\w}\de\imath c_z^d}
\ev{\imath\ov{\chi}_{\ba x}\imath\ov{\et}_x^a}\right|_{J=0}+
\de^{da}\de(z-x)\de_{\ga\ba}\de(\w-x)\right]
\nonumber\\
&&\hspace{-0.5cm}\left.
+\imath T_{\ba\al}^a\left[\left.\frac{\de^2}{\de\imath\ov{q}
_{\de v}\de\imath c_z^d}\ev{\imath\chi_{\ba x}\imath\ov{\et}_x^a}
\right|_{J=0}-\de^{da}\de(z-x)\de_{\de\ba}\de(v-x)\right]\ev{\imath
\ov{q}_{\al x}\imath q_{\ga\w}}\right\}. \hspace{2cm}
\label{eq:stid1}
\eea

We now introduce the following definitions:
\bea
\tilde{\G}_{\ov{q};\ov{c}cq\al\ga}^d(x,z,\w)&\equiv&\imath gT_{\al
\ba}^a\left.\frac{\de^2}{\de\imath q_{\ga\w}\de\imath c_z^d}\ev{
\imath\ov{\chi}_{\ba x}\imath\ov{\et}_x^a}\right|_{J=0}\nonumber\\
&=&\imath gT_{\al\ba}^a\left\{
-\ev{\imath\ov{\chi}_{\ba x}\imath\chi_\e}\ev{\imath\ov{q}_\e
\imath q_{\ga\w}\imath\Phi_\la}
\ev{ J_{\la}\imath J_\ka}
\ev{\ov{\et}_x^a\imath\et_\ta}\ev{\imath\ov{c}_\ta\imath c_z^d\imath\Phi_\ka}
\right.\nonumber\\&&\left.
+\ev{\imath\ov{\chi}_{\ba x}\imath\chi_\ka}\ev{\imath\ov{q}_\ka
\imath q_{\ga\w}\imath\ov{c}_\ta\imath c_z^d}\ev{\imath\ov{\et}_x^a
\imath\et_\ta}
\right\},\nonumber\\
\tilde{\G}_{q;\ov{c}c\ov{q}\de\al}^d(x,z,v)&\equiv&\left.
\frac{\de^2}{\de\imath\ov{q}_{\de v}\de\imath c_z^d}\ev{\imath\chi
_{\ba x}\imath\ov{\et}_x^a}\right|_{J=0}\imath gT_{\ba\al}^a
\nonumber\\
&=&\left\{
\ev{\ov{\et}_x^a\imath\et_\ta}\ev{\imath\ov{c}_\ta\imath c_z^d
\imath\Phi_\ka}\ev{\imath J_\ka\imath J_\la}\ev{\imath\ov{q}_
{\de v}\imath q_\e\imath\Phi_\la}\ev{\imath\ov{\chi}_\e\imath\chi_
{\ba x}}\right.\nonumber\\&&\left.
-\ev{\imath\ov{\et}_x^a\imath\et_\ta}\ev{\imath\ov{c}_\ta\imath 
c_z^d\imath\ov{q}_{\de v}\imath q_\ka}\ev{\imath\ov{\chi}_\ka\imath
\chi_{\ba x}}\right\}\imath gT_{\ba\al}^a.
\label{eq:kernels}
\eea

In the above, the sources/fields $J$ and $\Phi$ refer to either
$\vec{A}$ or $\si$, and the internal indices refer to all attributes
of the object in question (summed or integrated over).  A few more
comments concerning the above definitions and notations are in
order. Recall that in the ``standard'' \DS equations, the loop
expressions contain a propagator ($\ev{\imath\ov{\chi}_{\al
z}\imath\chi_{\ba x}}$ or any other propagator) and an associated
tee-level vertex, stemming directly from the interaction terms in the
Lagrangian.  However, as seen in \eq{eq:kernels}, in the case of the
(nonabelian) \ST identities one has a different structure which arises
from the Gauss-BRST transform (or generally from the BRST transform):
namely, one has to functionally differentiate objects such as
$\ev{\imath\ov{\chi}_{\ba x}\imath\ov{\et}_x^a}$, but in this case the
vertex arising from a direct interaction is missing. The resulting
expressions still have a (partial) meaning as loop integrals, even
though the tree-level vertex is absent.  These types of would-be loop
expressions, denoted with a tilde, are called \emph{quark-ghost
scattering kernels} and are common to the \ST identities of nonabelian
theories.\footnote{In Coulomb gauge \YM theory these objects have been
studied in detail in \cite{Watson:2008fb}, and in linear covariant
gauges, some well-known examples are the ghost-gluon and quark-ghost
scattering-like kernels, appearing in the identities for the
three-gluon and quark-gluon vertices (a nice presentation of this
subject can be found in Ref.~\cite{Marciano:1977su}).}

With these definitions, we return to the configuration space \ST
identity \eq{eq:stid1}, and rewritte it as
\bea
\lefteqn{0=\int d^4x\de(t-x_0)}\nonumber\\
&&\times\left\{
-\left(\imath\partial_x^0\ev{\imath\ov{q}_{\de v}\imath q_{\ga\w}
\imath\si_x^d}\right)\de(z-x)
+\left[\frac{\nabla_{ix}}{(-\nabla_x^2)}\ev{\imath\ov{q}_{\de v}
\imath q_{\ga\w}\imath A_{ix}^a}\right]\ev{\imath\ov{c}_x^a\imath 
c_z^d}\right.\nonumber\\&&\left.
+\ev{\imath\ov{q}_{\de v}\imath q_{\al x}}\left[\tilde{\G}_{\ov{q};
\ov{c}cq\al\ga}^d(x,z,\w)+\imath gT_{\al\ga}^d\de(z-x)\de(\w-x)
\right]\right.\nonumber\\&&\left.
+\left[\tilde{\G}_{q;\ov{c}c\ov{q}\de\al}^d(x,z,v)-\imath gT_{\de
\al}^d\de(z-x)\de(v-x)\right]\ev{\imath\ov{q}_{\al x}\imath q_{\ga
\w}}\right\}.
\label{eq:stid2}
\eea
\begin{figure}[t]
\centering
\includegraphics[width=0.6\linewidth]{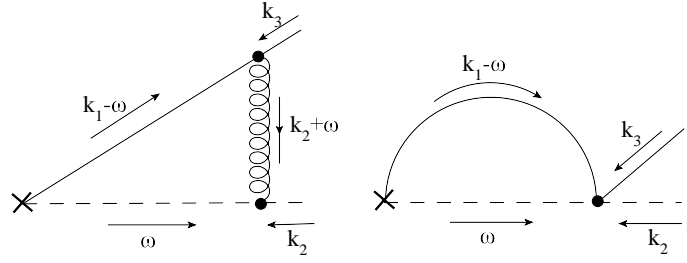}
\caption{\label{fig:quark-ghost-kernels} Quark-ghost scattering-like
kernels. Solid lines denote quarks, dashed lines denote ghosts and
spring lines denote gluons. All internal propagators are dressed.  The
cross indicates the absence of a tree-level vertex (see text for
details).}
\end{figure}

Using the formula \eq{eq:ftvertex} for the Fourier transform of the
vertex functions, one can write the \ST identity in momentum space
(momentum conservation is assumed)
\bea
k_3^0\G_{\ov{q}q\si\al\ba}^{d}(k_1,k_2,k_3)&=&\imath\frac{k_{3i}}
{\vec{k}_3^2}\G_{\ov{q}qA\al\ba i}^{a}(k_1,k_2,k_3)\G_{\ov{c}c}^
{ad}(-k_3)\nonumber\\&&
+\G_{\ov{q}q\al\de}(k_1)\left[\tilde{\G}_{\ov{q};\ov{c}cq}^{d}
(k_1+q_0,k_3-q_0;k_2)+\imath gT^d\right]_{\de\ba}\nonumber\\&&
+\left[\tilde{\G}_{q;\ov{c}c\ov{q}}^{d}(k_2+q_0,k_3-q_0;k_1)-
\imath gT^d\right]_{\al\de}\G_{\ov{q}q\de\ba}(-k_2)
\label{eq:stid3}
\eea
In the above, $\G_{\ov{c}c}$ is the proper ghost two-point function
and $q_0$ is the (arbitrary) energy injection scale that arises from
the time-dependence of the Gauss-BRST transform.  This expression is
very similar to the one derived in Ref.~\cite{Watson:2008fb} for the
\YM case.

We now write the quark-ghost kernels \eq{eq:kernels} in momentum
space. After performing the Fourier transform, we arrive at the
following expressions (represented diagrammatically in
\fig{fig:quark-ghost-kernels}):
\bea
\tilde{\G}_{\ov{q};\ov{c}cq}^d(p_1,p_2,p_3)&=&\imath gT^a\int\dk
{\w}W_{\ov{c}c}^{ab}(\w)W_{\ov{q}q}(p_1-\w) 
\bigg[\G_{\ov{c}cq\ov{q}}^{bd}(\w,p_2,p_1-\w,p_3)
\nonumber\\&&
-\G_{\ov{c}c\ka}^
{bdc}(\w,p_2,-p_2-\w)W_{\ka\la}^{ce}(p_2+\w)\G_{\ov{q}q\la}^e(p_1-
\w,p_3,p_2+\w)\bigg],
\nonumber\\
\tilde{\G}_{q;\ov{c}c\ov{q}}^d(p_1,p_2,p_3)&=&\int\dk{\w}\bigg[\G_
{\ov{c}c\ka}^{bdc}(\w,p_2,-p_2-\w)W_{\ka\la}^{ce}(p_2+\w)\G_{\ov{q}
q\la}^{e}(p_3,p_1-\w,p_2+\w)
\nonumber\\&&
-\G_{\ov{c}c\ov{q}q}^{bd}(\w,p_2,p_3,p_1-\w)\bigg]
W_{\ov{c}c}^{ab}(\w)W_{\ov{q}q}(\w-p_1)\imath gT^a,
\label{eq:kern0}
\eea
where the indices $\ka$, $\la$, refer to the gluonic field types $\si$
or $\vec{A}$ (with the associated spatial index).

  Just as in the \YM case, the above \ST identity, \eq{eq:stid3} is
used as a constraint, i.e. it helps to find meaningful truncations for
the vertex functions of the theory, which are then used to solve the
\DS equations. In particular, in the limit of the heavy quark mass,
this equation will be used to write the temporal quark-gluon vertex
$\G_{\ov{q}q\si}$ in terms of purely spatial, ghost or quark
propagators and proper functions. This will then be used to obtain an
exact solution of the quark gap equation and of the \BS equation for
quark-antiquark bound states.

Using the Feynman rules presented in chapter \ref{chap:g^2}, it is
easy to show that the identity \eq{eq:stid3} is trivially satisfied at
tree-level. Also, it can be verified that this identity is fulfilled
at one-loop perturbatively. In the next section, we will present this
derivation in detail.

\section{Perturbative analysis}

Let us now briefly show that the Slavnov-Taylor identity is fulfilled
at one-loop perturbative level. The proof is based on the
translational invariance of the loop-integrals, without explicitly
evaluating them.  At one-loop perturbative level, \eq{eq:stid3} reads:
\bea
\lefteqn{k_3^0\G_{\bar q q\si}^{(1)d}(k_1,k_2,k_3)
+k_3^i\G_{\bar q q Ai}^{(1)d}(k_1,k_2,k_3)
+\imath g T^a\ga^i\frac {k_{3i}}{\vec k_3^2}
\G_{\bar cc}^{(1)ad}(-\vec k_3)}
\nonumber\\
&&=
\imath g \left[\G_{\bar  qq}^{(1)}(k_1)T^d-T^d
\G_{\bar qq}^{(1)}(-k_2)\right]\nonumber\\
&&+\G_{\bar qq}^{(0)}(k_1)
\tilde\G_{\bar q;\bar c c q}^d(k_1+q_0,k_3-q_0,k_2)+
\tilde\G_{ q;\bar c c\bar q}^d(k_2+q_0,k_3-q_0,k_1)
\G_{\bar qq}^{(0)}(-k_2).
\label{eq:stid1loop}\nonumber\\
\eea
In the above, we have inserted the tree-level spatial quark-gluon
vertex, \eq{eq:treelevelquarkvertex2}, and the tree-level ghost
two-point function from Ref.~\cite{Watson:2007vc}
\be
\G_{\bar cc}^{ad (0)}(k)=\de^{ad}\imath\vec k^2. 
\ee

Using the expressions \eq{eq:qAvertex_1loop}, \eq{eq:qsivertex_1loop}
for the quark-gluon vertices, we obtain for the (nonperturbative)
combination of quark-gluon vertex functions, on the left-hand side of
\eq{eq:stid1loop}:
\bea
\lefteqn{k_3^0\G_{\bar q q\si}^{(1)d}(k_1,k_2,k_3)+
k_3^i\G_{\bar q q Ai}^{(1)d}(k_1,k_2,k_3)}\nonumber\\
&&\hspace{-0.2cm}=-\int\dk{\w}\bigg\{
\G_{\bar qqAk}^c(-k_1+\w,k_1,-\w)W_{\bar qq}(\w) 
\nonumber\\
&&\hspace{-0.2cm}\left[
k_{3}^0\G_{\bar q q\si}^{(0)d}(k_3,\w,-k_3-\w)+
k_{3i}\G_{\bar q qAi}^{(0)d}(k_3,\w,-k_3-\w)
\right]
\nonumber\\
&&\hspace{-0.2cm}\times W_{\bar q q}(k_3+\w)
\G_{\bar qqAj}^b(k_3+\w,k_2,k_1-\w)W_{AAjk}^{bc}(k_1-\w)
\nonumber\\
&&\hspace{-0.2cm}
-\G_{\bar qq\si}^c(-k_1+\w,k_1,-\w)W_{\bar qq}(\w)
\left[
k_{3}^0\G_{\bar q q\si}^{(0)d}(k_3,\w,-k_3-\w)+
k_{3i}\G_{\bar q qAi}^{(0)d}(k_3,\w,-k_3-\w)
\right]
\nonumber\\
&&\hspace{-0.2cm}\times W_{\bar q q}(k_3+\w)
\G_{\bar qq\si}^b(k_3+\w,k_2,k_1-\w)W_{\si\si}^{bc}(k_1-\w)
\nonumber\\
&&\hspace{-0.2cm}
-\G_{\bar qqAk}^a(-k_1+\w,k_1,-\w)W_{\bar qq}(\w)
\G_{\bar qq\si}^e(\w,k_2,-k_2-\w)W_{\si\si}^{ec}(k_2+\w)
W_{AAjk}^{ba}(-k_1+\w)\nonumber\\
&&\hspace{-0.2cm}\times\left[
k_3^0\G_{\si A\si j}^{(0)dbc}(k_2+\w,k_3,k_1-\w)
+k_{3i}\G_{\si AAij}^{(0)dbc}(k_2+\w,k_3,k_1-\w)
\right]
\nonumber\\
&&
\hspace{-0.2cm}
-\G_{\bar qq\si}^a(-k_1+\w,k_1,-\w)W_{\bar qq}(\w)
\G_{\bar qqAk}^e(\w,k_2,-k_2-\w)W_{AAjk}^{ec}(k_2+\w)
W_{\si\si}^{ba}(-k_1+\w)\nonumber\\
&&\hspace{-0.2cm}\times\left[
k_3^0\G_{\si A\si j}^{(0)dbc}(k_2+\w,k_3,k_1-\w)
+k_{3i}\G_{\si AAij}^{(0)dbc}(k_2+\w,k_3,k_1-\w)
\right]
\nonumber\\
&&\hspace{-0.2cm}
-\G_{\bar qqAl}^a(-k_1+\w,k_1,-\w)W_{\bar qq}(\w)
\G_{\bar qqAm}^e(\w,k_2,-k_2-\w)W_{AAmk}^{ec}(k_2+\w)
W_{AAjl}^{ba}(-k_1+\w)\nonumber\\
&&\hspace{-0.2cm}\times\left[
k_3^0\G_{\si AA jk}^{(0)dbc}(k_2+\w,k_3,k_1-\w)
+k_{3i}\G_{3 Aijk}^{(0)dbc}(k_2+\w,k_3,k_1-\w)
\right]
\nonumber\\
&&\hspace{-0.2cm}
-\G_{\bar qq\si}^a(-k_1+\w,k_1,-\w)W_{\bar qq}(\w)
\G_{\bar qq\si}^e(\w,k_2,-k_2-\w)W_{\si\si}^{ec}(k_2+\w)
W_{\si\si}^{ba}(-k_1+\w)\nonumber\\
&&\hspace{-0.2cm}
\times k_{3i}\G_{A\si\si i}^{(0)dbc}(k_2+\w,k_3,k_1-\w)
\bigg\}
\label{eq:stid_sum1}
\eea 
The lengthy expression above can be simplified if we make use of the
tree-level Yang-Mills Slavnov-Taylor identities derived in
Ref.~\cite{Watson:2008fb}.\footnote{In principle, the \YM identities
also contain ghost-gluon scattering-like kernels, but at tree-level
these factors do not give a contribution.}  For completeness we list
them all here (together with the tree-level quark Slavnov-Taylor
identity, which is trivial):
\bea
&&k_3^0\G_{\si A\si k}^{fed(0)}(k_1,k_2,k_3)
+k_{3i}\G_{\si AAki}^{fed(0)}
=-gf^{ade}\G_{\si Ak}^{fa(0)}(k_1)-gf^{adf}\G_{\si Ak}^{ea(0)}(k_2)
\nonumber\\
&&k_3^0\G_{A A\si jk}^{abc(0)}(k_1,k_2,k_3)
+k_{3i}\G_{ 3Aijk(0)}^{abc}=
-gf^{dcb}\G_{AAjk}^{ad(0)}(k_2)-gf^{dea}\G_{AAjk}^{bd(0)}(k_1)
\nonumber\\
&&k_{3i}\G_{\si\si Ai}^{abc(0)}(k_1,k_2,k_3)
=-gf^{dcb}\G_{\si\si}^{ad(0)}(k_2)-gf^{dca}\G_{\si\si}^{bd(0)}(k_1)
\nonumber\\
&&k_3^0\G_{\bar qq\si}^{(0)d}(k_1,k_2,k_3)
+k_{3i}\G_{\bar qqAi}^{(0)d}(k_1,k_2,k_3)=
\G_{\bar qq}^{(0)}(k_1) \imath g T^d-\imath g T^d\G_{\bar
  qq}^{(0)}(-k_2)
\hspace{2cm}
\label{eq:stid_ym}
\eea
Further, in order to derive the one-loop expression, in the full
nonperturbative equation \eq{eq:stid_sum1} we insert the tree-level
vertices and propagators given by \eq{eq:treelevelquarkprop},
\eq{eq:treelevelgluonprop}, \eq{eq:treelevelquarkvertex1},
\eq{eq:treelevelquarkvertex2}.  Putting all these together, we arrive
at the following simplified expression:
\bea
\lefteqn{k_3^0\G_{\bar q q\si}^{(1)d}(k_1,k_2,k_3)+
k_3^i\G_{\bar q q Ai}^{(1)d}(k_1,k_2,k_3)}\nonumber\\
&&=-\imath g^3 T^a T^a T^d \nonumber\\
&&\int\dk{\w}\bigg\{
\ga^k W_{\bar qq}(k_3+\w)\ga^j W_{AAjk}(k_1-\w)-
\ga^k W_{\bar qq}(\w)\ga^j W_{AAjk}(k_1-\w)\nonumber\\
&&+\ga^0 W_{\bar qq}(k_3+\w)\ga^0 W_{\si\si}(k_1-\w)-
\ga^0 W_{\bar qq}(\w)\ga^0 W_{\si\si}(k_1-\w)
\bigg\}\nonumber\\
&&=\frac{\imath}{2}T^dN_cg^3
\int\dk{\w}
\frac{k_{3j}t_{ij}(\vec\w)}{\w^2(\vec k_3-\vec\w)^2}\nonumber\\
&&\times\bigg\{\left[\ga^0(k_3-\w)_0-\ga^k(k_3-\w)_k\right]
\frac{\left[\ga^0(\w+k_2)_0-\ga^k(\w+k_2)_k-m\right]}
{(\w+k_2)^2-m^2}\ga^i\nonumber\\
&&+\ga^i
\frac{\left[\ga^0(\w+k_1)_0-\ga^k(\w+k_1)_k+m\right]}{(\w+k_1)^2-m^2}
 \left[\ga^0(k_3-\w)_0-\ga^k(k_3-\w)_k\right]
\bigg\}\hspace{2cm}
\label{eq:stid_vertex}
\eea
We first compare the first integral of \eq{eq:stid_vertex} with the
combination of one-loop quark two-point functions (without ghost
factors) on the r.h.s. of \eq{eq:stid1loop}.  Using the expression
\eq{eq:gapeqms} for the gap equation and recalling that the $\vec\pi$
and $\phi$ fields do not give a contribution at one-loop order in
perturbation theory, we deduce that these terms are equal and hence
they cancel in the \ST identity.  For the rest of the terms (the
combination of quark- ghost kernels on the r.h.s. and the remaining
ghost term on the l.h.s. of \eq{eq:stid1loop}) we get:
\bea
\lefteqn{\G_{\bar qq}^{(0)}(k_1)
\tilde\G_{\bar q;\bar c cq}^d(k_1+q_0,k_3-q_0,k_2)+
\tilde\G_{q;\bar c c\bar q}^d(k_2+q_0,k_3-q_0,k_1)
\G_{\bar qq}^{(0)}(-k_2)}\nonumber\\
&&-\imath g T^a\ga^i
\frac {k_{3i}}{\vec k_3^2}\G_{c}^{(1)ad}(-\vec k_3)
=\frac{\imath}{2}T^dN_cg^3
\int\dk{\w}
\frac{k_{3j}t_{ij}(\vec\w)}{\w^2(\vec k_3-\vec\w)^2}\nonumber\\
&&\left\{-(\ga^0k_{10}-\ga^k k_{1k}-m)
\frac{\left[\ga^0(\w+k_2)_0-\ga^k(\w+k_2)_k-m\right]}
{(\w+k_2)^2-m^2}\ga^i\right.\nonumber\\
&&+\left.\ga^i
\frac{\left[\ga^0(\w+k_1)_0
-\ga^k(\w+k_1)_k+m\right]}{(\w+k_1)^2-m^2}
 (-\ga^0 k_{20}+\ga^k k_{2k}-m)-2k_{3i}
\frac{\ga^k k_{3k}}{\vec k_3^2}
\right\}\hspace{1cm}
\label{eq:stid_1loop_qg}
\eea
After rearranging the terms, we find that the second integral in the
formula \eq{eq:stid_vertex} equals the above combination of
quark-ghost kernels and the ghost term. Collecting all the results, we
deduce that the l.h.s. of \eq{eq:stid1loop} equals the r.h.s., and
hence the \ST identity for the quark-gluon vertex is satisfied at
one-loop perturbative level.

As a useful check, we can also show that the divergent parts of the
quantities entering the formula \eq{eq:stid1loop} combine such that
the \ST identity is again satisfied.  For this we use the results
derived in Chapter~\ref{chap:g^2}, i.e.  the divergent part of the
quark gluon vertices \eq{eq:temporaldiv}, \eq{eq:spatialdiv}, and the
divergent part of the quark gap equation, obtained by inverting the
result \eq{eq:quarkproponeloop} for the quark propagator
\be
\G_{\bar q q}(k)=\imath \frac{C_F}{(4\pi)^2}\frac{1}{\e}
(\ga^0k_0-\ga^ik_i-4m)+\textrm{finite terms},
\label{eq:gapeqdiv}
\ee
together with the ghost two-point
function (previously calculated in \cite{Watson:2007mz})
\be
\G_c^{ab}(k)=-\imath\de^{ab}\vec
k^2\frac{4}{3}\frac{N_c}{(4\pi)^2}\frac{1}{\e}
+\textrm{finite terms}.
\label{eq:ghostqdiv}
\ee
Further, writing out the explicit form of the quark-ghost kernels
\eq{eq:kern0} and making a simple power analysis, we notice that these
factors do not contain divergences.\footnote{The convergence of the
quark-ghost kernels has been also shown in
Ref.~\cite{Leibbrandt:1997kh}, in the framework of split dimensional
regularization.}  Collecting the divergent factors
\eq{eq:temporaldiv}, \eq{eq:spatialdiv}, \eq{eq:gapeqdiv} and
\eq{eq:ghostqdiv}, it is straightforward to verify that that the \ST
identity is again satisfied.

%% file: hq.tex
\chapter{Heavy quarks}
\label{chap:hq} 
    
So far, we have investigated the \DS equations (along with the \ST
identities) for two- and three-point Green's functions at one-loop
perturbative level, where the coupling between quarks and gluons is
small.  In the second part of this thesis, we are concerned with the
nonperturbative (i.e., strong coupling) region of the theory, with the
aim to describe bound states of quarks, and interpret them in relation
to the linearly rising potential which confines them.

Traditionally, most of the nonperturbative studies are performed in
covariant gauges.  They concentrate on the light quark sector, where
one is concerned with dynamical chiral symmetry breaking (a review of
this subject can be found in Ref.~\cite{Alkofer:2000wg}, see also
Ref.~\cite{Tandy:1997qf} for the phenomenological implications for
meson and baryon states).  So far, the Landau gauge studies of the
heavy quark sector are unfortunately not conclusive, since the heavy
quark propagator contains spurious poles which prevent a solution of
the bound state \BS equation \cite{Burden:1997ja}. The difficulty
arises from the fact that nonperturbative \DS equations and heavy
quark limit are not automatically compatible, i.e., both the heavy
quark mass and the ultraviolet-cutoff scale of the loop integrals
(supposing that a ultraviolet-cutoff regularization is employed) tend
to be the largest scale in the problem.

An alternative approach is to reformulate the heavy quark sector of
QCD as an effective theory -- the Heavy Quark Effective Theory
[HQET]\footnote{ In \cite{Shifman:1987rj,Isgur:1989vq,
Eichten:1989zv,Georgi:1990um,Falk:1990yz} we collected the original
papers; extended reviews can be found, among others, in
Refs.~\cite{Neubert:1993mb, Mannel:1992fx} and in the textbook
\cite{Grozin:2004yc}.}.  In this approach, the main simplification
stems from the fact that the {\it effective} quark Lagrange function
is rewritten as an expansion in powers of $1/m$, where $m$ is the
heavy quark mass. Naturally, when the quarks are very heavy one can
keep only the first few terms in the mass expansion.  The effective
Lagrange function also enjoys two additional symmetries, which turn to
be very useful for practical applications \cite{Isgur:1989vq}: the
heavy flavor symmetry\footnote{ However, as we consider identical
quarks, this is not of particular interest for the present work.} and
the so called spin symmetry, which arises because in the heavy quark
limit the spin degrees of freedom do not couple to the gluon field.
Moreover, the apparent incompatibility of the heavy quark mass and the
ultraviolet cutoff has been recognized. To handle this problem, the
so-called matching procedure has been developed \cite{Neubert:1993mb}:
QCD operators are written as series in $1/m$ via HQET operators and
the coefficients in these series are determined by matching on-shell
matrix elements in both theories. These calculations have been
performed up to three loops in perturbation theory
\cite{Broadhurst:1994se, Grozin:2007fh, Grozin:2007ap}.

In the present chapter, the aim is to explore the heavy quark sector
of the theory by combining nonperturbative \DS equations with the
heavy quark mass expansion \cite{Popovici:2010mb}. We will employ the
full nonperturbative QCD functional formalism, involving the complete
quark fields, rather than HQET expressions that refer to the heavy
quark degrees of freedom.  The effect will be that, due to the
complexity of the equations, we will be obliged to restrict ourselves
to leading order in the mass expansion, and this means that we will
not be able to describe real quarks (which are not infinitely
heavy). However, this does not alter our goal to study the confining
potential, since all quarks should be confined, regardless of their
mass.

\section{Heavy quark mass expansion}

Let us start by writing out the explicit quark contribution to the
full QCD generating functional, \eq{eq:Z},
\bea
Z[\ov{\chi},\chi]&=&\int{\cal D}\Phi\exp{\left\{\imath\int d^4x
\ov{q}_\al(x)\left[\imath\ga^0D_0+\imath\s{\vec{\ga}}{\vec{D}}-m
\right]_{\al\ba}q_\ba(x)\right\}}
\nonumber\\&&\times
\exp{\left\{\imath\int d^4x\left[\ov{\chi}_\al(x)q_\al(x)+
\ov{q}_\al(x)\chi_\al(x)\right]+\imath{\cal S}_{YM}\right\}},
\label{eq:genfunc}
\eea
with the temporal and spatial components of the covariant derivative
(in the fundamental color representation) given by \eq{eq:covder}. The
\YM contribution to the generating functional is in the standard,
second order formalism \cite{Watson:2008fb,Watson:2007vc}, where the
auxiliary fields $\vec\pi$ and $\phi$ do not appear.

Now consider the following decomposition of the quark and antiquark
fields:
\bea
q_\al(x)&=&e^{-\imath mx_0}\left[h(x)+H(x)\right]_\al,
\nonumber\\
\ov{q}_\al(x)&=&e^{\imath mx_0}\left[\ov{h}(x)+\ov{H}(x)\right]_\al,
\label{eq:qdecomp1}
\eea
such that
\bea
&h_\al(x)=e^{\imath mx_0}\left[P_+q(x)\right]_\al,
&H_\al(x)=e^{\imath mx_0}\left[P_-q(x)\right]_\al \nonumber\\
&\ov{h}_\al(x)=e^{-\imath mx_0}\left[\ov{q}(x)P_+\right]_\al,
&\ov{H}_\al(x)=e^{-\imath mx_0}\left[\ov{q}(x)P_-\right]_\al.
\label{eq:qdecomp2}
\eea
In the above, the (spinor) projection operators are given
by\footnote{Recall that in the free Dirac theory the operators $P_\pm$
project on the positive and negative energy eigenstates (they select
the upper and lower components of the quark spinor).}
\be
P_\pm=\frac{1}{2}(\mathds{1}\pm\ga^0),\;\;\;\;P_++P_-=\mathds{1},
\;\;\;\;P_+P_-=0,\;\;\;\;P_\pm^2=P_\pm.
\ee
This decomposition is a particular case of the heavy quark transform
underlying HQET \cite{Neubert:1993mb}.  There, the starting point is
the observation that a heavy quark within a hadron is almost on-shell
and moves with the hadron velocity $v$.  Its 4-momentum can be written
\be
p^\mu=mv^\mu+k^\mu
\ee
 where $|k|\ll m|v|$ and $v^2=1$ (such that when $|k|=0$, $p^2=m^2$).
One then uses the general projectors
$P_\pm=(\mathds{1}\pm\slash\!\!\!{v})/2$ and the exponential terms are
generalized to $e^{\pm\imath mv\cdot x}$.  The case used here
corresponds to the rest frame of the quark, $v^\mu=(1,\vec{0})$.
Intuitively, this corresponds to a shift in the position of the ``zero
energy'' level, such that the energy of the state $m$ (the heavy quark
at rest) is the new zero level. In other words, instead of the true
energy $p_0=m+k_0$ of the heavy quark, the residual energy $k_0$ is
used.  However, within the context of the generating functional
\eq{eq:genfunc}, the decomposition \eq{eq:qdecomp1} can be initially
regarded simply as and arbitrary decomposition, which will later on
prove to be useful in Coulomb gauge.  In fact, this choice will result
in an important simplification: the spatial components of the \YM
Green's functions are absent at leading order in the mass expansion
and they apear only in next-to-leading order.  The projection
operators satisfy the following further relations
\be
P_+\ga^0P_+=P_+P_+,\;\;\;\;P_+\ga^0P_-=0,\;\;\;\;P_+\ga^iP_+=0
\ee
such that the following relations hold for the components of the quark
field:
\be
\ov{h}\ga^0h=\ov{h}h,\;\;\ov{H}\ga^0H=-\ov{H}H,\;\;\ov{h}\ga^0H=
\ov{H}\ga^0h=\ov{h}\ga^ih=\ov{H}\ga^iH=0.
\ee
Inserting the decomposition of the quark fields given by
\eq{eq:qdecomp1} into the generating functional \eq{eq:genfunc} and
using these relationships one obtains
\bea
\lefteqn{Z[\ov{\chi},\chi]=\int{\cal D}\Phi\exp{\left\{\imath\int d^4x
\left[
\ov{h}_\al(x)\left[\imath D_0\right]_{\al\ba}h_\ba(x)
\right.\right.}
}\nonumber\\
&&\!\!{\left.\left.
+\ov{h}_\al(x)\left[\imath\s{\vec{\ga}}{\vec{D}}\right]_{\al\ba}
H_\ba(x)
+\ov{H}_\al(x)\left[\imath\s{\vec{\ga}}{\vec{D}}\right]_{\al\ba}
h_\ba(x)
+\ov{H}_\al(x)\left[-2m-\imath D_0\right]_{\al\ba}H_\ba(x)\right]
\right\}}
\nonumber\\
&&\!\!\times
\exp{\left\{\imath\int d^4x\left[e^{-\imath mx_0}\ov{\chi}_\al(x)
\left[h(x)+H(x)\right]_\al+e^{\imath mx_0}\left[\ov{h}(x)+\ov{H}(x)
\right]_\al\chi_\al(x)\right]+\imath{\cal S}_{YM}\right\}}\!.
 \nonumber\\
\label{eq:genfunc1}
\eea
At this point, it is illuminating to discuss the difference between
the mass expansion (as used here) and HQET.  As already emphasized in
the introduction of this chapter, in this approach we work with the
full quark fields, i.e. we retain the source terms for the quark
fields $q, \bar q$ in order to derive the full gap, \BS and Faddeev
equations from $Z[\ov{\chi},\chi]$ (see below).  Hence, the source
term expression in the above is slightly modified by the appearance of
the exponential factors.  Since the Jacobian of the transformation is
field independent (and thus trivial), our generating functional has
not been altered, but merely rewritten in terms of different
integration variables.  This is in contrast to HQET where the quark
sources are replaced with sources for the projected fields $h$ and
$H$.  In principle, differentiation with respect to the sources of the
generating functional \eq{eq:genfunc1} would then lead to the Green's
functions of the projections $h,H$ of the spinor $q$. In order to
derive a Green's function with an external heavy quark, the components
$H$ (multiplied with twice the quark mass) are integrated out (as
below) \emph{and} in addition their sources are set to zero
\cite{Mannel:1992fx}.

Noticing that for the $h$-field (`large') components, the quark mass
parameter $m$ does not appear directly, we integrate out the
$H$-fields and get the following expression
\bea
Z[\ov{\chi},\chi]&=&\int{\cal D}\Phi\mbox{Det}\left[\imath D_0+2m
\right]\exp{\left\{\imath\int d^4x\left[
\ov{h}_\al(x)\left[\imath D_0\right]_{\al\ba}h_\ba(x)
\right.\right.}\nonumber\\&&{\left.\left.
+\left[\ov{h}(x)\imath\s{\vec{\ga}}{\vec{D}}+e^{-\imath mx_0}\ov{
\chi}(x)\right]_\al
\left[\imath D_0+2m\right]_{\al\ba}^{-1}
\left[\imath\s{\vec{\ga}}{\vec{D}}h(x)+e^{\imath mx_0}\chi(x)
\right]_\ba\right]\right\}}
\nonumber\\&&\times
\exp{\left\{\imath\int d^4x\left[e^{-\imath mx_0}\ov{\chi}_\al(x)
h_\al(x)+e^{\imath mx_0}\ov{h}_\al(x)\chi_\al(x)\right]+\imath
{\cal S}_{YM}\right\}}.
\label{eq:genfunc2}
\eea
Obviously, since we have integrated out a nontrivial component of the
original quark field, our expression is nonlocal and this is where the
heavy mass expansion is necessary.  The determinant arising from the
fermion functional integration depends on the temporal component of
the gauge field and it can be written in the following way:
\bea
\mbox{Det}\left[\imath D_0+2m\right]
&=&\exp{\ln\left[\imath  D_0+2m\right]}
= N \exp{\mbox{Tr} \ln 
\left[1+gT^a\si^a\frac{1}{\imath \pd_0+2m}\right]}\nonumber\\
&&\sim N+{\cal O}\left(1/m^2\right)
\eea
where $N$ is a (noninteracting) constant.  If we expand the logarithm
in power series we see that the first term, of ${\cal O}(1/m)$,
vanishes due to the fact that the generators $T^a$ are traceless
matrices, and the term at $O(1/m^2)$ is proportional to
$Tr(T^aT^b)=N_f$ (but, since we restrict to leading order in the mass
expansion, this term will not appear in our calculations).

 Leaving the $e^{\pm\imath mx_0}$ factors as they are, we can thus
write
\bea
\lefteqn{Z[\ov{\chi},\chi]=\int{\cal D}\Phi\exp\bigg\{\imath\int d^4x
\left[\right.
\ov{h}_\al(x)\left[\imath D_0\right]_{\al\ba}h_\ba(x)} \nonumber\\
&&
\hspace{-0.6cm}
+\frac{1}{2m}\left[\ov{h}(x)\imath\s{\vec{\ga}}{\vec{D}}+
e^{-\imath mx_0}\ov{\chi}(x)\right]_\al
\left[\imath\s{\vec{\ga}}{\vec{D}}h(x)+e^{\imath mx_0}
\chi(x)\left.\right]_\al\right]\bigg\}
\nonumber\\
&&
\hspace{-0.6cm}\times\exp
\left\{\imath\int d^4x\left[e^{-\imath mx_0}\ov{\chi}_\al(x)
h_\al(x)+e^{\imath mx_0}\ov{h}_\al(x)\chi_\al(x)\right]+
\imath{\cal S}_{YM}\right\}\!+\!{\cal O}\left(1/m^2\right)\!\!.
\label{eq:genfunc3}
\eea
Our generating functional is now local in the fields and arranged in
an expansion in the parameter $1/m$ (this will be referred to as the
mass expansion although strictly speaking, it is an expansion in the
\emph{inverse} mass).  However, locality does not mean that the above
expression can be directly applied.  Let us consider the classical
(full) quark field in the presence of sources:
\bea
\lefteqn{\frac{1}{Z}\int{\cal D}\Phi q_\al(x)\exp{\left\{\imath
{\cal S}\right\}}=\frac{1}{Z}\frac{\de Z}{\de\imath\ov{\chi}
_\al(x)}}\nonumber\\
&=&\frac{1}{Z}\int{\cal D}\Phi\left\{e^{-\imath mx_0}h_\al(x)+
\frac{e^{-\imath mx_0}}{2m}\left[\imath\s{\vec{\ga}}{\vec{D}}h(x)+
e^{\imath mx_0}\chi(x)\right]_\al\right\}\exp{\left\{\imath
{\cal S}\right\}}+{\cal O}\left(1/m^2\right). \nonumber\\
\eea
One sees immediately that even at ${\cal O}(1/m)$, the classical quark
field has components that involve interaction type terms
($\vec{D}h\sim\vec{A}h$) brought about by the truncation of the
nonlocality.  Of course, this is nothing more than the statement that
the $h$-field is nontrivially (and dynamically) related to the full
$q$-field.  It also means that if we want to use the nonperturbative
gap and \BS equations (i.e., those equations derived from the action
for the full quark fields $q$ and their sources $\chi$ and which are
the equations of QCD as opposed to HQET) then we cannot expect that
the mass expansion can realistically be extended far beyond the
leading order in order to do practical calculations.  As stated
previously though, the aim is to investigate the connection between
the \YM sector and the physical world made of quarks; the string
tension that represents our goal is not dependent on the quark mass
(both light and heavy quarks are confined in the same way, as far as
we know).  Therefore, we restrict our attention to the leading order
in the mass expansion as follows (and writing $D_0$ explicitly):
\bea
Z[\ov{\chi},\chi]&=&\int{\cal D}\Phi\exp{\left\{\imath\int d^4x
\ov{h}_\al(x)\left[\imath\partial_{0x}+gT^a\si^a(x)\right]_{\al\ba}
h_\ba(x)\right\}}
\nonumber\\&&\hspace{-0,5cm}\times
\exp{\left\{\imath\int d^4x\left[e^{-\imath mx_0}\ov{\chi}_\al(x)
h_\al(x)+e^{\imath mx_0}\ov{h}_\al(x)\chi_\al(x)\right]+
\imath{\cal S}_{YM}\right\}}+{\cal O}\left(1/m\right). \nonumber\\
\label{eq:genfunc4}
\eea

The standard machinery of functional methods is now employed.  Just as
for the usual QCD gap equation derived in Chapter~\ref{chap:ds}, we
start with the observation that up to boundary terms (which are
assumed to vanish) the integral of a total derivative vanishes, and
obtain for the quark field equation of motion:
\bea
0&=&\int{\cal D}\Phi\frac{\de}{\de\imath\ov{h}_\al(x)}\exp{\left\{
\imath{\cal S}\right\}}\nonumber\\
&=&\int{\cal D}\Phi\left\{\left[\imath\partial_{0x}+gT^a\si^a(x)
\right]_{\al\ba}h_\ba(x)+e^{\imath mx_0}\chi_\al(x)\right\}\exp{
\left\{\imath{\cal S}\right\}}+{\cal O}\left(1/m\right)\!.
\label{eq:feom}
\eea
The field equation of motion for the antiquark gives equivalent
results and can be neglected.  From the generating functional of
connected Green's functions $W[\ov{\chi},\chi]$, \eq{eq:genfunccon},
and using the bracket notation, \eq{eq:WJ}, for the derivatives of $W$
with respect to sources, we obtain the classical quark fields,
\eq{eq:qbarqclassic}.

Using the equation \eq{eq:eqgen7} and noticing that from
\eq{eq:genfunc4}, the following relations hold:
\bea
\frac{\de Z[\ov{\chi},\chi]}{\de\imath\ov{\chi}_\al(x)}&=&
\int{\cal D}\Phi e^{-\imath mx_0}h_\al(x)
\exp{\left\{\imath{\cal S}
\right\}}+{\cal O}\left(1/m\right),\nonumber\\
\frac{\de Z[\ov{\chi},\chi]}{\de\imath\chi_\al(x)}&=&
-\int{\cal D}\Phi e^{\imath mx_0}\ov{h}_\al(x)
\exp{\left\{\imath{\cal S}\right\}}+{\cal O}\left(1/m
\right)
\eea
(recalling the earlier discussion of neglecting the ${\cal O}
\left(1/m\right)$ terms), the field equation of motion, \eq{eq:feom},
can be written in terms of derivatives of $W$:
\bea
0&=&\left[\imath\partial_{0x}\right]_{\al\ba}e^{\imath mx_0}\ev
{\imath\ov{\chi}_\ba(x)}+\left[gT^a\right]_{\al\ba}e^{\imath mx_0}
\left[\ev{\imath\ro^a(x)\imath\ov{\chi}_\ba(x)}+\ev{\imath\ro^a(x)}
\ev{\imath\ov{\chi}_\ba(x)}\right]
\nonumber\\&&
+e^{\imath mx_0}\chi_\al(x)
+{\cal O}\left(1/m\right).
\eea
Factoring out the exponential terms gives then
\bea
0&=&\left[\imath\partial_{0x}-m\right]_{\al\ba}\ev{\imath\ov{\chi}_
\ba(x)}+\left[gT^a\right]_{\al\ba}\left[\ev{\imath\ro^a(x)\imath
\ov{\chi}_\ba(x)}+\ev{\imath\ro^a(x)}\ev{\imath\ov{\chi}_\ba(x)}
\right]\nonumber\\
&&+\chi_\al(x)+{\cal O}\left(1/m\right).
\label{eq:feom1}
\eea
To continue, we employ the Legendre transform \eq{eq:legendretr} in
order to construct the effective action for the full quark fields
(with the notations and conventions introduced in
Chapter~\ref{chap:ds}).  The field equation of motion can then be
rewritten in terms of derivatives of the effective action (which are
the proper Green's functions when sources are set to zero):
\be
\ev{\imath\ov{q}_\al(x)}=\left[\imath\partial_{0x}-m\right]_
{\al\ba}q_\ba(x)+\left[gT^a\right]_{\al\ba}\left[\ev{\imath\ro^a(x)
\imath\ov{\chi}_\ba(x)}+\si^a(x)q_\ba(x)\right]+{\cal O}\left(1/m
\right).
\label{eq:feom2}
\ee
With the field equation of motion written in the above forms, we can
now derive the Feynman rules for the quark components of the theory
(the \YM parts are already known \cite{Watson:2007vc}).  Following the
derivation presented in Chapter~\ref{chap:g^2}, we functionally
differentiate \eq{eq:feom1}, ignoring the interaction terms, and we
get the tree-level propagator in configuration space
\be
0=\left[\imath\partial_{0x}-m\right]_{\al\ba}\ev{\imath\chi_\ga(z)
\imath\ov{\chi}_\ba(x)}^{(0)}-\imath\de_{\ga\al}\de(z-x)+{\cal O}
\left(1/m\right).
\ee

Naively, one would write the solutions as
\be
W_{\ov{q}q\ba\al}^{(0)}(k)=\frac{-\imath\de_{\ba\al}}{\left[k_0-
m\right]}+{\cal O}\left(1/m\right),\;\;\;\;
W_{q\ov{q}\ga\ba}^{(0)}(k)=\frac{-\imath\de_{\ga\ba}}{\left[k_0+
m\right]}+{\cal O}\left(1/m\right).
\ee
These propagators will turn to have a couple of striking features,
which will be extensively discussed after deriving the nonperturbative
solution for the gap equation.  At this stage, we only emphasize that
the quark and antiquark propagators must be treated separately, and
this will turn to have important consequences for the bound state
equations.  Due to the mass expansion, we only have a single pole in
the complex $k_0$-plane, as opposed to the conventional quark
propagator, which possesses a pair of simple poles. Hence, in order to
define the Fourier transform, one must first define the Feynman
prescription for handling the poles in the energy integral. For the
quark propagator, we write
\be
W_{\ov{q}q\ba\al}^{(0)}(k)=\frac{-\imath\de_{\ba\al}}{\left[k_0-m+
\imath\e\right]}+{\cal O}\left(1/m\right).
\label{eq:quarkpropagator}
\ee
Let us state for the moment that for the antiquark propagator, the
following Feynman prescription is assigned (this will be explained in
the context of the \BS equation for bound states, see below):
\be
W_{q\ov{q}\ga\ba}^{(0)}(k)=\frac{-\imath\de_{\ga\ba}}{\left[k_0+m+
\imath\e\right]}+{\cal O}\left(1/m\right).
\label{eq:quarkantipropagator}
\ee

For the proper two-point function, we use functional derivatives of
\eq{eq:feom2} and we get directly in momentum space
\bea
\G_{\ov{q}q\al\ba}^{(0)}(k)&=&\imath\left[k_0-m\right]\de_{\al\ba}+
{\cal O}\left(1/m\right),\nonumber\\
\G_{q\ov{q}\al\ba}^{(0)}(k)&=&\imath\left[k_0+m\right]\de_{\al\ba}+
{\cal O}\left(1/m\right),
\label{eq:feyn1}
\eea
Since the two-point function requires no Feynman prescription and is
diagonal in the outer product of the fundamental color, flavor and
spinor spaces, we have the following relation between the quark and
antiquark two-point functions:
\be
\G_{\ov{q}q\al\ba}^{(0)}(k)=-\G_{q\ov{q}\al\ba}^{(0)}(-k)
\ee
 In addition, 
$W_{\ov{q}q}\G_{\ov{q}q}=\mathds{1}$ 
as usual.
For the three-point functions we obtain
\bea
\G_{\ov{q}q\si\al\ba}^{(0)a}(k_1,k_2,k_3)&=&\left[gT^a\right]_
{\al\ba}+{\cal O}\left(1/m\right),\nonumber\\
\G_{q\ov{q}\si\al\ba}^{(0)a}(k_1,k_2,k_3)&=&-\left[gT^a\right]_
{\ba\al}+{\cal O}\left(1/m\right).
\label{eq:feyn2}
\eea
Notice the ordering of the indices for the $\G_{q\ov{q}\si}$ vertex.
Importantly, the tree-level spatial quark-gluon vertex does not appear
at leading order in the mass expansion:
\be
\G_{\ov{q}qA\al\ba i}^{(0)a}=\G_{q\ov{q}A\al\ba i}^{(0)a}\sim{\cal O}
\left(1/m\right).
\label{eq:spglsupp}
\ee
Let us again stress that, because of our insistence of using the full
quark sources, all the nonperturbative equations involving the quarks
(the \DS equations for two-point and three-point functions, the \ST
identities, the \BS equation and the Faddeev equation) will not alter
their form at leading order in the mass expansion, the only
alterations being at the level of the tree-level factors.

\section{Truncation scheme}

In order to solve the nonperturbative system we must further specify
our truncation scheme. In the context of the heavy mass expansion, we
propose to consider only the dressed two-point functions of the \YM
sector (i.e., the nonperturbative gluon propagators) and to set all
the pure \YM vertices and higher $n$-point functions occurring in the
quark equations to zero.

Since the tree-level spatial quark-gluon vertex is suppressed by the
mass expansion, in our approximation any diagram containing this
vertex will not contribute at one-loop order in perturbation theory.
Further, the fully temporal gluon Green's functions
$\G_{\si\si\si},\G_{\si\si\si\si},\ldots$ are zero at tree-level
\cite{Watson:2007vc}, and this means that the leading order
perturbative corrections containing this vertices again vanish. This
implies that that the number of loop diagrams arising because of the
\YM vertices is heavily restricted and they first contribute to the
next to leading order in perturbation theory. Physically, the most
important point that will emerge is that when we set the \YM vertices
to zero, we exclude the non-Abelian part of the charge screening
mechanism of the quark color charge and any potential glueball states.
On the other hand, the charge screening mechanism and glueball
contributions of the gluon field (i.e., the color string) is
implicitly encoded in the nonperturbative form of the temporal gluon
propagator. We will come back to the physical implications of this
truncation after deriving the confining potential between a quark and
an antiquark from the \BS equation.

With the truncation scheme as outlined above, the \YM sector reduces
to the inclusion of a single object: the temporal gluon propagator
which is written as \cite{Watson:2007vc}
\be
W_{\si\si}^{ab}(k)=\de^{ab}\frac{\imath}{\vec{k}^2}D_{\si\si}(\vec
{k}^2).
\label{eq:gtemp}
\ee
There are three important features to this propagator.  Firstly, there
are indications on the lattice that the dressing function $D_{\si\si}$
is largely independent of energy \cite{Quandt:2008zj}, justifying the
energy independence of the above form. As we already emphasized in the
first chapter, within the first order formalism one can justify that
the temporal gluon propagator must have some part that is constant in
the energy in order to cancel closed ghost loops and resolve the
Coulomb gauge energy divergences\footnote{See also the more formal
considerations of Ref.~\cite{Cucchieri:2000hv}.}.  Second, the lattice
analysis indicates that the dressing function $D_{\si\si}$ is infrared
divergent and is likely to behave as $1/ \vec{k}^2$ for vanishing
$\vec{k}^2$.  Since we are interested mainly in the relationship
between $D_{\si\si}$ (as the input of the \YM sector) and the string
tension, we will not need the specific form until towards the end.
Third, the product $g^2D_{\si \si}$ is a renormalization group
invariant quantity and thus a good candidate for being related to the
physical string tension \cite{Zwanziger:1998ez,Watson:2008fb}.

\section{Gap equation}
\subsection{Nonperturbative treatment}

Let us begin by considering the \DS equation for the quark two-point
proper function.  In full, second order formalism, QCD it is given
by\footnote{This can be directly inferred from the expression
\eq{eq:gapeqms}, in the first order formalism, by setting the
auxiliary fields $\vec\pi$, $\phi$ to zero.}:
\bea
\G_{\ov{q}q\al\de}(k)&=&\G_{\ov{q}q\al\de}^{(0)}(k)\nonumber\\
&&+\int\dk{\w}\left\{\G_{\ov{q}q\si\al\ba}^{(0)a}(k,-\w,\w-k)
W_{\ov{q}q\ba\ga}(\w)\G_{\ov{q}q\si\ga\de}^{b}(\w,-k,k-\w)
W_{\si\si}^{ab}(k-\w)
\right.\nonumber\\ &&
\left.+\G_{\ov{q}qA\al\ba
i}^{(0)a}(k,-\w,\w-k)W_{\ov{q}q\ba\ga}(\w)
\G_{ \ov{q}qA\ga\de j}^{b}(\w,-k,k-\w)
W_{AAij}^{ab}(k-\w) \right\}\nonumber\\
\label{eq:gap}
\eea
(the spatial gluon propagator $W_{AA}$ will be unimportant here
because of the suppression \eq{eq:spglsupp} of the spatial quark-gluon
vertex).  The gap equation is supplemented by the \ST identity,
\eq{eq:stid3}, derived in Chapter~\ref{chap:st} .

In order to use the \ST identity as input for solving the gap
equation, we first apply our truncation scheme in the context of the
heavy mass expansion at leading order.  Starting with the dressed
spatial quark-gluon vertex, consider the terms that contribute to the
\DS equation shown schematically in \fig{fig:quark-gluon-vertex}.
According to the truncation scheme, we set all \YM vertices to zero,
meaning that diagrams (c) and (e-i) are excluded.  This then leaves us
with the tree-level term (a) and the quark contributions (b), (d).
However, all of these involve at least one tree-level spatial
quark-gluon vertex, which is not present at leading order in the mass
expansion.  Thus, we obtain the nonperturbative result that
\be
\G_{\ov{q}qA\al\ba i}^{a}(k_1,k_2,k_3)\sim{\cal O}\left(1/m\right).
\label{eq:qbqa}
\ee
Similarly, the ghost-quark kernels of the \ST identity, given their
definition, \eq{eq:kern0}, involve \YM vertices and can be neglected.
Thus, in our truncation scheme and at leading order in the mass
expansion, the \ST identity reduces to
\be
k_3^0\G_{\ov{q}q\si\al\ba}^{d}(k_1,k_2,k_3)=
\G_{\ov{q}q\al\de}(k_1)\left[\imath gT^d\right]_{\de\ba}-\left[
\imath gT^d\right]_{\al\de}\G_{\ov{q}q\de\ba}(-k_2)+{\cal O}
\left(1/m\right).
\ee
Clearly, the \ST identity under truncations results in an Abelian type
Ward identity.  Moreover, since the temporal quark-gluon vertex is
simply multiplied by the temporal gluon energy (an essential feature
of Coulomb gauge \ST identities, as opposed to covariant gauges) and
the quark proper two-point function is color diagonal, we can
immediately write the solution:
\be
\G_{\ov{q}q\si\al\ba}^{d}(k_1,k_2,k_3)=\frac{\imath g}{k_3^0}\left
\{T^d\left[\G_{\ov{q}q}(k_1)-\G_{\ov{q}q}(-k_2)\right]\right\}_{\al
\ba}+{\cal O}\left(1/m\right).
\label{eq:qbqs}
\ee
The above solution is trivially satisfied at tree-level.  The apparent
singularity arising for vanishing gluon energy ($k_3^0=0$, but
$\vec{k}_3\neq0$) must somehow be canceled by the difference of proper
quark two-point functions. Since the spatial momentum configuration is
arbitrary, this leads to the requirement that
$\G_{\ov{q}q}(k)\rightarrow\G_{\ov{q}q}(k_0) +{\cal
O}\left(1/m\right)$. Later on we shall see that our solution does
indeed satisfy this condition.  The demand that the nonperturbative
vertex solution to the Coulomb gauge \ST identity be free of kinematic
divergences (here, simply the $1/k_3^0$ factor) is a variation of the
familiar covariant gauge situation considered in
Ref.~\cite{Ball:1980ay}.

Inserting the results, \eq{eq:qbqs} and \eq{eq:qbqa}, for the
vertices, using the Feynman rules given by \eq{eq:feyn1},
\eq{eq:feyn2}, and the temporal gluon propagator given by
\eq{eq:gtemp} and resolving the color structure, the nonperturbative
gap equation, \eq{eq:gap}, under truncation and at leading order in
the mass expansion thus reads
\bea
\G_{\ov{q}q\al\de}(k_0)&=&
\!\!\imath\left[k_0-m\right]\de_{\al\de}-g^2C_F
\int\frac{\dk{\w}D_{\si\si}(\vec{k}-\vec{\w})}{(k_0-\w_0)(\vec{k}-
\vec{\w})^2}W_{\ov{q}q\al\ba}(\w_0)\left[\G_{\ov{q}q}(\w_0)-\G_{\ov
{q}q}(k_0)\right]_{\ba\de}\nonumber\\
&&\!\!+{\cal O}\left(1/m\right).
\label{eq:gap1}
\eea
There exists one particularly simple solution to this equation, given
by
\bea
\G_{\ov{q}q\al\ba}(k)&=&
\imath\de_{\al\ba}\left[k_0-m-{\cal I}_r\right]+{\cal O}
\left(1/m\right) \label{eq:gapsol}\\
W_{\ov{q}q\al\ba}(k_0)&=&\frac{-\imath\de_{\al\ba}}
{\left[k_0-m-{\cal I}_r+\imath\e\right]}+{\cal O}\left(1/m\right),
\label{eq:quarkpropnonpert}
\eea
with the constant
[$\dk{\vec{\w}}=d^3\vec{\w}/(2\pi)^3$]
\be
 {\cal I}_r =\frac{1}{2}g^2C_F
\int_r\frac{\dk{\vec{\w}}D_{\si\si}(\vec{\w})}{\vec{\w}^2}
+{\cal O}\left(1/m\right).
\label{eq:const}
\ee
A short comment regarding the ordering of the limits in the spatial
and temporal integrals and the potential divergences in the constant $
{\cal I}_r$ is in order.  Here, we state that the temporal integral is
performed first under the condition that the spatial integral is
somehow regularized, i.e. finite (the implicit regularization is
signaled by the subscript ``$r$'')\footnote{This regularization will
be assumed throughout the rest of this work.}.  Extracting the
expression for the constant $ {\cal I}_r$ from the gap equation under
truncation, \eq{eq:gap1}, we insert the solution given by
\eq{eq:gapsol}, \eq{eq:quarkpropnonpert} (assuming that the spatial
integral is regularized), and obtain
\bea
{\cal I}_r&=&g^2C_F\int_r\frac{\dk{\vec{\w}}D_{\si\si}(\vec{k}-
\vec{\w})}{(\vec{k}-\vec{\w})^2}\frac{\imath}{2\pi}\lim_{R
\rightarrow\infty}\int_{-R}^{R}\frac{d\w_0}{\left[\w_0-m-{\cal I}_r+
\imath\e\right]}+{\cal O}\left(1/m\right)\nonumber\\
&=&\frac{1}{2}g^2C_F\int_r\frac{\dk{\vec{\w}}D_{\si\si}(\vec{k}-
\vec{\w})}{(\vec{k}-\vec{\w})^2}+{\cal O}\left(1/m\right).\label
{eq:const_ren}
\eea
Clearly, the effect of performing the temporal integral first is that
the regularized constant ${\cal I}_r$ in the denominator factor can be
shifted away and hence becomes irrelevant.  Further, by shifting the
integration variable in the spatial integral we find that this
integral involves no external scale, and thus we arrive at our above
result, \eq{eq:const}.

Inserting the solution \eq{eq:gapsol} into the \ST identity, we also
have that for the vertex
\be
\G_{\ov{q}q\si\al\ba}^{d}(k_1,k_2,k_3)=\left[gT^d\right]_{\al\ba}+
{\cal O}\left(1/m\right),
\label{eq:vsol}
\ee
and this means that the dressed temporal quark-gluon vertex is trivial
and the gap equation reduces to the rainbow truncation.

Let us now briefly discuss the physical interpretation of these
results.  Firstly, it might be the case that there exist other
solutions to the truncated gap equation. However, as will be shown in
the next section, the above solution can also be derived from a
semi-perturbative type of expansion.  One has to bear in mind that in
principle the fully nonperturbative solution might not be the same as
the resummed perturbative solution.
  
Secondly, as already outlined when defining the Feynman prescription,
the quark propagator has a single pole, so cannot represent physical
propagation (which requires a covariant double pole) and this arises
obviously from the truncation of the mass expansion.  From
\eq{eq:quarkpropnonpert} it then follows that the closed quark loops
(virtual quark-antiquark pairs) vanish due to the energy integration,
which implies that the theory is quenched in the heavy mass limit (see
also Ref.~\cite{Neubert:1993mb} for an alternate discussion on this
topic):
\be
\int\frac{dk_0}{\left[k_0-m- {\cal I}_r+\imath\e\right]
\left[k_0+p_0-m- {\cal I}_r+\imath\e\right]}=0.
\label{eq:tempint}
\ee
The above result can be interpreted as being a consequence of the
breaking of time reversal symmetry of the full Dirac equation (quark
and antiquark moving either forwards or backwards in time according to
causality), as a result of the mass expansion. Time reversal symmetry
breaking means that there is only the quark (or only the antiquark)
moving forward in time, corresponding to the above Feynman
prescription, and, in this sense, the corresponding antiquark (or
quark) is not present.  On the other hand, the closed quark loop
involves a quark going backwards in time (similarly for an antiquark
loop), which is prohibited according to our argumentation, so that
such closed quark loop integrals vanish at leading order.  Clearly,
the breaking of the time reversal symmetry is also reflected in the
presence of only a single pole in the heavy quark propagator.

Thirdly, the quark propagator \eq{eq:quarkpropnonpert} is diagonal in
the outer product of the fundamental color, flavor and spinor spaces
as a consequence of the mass expansion -- physically this corresponds
to the decoupling of the spin from the heavy quark system.  In fact,
$W_{\ov{q}q}^{(0)}$ is identical to the heavy quark tree-level
propagator \cite{Neubert:1993mb} up to the appearance of the mass
term, and this is due to the fact that in HQET one uses the sources
for the large $h$-fields directly, while we retain the sources of the
full quark fields.  Also, note that the kinetic term of the
(tree-level) propagator would read $-\vec{k}^2/2m$ in the denominator
factor and hence appears at higher order in the mass expansion.  Such
terms are obviously important to the UV properties of the loop
integrals but do not play any role in the infrared limit considered
here.

Finally, the fact that the solution involves potentially divergent
constants is not a comfortable situation but does not necessarily
contradict the physics, since the position of the pole has no physical
meaning (the quark can never be on-shell). The poles in the quark
propagator are situated at infinity (thanks to ${\cal I}_r$ as the
regularization is removed) meaning that either one requires infinite
energy to create a quark from the vacuum or, if a a hadronic system is
considered, only the relative energy is important. Indeed, it was
shown some time ago \cite{Adler:1984ri} that the divergence of the
absolute energy has no physical meaning and only the relative energy
(derived from the \BS equation, see below) must be considered.
Further, the divergences appearing in the quark propagator have no
interpretation with regards to renormalization, at least within the
context of the mass expansion to leading order.  The mass parameter
cannot be renormalized simply because one cannot construct an
appropriate counterterm in the action.  Also, the quark field
renormalization is trivial at leading order, as one sees from the
explicit form of the temporal quark-gluon vertex, \eq{eq:vsol}.

Having discussed the quark propagator, let us now discuss the
antiquark propagator.  Recall that at tree-level, we used a different
Feynman prescription for the two denominator factors and this gives
rise to some rather interesting physical consequences.  As previously
discussed, the heavy mass expansion employed here breaks the charge
conjugation symmetry relating particle and antiparticle, so we cannot
expect that the two propagators are necessarily equivalent.  Starting
with the gap equation for full QCD, \eq{eq:gap}, we reverse the
ordering of the quark and antiquark functional derivatives that form
the quark Green's functions (still within the context of the full
quark fields and sources) and rearrange the ordering to get the gap
equation for the antiquark propagator:
\bea
\lefteqn{-\G_{q\ov{q}\de\al}(-k)=
-\G_{q\ov{q}\de\al}^{(0)}(-k)}\nonumber\\
&&-\int\dk{\w}
\left\{\G_{q\ov{q}\si\de\ga}^{b}(-k,\w,k-\w)W_{q\ov{q}\ga\ba}(-\w)
\G_{q\ov{q}\si\ba\al}^{(0)a}(-\w,k,\w-k)W_{\si\si}^{ab}(k-\w)
\right.\nonumber\\
&&+\G_{q\ov{q}A\de\ga j}^{b}(-k,\w,k-\w)W_{q\ov{q}\ga\ba}(-\w)
\G_{q\ov{q}A\ba\al i}^{(0)a}(-\w,k,\w-k)W_{AAij}^{ab}(k-\w)
\left.\right\}.
\label{eq:bgap}
\eea
Applying our truncation scheme reduces the above to
\bea
\lefteqn{\G_{q\ov{q}\de\al}(-k)=\G_{q\ov{q}\de\al}^{(0)}(-k) }
\nonumber\\
&&+\int\dk{\w}\G_{q\ov{q}\si\de\ga}^{b}(-k,\w,k-\w)
W_{q\ov{q}\ga\ba}(-\w)\G_{q\ov{q}
\si\ba\al}^{(0)a}(-\w,k,\w-k)W_{\si\si}^{ab}(k-\w)+{\cal O}
\left(1/m\right). \nonumber\\
\label{eq:bglap1}
\eea
In similar fashion, we have the \ST identity for the antiquark-gluon
vertex:
\be
-k_3^0\G_{q\ov{q}\si\ba\al}^{d}(k_2,k_1,k_3)=
+\G_{q\ov{q}\ba\de}(k_2)\left[\imath gT^d\right]_{\de\al}^T
-\left[\imath gT^d
\right]_{\ba\de}^T\G_{q\ov{q}\de\al}(-k_1)+{\cal O}\left(1/m\right).
\ee
Before we give the solution to \eq{eq:bglap1}, let us remark that in
order to construct a quark-antiquark pair in the \BS equation (which
has a physical interpretation of a bound state equation), one is not
considering a virtual quark-antiquark pair but rather a system
composed of two separate unphysical particles.\footnote{It is known
that in Coulomb gauge, Gauss' law forbids the creation of a colored
state in isolation (the total color charge is conserved and vanishing
\cite{Reinhardt:2008pr}) and so, the Feynman prescription for the
quark (or the antiquark) propagator has no physical meaning in
isolation.}  Since the quark and antiquark lines of the \BS equation
are never connected by a primitive vertex (unlike the closed quark
loop discussed above), we are allowed to assign the following Feynman
prescription for the antiquark propagator:
\bea
W_{q\ov{q}\al\ba}(k)&=&
\frac{-\imath\de_{\al\ba}}
{\left[k_0+m-{\cal I}_r+\imath\e\right]}+{\cal O}\left(1/m\right),
\label{eq:bgapsol0}\\
\G_{q\ov{q}\al\ba}(k)&=&
\imath\de_{\al\ba}\left[k_0+m-{\cal I}_r\right]
+{\cal O}\left(1/m\right)
\label{eq:bgapsol}
\eea
with the corresponding solution for the vertex
\be
\G_{q\ov{q}\si\al\ba}^{d}(k_1,k_2,k_3)=
-\left[gT^d\right]_{\ba\al}+
{\cal O}\left(1/m\right).
\label{eq:bvsol}
\ee
In the above, notice that the sign of the loop correction has changed
and this will result (in the context of the bound state studies for
mesons and diquarks) in a physical interpretation for the
quark-antiquark \BS equation as a whole.

\subsection{Semiperturbative treatment}

After solving the gap (and anti-gap) equation, we have seen that the
solutions for the proper two-point function leads to a
(nonperturbatively) bare temporal quark-gluon (and antiquark-gluon)
vertex. In this section, we will reconsider this vertex within the
context of a semiperturbative analysis. This will then introduce a
technical feature crucial for considering the \BS equation
nonperturbatively.

\begin{figure}[t]
\centering\includegraphics[width=0.45\linewidth]{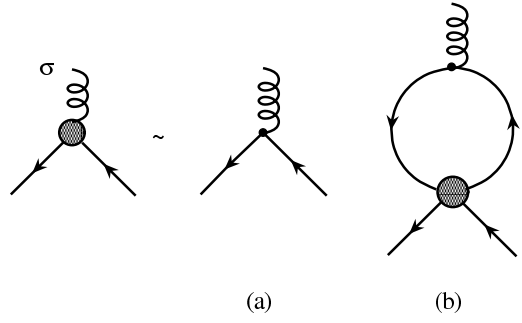}
\caption{\label{fig:qbqs}Diagrams that contribute under our truncation
scheme to the \DS equation for the \emph{temporal} quark-gluon vertex
(without prefactors or signs).  Internal propagators are fully dressed
and blobs represent dressed proper vertex and (reducible) kernels.
Internal propagators represented by solid lines represent the quark
propagator.}
\end{figure}

\begin{figure}[t]
\centering\includegraphics[width=0.35\linewidth]{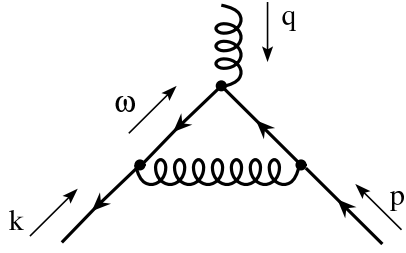}
\caption{\label{fig:qbqslad}Ladder type loop correction to the
temporal quark-gluon vertex.  Internal propagators are fully dressed:
solid lines represent the quark propagator and springs denote the
temporal gluon propagator.}
\end{figure}

\begin{figure}[t]
\centering\includegraphics[width=0.3\linewidth]{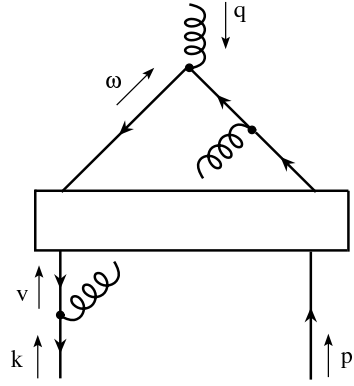}
\caption{\label{fig:qbqscbox}Generic crossed box type loop correction
to the temporal quark-gluon vertex.  Internal propagators are fully
dressed: solid lines represent the quark propagator and springs denote
the temporal gluon propagator.  The box represents any combination of
interactions allowed under our truncation.}
\end{figure}

 Under our truncation scheme, the nonperturbative \DS equation for the
temporal quark-gluon vertex involves the diagrams shown in
Fig.~\ref{fig:qbqs}.  Semiperturbative expansion means in this case
that in the loop expansion all internal propagators are taken to be
dressed, but all internal vertices are tree-level.  Clearly, the goal
is to show that all the loop corrections (contained in diagram (b) of
Fig.~\ref{fig:qbqs}) vanish. For this, it suffices to consider two
types of diagram, given in Figs.~\ref{fig:qbqslad} and
\ref{fig:qbqscbox}.

The easiest to start with is Fig.~\ref{fig:qbqslad}, where a single
ladder exchange correction to the temporal quark-gluon vertex is
considered.  This diagram (neglecting the overall color and
prefactors) gives rise to the following scalar integral:
\be
\int_r\frac{\dk{\vec{\w}}D_{\si\si}(\vec{k}-\vec{\w})}{(\vec{k}-
\vec{\w})^2}\frac{1}{2\pi}\int_{-\infty}^{\infty}\frac{d\w_0}{\left
[\w_0-m-{\cal I}_r+\imath\e\right]\left[\w_0+q_0-m-{\cal I}_r
+\imath\e\right]}.
\ee
Now we apply the following identity (for finite, real $a,b$; the case
$a=b$ is trivial):
\be
\int_{-\infty}^{\infty}\frac{dz}{\left[z-a+\imath\e\right]
\left[z-b+\imath\e\right]}=\frac{1}{(a-b)}\int_{-\infty}^{\infty}
dz\left\{\frac{1}{\left[z-a+\imath\e\right]}-\frac{1}{\left[z-b+
\imath\e\right]}\right\}=0.
\label{eq:intid}
\ee
Thus we see that where the spatial integral is regularized, the
temporal integral vanishes.  It is simple to see that the planar
one-loop diagrams with two or more external temporal gluon legs (which
under the truncation scheme considered here connect only to the
internal quark line) and one internal temporal gluon will also vanish.

Now let us consider a generic crossed box (nonplanar) type of diagram,
illustrated in Fig.~\ref{fig:qbqscbox}.  Considering only the temporal
double integral components of the explicit internal quark propagators,
we have the following form
\bea
&&\int_{-\infty}^{\infty}\frac{d\w_0\,dv_0}{\left[v_0-a_1+\imath\e
\right]\left[\w_0-a_2+\imath\e\right]\left[\w_0+q_0-a_3+\imath\e
\right]\left[\w_0-v_0-p_0-a_4+\imath\e\right]}\nonumber\\&&
=\int_{-\infty}^{\infty}\frac{d\w_0}{\left[\w_0-a_2+\imath\e\right]
\left[\w_0+q_0-a_3+\imath\e\right]\left[\w_0-p_0-a_1-a_4+2\imath\e
\right]}\nonumber\\&&
\times\int_{-\infty}^{\infty}dv_0\left\{\frac{1}{\left[v_0-a_1+
\imath\e\right]}-\frac{1}{\left[v_0-\w_0+p_0+a_4-\imath\e\right]}
\right\}\nonumber\\&&
=-2\pi\imath\int_{-\infty}^{\infty}\frac{d\w_0}{\left[\w_0-a_2+
\imath\e\right]\left[\w_0+q_0-a_3+\imath\e\right]\left[\w_0-p_0-a_1
-a_4+2\imath\e\right]}\nonumber\\&&
=0
\eea
where in the last line, we have used a variation of the identity
\eq{eq:intid}.  Thus we have the result that the generic crossed box
type of diagram shown in Fig.~\ref{fig:qbqscbox} also vanishes.

With the result that both the single ladder type exchange diagram and
the generic crossed box diagrams considered so far vanish, it is easy
to see that any vertex dressing diagram will vanish (including all
subdiagrams such as internal vertex corrections and so on), since all
associated diagrams are merely variations or combinations of these two
under our truncation scheme.  We stress that this result is a
consequence of the fact that the energy and Feynman prescription of
the denominator factors follow the quark line through the diagram so
that eventually the identity, \eq{eq:intid}, can be used.  It is also
precisely the reason why all closed quark loops vanish, according to
the relation \eq{eq:tempint}.

Thus, the semiperturbative expansion confirms our previous result that
the temporal quark-gluon vertex remains bare to all orders.  Clearly,
the result also applies to the antiquark-gluon vertex.  With the
corresponding simple forms for the self-energy integrals of the gap
and anti-gap equations (which as we recall, reduce to the rainbow
truncation), the results for the quark propagator functions are also
confirmed.  Notice though that whilst the all orders semiperturbative
result must match the nonperturbative result, the converse is not
necessarily true.  It remains the case that there may exist further
solutions, but these must be purely nonperturbative in character if
they exist.

To summarize, in this section we have introduced a set of valuable
identities for the energy integrals, which confirm the nonperturbative
case previously studied.  This is very useful for further
investigations, since we are allowed to apply them with confidence to
study the bound state equations in the next section.

\section{Bound state equations}
\label{eq:secbstates}

Bound states of quarks appear as free-particle poles in their
respective $n$-point Green's functions. For example, the
quark-antiquark, and three-quark Green's functions exhibit meson and
baryon poles, respectively. In this section, we will consider the
corresponding bound state equations in the limit of the heavy quark
mass -- named \BS equation for mesons and Faddeev equation for baryons
-- and interpret them in connection with the linearly rising potential
which confines the quarks.  As has been emphasized, due to the fact
that we include the full quark sources and fields in our generating
functional, we are allowed to use the full functional
(nonperturbative) equations as a starting point and then subsequently
apply our mass expansion and truncation scheme.  In this context, we
will also seek for a relationship between \YM sector of the theory and
nonperturbative external physical scale (i.e., the string tension).

\subsection{\BS equation for mesons and diquarks}

\begin{figure}[t]
\centering\includegraphics[width=0.5\linewidth]{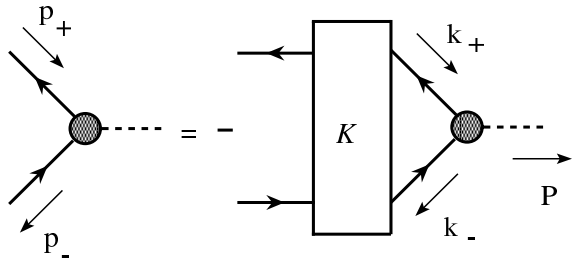}
\caption{\label{fig:fbse}Homogeneous \BS equation for quark-antiquark
bound states.  Internal propagators are fully dressed and solid lines
represent the quark propagator.  The box represents the \BS kernel $K$
and filled blobs represent the \BS vertex function $\G$ with the
(external) bound state leg given by a dashed line.  See text for
details.}
\end{figure}

Let us now consider the homogeneous \BS equation for quark-antiquark
bound states.  In full QCD this equation reads
\be
\G_{\al\ba}(p;P)=-\int\dk{k}K_{\al\ba;\de\ga}(p,k;P)\left[W_{\ov{q}
q}(k_+)\G(k;P)W_{\ov{q}q}(k_-)\right]_{\ga\de}
\ee
In the above, the minus sign arises from the definitions of the
Legendre transform and Green's functions in Coulomb gauge. This will
be shown in Chapter~\ref{chap:nGreen}, where the 4-point Green's
function (which includes the homogeneous \BS equation) will be
explicitly derived.  The momenta of the quarks are given by $k_+=k+\xi
P$, $k_-=k+(\xi-1)P$ (similarly for $p_\pm$), $P$ is the pole
4-momentum of the bound state (assuming that a solution exists),
$\xi=[0,1]$ is the so-called momentum sharing fraction that dictates
how much of the total meson momentum is carried by each quark
constituent.  $K$ represents the \BS kernel and $\G$ is the \BS vertex
function for the particular bound state that one is considering and
whose indices explicitly denote only its quark content. Later on, we
will explicitly investigate what color, flavor and spin structure the
solutions may have. We also mention that physically, the results
should be independent of $\xi$ and this has been numerically observed
in phenomenological studies \cite{Alkofer:2002bp}.  The \BS equation
is shown pictorially in Fig.~\ref{fig:fbse}.

To solve the \BS equation, we firstly address the problem of
constructing the kernel $K$.  For technical reasons the most widely
studied system is based on the ladder kernel which is either
constructed via the interchange of a single gluon (for example
\cite{Maris:1999nt}) or as a phenomenological potential (see for
example Ref.~\cite{Adler:1984ri}).  However, there has been much
recent attention focused on the construction of more sophisticated
kernels.  One key element of the construction is the axialvector
Ward--Takahashi identity [AXWTI], which relates the gap equation to
the \BS kernel and which ensures that chiral symmetry and its
spontaneous breaking are consistently implemented (e.g.,
Refs.~\cite{Adler:1984ri,Bender:1996bb,Watson:2004kd}).  Here, we
shall show that the ladder \BS kernel is \emph{exact} at leading order
in the heavy mass expansion and under our truncation scheme.

\begin{figure}[t]
\centering\includegraphics[width=0.35\linewidth]{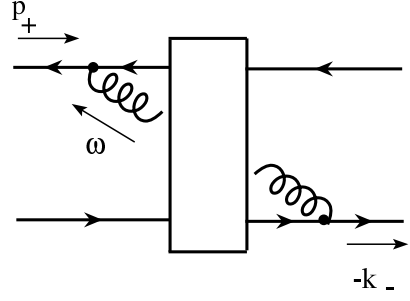}
\caption{\label{fig:bscbox}Generic crossed box type of diagram that
contributes to the \BS kernel. Internal propagators are fully dressed,
whereas vertices are tree-level.  The upper (lower) solid line denotes
the quark (antiquark) propagator; springs denote the temporal gluon
propagator and the box represents any combination of nontrivial
interactions allowed under our truncation scheme.  See text for
details.}
\end{figure}

To construct the kernel $K$, we follow the semiperturbative analysis
of the previous section. As before, at leading order in the mass
expansion and under our truncation scheme we only have the temporal
quark-gluon and antiquark-gluon vertices, both of which have been
shown to be given by their tree-level forms. Consider now the generic
semiperturbative crossed box contribution, which contains all possible
nontrivial contributions allowed within our truncation scheme.  This
diagram is depicted in Fig.~\ref{fig:bscbox}.  Such a diagram has at
least the following terms in the temporal integral (as before, we
assume that the spatial integral is regularized and finite so that we
are able to firstly perform the temporal integral without
complication)
\be
\int\frac{d\w_0}{\left[\w_0+p_{+}^0-m-{\cal I}_r+\imath\e\right]
\ldots\left[\w_0-k_{-}^0+m-{\cal I}_r+\imath\e\right]}.
\ee
The first factor corresponds to the explicit quark (upper) propagator,
the last factor to the explicit antiquark (lower) propagator.  The
dots represent the multiple internal propagator factors which carry
the same dependence on the integration energy $\w_0$ $and$ have the
same relative sign for the Feynman prescription term, i.e.,
$\w_0+\imath\e$, regardless of whether they originate from internal
quark or antiquark propagators.  Therefore, this type of integral can
always be reduced to the difference of integrals over a simple pole
and with the same sign for carrying out the analytic integration, just
as in \eq{eq:intid}.  Thus, all generic crossed box diagrams in the
\BS kernel are zero and one is left with simply the ladder
contribution to the kernel.  In fact, this result can also be inferred
from the AXWTI: this identity connects the self-energy term of the gap
equation and the \BS kernel and since it has been shown explicitly
that the self-energy integral reduces to the rainbow truncation, the
corresponding \BS kernel is simply given by ladder exchange.

Having analyzed the kernel $K$, let us now inspect the quark and
antiquark pieces of the \BS equation.  Recalling that the antiquark
propagator must be treated as distinct from the quark propagator, in
Fig.~\ref{fig:fbse} we must explicitly specify which line is assigned
to the quark propagator and which one, to the antiquark propagator.
For the quark-antiquark system under consideration (later on we will
analyze the quark-quark, or diquark system), the \BS equation more
properly reads
\be
\G_{\al\ba}(p;P)=-\int\dk{k}K_{\al\ba;\de\ga}(p,k;P)\left[W_{\ov{q}
q}(k_+)\G(k;P)(-)W_{q\ov{q}}^T(-k_-)\right]_{\ga\de}
\label{eq:bs1}
\ee
where we have explicitly identified the antiquark propagator
contribution (it corresponds to the lower line of Fig.~\ref{fig:fbse})
by reordering the functional derivatives,
i.e. $W_{q\ov{q}}(k_-)=-W_{q\ov{q}}^T(-k_-)$. The above equation is
still valid in full QCD. Further, we explicitly replace the tree-level
form for the kernel, and we arrive at the following expression \BS
equation for the quark-antiquark system, at leading order in the mass
expansion and within our truncation scheme:
\bea
\G_{\al\ba}(p;P)&=&-\int\dk{k}
\left[\G_{\ov{q} q\si}^a(p_+,-k_+,k-p)
W_{\ov{q}q}(k_+)\right]_{\al\la}\nonumber\\
&\times&\!\!\!\!\!
\left[(-)W_{q\ov{q} }^T(-k_-)(-)
\G_{q\ov{q}\si}^b(-p_-,k_-,p-k)\right]_{\ba\ka}
W_{\si\si}^{ab}(p-k) \G_{\la\ka}(k;P)
+{\cal O}\left(1/m\right)\!. \nonumber\\
\eea
Notice the antiquark contribution within the kernel which absorbs the
explicit minus sign from \eq{eq:bs1}.  Also, recall that we are
implicitly considering only the flavor non-singlet case.  The above
equation can be trivially rewritten as:
\bea
\G_{\al\ba}(p;P)&=&-\int\dk{k}\G_{\ov{q}q\si\al\ga}^{a}(p_+,-k_+,
k-p)W_{\si\si}^{ab}(p-k)\G_{q\ov{q}\si\ba\de}^{bT}(-p_-,k_-,p-k) \nonumber\\
&&\times W_{\ov{q}q\ga\la}(k_+)W_{q\ov{q}\de\ka}^T(-k_-)\G_{\la\ka}(k;P)
+{\cal O}\left(1/m\right). \hspace{4cm}
\eea
Inserting the nonperturbative results for the propagators and vertices
so far, Eqs.~(\ref{eq:quarkpropnonpert}), (\ref{eq:bgapsol}),
(\ref{eq:vsol}), (\ref{eq:bvsol}) and taking the form, \eq{eq:gtemp},
for the temporal gluon propagator, we obtain the equation
\bea
\lefteqn{\G_{\al\ba}(p;P)=
g^2\int_r\frac{\dk{\vec{k}}D_{\si\si}(\vec{p}-\vec
{k})}{(\vec{p}-\vec{k})^2}}\nonumber\\
&&\times\frac{\imath}{2\pi}\int_{-\infty}^\infty
\frac{dk_0}{\left[k_+^0-m-{\cal C}_r+\imath\e\right]
\left[k_-^0-m+{\cal I}_r-\imath\e\right]}
\left[T^a\G(k;P)T^a\right]_{\al\ba}+
{\cal O}\left(1/m\right).\hspace{1cm}
\eea
We see immediately that the flavor and spin structure of the meson
decouples from the problem --- this is well a known property of the
heavy mass expansion.  The color structure will be discussed shortly.
Since the external energy $p_0$ does not enter the right-hand side of
the above equation, we further have that the \BS vertex
$\G_{\al\ba}(p;P)$ must be independent of the quark energy (and only
implicitly dependent on the bound state energy $P_0$). Thus we can
write
\bea
\G_{\al\ba}(\vec{p};P)&=&g^2\int_r\frac{\dk{\vec{k}}
D_{\si\si}(\vec{p}-\vec{k})}{(\vec{p}-\vec{k})^2}
\left[T^a\G(\vec{k};P)T^a\right]_{\al\ba}\nonumber\\
&&\times\frac{\imath}{2\pi}\int_{-\infty}^\infty
\frac{dk_0}{\left[k_+^0-m-{\cal I}_r+\imath\e\right]
\left[k_-^0-m+{\cal I}_r-\imath\e\right]}
+{\cal O}\left(1/m\right)\nonumber\\
&=&-g^2\int_r\frac{\dk{\vec{k}}D_{\si\si}(\vec{p}-\vec{k})}{(\vec
{p}-\vec{k})^2}\frac{\left[T^a\G(\vec{k};P)T^a\right]_{\al\ba}}{
\left[P_0-2{\cal I}_r+2\imath\e\right]}+{\cal O}
\left(1/m\right). \hspace{3cm}
\label{eq:fpaux}
\eea
Thus, at leading order in the mass expansion, inserting the expression
\eq{eq:const} for ${\cal I}_r$, it is now clear that
\bea
\left[P_0-g^2C_F\int_{r}\frac{\dk{\vec{\w}}D_{\si\si}(\vec{\w})}
{\vec{\w}^2}\right]\G_{\al\ba}(\vec{p};P)&=&-g^2\int_{r}\frac{\dk{
\vec{k}}D_{\si\si}(\vec{p}-\vec{k})}{(\vec{p}-\vec{k})^2}\left[T^a
\G(\vec{k};P)T^a\right]_{\al\ba}\nonumber\\
&&+{\cal O}\left(1/m\right).
\eea
We notice that the explicit quark mass contributions of the
self-energy expressions cancel.  This is a feature of the
quark-antiquark \BS equation --- it does not make any reference to the
origins of its constituents (this explains, for example, why the pion
can be a massless bound state of massive constituents).  Physically,
one can visualize that the quark and antiquark are moving with equal
and opposite velocities such that the center of mass is stationary.
This is related explicitly to the choice of Feynman prescription for
the constituent quark and antiquark.  Were the Feynman prescription
for the antiquark chosen to coincide with that of the quark, the
right-hand side of \eq{eq:fpaux} would simply vanish and there would
be certainly no physical quark-antiquark state.  The Feynman
prescription for the antiquark corresponds precisely to a particle
moving with the opposite velocity.  Also, at leading order in the mass
expansion, the momentum sharing parameter, $\xi$, has dropped out, and
therefore the results are independent of $\xi$.  Shifting the
integration momenta, we can write
\be
P_0\G_{\al\ba}(\vec{p};P)=g^2\int_{r}\frac{\dk{\vec{\w}}D_{\si\si}
(\vec{\w})}{\vec{\w}^2}\left\{C_F\G_{\al\ba}(\vec{p};P)-\left[T^a\G
(\vec{p}-\vec{\w};P)T^a\right]_{\al\ba}\right\}+{\cal O}\left(1/m
\right).
\ee
To see the physical meaning of this equation, we rewrite the \BS 
vertex function as a Fourier transform:
\be
\G_{\al\ba}(\vec{p};P)=\int d\vec{y}e^{-\imath\vec{p}\cdot\vec{y}}
\G_{\al\ba}(\vec{y})
\ee
(in the homogeneous \BS equation, the total momentum $P$ denotes 
the solution and is not a variable).  We also assign the following  color 
structure
\be
\left[T^a\G(\vec{y})T^a\right]_{\al\ba}=C_M\G_{\al\ba}(\vec{y})
\ee
where $C_M$ is yet to be identified.  Then, the \BS equation 
reduces to
\bea
\int d\vec{y}e^{-\imath\vec{p}\cdot\vec{y}}P_0\G_{\al\ba}(\vec{y})&=&
\int d\vec{y}e^{-\imath\vec{p}\cdot\vec{y}}g^2\int_{r}\frac{\dk{
\vec{\w}}D_{\si\si}(\vec{\w})}{\vec{\w}^2}\left\{C_F\G_{\al\ba}(
\vec{y})-e^{\imath\vec{\w}\cdot\vec{y}}C_M\G_{\al\ba}(\vec{y})
\right\}\nonumber\\
&&+{\cal O}\left(1/m\right)
\eea
with the simple solution
\be
P_0=g^2\int_{r}\frac{\dk{\vec{\w}}D_{\si\si}(\vec{\w})}{\vec{\w}^2}
\left\{C_F-e^{\imath\vec{\w}\cdot\vec{y}}C_M\right\}+{\cal O}
\left(1/m\right).
\label{eq:p0sol}
\ee
As already mentioned, because the total color charge of the system is
conserved and vanishing \cite{Reinhardt:2008pr}, neither the quark or
antiquark can exist as an independent asymptotic physical state.
Thus, the $\bar qq$ system is either confined, such that the bound
state energy $P_0$ increases linearly as the separation between the
quark and antiquark increases, or the system cannot be physically
created, such that the energy $P_0$ is infinite when the hypothetical
regularization is removed.  Whether the system is confining or
disallowed can only depend on the color structure, since the temporal
gluon propagator dressing function would be common to both situations.

An infrared confining solution is characterized in configuration space
by the solution
\be
P_0=\si|\vec{y}|\textrm{~for large~} |\vec{y}|,
\ee
 where $\si \sim 1GeV/fm$ is called the string tension, such that as
the separation between the quark and antiquark increases, the energy
of the system should increase linearly without bound and infinite
energy input is required to fully separate them\footnote{The small
$|\vec{y}|$ (and large $|\vec{\w}|$) properties are of no concern
here.}  (at least in the absence of unquenching).  The Fourier
transform integral needed to obtain the above form of the solution is
\be
\int\frac{\dk{\vec{\w}}}{\vec{\w}^4}\left[1-e^{\imath\vec{\w}\cdot
\vec{y}}\right]=\frac{|\vec{y}|}{8\pi}.
\ee
This implies that the temporal gluon propagator dressing function
diverges like $1/\vec{\w}^2$  and in addition the condition
\be
C_F=C_M
\ee
must be satisfied. Moreover, with the these conditions, the spatial
integral in \eq{eq:p0sol} becomes automatically convergent and hence
the energy of the system is well-defined, as the regularization is
removed (since we are interested in the low $|\vec\w|$ regime, it
becomes clear that the regularization here would be infrared in
character).  Using the Fierz identity for the generators,
\eq{eq:fierz}, we get the condition
\be
C_F\G_{\al\ga}(\vec{y})\equiv C_M\G_{\al\ga}(\vec{y})=\frac{1}{2}
\de_{\al\ga}\G_{\ba\ba}(\vec{y})-\frac{1}{2N_c}\G_{\al\ga}(\vec{y}),
\ee
or with the definition \eq{eq:casimir},
\be
\G_{\al\ga}(\vec{y})=\de_{\al\ga}\G(\vec{y}).
\ee
In other words, the quark-antiquark \BS equation can only have a
finite solution for \emph{color singlet} states where the divergent
constant integral coming from the unphysical quark self-energy
cancels; otherwise the energy of the system is divergent.

Assuming that in the infrared (as is indicated by the lattice data
\cite{Quandt:2008zj} or by the above argument about the non-existence
of asymptotic quark states), $D_{\si\si}=X/\vec{\w}^2$ where $X$ is
some combination of constants (and further knowing that $g^2X$ is a
renormalization group invariant
\cite{Zwanziger:1998ez,Watson:2008fb}), then
\be
P_0\equiv\si|\vec{y}|=\frac{g^2C_FX}{8\pi}|\vec{y}|+{\cal O}
\left(1/m\right).
\ee
The above result shows that there exists a direct connection between
the string tension and the nonperturbative \YM sector of QCD (i.e.,
the temporal gluon propagator) at least under the truncation scheme
considered here.

Let us now shortly discuss what are the possible consequences of
including the pure \YM vertices in our approach.  Recall that the
confining potential follows from the rainbow-ladder \BS kernel (i.e.,
dressed temporal gluon propagator and tree-level quark-gluon
vertices).  This suggests that the dressing function $D_{\si\si}$ does
implicitly contain all nonperturbative effects associated with the
dynamical dressing of the color charge (including, for example,
potential glueball states), whereas the quark-gluon vertices
correspond to the naked quark color charge.  Pictorially, one can
visualize this as a dressed color string confining two naked color
sources.  Consequently, we anticipate that the effect of including the
non-Abelian corrections to the formalism presented here would not
result in the removal of the linearly rising bound state energy (this
would correspond to the cancellation of the ladder exchange). Instead,
one expects a lowering of the string tension $\si$ by shifting the
pole position by some finite amount.  Physically, this corresponds to
the screening of the quark color charge --- this statement has been
phrased by Zwanziger as ``no confinement without Coulomb
confinement''.

Let us now turn to the diquark \BS equation.  From a technical point
of view, the difference between this and the previously considered
quark-antiquark system is rather simple, but as we shall see it leads
to a completely different physical result.  Since the quark and the
antiquark propagators share the same Feynman prescription relative to
their energy, the result that the crossed box contributions to the \BS
kernel vanish extends to the diquark case.  This means that we can
immediately write down the \BS equation for diquarks, at leading order
in the mass expansion and within our truncation scheme:
\bea
\G_{\al\ba}(p;P)&=&-\int\dk{k}\G_{\ov{q}q\si\al\ga}^{a}(p_+,-k_+,
k-p)W_{\si\si}^{ab}(p-k)\G_{\ov{q}q\si\ba\de}^{b}(-p_-,k_-,p-k)
\nonumber\\
&&\times W_{\ov{q}q\ga\la}(k_+)W_{\ov{q}q\de\ka}(-k_-)
\G_{\la\ka}(k;P)+{\cal O}\left(1/m\right).
\eea
Again, the indices of the \BS vertex function correspond to the quark
content of the diquark and since the flavor and spin content decouple
from the system, we shall only be interested in the color content of
the diquark.  Expanding this out as before, we get the analogous
result
\bea
\left[\!P_0-2m-g^2C_F\!\! \int_{r}\frac{\dk{\vec{\w}}
D_{\si\si}(\vec{\w})}
{\vec{\w}^2}\right]\!\!\G_{\al\ba}(\vec{p};P)
&\!\!\!\!\!=\!\!\!\!\!&\!g^2 \!\!\int_{r}\frac{\dk{\vec
{\w}}D_{\si\si}(\vec{\w})}{\vec{\w}^2}\left[T^a\right]_{\al\la}\!\!
\left[T^a\right]_{\ba\ka}\G_{\la\ka}(\vec{p}-\vec{\w};P)\nonumber\\
&&+{\cal O}\left(1/m\right).
\eea
Fourier transforming as previously, and writing
\be
\left[T^a\right]_{\al\la}\left[T^a\right]_{\ba\ka}\G_{\la\ka}(\vec{
y})=C_D\G_{\al\ba}(\vec{y})
\ee
gives the solution
\be
P_0=2m+g^2\int_{r}\frac{\dk{\vec{\w}}D_{\si\si}(\vec{\w})}{\vec{\w}
^2}\left\{C_F+e^{\imath\vec{\w}\cdot\vec{y}}C_D\right\}+{\cal O}
\left(1/m\right).
\ee
The dependence of the solution on the quark mass simply indicates that
in contrast to the quark-antiquark system, there are now two co-moving
quarks. For the anti-diquark system the solution is identical to the
above, but with \emph{minus} twice the mass -- their velocities are
simply reversed.  The diquark is antisymmetric under interchange of
the two quark legs and this means that the color structure must be
antisymmetric.  Similar to the quark-antiquark system, the system can
only have a confining (finite) energy solution, i.e., if $C_D=-C_F$,
or no finite solution at all.  Using the Fierz identity \eq{eq:fierz},
the color condition then reads
\be
-C_F\G_{\al\ba}(\vec{y})\equiv C_D\G_{\al\ba}(\vec{y})=\frac{1}{2}
\G_{\ba\al}(\vec{y})-\frac{1}{2N_c}\G_{\al\ba}(\vec{y}).
\ee
Demanding the diquark color antisymmetry and with the definition 
\eq{eq:casimir} this becomes
\be
N_c^2-N_c-2=0\;\;\;\;\Rightarrow\;\;\;\;N_c=-1,2.
\ee
This means that in $SU(N_c=2)$ there exists a confined, antisymmetric
bound state of two quarks -- the $SU(2)$ baryon -- and otherwise no
physical states are allowed. In the next section, we will consider the
$SU(3)$ baryon, and explicitly demonstrate how the confinement
potential is obtained in this case.

\subsection{Faddeev equation for baryons}

\begin{figure}
\centering
\includegraphics[width=1.0\linewidth, height=0.11
\textheight]{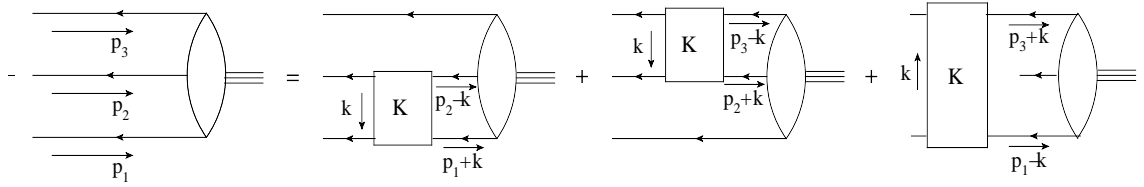}
\caption{\label{fig:faddeev} Faddeev equation for three quark bound
states. Solid lines represent the quark propagator, the box represents
the diquark kernel $K$ and the ellipse represents the Faddeev vertex
function with the bound state leg depicted by a triple-line. See text
for details.}
\end{figure}

Baryons appear as three-quark bound-states in the Faddeev equation,
which generally can be written as:
\be
\G=K^{(3)}\G
\ee
where $\G$ is the quark-baryon vertex and $K^{(3)}$ is the three-body
kernel, containing an irreducible term $K^{(3)}_{ir}$ and the sum of
permuted diquark kernels:
\be
K^{(3)}=K^{(3)}_{ir}+\sum_{i=1}^{3}K^{(2)}_{i}
\ee
The Faddeev equation \cite{Faddeev:1960su} and its subsequent
developments \cite{Taylor:1966zza,Boehm:1976ya} (for an extended
review see \cite{Loring:2001kv} and the textbook \cite{Gloeckle})
provide a general formulation of the relativistic three-body
problem. It is a bound state equation (the direct analogue of the
homogeneous two-body \BS equation) and it has been efficiently applied
in QCD to study baryon states, via the Green's functions of the
theory. Typically, these studies are performed in Landau gauge and,
due to the complexity of the equations, they have been mainly
restricted to rainbow-ladder truncation, where the kernel is reduced
to the single exchange of a dressed gluon. Within this approximation
and by employing phenomenological ans\"atze for the \YM part of the
theory, the nucleon and $\Delta$ properties have been studied
\cite{Hellstern:1997pg,Oettel:2000jj,Eichmann:2009qa,
Nicmorus:2010mc}. Other simplifications include the three-body
spectator formalism \cite{Loring:2001kv}, a Salpeter-type equation
with instantaneous interaction \cite{Stadler:1997iu} or the diquark
correlations \cite{Anselmino:1992vg}.

In the following, we investigate the Faddeev equation for three-quark
bound states, by employing only the permuted two quark kernels
$K^{(2)}_{i}$ (which coincide with kernel appearing in the \BS
equation for diquark states) and neglecting the three-quark
irreducible diagrams, i.e., genuine three-body forces
\cite{Popovici:2010ph}. This approximation is also motivated by the
fact that in the quark-diquark model the binding energy is assumed to
be mainly provided by the two-quark correlations
\cite{Reinhardt:1989rw}. In this truncation, the Faddeev equation
reads (see also \fig{fig:faddeev}):
\bea
\lefteqn{\G_{\al\ba\ga}(p_1,p_2,p_3;P)=}\nonumber\\
&&-\int\dk{k}
\left\{K_{\ba\al;\al^{'}\ba^{'}}(k)
W_{\bar qq\al^{'}\al^{''}}(p_1+k)W_{\bar qq\ba^{'}\ba^{''}}(p_2-k)
\G_{\al^{''}\ba^{''}\ga}(p_1+k,p_2-k,p_3;P)\right.\nonumber\\
&&+K_{\ga\ba;\ba^{'}\ga^{'}}(k)
W_{\bar qq\ba^{'}\ba^{''}}(p_2+k)W_{\bar qq\ga^{'}\ga^{''}}(p_3-k)
\G_{\al\ba^{''}\ga^{''}}(p_1,p_2+k,p_3-k;P)\nonumber\\
&&+\left. K_{\al\ga;\ga^{'}\al^{'}}(k)
W_{\bar qq\ga^{'}\ga^{''}}(p_3+k)W_{\bar qq\al^{'}\al^{''}}(p_1-k)
\G_{\al\ba^{''}\ga^{''}}(p_1-k,p_2,p_3+k;P)
\right\}
\label{eq:faddeev.gen}
\eea
where $p_1,p_2,p_3$ are the momenta of the quarks, $P=p_1+p_2+p_3$ is
the pole 4-momentum of the bound baryon state and $\G$ is the
so-called quark-baryon Faddeev vertex for the particular bound state
under consideration and whose indices denote explicitly only its quark
content. Due to the fact that in the heavy mass limit the spin degrees
of freedom decouple from the system, at leading order in the mass
expansion the Faddeev baryon amplitude $\G_{\al\ba\ga}$ becomes a
Dirac scalar, similar to the heavy quark propagator
\eq{eq:quarkpropnonpert}. The explicit momentum dependence of the
kernels $K$ is abbreviated for notational convenience. As in the
homogeneous \BS equation, the integral equation depends only
parametrically on the total four momentum $P$.

As discussed in the case of the \BS equation, the kernel $K$ reduces
to the ladder approximation (constructed via gluon exchange) and it
reads
\be
K_{\ba\al;\al^{'}\ba^{'}}(k)= 
\G_{\bar qq\si \al\al^{'}}^{a}W_{\si\si}^{ab}(\vec k) 
\G_{\bar qq\si\ba\ba^{'}}^{b}=g^2
T_{\al\al^{'}}^{a}W_{\si\si}^{ab}(\vec k) T_{\ba\ba^{'}}^{b}
\label{eq:kernel}
\ee
with the temporal quark-gluon vertex and the temporal gluon propagator
given by \eq{eq:gtemp} and \eq{eq:feyn1}, respectively. Similar to the
\BS equation for meson bound states, the energy independence of this
propagator will turn to be crucial in the derivation of the confining
potential.

Let us now investigate the energy dependence of the equation
\eq{eq:faddeev.gen}. As shown in the previous section, the \BS kernel
was energy independent, and thus it was straightforward to show that
the \BS vertex itself did not contain an energy dependent part. This
observation was then used to calculate the confining potential from
the \BS equation, via a simple analytical integration over the
relative energy variable. Unfortunately this approach cannot be
extended to baryon states: despite the instantaneous kernel, a
relative energy dependence still remains and thus one cannot assume an
energy-independent Faddeev vertex. Therefore, in order to proceed, we
make the following separable ansatz for the Faddeev vertex:
\be
\G_{\al\ba\ga}( p_1, p_2,  p_3;P)=\Psi_{\al\ba\ga}
\G_{t}( p_1^0, p_2^0, p_3^0;P)
\G_{s}( \vec p_1,\vec  p_2, \vec  p_3;P)
\label{eq:qqqbaryon}
\ee
where we have introduced a purely antisymmetric (in the quark legs)
color factor $\Psi$ (the possible baryon color index is omitted) and
the symmetric (Dirac scalar) temporal and spatial components $\G_t$
and $\G_s$, respectively.

Inserting the explicit form of the kernel \eq{eq:kernel} and the
quark-baryon vertex ansatz \eq{eq:qqqbaryon}, the Faddeev equation
\eq{eq:faddeev.gen} can be explicitly written as (for simplicity we
drop the label $P$ in the arguments of the vertex functions):
\bea
\lefteqn{\G_{\al\ba\ga}(p_1,p_2,p_3)=-g^2
T_{\al\tau}^{a} T_{\ba\kappa}^{a}\Psi_{\tau\kappa\ga}
\int \dk{k}
W_{\si\si}(\vec k)} \nonumber\\
&&\times W_{\bar qq}(p_1+k)W_{\bar qq}(p_2-k)
\G_{t}( p_1^0+k_0, p_2^0-k_0, p_3^0)
\G_{s}( \vec p_1+\vec k,\vec p_2-\vec k, \vec  p_3)
+\mbox{c.p.},\hspace{1cm}
\label{eq:faddeev.gen1}
\eea
where the explicit color structure has been extracted 
($W_{\si\si}^{ab}=\de^{ab}W_{\si\si}, 
W_{\bar qq \al\ba}=\de_{\al\ba} W_{\bar qq}$).

Analogous to the \BS equation, we use the Fierz identity for the
generators, \eq{eq:fierz}, to write the color structure as
\be
T_{\al\al^{'}}^{a} T_{\ba\ba^{'}}^{a}\Psi_{\al^{'}\ba^{'}\ga}=
-C_{B} \Psi_{\al\ba\ga}
\ee
with $C_B=(N_c+1)/2N_c$, where $N_c$ is the number of colors, yet to
be identified (i.e., the baryon is not assumed to be a color singlet).

In the next step we perform the Fourier transform for the spatial part
of the equation, recalling that the heavy quark propagator is only a
function of energy. We define the coordinate space vertex function via
its Fourier transform
\be
\G_{s}(\vec{p_1},\vec{p_2},\vec{p_3})=
\int d\vec x_1d \vec x_2  d\vec x_3
e^{-\imath\vec p_1\cdot\vec x_1-\imath\vec p_2\cdot\vec x_2
-\imath\vec p_3\cdot\vec x_3}
\G_{s}(\vec{x}_1,\vec{x}_2,\vec{x}_3)
\ee
such that 
\bea
\lefteqn{\int\dk{\vec k} W_{\si\si}(\vec k)
\G_{s}( \vec p_1+\vec k,\vec p_2-\vec k, \vec  p_3)=}\nonumber\\
&&\int \dk{\vec x_1}\dk{\vec x_2} \dk{\vec x_3}
e^{-\imath\vec p_1\cdot\vec x_1-\imath\vec p_2\cdot\vec x_2
-\imath\vec p_3\cdot\vec x_3}
W_{\si\si}(\vec x_2-\vec x_1)\G_{s}(\vec{x}_1,\vec{x}_2,\vec{x}_3).
\eea
Clearly, the component $\G_s$ trivially simplifies (as before, we have
separated the temporal and spatial integrals, under the assumption the
spatial integral is regularized and finite) and the equation
\eq{eq:faddeev.gen1} reduces to [$\dk{k_0}=d k_{0}/(2\pi)$]:
\bea
\lefteqn{\G_{t}(p_1^0,p_2^0,p_3^0)
\!=\!g^2C_B W_{\si\si}(\vec x_2-\vec x_1)
\!\!\int\!\!\dk{k_0}
W_{\bar qq}(p_1^0+k_0)W_{\bar qq}(p_2^0-k_0)
\G_{t}(p_1^0+k_0,p_2^0-k_0,p_3^0) }\nonumber\\
&&+\mbox{~~c.p.}.\hspace{12,8cm}
\label{eq:faddeev.coordinate}
\eea
At this point we make a further simplification, motivated by the
symmetry of the three-quark system: we restrict to a particular
geometry, namely to equal quark separations, i.e. 
$|\vec r|= |\vec x_2-\vec x_1|=| \vec x_3-\vec x_2| 
=| \vec x_1-\vec x_3|$.  By inserting the explicit form of the
quark propagators, \eq{eq:quarkpropnonpert}, we have
\bea
\lefteqn{
\G_{t}(p_1^0,p_2^0,p_3^0)= -g^2C_B W_{\si\si}(|\vec r|)
\int\dk{k_0}
\frac{\G_{t}(p_1^0+k_0,p_2^0-k_0,p_3^0)}
{\left[p_1^0+k_0-m-{\cal I}_r+\imath\e\right]
\left[p_2^0-k_0-m-{\cal I}_r+\imath\e\right]}}\nonumber\\
&&+\mbox{~c.p.}. \hspace{12,7cm}
\label{eq:faddeev.simpl}
\eea
With the assumption that the vertex $\G_t$ is symmetric under
permutation of quark legs, an ansatz that satisfies this equation is:
\be
\G_{t}(p_1^0,p_2^0,p_3^0)= \sum_{i=1,2,3}
\frac{1}{2P_0-3(p_i^0 +m+{\cal I}_r)+\imath\e}.
\label{eq:gammat}
\ee
Since the explicit derivation is rather technical, we only give here
the solution and defer the details to the
Appendix~\ref{chap:qbvertex}.

Notice that in the expression \eq{eq:gammat} there are simple poles
(in the energy) present. These poles however do not occur for finite
energies and cannot be physical. As discussed, this is also the case
for the quark propagator. Intuitively, when a single heavy quark is
pulled apart from the system, the $qqq$ state becomes equivalent
(i.e., it has the same color quantum numbers) to the $\bar qq$ system
in the sense that the remaining two quarks form a diquark which for
$N_c=3$ would be a color antitriplet configuration, and hence the
physical interpretation of the vertex \eq{eq:gammat} can be directly
related to the heavy quark propagator \eq{eq:quarkpropnonpert}: the
presence of the single pole in \eq{eq:gammat} simply means that this
cannot have the meaning of physical propagation (this would require a
covariant double pole).  Moreover, the divergent constant ${\cal I}_r$
appearing in the absolute energy does not contradict the physics --
the only relevant quantity is the relative energy of the three quark
system.

With this ansatz at hand, we return to the formula
\eq{eq:faddeev.simpl}, insert the definitions \eq{eq:gtemp} and
\eq{eq:const} for $W_{\si\si}(\vec x)$ and ${\cal I}_r$, and arrive at
the following solution for the bound state energy $P_0$, in the case
of equal quark separations:
\be
P_0=3
m+\frac{3}{2}g^2\int\dk{\vec\w}\frac{D_{\si\si}(\vec\w)}{\vec\w^2}
\left[C_F- 2 C_B e^{\imath\vec\w\cdot \vec r}\right].
\label{eq:P0solution}
\ee
The following reasoning is similar to our discussion from the case of
the bound states of a meson or diquark system.  Since the quarks can
not be prepared as isolated states, the only possibilities for the
$qqq$ state are either that the system is confined (i.e., the bound
state energy $P_0$ increases with the separation), or the system is
physically not allowed (i.e., the energy $P_0$ is infinite). From the
formula \eq{eq:P0solution} and knowing that $D_{\si\si}(\vec\w)$ is
infrared enhanced, it is clear that in order to have an infrared
confining solution (corresponding to a convergent three-momentum
integral), the condition
\be
C_B=\frac{C_F}{2}
\ee
must be satisfied. This is fulfilled for $N_c=3$ colors, implying that
$\Psi_{\al\ba\ga}=\e_{\al\ba\ga}$ and that the baryon is a color
singlet (confined) bound state of three quarks; otherwise, for $N_c\ne
3$ the energy of the the system is infinite for any separation $|\vec
r|$.

As in the preceding section, with the assumption that in the infrared
$D_{\si\si}(\vec\w)= X/\vec \w^2$ (as indicated by the lattice data
\cite{Burgio:2008jr,Quandt:2008zj,
Nakagawa:2009zf,Cucchieri:2000gu,Langfeld:2004qs} and by the
variational calculations in the continuum \cite{Epple:2006hv}), where
$X$ is some combination of constants, it is straightforward to perform
the integration on the right hand side of \eq{eq:P0solution}, with the
result that for large separation $|\vec r|$:
\be
P_0=3m+\frac{3}{2}\frac{g^2C_F X}{8\pi}|\vec r|.
\label{eq:bound_state}
\ee
This mimics the previous findings for $\bar qq$ and $qq$ systems,
namely that there exists a direct connection between the string
tension and the nonperturbative \YM Green's functions (at least under
truncation). In this case, the standard term ``string tension'' refers
to the coefficient of the three-body linear confinement term
$\si_{3q}|\vec r|$. Also, comparing with the result of the previous
section, we find that the string tension corresponding to the $qqq$
system is $3/2$ times that of the $\bar qq$. To our knowledge, no
direct comparison between the string tensions of the two systems has
been made and hence this relation would be interesting to investigate
on the lattice. The appearance of three times the quark mass stems
from the presence of the mass term in the heavy quark propagator
\eq{eq:quarkpropnonpert} which enters the Faddeev equation.

%% file: nGreen.tex
\chapter{Higher order Green's functions}
\label{chap:nGreen}

We have seen that in general, the underlying equation for the
description of meson bound states is the two-body homogeneous \BS
equation.  In the rainbow-ladder approximation, this equation has been
successfully used to describe the properties of light mesons (see, for
example \cite{Alkofer:2002bp, Maris:1999nt} and for a recent review
\cite{Fischer:2006ub}), where the driving mechanism is the chiral
symmetry breaking.  Beyond this approximation, models with dressed
vertex contributions
\cite{Watson:2004kd,Williams:2009wx,Bender:2002as,
Bhagwat:2004hn,Matevosyan:2006bk,Bender:1996bb} and unquenching
effects \cite{Fischer:2005en,Watson:2004jq,Fischer:2009jm} have been
considered, and more sophisticated numerical methods to solve both the
homogeneous and the inhomogeneous \BS equation have been recently
developed \cite{Blank:2010bp}.  In spite of this success, an exact
derivation of the meson or diquark bound state energies via Green's
functions techniques has not been yet reported.  The difficulty stems
from the fact that the (irreducible) interaction kernel contains
higher order vertex functions which in general can not be calculated
exactly.

In this chapter, we continue our investigations of the quark-antiquark
and diquark states by using Green's functions techniques.  As before,
our study is based on the heavy mass expansion underlining HQET and
with the truncation of the \YM sector to include only dressed
two-point functions.  By means of functional methods, we explicitly
derive the (fully amputated) quark 4-point Green's functions and give
an exact, analytical solution. This will enable us to verify that
bound states are related to the occurrence of the poles in the Green's
functions and hence we will be able to provide a direct connection
between the homogeneous \BS equation considered in the previous
chapter and the singularities of the Green's function, at least within
the scheme considered here.

\section{4-point Green's functions for quark-antiquark systems}

\begin{figure}
\centering\includegraphics[width=1\linewidth]{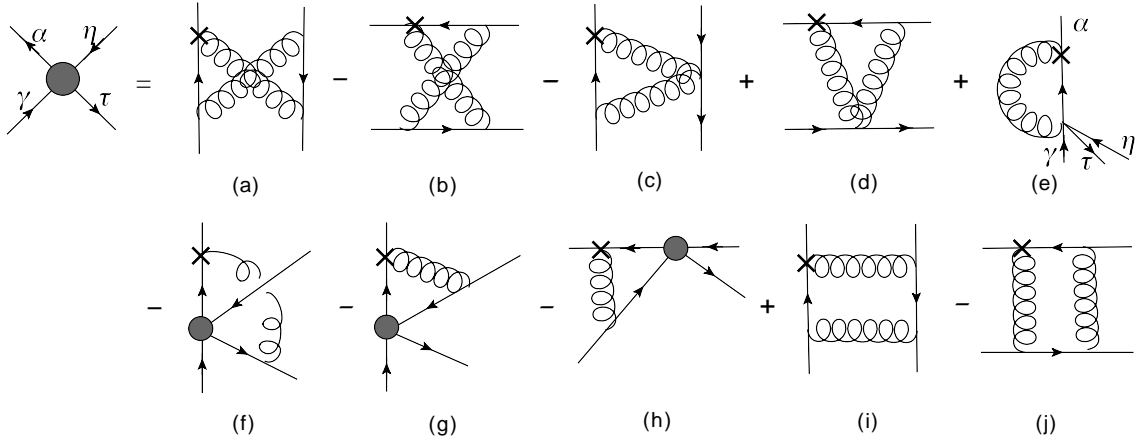}
\caption{\label{fig:1PI}
Diagrammatic representation of the one particle irreducible 4-point
quark-antiquark Green's function. Blobs represent dressed proper (1PI)
4-point vertex, solid lines represent the quark propagator, springs
denote either spatial ($\vec A$) or temporal ($\si$) gluon propagator
and cross denotes the tree level quark-gluon vertex. Internal
propagators are fully dressed.}
\end{figure}

In this section we derive the \DS equations for the one-particle
irreducible and for the fully amputated 4-point quark-antiquark
Green's functions (from which later on the bound state energy for the
heavy quark system will be extracted). The presentation follows
\cite{Popovici:2011yz}.

Let us start by deriving the \DS equation for the one-particle
irreducible (1PI) 4-point function.  As an illustration of the
functional differentiation techniques, we present the explicit
derivation of the first term in this function and notice that the rest
of the terms follow from an identical calculation.  Just as for the
gap equation, we first take the functional derivative of the quark
\eq{eq:qeom} with respect to $\imath q_{\ga z}$, and obtain the result
\eq{eq:rhoqq}; further, we functionally differentiate with respect to
$\imath \bar q_{\tau w}$ and, using the product rule, we obtain:
\bea
\frac{\delta^2}{\delta\imath \bar q_{\tau w}\delta\imath q_{\ga z}}
\ev{\imath\rho_{y}^{a}\imath\bar\chi_{\ba x}}
&=&-S[\la,\ka]\left\{
\left[\frac{\delta}{\delta\imath \bar q_{\tau w}}
\ev{\imath\rho_{y}^{a}\imath J_{\la}}\right]
\ev{\imath\Phi_{\la}\imath q_{\ga z}\imath\Phi_{\ka}}
\ev{\imath J_{\ka} \imath\bar\chi_{\ba x}}\right.\nonumber\\
&&+
\ev{\imath\rho_{y}^{a}\imath J_{\la}}
\ev{\imath\Phi_{\la}\imath \bar q_{\tau w}
\imath q_{\ga z}\imath\Phi_{\ka}}
\ev{\imath J_{\ka} \imath\bar\chi_{\ba x}}\nonumber\\
&&+\left.
\ev{\imath\rho_{y}^{a}\imath J_{\la}}
\ev{\imath\Phi_{\la}\imath q_{\ga z}\imath\Phi_{\ka}}
\left[\frac{\delta}{\delta\imath \bar q_{\tau w}}
\ev{\imath J_{\ka} \imath\bar\chi_{\ba x}}\right]\right\}
\eea
As explained above, we only retain the first term in the product (and
denote the rest with dots). Again, we make use of the formula
\eq{eq:mfd4} and obtain:
\bea
\frac{\delta^2}{\delta\imath \bar q_{\tau w}\delta\imath q_{\ga z}}
\ev{\imath\rho_{y}^{a}\imath\bar\chi_{\ba x}}
&=&S[\la,\ka,\mu,\nu]
\nonumber\\
&&\hspace{-2,7cm}\times\ev{\imath\rho_{y}^{a}\imath J_{\mu}}
\ev{\imath\Phi_{\mu}\imath \bar q_{\tau w}\imath\Phi_{\nu}}
\ev{\imath J_{\nu}^{a}\imath J_{\la}}
\ev{\imath\Phi_{\la}\imath q_{\ga z}\imath\Phi_{\ka}}
\ev{\imath J_{\ka} \imath\bar\chi_{\ba x}}+\dots .
\eea
A last functional derivative with respect to the quark field $q_{\eta
t}$ gives
\bea
 \lefteqn{
\frac{\delta^3}{
\delta\imath \bar q_{\tau w}\delta\imath q_{\ga z}
\delta\imath  q_{\eta t}}
\ev{\imath\rho_{y}^{a}\imath\bar\chi_{\ba x}}
=S[\la,\ka,\mu,\nu]}\nonumber\\
&&\times\left\{
\left[\frac{\delta}{\delta\imath  q_{\eta t}}
\ev{\imath\rho_{y}^{a}\imath J_{\mu}}
\right]
\ev{\imath\Phi_{\mu}\imath \bar q_{\tau w}\imath\Phi_{\nu}}
\ev{\imath J_{\nu}^{a}\imath J_{\la}}
\ev{\imath\Phi_{\la}\imath q_{\ga z}\imath\Phi_{\ka}}
\ev{\imath J_{\ka} \imath\bar\chi_{\ba x}}\right. \nonumber\\
&&+
\ev{\imath\rho_{y}^{a}\imath J_{\mu}}
\ev{\imath\Phi_{\mu}\imath  q_{\eta  t}\imath 
\bar q_{\tau w}\imath\Phi_{\nu}}
\ev{\imath J_{\nu}^{a}\imath J_{\la}}
\ev{\imath\Phi_{\la}\imath q_{\ga z}\imath\Phi_{\ka}}
\ev{\imath J_{\ka} \imath\bar\chi_{\ba x}}\nonumber\\
&&+
\ev{\imath\rho_{y}^{a}\imath J_{\mu}}
\ev{\imath\Phi_{\mu}\imath \bar q_{\tau w}
\imath\Phi_{\nu}}
\left[\frac{\delta}{\delta\imath  q_{\eta t}}
\ev{\imath J_{\nu}^{a}\imath J_{\la}}
\right]
\ev{\imath\Phi_{\la}\imath q_{\ga z}\imath\Phi_{\ka}}
\ev{\imath J_{\ka} \imath\bar\chi_{\ba x}}\nonumber\\
&&+
\ev{\imath\rho_{y}^{a}\imath J_{\mu}}
\ev{\imath\Phi_{\mu}\imath \bar q_{\tau w}\imath\Phi_{\nu}}
\ev{\imath J_{\nu}^{a}\imath J_{\la}}
\ev{\imath\Phi_{\la}\imath q_{\eta t}\imath q_{\ga z}
\imath\Phi_{\ka}}
\ev{\imath J_{\ka} \imath\bar\chi_{\ba x}}\nonumber\\
&&+\left.
\ev{\imath\rho_{y}^{a}\imath J_{\mu}}
\ev{\imath\Phi_{\mu}\imath \bar q_{\tau w}\imath\Phi_{\nu}}
\ev{\imath J_{\nu}^{a}\imath J_{\la}}
\ev{\imath\Phi_{\la}\imath q_{\ga z}\imath\Phi_{\ka}}
\!\!\left[\frac{\delta}{\delta\imath  q_{\eta t}}
\ev{\imath J_{\ka} \imath\bar\chi_{\ba x}}
\right]\!\right\}\!\!+\!\dots .
\eea

Again, we take only the first term from the above sum and as before we
use the formula \eq{eq:mfd4}. We obtain:
\bea
\frac{\delta^3}{\delta\imath \bar q_{\tau w}\delta\imath q_{\ga z}
  \delta\imath  q_{\eta t}}
\ev{\imath\rho_{y}^{a}\imath\bar\chi_{\ba x}}
&=&-S[\la,\ka,\mu,\nu,\e,\de]
\ev{\imath\rho_{y}^{a}\imath J_{\e}}
\ev{\imath\Phi_{\e}\imath \bar q_{\eta t}\imath\Phi_{\de}}
\ev{\imath J_{\de}^{a}\imath J_{\mu}}\nonumber\\
&&\hspace{-1,7cm}\times
\ev{\imath\Phi_{\mu}\imath \bar q_{\tau w}\imath\Phi_{\nu}}
\ev{\imath J_{\nu}^{a}\imath J_{\la}}
\ev{\imath\Phi_{\la}\imath q_{\ga z}\imath\Phi_{\ka}}
\ev{\imath J_{\ka} \imath\bar\chi_{\ba x}}+\dots .
\eea
Identifying the various fields and sources, we arrive at the the first
term in the expression below for the 4-point Green's function (all the
other terms are derived by a similar calculation and we omit the terms
which will vanish when the sources are set to zero):
\bea
\lefteqn{\ev{\imath\ov{q}_{\al}\imath q_{\ga}
\imath\ov{q}_{\tau}\imath  q_{\eta}}=
[g\ga^0T^a]_{\al\ba}\int dy\, \de(x-y)} \nonumber\\
&&\times\left\{ 
\left[
\ev{\imath\ov\chi_{\ba}\imath\chi_{\ka}}
\ev{\imath\ov{q}_{\ka}\imath q_{\ga}\imath \si^{c}_{\la}}
\ev{\imath\rho^c_{\la}\imath\rho^d_{\nu}}
\right]
\left[
\ev{\imath\ov{q}_{\tau}\imath q_{\mu}\imath \si^{d}_{\nu}}
\ev{\imath\ov\chi_{\mu}\imath\chi_{\de}}
\ev{\imath\ov{q}_{\de}\imath q_{\eta}\imath \si^{b}_{\e}}
\ev{\imath\rho^b_{\e}\imath\rho^a_y}
\right]\right.\nonumber\\
&&-
\left[
\ev{\imath\ov\chi_{\ba}\imath\chi_{\ka}}
\ev{\imath\ov{q}_{\ka}\imath q_{\ga}\imath \si^{c}_{\la}}
\ev{\imath\rho^c_{\la}\imath\rho^d_{\nu}}
\ev{\ov{q}_{\tau}\imath q_{\eta}\imath \si^{d}_{\nu}\imath
  \si^{b}_{\mu}}
\ev{\imath\rho^b_{\mu}\imath\rho^a_y}
\right]\nonumber\\
&&+
\left[
\ev{\imath\ov\chi_{\ba}\imath\chi_{\ka}}
\ev{\imath\ov{q}_{\ka}\imath q_{\ga}\imath \si^{c}_{\la}}
\ev{\imath\rho^c_{\la}\imath\rho^d_{\de}}
\right]
\left[
\ev{\imath\ov{q}_{\tau}\imath q_{\nu}\imath \si^{b}_{\mu}}
\ev{\imath\ov\chi_{\nu}\imath\chi_{\e}}
\ev{\imath\ov{q}_{\e}\imath q_{\eta}\imath \si^{d}_{\de}}
\ev{\imath\rho^b_{\mu}\imath\rho^a_y}
\right]\nonumber\\
&&-
\left[
\ev{\imath\ov\chi_{\ba}\imath\chi_{\ka}}
\ev{\imath\ov{q}_{\ka}\imath q_{\ga}
\imath\ov{q}_{\la}\imath q_{\eta}}
\ev{\imath\ov{q}_{\tau}\imath q_{\nu}\imath \si^{b}_{\mu}}
\ev{\imath\ov\chi_{\nu}\imath\chi_{\la}}
\ev{\imath\rho^b_{\mu}\imath\rho^a_y}
\right]\nonumber\\
&&-
\left[
\ev{\imath\ov\chi_{\ba}\imath\chi_{\de}}
\ev{\imath\ov{q}_{\de}\imath q_{\eta}\imath \si^{c}_{\e}}
\ev{\imath\rho^c_{\e}\imath\rho^d_{\ka}}
\right]
\left[
\ev{\imath\ov{q}_{\tau}\imath q_{\nu}\imath \si^{b}_{\mu}}
\ev{\imath\ov\chi_{\nu}\imath\chi_{\la}}
\ev{\imath\ov{q}_{\la}\imath q_{\gamma}\imath \si^{d}_{\ka}}
\ev{\imath\rho^b_{\mu}\imath\rho^a_y}
\right]\nonumber\\
&&-
\left[
\ev{\imath\ov\chi_{\ba}\imath\chi_{\ka}}
\ev{\imath\ov{q}_{\ka}\imath q_{\ga}
\imath\ov{q}_{\tau}\imath q_{\la}}
\ev{\imath\ov\chi_{\la}\imath\chi_{\de}}
\ev{\imath\ov{q}_{\de}\imath q_{\eta}\imath \si^{b}_{\e}}
\ev{\imath\rho^b_{\e}\imath\rho^a_y}
\right]\nonumber\\
&&+
\left[
\ev{\imath\ov\chi_{\ba}\imath\chi_{\ka}}
\ev{\imath\ov{q}_{\ka}\imath q_{\ga}
\imath\ov{q}_{\tau}\imath q_{\eta}\si^{b}_{\la}}
\ev{\imath\rho^b_{\la}\imath\rho^a_y}
\right]\nonumber\\
&&+
\left[
\ev{\imath\ov\chi_{\ba}\imath\chi_{\de}}
\ev{\imath\ov{q}_{\de}\imath q_{\eta}\imath \si^{c}_{\e}}
\ev{\imath\rho^c_{\e}\imath\rho^d_{\ka}}
\ev{\imath\ov{q}_{\tau}\imath q_{\ga}
\imath\si^d_{\ka}\imath\si^b_{\la}}
\ev{\imath\rho^b_{\la}\imath\rho^a_y}
\right]\nonumber\\
&&-
\left[
\ev{\imath\ov\chi_{\ba}\imath\chi_{\nu}}
\ev{\imath\ov{q}_{\nu}\imath q_{\mu}
\imath\ov{q}_{\tau}\imath q_{\eta}}
\ev{\imath\ov\chi_{\mu}\imath\chi_{\ka}}
\ev{\imath\ov{q}_{\ka}\imath q_{\ga}\imath \si^{b}_{\la}}
\ev{\imath\rho^b_{\la}\imath\rho^a_y}
\right]\nonumber\\
&&-
\left.
\left[
\ev{\imath\ov\chi_{\ba}\imath\chi_{\de}}
\ev{\imath\ov{q}_{\de}\imath q_{\eta}\imath \si^{c}_{\e}}
\ev{\imath\rho^c_{\e}\imath\rho^d_{\nu}}
\right]
\!\!
\left[
\ev{\imath\ov{q}_{\tau}\imath q_{\mu}\imath \si^{d}_{\nu}}
\ev{\imath\ov\chi_{\mu}\imath\chi_{\ka}}
\ev{\imath\ov{q}_{\ka}\imath q_{\ga}\imath \si^{d}_{\la}}
\ev{\imath\rho^d_{\la}\imath\rho^a_y}
\right]
\right\} \nonumber\\
&&+\dots
\label{eq:1pi_4quark}
\eea
where the dots represent the $\vec A$ vertex terms which are not
considered here and we have already replaced the tree-level temporal
quark-gluon vertex with its expression \eq{eq:treelevelquarkvertex1}.
Also, notice the minus sign in the fourth term, which corresponds to
the minus sign arising in the \BS equation considered in
Chapter~\ref{chap:hq}.  This equation is diagrammatically represented
in \fig{fig:1PI}.

The next step is to derive the \DS equation for the connected 4-point
quark-antiquark Green's function, which is related to the 1PI Green's
function via Legendre transform.  Starting with the identity
\be
\ev{\imath q_{\ba}\imath\ov{q}_{\ga}}
\ev{\imath\chi_{\ga}\imath\ov\chi_{\al}}=\de_{\ba\al},
\ee
we first take a functional derivative with respect to the source
$\imath J_{\de}$ and, with the help of the formula \eq{eq:mfd11}, we
derive the relation:
\be
0=\ev{\imath q_{\ba}\imath \ov q_{\rho}}
\ev{\imath \chi_{\rho}\imath\ov\chi_{\al} \imath J_{\de}}
+\imath S[\ka]
\ev{\imath J_{\de}\imath J_{\ka}}
\ev{\imath \Phi_{\ka}\imath q_{\ba}\imath \bar q_{\ga}}
\ev{\imath \chi_{\ga}\imath \ov\chi_{\al}}.
\label{eq:mfd5}
\ee
Identifying $J_{\de}$ with a gluon source, we have that
\be
\ev{\imath\ov\chi_{\al}\imath\chi_{\rho}\imath J_{\de}}=
(-\imath) 
\ev{\imath\ov\chi_{\al}\imath\chi_{\ga}}
\ev{\imath\ov{q}_{\ga}\imath q_{\ba}\imath \phi_{\ka}}
\ev{\imath\ov\chi_{\ba}\imath\chi_{\rho}}
\ev{\imath J_{\ka}\imath J_{\de}}.
\label{eq:mfd6}
\ee
Taking a further functional derivative of \eq{eq:mfd5} with respect to
the quark source $\imath\ov\chi_{\la}$ (in this case, $J_{\de}$ is
identified as quark source), we arrive at the following expression
\bea
\ev{\imath\ov\chi_{\al}\imath\chi_{\de}
\imath\ov\chi_{\la}\imath\chi_{\eta}}&=&
-\imath
\ev{\imath\ov\chi_{\al}\imath\chi_{\de}\imath J_{\rho}}
\ev{\imath\ov{q}_{\e}\imath q_{\ba}\imath \phi_{\rho}}
\ev{\imath\ov\chi_{\la}\imath\chi_{\e}}
\ev{\imath\ov\chi_{\ba}\imath\chi_{\eta}}
\nonumber\\
&&+\imath
\ev{\imath\ov\chi_{\al}\imath\chi_{\ga}}
\ev{\imath\ov{q}_{\ga}\imath q_{\ba}\imath \phi_{\ka}}
\ev{\imath\ov\chi_{\la}\imath\chi_{\de}\imath J_{\ka}}
\ev{\imath\ov\chi_{\ba}\imath\chi_{\eta}}\nonumber\\
&&-
\ev{\imath\ov\chi_{\al}\imath\chi_{\ga}}
\ev{\imath\ov{q}_{\ga}\imath q_{\ka}
\imath\ov{q}_{\e}\imath q_{\ba}}
\ev{\imath\ov\chi_{\la}\imath\chi_{\e}}
\ev{\imath\ov\chi_{\ka}\imath\chi_{\de}}
\ev{\imath\ov\chi_{\ba}\imath\chi_{\eta}}.\hspace{1cm}
\label{eq:connected_4quark0}
\eea
We now replace the the quark-gluon vertices with the expressions
\eq{eq:mfd6} and obtain for the 4-point quark-antiquark connected
Green's functions, written in terms of 1PI Green's functions:
\bea
\lefteqn{
\ev{\imath\ov\chi_{\al}\imath\chi_{\de}
\imath\ov\chi_{\la}\imath\chi_{\eta}}=}
\nonumber\\
&&-
\left[
\ev{\imath\ov\chi_{\al}\imath\chi_{\tau}}
\ev{\imath\ov{q}_{\tau}\imath q_{\mu}\imath \phi_{\nu}}
\ev{\imath\ov\chi_{\mu}\imath\chi_{\de}}
\right]
\left[
\ev{\imath\ov\chi_{\la}\imath\chi_{\e}}
\ev{\imath\ov{q}_{\e}\imath q_{\ba}\imath \phi_{\rho}}
\ev{\imath\ov\chi_{\ba}\imath\chi_{\eta}}
\right]
\ev{\imath J_{\rho}\imath J_{\nu}}\nonumber\\
&&+
\left[
\ev{\imath\ov\chi_{\al}\imath\chi_{\ga}}
\ev{\imath\ov{q}_{\ga}\imath q_{\ba}\imath \phi_{\ka}}
\ev{\imath\ov\chi_{\ba}\imath\chi_{\eta}}
\right]
\left[
\ev{\imath\ov\chi_{\la}\imath\chi_{\tau}}
\ev{\imath\ov{q}_{\tau}\imath q_{\mu}\imath \phi_{\nu}}
\ev{\imath\ov\chi_{\mu}\imath\chi_{\de}}
\right]
\ev{\imath J_{\ka}\imath J_{\nu}}\nonumber\\
&&+
\left[
\ev{\imath\ov\chi_{\al}\imath\chi_{\ga}}
\ev{\imath\ov\chi_{\la}\imath\chi_{\e}}
\ev{\imath\ov{q}_{\ga}
\imath q_{\ka}\imath\ov{q}_{\e}\imath q_{\ba}}
\ev{\imath\ov\chi_{\ka}\imath\chi_{\de}}
\ev{\imath\ov\chi_{\ba}\imath\chi_{\eta}}
\right].
\label{eq:connected_4quark}
\eea

\begin{figure}[t]
\centering\includegraphics[width=0.75\linewidth]{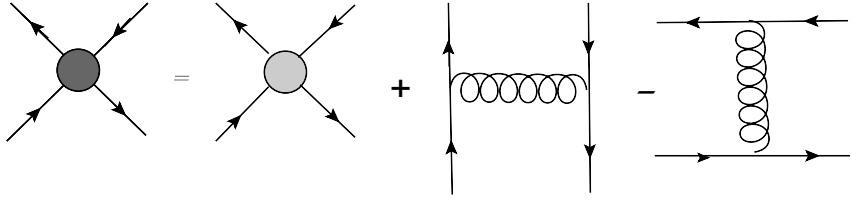}
\caption{\label{fig:1PI_amputated}
Relation between the one particle irreducible (filled blob) and
amputated (dashed blob) 4-point Greens function for the
quark-antiquark system. Internal propagators are fully dressed.}
\end{figure}

The relation \eq{eq:connected_4quark} can be further simplified by
introducing the fully amputated Green's function, i.e.  dividing by
the quark propagators (cut the quark legs), as shown in
\fig{fig:1PI_amputated}. The resulting expression for the 1PI Green's
function (as function of the amputated Green's function), is then
replaced in \eq{eq:1pi_4quark}.

\section{Solution in the heavy mass limit}

In the heavy mass limit and under our truncation, introduced at the
beginning of the previous chapter, Eqs.~(\ref{eq:1pi_4quark},
\ref{eq:connected_4quark}) simplify dramatically, allowing us to
derive exact solutions for both the 1PI and the fully amputated
4-point quark Green's functions.  In this section, the consecutive
steps required to arrive at the simplified form of these equations
will be discussed in detail. Furthermore, the physical implications of
the corresponding solutions will be discussed.

\subsection{One particle irreducible Green's function}

Let us start with the 1PI Green's function, \eq{eq:1pi_4quark}, and
the corresponding diagrammatic representation, \fig{fig:1PI}, and
apply our truncation scheme in the heavy quark limit, at leading order
in the mass expansion.

\begin{figure}[t]
\centering\includegraphics[width=0.35\linewidth]{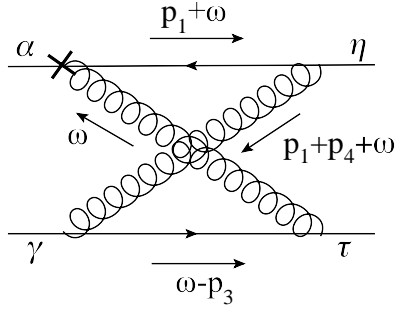}
\caption{\label{fig:diagr-b}
Crossed ladder diagram that contributes to the 1PI 4-point Green's
function. The upper line denotes the quark propagator, the lower one,
the antiquark propagator and springs denote the temporal gluon
propagator.}
\end{figure}

For the quark-antiquark system, we consider the flavor non-singlet
Green's function in the $s$-channel, recalling that the quark and the
antiquark are regarded as two distinct flavors. Hence, the diagrams
(a), (c) and (i) are excluded.  The diagram (b) (crossed ladder type
exchange diagram) explicitly reads (see also \fig{fig:diagr-b}):
\bea
\lefteqn{\int\dk{\w}\left[
\G_{\bar qq\si \al\de}^{(0)a}(p_1,-p_1-\w,\w)
W_{\bar  qq \de\phi}(p_1+\w)
\G_{\bar qq\si \phi\eta}^{b}(p_1+\w,p_4,-p_1-p_4-\w)\right]}
\nonumber\\
&&\times\left[\G_{q\bar q\si\tau\mu}^{c}(p_3,\w-p_3,-\w) 
W_{\bar qq\mu\la}(p_3-\w)
\G_{q\bar q\si\la\ga}^{d}(p_3-\w,p_2,p_1+p_4+\w)\right]
\nonumber\\
&&\times W_{\si\si}^{ac}(-\vec \w)
W_{\si\si}^{bd}(\vec p_1+\vec p_4+\vec \w)
\label{eq:diagr-b}
\eea
Isolating the energy integral over the quark propagators and using the
fact that $W_{\bar qq}(k)=-W_{q\bar q}^{T}(-k)$ (before the
truncation), we find that the energy integral vanishes, due to the
fact that both quark and antiquark propagators have the same Feynman
prescription, just as the crossed box contributions from the kernel of
the \BS equation discussed in Section~\ref{eq:secbstates} of
Chapter~\ref{chap:hq}:
\bea
\lefteqn{\int\dk{\w_{0}}W_{\bar  qq }(p_1+\w) W_{q\bar q}(\w-p_3)=}
\nonumber\\
&&\frac{1}{-(p_1^0+p_3^0)+2m}\int\dk{\w_{0}}
\left\{
\frac{1 }{\w_0+p_1^0-m-{\cal I}_r+\imath\e}
-\frac{1}{\w_0-p_3^0+m-{\cal I}_r+\imath\e}
\right\}=0.\nonumber\\
\label{eq:diagr-b-energy}
\eea
In the above, the first factor corresponds to the explicit quark
propagator, and the second factor, to the explicit antiquark
propagator (upper and lower line in the diagram \fig{fig:diagr-b},
respectively).

Let us now analyze the diagram (d), containing a quark-2 gluon vertex.
This vertex can be obtained as a solution of the corresponding \ST
identity.  The derivation is identical to the \ST identity for the
quark-gluon vertex presented in Chapter~\ref{chap:st}. Starting with
\eq{eq:stid0}, we functionally differentiate with respect to the
quark, antiquark and gluon fields. In principle, the resulting
equation contains a large number of terms, however most of them
simplify in the heavy mass limit and under truncation. To be specific,
after taking the functional derivatives there are four categories of
terms entering the equation. Firstly, the terms multiplied by a
spatial quark-gluon vertex do not contribute, since this vertex is
suppressed at leading order in the mass expansion, according to
\eq{eq:qbqa} (see also the related discussion). Secondly, the terms
containing a 4-point function $\G_{\bar qqA\si }$ are also of ${\cal
O}(1/m)$, due to the fact that in the corresponding \DS equation at
least one vertex must be at tree-level, and we are at liberty to chose
this to be $\G_{\bar qqA}$, which is truncated out by the mass
expansion. Thirdly, the ghost kernels arising from the functional
derivatives also vanish, since they only interact with the spatial \YM
sector of the theory. Hence, under truncation and in the heavy mass
limit, only the terms that involve a temporal quark-gluon vertex will
survive. Explicitly, the equation \eq{eq:stid0}, from which the \ST
identity for the quark-2 gluon vertex is derived, reduces to:
\bea
0&=&\int d^4x\de(t-x_0)\nonumber\\
&&\times\left\{
-\frac{\imath}{g}\left(\partial_x^0\ev{\imath\si_x^d}\right)
\de(z-x)
+f^{abd}\ev{\imath\si_x^a}\imath\si_x^b \de(z-x)\right.
 \nonumber\\
&&\left.-\imath T_{\al\ba}^d
\ev{\imath q_{\al x}}\imath q_{\ba x}\de(z-x)
-\imath T_{\ba\al}^d\imath\ov{q}_{\ba x}
\ev{\imath\ov{q}_{\al x}}\de(z-x)\right\}. \nonumber\\
\label{eq:stid0tr}
\eea
We now functionally differentiate with respect to $\imath q_{\e
y},\imath \bar q_{\rho t}$ and $\imath\si_{w}^{e}$ and arrive at the
following expression:
\bea
\lefteqn{0=\int dx_{0}\de(t-x_0)} \nonumber\\
&&\times\left\{
-\frac{1}{g}\pd_z^0
\ev{\imath\bar  q_{\rho t}\imath q_{\e y}\imath\si_x^d
 \imath\si_w^e} \de(z-x)
-\imath f^{aed}\ev{\imath\bar  q_{\rho t}\imath q_{\e y}
\imath\si_w^a} \de(z-x) \de(x-w)
\right.\nonumber\\
&&\left.+T^d_{\al\e}\ev{\imath\bar  q_{\rho t}\imath q_{\al y}
\imath\si_w^e}\de(z-x) \de(x-y)
-T^d_{\rho\al}\ev{\imath\bar  q_{\al z}\imath q_{\e y}
\imath\si_w^e}\de(z-x) \de(x-t)
\right\}\nonumber\\
\label{eq:stid5pt}
\eea
Replacing the temporal quark-gluon vertex with the expression
\eq{eq:feyn1}, and using the identity \eq{eq:com} for the generators,
it is straightforward to see that the color structure cancels and
hence $\G_{\bar qq\si\si}$ is zero (and correspondingly, the diagram
(d) vanishes).

 Let us for the moment discard the diagrams (f) and (g), which include
the 1PI 4-point quark Green's function, and the diagram (e),
containing a 4 quark-gluon vertex.  Then we are only left with the
diagram (h) and the rainbow-ladder term (j). This simplification
enables us to derive a solution for the corresponding (truncated)
equation for the 1PI 4-point quark Green's function. With this result
at hand, we will then return to the diagrams (f), (g) and (e), and
explicitly show that they cancel (and hence our assumption is
justified).  With these observations, \eq{eq:1pi_4quark} reduces to:
\bea
\lefteqn{
\ev{\imath\ov{q}_{\al}\imath q_{\ga}
\imath\ov{q}_{\tau}\imath  q_{\eta}}
=- \left[g T^{a}\ga^{0}\right] _{\al\ba}\left\{
\ev{\imath\ov\chi_{\ba}\imath\chi_{\nu}}
\ev{\imath\ov{q}_{\nu}\imath q_{\mu}
\imath\ov{q}_{\tau}\imath q_{\eta}}
\ev{\imath\ov\chi_{\mu}\imath\chi_{\ka}}
\ev{\imath\ov{q}_{\ka}\imath q_{\ga}\imath \si^{b}_{\la}}
\ev{\imath\rho^b_{\la}\imath\rho^a}
\right.
}\nonumber\\
&&+
\left.\left[
\ev{\imath\ov\chi_{\ba}\imath\chi_{\de}}
\ev{\imath\ov{q}_{\de}\imath q_{\eta}\imath \si^{c}_{\e}}
\ev{\imath\rho^c_{\e}\imath\rho^d_{\ka}}
\right]
\left[
\ev{\imath\ov{q}_{\tau}\imath q_{\nu}\imath \si^{b}_{\mu}}
\ev{\imath\ov\chi_{\nu}\imath\chi_{\la}}
\ev{\imath\ov{q}_{\la}\imath q_{\gamma}\imath \si^{d}_{\ka}}
\ev{\imath\rho^b_{\mu}\imath\rho^a}
\right]\right\}. \nonumber\\
\label{eq:1pi_4quark1}
\eea

\begin{figure}[t]
\centering\includegraphics[width=0.9\linewidth]{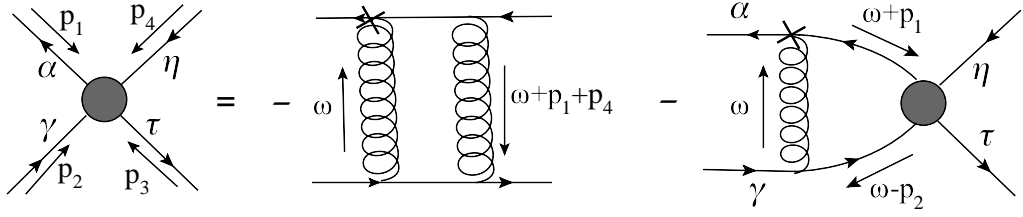}
\caption{\label{fig:1PI_mr}
\DS equation for the 1PI 4-point Green's function in the
$s$-channel. Same conventions as in \fig{fig:1PI} apply.}
\end{figure}
It is convenient to express the resulting equation in momentum space.
We define the momentum space Green's functions via their respective
Fourier transform (in order to avoid proliferation of indices, we
introduce the convention that $x_1, k_1$ correspond to the index $\al$
etc.)
\be
\ev{\imath\ov{q}_{\al}\imath q_{\ga}
\imath\ov{q}_{\tau}\imath q_{\eta}}=
\int \dk{k_1}\dk{k_2}\dk{k_3}\dk{k_4}
e^{-\imath k_1x_1-\imath k_2x_2-\imath k_3x_3-\imath k_4x_4}
\G^{(4)}_{\al\ga\tau\eta}(k_1,k_2,k_3,k_4)
\label{eq:g4fourier}
\ee
and arrive at the following \DS equation for the 1PI 4-point quark
Green's function in the $s$-channel (shown diagrammatically in
\fig{fig:1PI_mr}):
\bea
\lefteqn{
 \G^{(4)}_{\al\ga\tau\eta}(p_1,p_2,p_3,p_4)=}\nonumber\\
&&-\int\dk{\w}\left[\G_{\bar qq\si \al\de}^{(0)a}(p_1,-p_1-\w,\w)
W_{\bar  qq \de\phi}(p_1+\w)
\G_{\bar qq\si \phi\eta}^{b}(p_1+\w,p_4,-p_1-p_4-\w)\right]
\nonumber\\
&&\times\left[\G_{\bar qq\si\tau\mu}^{c}(p_3,p_2-\w,p_1+p_4+\w)
W_{\bar qq\mu\la}(\w-p_2)
\G_{\bar qq\si\la\ga}^{d}(\w-p_2,p_2,-\w)\right]\nonumber\\
&&\times W_{\si\si}^{ad}(-\vec \w)
W_{\si\si}^{bc}(\vec p_1+\vec p_4+\vec \w) \nonumber\\
&&-\int\dk{\w}\G_{\bar qq\si\al\de}^{(0)a} (p_1,-p_1-\w,\w) 
W_{\bar qq \de\phi}(p_1+\w)
W_{\bar qq\mu\la}(\w-p_2)
\G_{\bar qq\si \la\ga}^{b} (\w-p_2,p_2,-\w)\nonumber\\
&&\times W_{\si\si}^{ab}(-\vec \w)
\G^{(4)}_{\phi\mu\tau\eta}(p_1+\w,p_2-\w,p_3,p_4). 
\label{eq:4pointDS1}
\eea

In order to proceed, we make the following assumption for the function
$\G^{(4)}$:
\be
\G ^{(4)}(p_1,p_2,p_3,p_4)=\G ^{(4)}(P_0;\vec p_1+\vec p_4),
\label{eq:ansatzG4}
\ee
with $P_0=p_1^0+p_2^0$.  This implies that in the above equation the
4-point function $\G^{(4)}_{\phi\mu\tau\eta}(p_1+\w,p_2-\w,p_3,p_4) $
does not depend on the integration variable $\w_0$, and hence we can
separate the energy and three-momentum integrals and perform the
energy integration over the quark propagators. Moreover, with this
ansatz we are allowed to Fourier transform the resulting spatial
integral back to coordinate space, as shall be explained shortly
below.

After inserting the expressions \eq{eq:quarkpropnonpert},
\eq{eq:bgapsol}, for the quark and antiquark propagators and
completing the energy integration, \eq{eq:4pointDS1} simplifies to:
\bea
\lefteqn{
\left[ p_1^0+p_2^0-2 {\cal I}_{r} +2\imath\e\right]
\G^{(4)}_{\al\ga\tau\eta}(P_0;\vec p_1+\vec p_4) }\nonumber\\
&&=\imath\frac{g^4}{4}\left[\left(N_c-\frac{2}{N_c}\right)
\de_{\al\ga}\de_{\tau\eta}
+\frac{1}{N_c^2}\de_{\al\eta}\de_{\ga\tau}\right]
\int\dk{\vec\w}W_{\si\si}(\vec\w)W_{\si\si}(\vec k+\vec\w)
\nonumber\\
&&+\imath
\frac{g^2}{2}\left[\de_{\al\ga}\de_{\mu\phi}-
\frac{1}{N_c}\de_{\al\phi}\de_{\mu \ga}\right]
\int\dk{\vec\w}W_{\si\si}(\vec\w)
\G^{(4)}_{\phi\mu\tau\eta}(P_0;\vec p_1+\vec p_4+\vec \w).
\hspace{1cm} 
\eea

Let us now make the following color decomposition for the function
$\G^{(4)}$:
\be
\G^{(4)}_{\al\ga\tau\eta}
=\de_{\al\ga}\de_{\tau\eta}\G_{1}+\de_{\al\eta}\de_{\ga\tau}\G_{2}.
\label{eq:g4color1PI}
\ee
where (for a given flavor structure) $\G^{(4)}_{1}$ and $\G^{(4)}_{2}$
are Dirac scalar functions.

At this point, it is convenient to Fourier transform back to
coordinate space.  In general, since \eq{eq:4pointDS1} might in
principle contain momentum-dependent vertex functions, as well as
mixing of energy and three-momentum variables, this transformation
could not be carried out. However, in our case momentum-dependent
vertices are absent and moreover, with the ansatz \eq{eq:ansatzG4},
the energy and tree-momentum integrals have separated such that the
spatial integral is performed only over two functions (spatial gluon
propagator and the spatial component of the quark 4-point
function). Hence the Fourier transform simplifies to the usual
convolution product:
\bea
\int\dk{\vec \w} W_{\si\si}(\vec \w)\G^{(4)}(\vec q+\vec \w)=
\int d \vec x e^{-\imath\vec q\cdot\vec x}
W_{\si\si}(\vec x)\G^{(4)}(-\vec x),
\label{eq:convolution}
\eea
Using the Fierz identity for the generators, \eq{eq:fierz}, and
sorting out the color factors, it is straightforward to obtain for the
components $\G_1^{(4)}$, $\G_2^ {(4)}$:
\bea
\G_1^{(4)}(P_0;-\vec x)&=& \imath \left(\frac{g^2}{2N_c}\right)^2
 \frac{W_{\si\si}(\vec x) W_{\si\si}(-\vec x) }
{P_0-2{\cal I}_r+\imath \frac{g^2}{2N_c}
W_{\si\si}(\vec x)+2\imath\e}\nonumber\\
&&\times
\frac{ N_c\left[(P_0-2{\cal I}_r)(N_c^2-2)+\imath g^2 C_F 
W_{\si\si}(\vec x)\right]}
{P_0-2{\cal I}_r-\imath g^2C_FW_{\si\si}(\vec x) +2\imath\e}
\nonumber\\
\G_2^{(4)}(P_0;-\vec x)&=&
\imath\left(\frac{g^2}{2N_c}\right)^2
\frac{W_{\si\si}(\vec x) W_{\si\si}(-\vec x)}
{P_0-2{\cal I}_r+\imath \frac{g^2}{2N_c}
W_{\si\si}(\vec x) +2\imath\e},\hspace{3cm}
\eea
where $\vec x$ is the separation associated with the momentum $\vec
p_1+\vec p_4$.  Inserting the above results into the decomposition
\eq{eq:g4color1PI}, we find the final formula for the 1PI quark
Green's function:
\bea
\G^{(4)}_{\al\ga\tau\eta} (P_0;-\vec x)&=&
\imath\left(\frac{g^2}{2N_c}\right)^2
\frac{W_{\si\si}(\vec x)^2}
{P_0-2{\cal I}_r+\imath 
\frac{g^2}{2N_c}W_{\si\si}(\vec x) +2\imath\e}\nonumber\\
&&\hspace{-0,8cm}\times\left\{
\de_{\al\ga}\de_{\tau\eta}
\frac{(P_0-2{\cal  I}_r)N_c(N_c^2-2)
+\imath g^2N_c C_F W_{\si\si}(\vec x)}
{P_0-2{\cal I}_r-\imath g^2C_FW_{\si\si}(\vec x) +2\imath\e}
+\de_{\al\eta}\de_{\ga\tau}
\right\}\!.
\label{eq:sol1PI}
\eea
where we recall that $P_0=p_1^0+p_2^0$.
\begin{figure}[t]
\centering\includegraphics[width=0.3\linewidth]{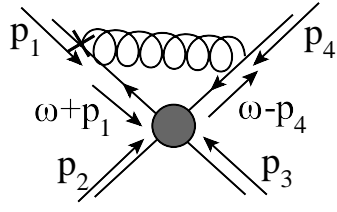}
\caption{\label{fig:diagr-g}
Momentum routing for the diagram (g). See text for details.}
\end{figure}

Having derived the solution \eq{eq:sol1PI} for the 1PI Green's
function, we return to the diagrams (f), (g) and (e) and show that
they do not contribute to the final result. To see this, we use our
usual trick and consider the energy integral. In the case of the
diagram (g), this reads (with the momentum routing from
\fig{fig:diagr-g}):
\bea
\lefteqn{ \int \dk{\w_0} W_{\bar qq}(p_1+\w) W_{\bar qq}(\w-p_4)
\G^{(4)}_{E}(p_1+\w,p_2,p_3,\w-p_4) }\nonumber\\
&&\sim\int \dk{\w_0}
\frac{1}{\left[p_1^0+\w_0+m-{\cal I}_r+\imath\e\right]
\left[\w^0-p_4^0+m-{\cal I}_r+\imath\e\right]}
\G^{(4)}_{E}(\w_0+P_0) \nonumber\\
&&=\int \dk{\w_0}
\frac{1}{\left[p_1^0+\w_0+m-{\cal I}_r+\imath\e\right]
\left[\w^0-p_4^0+m-{\cal I}_r+\imath\e\right]}\nonumber\\
&&\times
\frac{P_0+\w_0+\al}
{\left[P_0+\w_0+C_1+2\imath\e\right]\left[P_0
+\w_0+C_2+2\imath\e\right]}
\eea
where $\al$, $C_1$ and $C_2$ are energy independent constants that
appear in the expression \eq{eq:sol1PI} for the 1PI Green's
function. The above formula can be trivially rewritten as
\bea
\lefteqn{\int \dk{\w_0}
\frac{1}{\left[p_1^0+\w_0+m-{\cal I}_r+\imath\e\right]
\left[\w^0-p_4^0+m-{\cal I}_r+\imath\e\right]}}\nonumber\\
&&\times
\left\{
\frac{1}{P_0+\w_0+C_1+2\imath\e}
+\frac{\al-C_1}
{\left[P_0+\w_0+C_1+2\imath\e\right]
\left[P_0+\w_0+C_2+2\imath\e\right]}
\right\}.
\eea
Clearly, both terms in the above sum (having the same Feynman
prescription) can be reduced to differences of integrals over a simple
pole, and with the same sign for performing the integration in the
complex plane and this vanishes (same as for the kernel of the \BS
equation and the diagram (b) from above).  An identical calculation
for the integral (f) leads us to the fact that this integral is also
vanishing.
\begin{figure}[t]
\centering\includegraphics[width=0.75\linewidth]{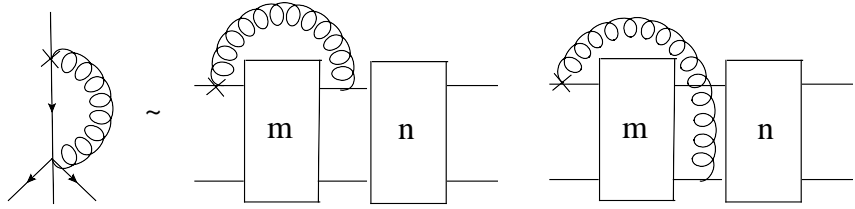}
\caption{\label{fig:diagr-e}
Perturbative expansion of the diagram (e). Boxes comprise $m$ and $n$
gluon legs, respectively, with $m, n\ge 1$.  See text for details.}
\end{figure}

Finally, we are now in the position to show that the diagram (e),
containing the 4 quark-gluon vertex, is also vanishing.  The
argumentation is based on our previous findings, namely that the
diagrams (f) and (g), containing the 1PI quark Green's functions, are
zero. We first observe that the diagram (e) can be written as a
combination of diagrams of the form shown in \fig{fig:diagr-e}, where
the boxes contain an arbitrary number of gluon legs (ladder
resummation).\footnote{To see this, it is enough to analyze the first
few terms in perturbation theory of the 4 quark-gluon vertex, which
are then included into the diagram (e).} On the other hand, as a
result of the \DS equation for the 1PI quark Green's functions, the
diagrams (f) and (g) can also be written as a ladder resummation,
which exactly coincide with the two terms in the diagram
\fig{fig:diagr-e}.  Hence, the perturbative series of diagram (e) has
been reorganized such that although the function $\G_{\bar qq\bar
qq\si}$ itself does not vanish, this 5-point interaction vertex and
and the gluon line on top of it provide the cancellation at every
order perturbatively. In turn, this implies that our original
assumption is correct and the solution \eq{eq:sol1PI} is valid
nonperturbatively.

\subsection{Amputated Green's function}

In the following, we return to the \DS equation for the fully
amputated 4-point quark-antiquark Green's function in the $s$-channel
and, with the simplifications outlined in the previous section, we
will derive a solution to this equation. We will verify that this is
consistent with the 1PI Green's function obtained in the previous
section and moreover, we will analyze the position of the poles and
compare them with the \BS equation for physical states.

\begin{figure}[t]
\centering\includegraphics[width=0.7\linewidth]{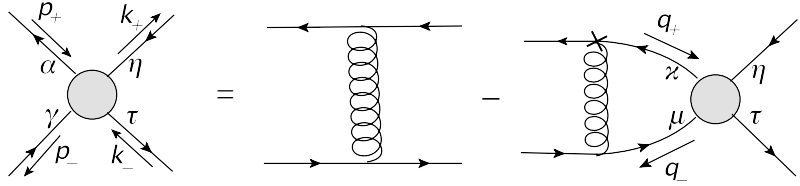}
\caption{\label{fig:amputated_mr}
\DS equation for the fully amputated quark-antiquark 4-point Green's
function in the $s$-channel. }
\end{figure}

The \DS equation for the fully amputated 4-point quark-antiquark
Green's function is obtained from the formula \eq{eq:1pi_4quark1}, by
replacing the 1PI Green's function with the expression
\eq{eq:connected_4quark} and cutting the legs. Thus this equation
(shown diagrammatically in \fig{fig:amputated_mr}) is equivalent to
the \DS equation \eq{eq:4pointDS1}, derived for 1PI Green's
functions. It is given by:
\bea
\lefteqn{
 G^{(4)}_{\al\ba;\de\ga}(p_+,p_-;k_+,k_-)=
W_{\si\si}^{ab}(\vec p-\vec k)\left[\G_{\bar qq\si}^a\right]_{\al\ga}
\left[\G_{\bar q q\si}^{b}\right]_{\de\ba}
}\nonumber\\
&&-\int\dk{q}\left[\G_{\bar qq\si}^a W_{\bar qq}(q_+)\right]_{\al\ka}
\left[W_{q\bar q}^{T}(-q_-)\G_{q\bar q\si}^{Tb}\right]_{\ba\tau}
W_{\si\si}^{ab}(\vec p-\vec q)
G^{(4)}_{\ka\tau;\de\ga}(q_+,q_-;k_+,k_-). \nonumber\\
\label{eq:4pointDS2}
\eea
In the above, the momenta of the quarks are given by $p_+=p+\xi P$,
$p_-=p-(1-\xi) P$ (similarly for $k$ and $q$), and $\xi$ is the
momentum sharing fraction.  $P$ indicates the dependence on the total
four momentum, which will become important for the investigation of
the bound-state contributions to the Green's function.
\eq{eq:4pointDS1} is an inhomogeneous integral equation which --
unlike the \BS equation -- contains both a resonant component (later
on used to reproduce the bound state confining energy of the $\bar qq$
system), \emph {and} a nonresonant term.

The right hand side of equation \eq{eq:4pointDS1} does not depend on
the external energy $p_0$, implying that the 4-point function
$G^{(4)}$ has to be independent on the relative energy $q_0$, and we
further assume that $G^{(4)}$ depends only on the relative momentum
$\vec p-\vec k$.  Accordingly, we replace the quark and antiquark
propagators with the expressions \eq{eq:quarkpropnonpert},
\eq{eq:bgapsol}, and as before we perform the energy integration. We
arrive at the following expression:
\bea
G^{(4)}_{\al\ba;\de\ga}(P_0;\vec p-\vec k)&=&
g^2 T^{a}_{\al\ga}T^{b}_{\de \ba }W_{\si\si}^{ab}(\vec p-\vec k)
\nonumber\\
&&+g^2T^{a}_{\al\ka}T^{a}_{\tau\ba}
\frac{\imath}{P_0-2 {\cal I}_r+2\imath\e}
\int\dk{\vec q}W_{\si\si}(\vec p-\vec q)
G^{(4)}_{\ka\tau;\de\ga}(P_0;\vec q-\vec k). 
\nonumber\\
\label{eq:DS4point}
\eea
In the above, we have also replaced the vertex functions with their
tree-level expressions \eq{eq:vsol}, \eq{eq:bvsol}.  Having integrated
out the energy, it is convenient to rewrite the above formula back
into coordinate space.  Using the definition \eq{eq:g4fourier} and the
relation \eq{eq:convolution}, the equation \eq{eq:DS4point} simplifies
to:
\be
G^{(4)}_{\al\ba;\de\ga}(P_0;\vec x)
=g^2T^a_{\al\ga}T^{a}_{\de \ba}W_{\si\si}(\vec x)+
g^2T^{a}_{\al\ka}T^{a}_{\tau\ba}\frac{\imath}{P_0-2{\cal I}_r
+2\imath\e}W_{\si\si}(\vec x)
G^{(4)}_{\ka\tau;\de\ga}(P_0;\vec x).
\ee

Again, we decompose the function $G^{(4)}$ :
\be
G^{(4)}_{\al\ba;\de\ga}
=\de_{\al\ba}\de_{\ga\de}G^{(4)}_{1}
+\de_{\al\ga}\de_{\ba\de}G^{(4)}_{2},
\label{eq:g4color}
\ee
(for a given flavor structure, $G^{(4)}_{1}$ and $G^{(4)}_{2}$ are
Dirac scalar functions), use the Fierz identity, \eq{eq:fierz}, to
sort out the color factors, and obtain the following results for the
components $G^{(4)}_1$, $G^{(4)}_2$:
\bea
G_1^{(4)}(P_0;\vec x)&=&\left(\frac{g^2}{2}\right)
 \frac{(P_0-2{\cal  I}_r) W_{\si\si}(\vec x) }
{P_0-2{\cal I}_r+\imath
\frac{g^2}{2N_c}W_{\si\si}(\vec x)+2\imath\e}\\
&&\times\frac{(P_0-2{\cal  I}_r)}
{P_0-2{\cal I}_r-\imath \frac{g^2}{2}
\left(N_c-\frac 1N_c\right)W_{\si\si}(\vec x) +2\imath\e}
\nonumber\\
G_2^{(4)}(P_0;\vec x)&=&-\left(\frac{g^2}{2N_c}\right)
\frac{(P_0-2{\cal  I}_r) W_{\si\si}(\vec x) }
{P_0-2{\cal I}_r+\imath \frac{g^2}{2N_c}
W_{\si\si}(\vec x) +2\imath\e}\hspace{3cm}
\eea
Replacing these results in the formula \eq{eq:g4color}, we get the
final result for the function $G^{(4)}$:
\bea
G^{(4)}(P_0;\vec x)&=&\frac{g^2}{2}
\frac{\left(P_0-2{\cal I}_r\right) 
W_{\si\si}(\vec x)}{P_0-2{\cal I}_r+\imath
\frac{g^2}{2N_c}W_{\si\si}(\vec x) +2\imath\e}\nonumber\\
&&\times\left[
\de_{\al\ba}\de_{\ga\de}
\frac{(P_0-2{\cal  I}_r) }
{P_0-2{\cal I}_r-\imath \frac{g^2}{2} W_{\si\si}(\vec x) C_F 
+2\imath\e}-
\de_{\al\ga}\de_{\ba\de}\frac{1}{N_c} \right]
\label{eq:4pointfinal}
\eea

A few comments regarding the structure of the above equation are here
in order. Firstly, a direct calculation shows that our result for the
amputated 4-point function is related to the result \eq{eq:sol1PI} for
the 1PI Green's function, via the formula \eq{eq:connected_4quark} (or
alternatively, \fig{fig:1PI_amputated}).

Also, notice that despite the truncation, the poles in the 1PI and
amputated Green's functions are identical. Moreover, in this approach
the physical and nonphysical poles disentangle automatically, as
opposite to the ``standard'' QCD setting, where this separation does
not occur. Using the form \eq{eq:gtemp} for the temporal gluon
propagator, we notice that the bound state (infrared confining) energy
$P_0=\si|\vec x|$ emerges as the pole of the resonant component (first
term in the bracket of \eq{eq:4pointfinal}), for arbitrary number of
colors. Hence, this provides an explicit analytical dependence of the
4-point Green's function on the $\bar qq$ bound state energy, which
results from the \BS equation presented in Chapter~\ref{chap:hq}.  The
term multiplying the bracket has a pole for $N_c=0$ and hence this
cannot represent a physical state.  This (common) pole in the
nonresonant term can be shifted to infinity (as the regularization of
$I_r$ is removed) and, as discussed in the previous chapter, is
nonphysical just like the poles in the quark propagator or in the
baryon vertex. In the case of the 4-point Green's function, this can
be simply absorbed in the normalization. Also, the appearance of this
spurious pole does not contradict the physical results, since the
bound state energy, stemming from the first pole of
\eq{eq:4pointfinal}, is the only relevant quantity.

\section{4-point Green's functions for diquarks}

Let us now consider the diquark 4-point Green's function. This can be
easily obtained from the equation \eq{eq:1pi_4quark} for
quark-antiquark systems, by interchanging the quark legs and inserting
the appropriate minus signs. We obtain (see also the diagrammatic
representation from \eq{fig:1PIdiquarks}):
\bea
\lefteqn{
\ev{\imath\ov{q}_{\al}\imath\ov{q}_{\tau}
\imath q_{\ga}\imath  q_{\eta}}=
[g\ga^0T^a]_{\al\ba}\int dy\, \de(x-y) }\nonumber\\
&&\times\left\{ 
\left[
\ev{\imath\ov\chi_{\ba}\imath\chi_{\ka}}
\ev{\imath\ov{q}_{\ka}\imath q_{\ga}\imath \si^{c}_{\la}}
\ev{\imath\rho^c_{\la}\imath\rho^d_{\nu}}
\ev{\ov{q}_{\tau}\imath q_{\eta}\imath \si^{d}_{\nu}\imath
  \si^{b}_{\mu}}
\ev{\imath\rho^b_{\mu}\imath\rho^a_y}
\right]
\right.\nonumber\\
&&
-\left[
\ev{\imath\ov\chi_{\ba}\imath\chi_{\ka}}
\ev{\imath\ov{q}_{\ka}\imath q_{\ga}\imath \si^{c}_{\la}}
\ev{\imath\rho^c_{\la}\imath\rho^d_{\nu}}
\right]
\left[
\ev{\imath\ov{q}_{\tau}\imath q_{\mu}\imath \si^{d}_{\nu}}
\ev{\imath\ov\chi_{\mu}\imath\chi_{\de}}
\ev{\imath\ov{q}_{\de}\imath q_{\eta}\imath \si^{b}_{\e}}
\ev{\imath\rho^b_{\e}\imath\rho^a_y}
\right]
\nonumber\\
&&-
\left[
\ev{\imath\ov\chi_{\ba}\imath\chi_{\ka}}
\ev{\imath\ov{q}_{\ka}\imath q_{\ga}\imath \si^{c}_{\la}}
\ev{\imath\rho^c_{\la}\imath\rho^d_{\de}}
\right]
\left[
\ev{\imath\ov{q}_{\tau}\imath q_{\nu}\imath \si^{b}_{\mu}}
\ev{\imath\ov\chi_{\nu}\imath\chi_{\e}}
\ev{\imath\ov{q}_{\e}\imath q_{\eta}\imath \si^{d}_{\de}}
\ev{\imath\rho^b_{\mu}\imath\rho^a_y}
\right]\nonumber\\
&&-
\left[
\ev{\imath\ov\chi_{\ba}\imath\chi_{\ka}}
\ev{\imath\ov{q}_{\ka}\imath\ov{q}_{\la}
\imath q_{\ga}\imath q_{\eta}}
\ev{\imath\ov{q}_{\tau}\imath q_{\nu}\imath \si^{b}_{\mu}}
\ev{\imath\ov\chi_{\nu}\imath\chi_{\la}}
\ev{\imath\rho^b_{\mu}\imath\rho^a_y}
\right]\nonumber\\
&&+
\left[
\ev{\imath\ov\chi_{\ba}\imath\chi_{\de}}
\ev{\imath\ov{q}_{\de}\imath q_{\eta}\imath \si^{c}_{\e}}
\ev{\imath\rho^c_{\e}\imath\rho^d_{\ka}}
\right]
\left[
\ev{\imath\ov{q}_{\tau}\imath q_{\nu}\imath \si^{b}_{\mu}}
\ev{\imath\ov\chi_{\nu}\imath\chi_{\la}}
\ev{\imath\ov{q}_{\la}\imath q_{\gamma}\imath \si^{d}_{\ka}}
\ev{\imath\rho^b_{\mu}\imath\rho^a_y}
\right]\nonumber\\
&&-
\left[
\ev{\imath\ov\chi_{\ba}\imath\chi_{\ka}}
\ev{\imath\ov{q}_{\ka}\imath\ov{q}_{\tau}\imath q_{\ga}
\imath q_{\la}}
\ev{\imath\ov\chi_{\la}\imath\chi_{\de}}
\ev{\imath\ov{q}_{\de}\imath q_{\eta}\imath \si^{b}_{\e}}
\ev{\imath\rho^b_{\e}\imath\rho^a_y}
\right]\nonumber\\
&&+
\left[
\ev{\imath\ov\chi_{\ba}\imath\chi_{\ka}}
\ev{\imath\ov{q}_{\ka}\imath\ov{q}_{\tau}\imath q_{\ga}
\imath q_{\eta}\si^{b}_{\la}}
\ev{\imath\rho^b_{\la}\imath\rho^a_y}
\right]\nonumber\\
&&-
\left[
\ev{\imath\ov\chi_{\ba}\imath\chi_{\de}}
\ev{\imath\ov{q}_{\de}\imath q_{\eta}\imath \si^{c}_{\e}}
\ev{\imath\rho^c_{\e}\imath\rho^d_{\ka}}
\ev{\imath\ov{q}_{\tau}\imath q_{\ga}\imath\si^d_{\ka}
\imath\si^b_{\la}}
\ev{\imath\rho^b_{\la}\imath\rho^a_y}
\right]\nonumber\\
&&-
\left[
\ev{\imath\ov\chi_{\ba}\imath\chi_{\nu}}
\ev{\imath\ov{q}_{\nu}\imath\ov{q}_{\tau}\imath q_{\mu}
\imath q_{\eta}}
\ev{\imath\ov\chi_{\mu}\imath\chi_{\ka}}
\ev{\imath\ov{q}_{\ka}\imath q_{\ga}\imath \si^{b}_{\la}}
\ev{\imath\rho^b_{\la}\imath\rho^a_y}
\right]\nonumber\\
&&+
\left.
\left[
\ev{\imath\ov\chi_{\ba}\imath\chi_{\de}}
\ev{\imath\ov{q}_{\de}\imath q_{\eta}\imath \si^{c}_{\e}}
\ev{\imath\rho^c_{\e}\imath\rho^d_{\nu}}
\right]
\left[
\ev{\imath\ov{q}_{\tau}\imath q_{\mu}\imath \si^{d}_{\nu}}
\ev{\imath\ov\chi_{\mu}\imath\chi_{\ka}}
\ev{\imath\ov{q}_{\ka}\imath q_{\ga}\imath \si^{d}_{\la}}
\ev{\imath\rho^d_{\la}\imath\rho^a_y}
\right]
\right\}\nonumber\\
&&+\dots . 
\label{eq:1pi_4diquark}
\eea
\begin{figure}
\centering\includegraphics[width=1.0\linewidth]{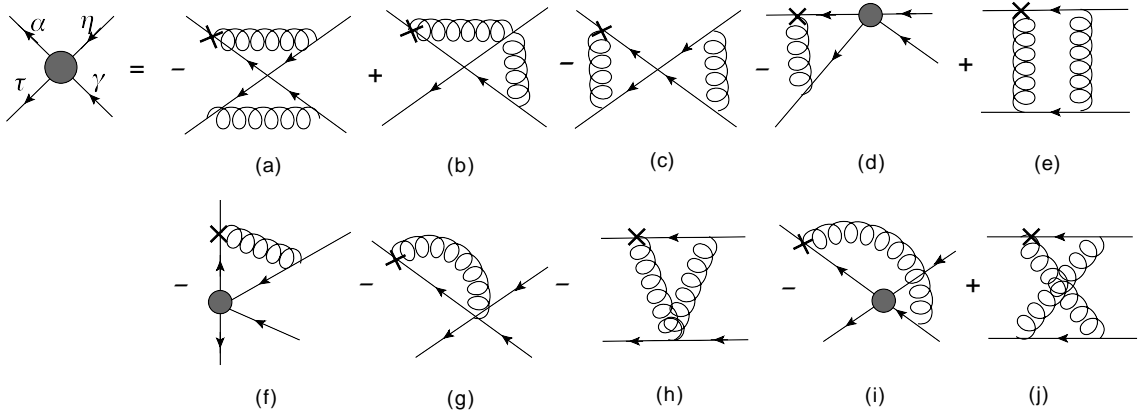}
\caption{\label{fig:1PIdiquarks}
Diagrammatic representation of the one particle irreducible 4-point
diquark. Green's function. Same conventions as in \fig{fig:1PI}
apply.}
\end{figure}
As in the $\bar qq$ case, the dots represent the $\vec A$ vertex terms
which are not considered here and the tree-level temporal quark-gluon
vertex has been replaced with its expression
\eq{eq:treelevelquarkvertex1}.  Also, notice that the diquark is
antisymmetric under the exchange of two quark legs. In this case we
explicitly take into account the flavor structure, i.e. we consider
equal (heavy) mass quarks but with different flavors.

\begin{figure}[t]
\centering\includegraphics[width=0.25\linewidth]{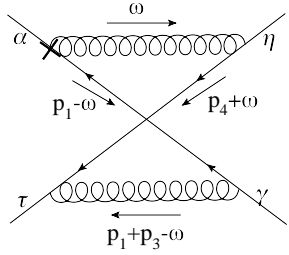}
\caption{\label{fig:diagr-a-diq}
Crossed ladder diagram that contributes to the 1PI 4-point Green's
function in the diquark channel.}
\end{figure}

As before, we analyze the diagrammatic representation from
\fig{fig:1PIdiquarks} and show that the same type of cancellations
occur. Starting with the diagram (a), we notice that this is a crossed
ladder type exchange diagram (see also \fig{fig:diagr-a-diq}):
\bea
\lefteqn{
\int\dk{\w}\left[\G_{\bar qq\si \al\ba}^{(0)a}(p_1,\w-p_1,-\w)
W_{\bar  qq \ba\ka}(p_1-\w)
\G_{\bar qq\si \ka\ga}^{d}(p_1-\w,p_3,\w-p_1-p_3)\right]}
\nonumber\\
&&\times\left[\G_{q\bar q\si\tau\mu}^{c}(p_2,p_4+\w,p_1+p_3-\w)
 W_{\bar qq\mu\de}(-p_4-\w)
\G_{q\bar q\si\de\eta}^{b}(-p_4-\w,p_4,\w)\right]\nonumber\\
&&\times W_{\si\si}^{ab}(\vec \w)
W_{\si\si}^{dc}(\vec p_1+\vec p_3-\vec \w).\hspace{9cm}
\label{eq:diagr-a-diq}
\eea
It has been already shown that the integral over the quark propagators
(with the same Feynman prescription) vanishes and thus the diagram (a)
is zero.  A similar type of integral arises in the diagram (j), and
hence this term is also not giving a contribution. Further, the
diagrams (b) and (h) are zero, since the corresponding quark-2 gluon
vertex vanishes according to the \ST identity, \eq{eq:stid5pt}.  As in
the case of the $\bar qq$ system, let us for the moment assume that
the integrals (f), (i), containing the diquark 4-point vertex, and
(g), containing the 5-point functions $\G_{\bar q\bar q qq\si}$ are
also zero, and solve the \DS equation with the remaining terms. Having
derived the solution, we will then return to the diagrams (f),(i) and
(g) and show that they vanish.

Putting all these together, the \DS equation for the diquark 4-point
function with the remaining diagrams, i.e. diagrams (c), (d) and (e)
explicitly reads:
\bea
\lefteqn{\G_{\bar qq \bar qq\si \al\tau\ga\eta}(p_1,p_2,p_3,p_4)=}
\nonumber\\
&&-\int\dk{\w}\left[\G_{\bar qq\si \al\ba}^{(0)a}(p_1,-\w-p_1,\w)
W_{\bar  qq \ba\ka}(p_1+\w)
\G_{\bar qq\si \ka\ga}^{c}(p_1+\w,p_3,-\w-p_1-p_3)\right]
\nonumber\\
&&\times\left[\G_{q\bar q\si\tau\nu}^{b}(p_2,\w-p_2,-\w) 
W_{\bar qq\nu\e}(p_2-\w)
\G_{q\bar q\si\e\eta}^{d}(p_2-\w,p_4,p_1+p_3+\w)\right]
\nonumber\\
&&\times W_{\si\si}^{ab}(-\vec \w)
W_{\si\si}^{dc}(\vec p_1+\vec p_3+\vec \w) \nonumber\\
&&+\int\dk{\w}\left[\G_{\bar qq\si \al\ba}^{(0)a}(p_1,-\w-p_1,\w)
W_{\bar  qq \ba\de}(p_1+\w)
\G_{\bar qq\si \de\eta}^{d}(p_1+\w,p_4,-\w-p_1-p_4)\right]
\nonumber\\
&&\times\left[\G_{q\bar q\si\tau\nu}^{b}(p_2,\w-p_2,-\w) 
W_{\bar qq\nu\la}(p_2-\w)
\G_{q\bar q\si\la\ga}^{c}(p_2-\w,p_3,p_1+p_4+\w)\right]
\nonumber\\
&&\times W_{\si\si}^{ab}(-\vec \w)
W_{\si\si}^{dc}(\vec p_1+\vec p_4+\vec \w) \nonumber\\
&&-\int\dk{\w}\left[\G_{\bar qq\si \al\ba}^{(0)a}(p_1,-\w-p_1,\w)
W_{\bar  qq \ba\ka}(p_1+\w)\right]\nonumber\\
&&\times\left[\G_{\bar q q\si\tau\nu}^{b}(p_2,\w-p_2,-\w) 
W_{\bar qq\nu\la}(p_2-\w)\right]
\G_{\bar qq \bar qq\si \ka\la\ga\eta}(p_1+\w,p_2-\w,p_3,p_4)
W_{\si\si}^{ab}(\vec \w)\nonumber\\
\label{eq:1PImomdiq}
\eea
After sorting out the color factors and applying the usual separation
of the energy and three-momentum integrals, we arrive at the following
formula:
\bea
\lefteqn{
\G^{(4)}_{\al\tau\ga\eta}(p_1,p_2,p_3,p_4) =
\frac{1}{ P^0-2m-2 {\cal I}_{r} +2\imath\e}}\nonumber\\
&&\left\{
-\imath\left(\frac{g^2}{2N_c}\right)^2
\de_{\al\ga}^f\de_{\tau\eta}^f
\left[\left(N_c^2+1\right)\de_{\al\ga}\de_{\tau\eta}
-2N_c\de_{\al\eta}\de_{\tau \ga}\right]
\int\dk{\vec\w}W_{\si\si}(\vec\w)
W_{\si\si}(\vec p_1+\vec p_3+\vec\w)\right.\nonumber\\
&&+\imath\left(\frac{g^2}{2N_c}\right)^2
\de_{\al\eta}^f\de_{\tau \ga }^f
\left[\left(N_c^2+1\right)\de_{\al\eta}\de_{\tau \ga}
-2N_c\de_{\al\ga}\de_{\tau\eta}\right]
\int\dk{\vec\w}W_{\si\si}(\vec\w)W_{\si\si}(\vec p_1+\vec p_4+\vec\w)
\nonumber\\
&&+
\frac{g^2}{2N_c} \de_{\al\la}^f\de_{\tau\ka}^f
\left[N_c\de_{\al\ka}\de_{\tau\la}-\de_{\al\la}\de_{\tau\ka}\right]
\nonumber\\
&&\times\int\dk{\w_0} 
\left[
\frac{1}{\w_0+ p_1^0-m- {\cal I}_{r} +\imath\e}
-\frac{1}{\w_0- p_2^0+m+ {\cal I}_{r} -\imath\e}
\right]\nonumber\\
&&\times\left.\int\dk{\vec\w}W_{\si\si}(\vec\w)
\G^{(4)}_{\la\ka\ga\eta}(p_1+\w,p_2-\w,p_3,p_4) \right\}
\label{eq:1PIsimpl-di}
\eea

Since the diquarks we are not restricted to flavor singlet, we make
the following flavor decomposition:
\be
\G^{(4)}_{\al\tau\ga\eta}
=\de_{\al\ga}^{f}\de_{\tau\eta}^{f}\G^{(c)}_{\al\tau\ga\eta}
+\de_{\al\eta}^{f}\de_{\ga\tau}^{f}\G^{(p)}_{\al\tau\ga\eta}
\label{eq:g4flavour1PIdi}
\ee
(the superscripts $c$ and $p$ stand for crossed and parallel
configurations, respectively).  As before, in order to carry on the
energy integration, and in the view of Fourier transforming to
coordinate space the resulting spatial integral, we make the following
ansatz for the components $\G^{(c)}$ and $\G^{(p)}$:
\bea
\G^{(c)}( p_1+\w, p_2-\w, p_3,p_4)&=&
\G^{(c)}(P_0;\vec p_1+\vec p_3+\vec\w)\\
\G^{(p)}( p_1+\w, p_2-\w, p_3,p_4)&=&
\G^{(c)}(P_0;\vec p_1+\vec p_4+\vec\w) ,
\eea 
with $P_0=p_1^0+p_2^0$.

With these notations, \eq{eq:1PIsimpl-di} decouples to:
\bea
\lefteqn{\G^{(c)}_{\al\tau\ga\eta}(P_0;\vec p_1+\vec p_3) 
=(-\imath)\frac{g^2}{2N_c}
\frac{1}{ P^0-2m-2 {\cal I}_{r} +2\imath\e}}\nonumber\\
&&\times\left\{
\frac{g^2}{2N_c}\left[\left(N_c^2+1\right)
\de_{\al\ga}\de_{\tau\eta}
-2N_c\de_{\al\eta}\de_{\ga\tau}\right]
\int\dk{\vec\w}W_{\si\si}(\vec\w)
W_{\si\si}(\vec p_1+\vec p_3+\vec\w)\right.\nonumber\\
&&+\left.
\left[N_c\de_{\al\la}\de_{\tau \ka}-\de_{\al\ka}
\de_{\tau\la}\right]
\int\dk{\vec\w}W_{\si\si}(\vec\w)
\G^{(c)}_{\ka\la\ga\eta}(P_0;\vec p_1+\vec p_3+\vec\w)\right\}
\hspace{3cm}
\label{eq:1PIsimpl-di1}
\eea
\bea
\lefteqn{\G^{(p)}_{\al\tau\ga\eta}(P_0;\vec p_1+\vec p_4) 
=\imath\frac{g^2}{2N_c}
\frac{1}{ P^0-2m-2 {\cal I}_{r} +2\imath\e}}\nonumber\\
&&\times\left\{
\frac{g^2}{2N_c}\left[\left(N_c^2+1\right)\de_{\al\eta}\de_{\tau\ga}
-2N_c\de_{\al\ga}\de_{\tau\eta}\right]
\int\dk{\vec\w}W_{\si\si}(\vec\w)
W_{\si\si}(\vec p_1+\vec p_4+\vec\w)\right.\nonumber\\
&&-\left.
\left[N_c\de_{\al\la}\de_{\tau \ka}-\de_{\al\ka}\de_{\tau\la}\right]
\int\dk{\vec\w}W_{\si\si}(\vec\w)
\G^{(p)}_{\ka\la\ga\eta}(P_0;\vec p_1+\vec p_4+\vec\w) \right\}
\hspace{3cm}
\label{eq:1PIsimpl-di2}
\eea

Further, we make the color decomposition:
\be
\G^{(c,p)}_{\al\tau\ga\eta}
=\de_{\al\ga}\de_{\tau\eta}\G^{(c,p1)}
+\de_{\al\eta}\de_{\tau \ga}\G^{(c,p2)},
\label{eq:g4color1PIdi}
\ee
where $\G^{(c,p1)}$, $\G^{(c,p2)}$ are Dirac scalar functions.
Inserting this into Eqs.~(\ref{eq:1PIsimpl-di1},
\ref{eq:1PIsimpl-di2}), we obtain the following set of equations:
\bea
\lefteqn{\G^{(c1)}(P_0;\vec p_1+\vec p_3) 
=(-\imath)\frac{g^2}{2N_c}
\frac{1}{ P^0-2m-2 {\cal I}_{r} +2\imath\e}}\nonumber\\
&&\times\left\{
\frac{g^2}{2N_c}\left(N_c^2+1\right)\int\dk{\vec\w}
W_{\si\si}(\vec\w)
W_{\si\si}(\vec p_1+\vec p_3+\vec\w)\right.\nonumber\\
&&+\left. \int\dk{\vec\w}W_{\si\si}(\vec\w)
\left[N_c\G^{(c2)}(P_0;\vec p_1+\vec p_3+\vec\w)
-\G^{(c1)}(P_0;\vec p_1+\vec p_3+\vec\w) \right] \right\}
\label{eq:1PIcomp-di1}\\
\lefteqn{\G^{(p1)}(P_0;\vec p_1+\vec p_4)
 =(-\imath)\frac{g^2}{2N_c}
\frac{1}{ P^0-2m-2 {\cal I}_{r} +2\imath\e}}\nonumber\\
&&\times\left\{
g^2\int\dk{\vec\w}W_{\si\si}(\vec\w)
W_{\si\si}(\vec p_1+\vec p_4+\vec\w)\right.\nonumber\\
&&+\left. \int\dk{\vec\w}W_{\si\si}(\vec\w)
\left[N_c\G^{(p2)}(P_0;\vec p_1+\vec p_4+\vec\w)
-\G^{(p1)}(P_0;\vec p_1+\vec p_4+\vec\w) \right] \right\}
\label{eq:1PIcomp-di2}\\
\lefteqn{\G^{(c2)}(P_0;\vec p_1+\vec p_3) 
=\imath\frac{g^2}{2N_c}
\frac{1}{P^0-2m-2 {\cal I}_{r} +2\imath\e}}\nonumber\\
&&\times\left\{
g^2\int\dk{\vec\w}W_{\si\si}(\vec\w)
W_{\si\si}(\vec p_1+\vec p_3+\vec\w)\right.\nonumber\\
&-&\left. \int\dk{\vec\w}W_{\si\si}(\vec\w)
\left[N_c\G^{(c1)}(P_0;\vec p_1+\vec p_3+\vec\w)
-\G^{(c2)}(P_0;\vec p_1+\vec p_3+\vec\w) \right] \right\}
\label{eq:1PIcomp-di3}\\
\lefteqn{\G^{(p2)}(P_0;\vec p_1+\vec p_4) 
=\imath\frac{g^2}{2N_c}
\frac{1}{ P^0-2m-2 {\cal I}_{r} +2\imath\e}}\nonumber\\
&&\times\left\{
\frac{g^2}{2N_c}\left(N_c^2+1\right)\int\dk{\vec\w}
W_{\si\si}(\vec\w)
W_{\si\si}(\vec p_1+\vec p_4+\vec\w)\right.\nonumber\\
&-&\left. \int\dk{\vec\w}W_{\si\si}(\vec\w)
\left[N_c\G^{(p1)}(P_0;\vec p_1+\vec p_4+\vec\w)
-\G^{(p2)}(P_0;\vec p_1+\vec p_4+\vec\w) \right] \right\}
\label{eq:1PIcomp-di4}
\hspace{2cm}
\eea

In the next step, we transform to coordinate space and rearrange the
terms, such that we obtain:
\bea
&&\hspace{-0,7cm}\G^{(c1)}(P_0;\vec y)+\G^{(c2)}(P_0;\vec y) 
=(-\imath)
\left(\frac{g^2}{2N_c}\right)^2
\frac{\left(N_c-1\right)^2W_{\si\si}(\vec y)^2}
{ P_0-2m-2 {\cal I}_{r} -\imath\frac{g^2}{2N_c}(1-N_c) 
W_{\si\si}(\vec y)+2\imath\e}\nonumber\\
\label{eq:1PIcomp-di1fin}\\
&&\hspace{-0,7cm}\G^{(c1)}(P_0;\vec y)-\G^{(c2)}(P_0;\vec y) 
=(-\imath)
\left(\frac{g^2}{2N_c}\right)^2
\frac{\left(N_c+1\right)^2W_{\si\si}(\vec y)^2}
{ P_0-2m-2 {\cal I}_{r}-\imath\frac{g^2}{2N_c}(1+N_c) 
W_{\si\si}(\vec y)+2\imath\e}\nonumber\\
\label{eq:1PIcomp-di2fin}\\
&&\hspace{-0,7cm}\G^{(p1)}(P_0;\vec x)+\G^{(p2)}(P_0;\vec x) 
=(+\imath)
\left(\frac{g^2}{2N_c}\right)^2
\frac{\left(N_c-1\right)^2W_{\si\si}(\vec x)^2}
{ P_0-2m-2 {\cal I}_{r} -\imath\frac{g^2}{2N_c}(1-N_c)
 W_{\si\si}(\vec x)+2\imath\e}\nonumber\\
\label{eq:1PIcomp-di3fin}\\
&&\hspace{-0,7cm}\G^{(p1)}(P_0;\vec x)-\G^{(p2)}(P_0;\vec x) 
=(-\imath)
\left(\frac{g^2}{2N_c}\right)^2
\frac{\left(N_c+1\right)^2W_{\si\si}(\vec x)^2}
{ P_0-2m-2 {\cal I}_{r}-\imath\frac{g^2}{2N_c}(1+N_c) 
W_{\si\si}(\vec x)+2\imath\e}\nonumber\\
\label{eq:1PIcomp-di4fin}
\eea
In the above, $\vec x$ and $\vec y$ represent the separations
corresponding to the momentum $\vec p_1+\vec p_4$ and $\vec p_1+\vec
p_3$, respectively.  The final expression for the diquark 4-point
Green's function reads:
\bea
\G_{\al\tau\ga\eta}(P_0;\vec x, \vec y)&=&
\de_{\al\ga}^f\de_{\tau\eta}^f
\left\{
\frac{1}{2}\left(\de_{\al\ga}\de_{\tau\eta}
+\de_{\al\eta}\de_{\tau \ga}\right)
\left[\G^{(c1)}(P_0;\vec y)+\G^{(c2)}(P_0;\vec y)\right]\right. 
\nonumber\\
&&+\left.\frac{1}{2}\left(\de_{\al\ga}\de_{\tau\eta}
-\de_{\al\eta}\de_{\tau \ga}\right)
\left[\G^{(c1)}(P_0;\vec y)-\G^{(c2)}(P_0;\vec y)\right]
\right\}\nonumber\\
&&+\de_{\al\eta}^f\de_{\tau \ga}^f
\left\{
\frac{1}{2}\left(\de_{\al\ga}\de_{\tau\eta}
+\de_{\al\eta}\de_{\tau \ga}\right)
\left[\G^{(p1)}(P_0;\vec x)+\G^{(p2)}(P_0;\vec x)\right]\right. 
\nonumber\\
&&+\left.\frac{1}{2}\left(\de_{\al\ga}\de_{\tau\eta}
-\de_{\al\eta}\de_{\tau \ga}\right)
\left[\G^{(p1)}(P_0;\vec x)-\G^{(p2)}(P_0;\vec x)\right]
\right\}
\label{eq:diqsolution}
\eea
with the components given by Eqs.~(\ref{eq:1PIcomp-di1fin} --
\ref{eq:1PIcomp-di4fin}).

As in the case of the $\bar qq$ systems, with this solution we return
to the diagrams (f), (i) and (g). Writing out the explicit form of the
energy integrals we notice that their form is identical to the
quark-antiquark case, since the $\e$ prescription is similar,
regardless of the internal quark (or antiquark) propagator. Thus,
these diagrams are also vanishing.

The above equation can be rewritten such that the pole structure
becomes manifest. Introducing the notations
\bea
f_{+}(\vec y)&=&(-\imath)
\left(\frac{g^2}{2N_c}\right)^2
\frac{\left(N_c-1\right)^2W_{\si\si}(\vec y)^2}
{ P_0-2m-2 {\cal I}_{r} -\imath\frac{g^2}{2N_c}(1-N_c) 
W_{\si\si}(\vec  y)+2\imath\e}\\
 f_{-}(\vec y) &=&(-\imath)
\left(\frac{g^2}{2N_c}\right)^2
\frac{\left(N_c+1\right)^2W_{\si\si}(\vec y)^2}
{ P_0-2m-2 {\cal I}_{r}-\imath\frac{g^2}{2N_c}(1+N_c)
 W_{\si\si}(\vec y)+2\imath\e},
\eea
we have that
\bea
\G_{\al\tau\ga\eta}(P_0;\vec x, \vec y)&=&
\frac{1}{2}\left\{
\left(\de_{\al\ga}\de_{\tau\eta}+\de_{\al\eta}\de_{\tau \ga}\right)
\left[ \de_{\al\ga}^f\de_{\tau\eta}^f  f_{+}(\vec y)- 
\de_{\al\eta}^f\de_{\tau\ga}^f f_{+}(\vec x)\right]\right.
 \nonumber\\
&&+\left.
\right(\de_{\al\ga}\de_{\tau\eta}-\de_{\al\eta}\de_{\tau \ga}\left)
\left[ \de_{\al\ga}^f\de_{\tau\eta}^f  f_{-}(\vec y)
+ \de_{\al\eta}^f\de_{\tau\ga}^f f_{-}(\vec x)\right]
\right\}
\label{eq:diqsolution1}
\eea

Analyzing the pole structure of the above equation we notice that, as
in the case of the $\bar qq$ systems, we have two different pole
conditions:
\be
P_0-2m-2 {\cal I}_{r}-\imath\frac{g^2}{2N_c}(1\pm N_c) W_{\si\si}(\vec x)=0
\ee
From the first equation (corresponding to the color antisymmetric term
in the above formula), we find that $N_c=2,-1$, and from the second
equation, we obtain $N_c=-2,1$. Hence, the only physical solution,
with $N_c=2$, corresponds to color antisymmetric and flavor symmetric
configuration, in agreement with our findings from the previous
chapter that bound states exist only for color antisymmetric $SU(2)$
baryons (in that case flavor symmetry was implicit).  Notice also that
the poles in the case of the diquark system disconnect, as opposite to
the meson case, where they multiply.

%% file: conclusions.tex
\chapter{Summary and conclusions}
\label{chap:concl}

This thesis has been constructed from three major building blocks: the
general derivation of the quark \DS equations in Coulomb gauge first
order formalism, the perturbative studies at one-loop order, and the
nonperturbative investigations in the limit of the heavy quark mass.
We will present our summary and conclusions separately for each part.

\section{Functional derivation of \DS equations}

Starting with the QCD Lagrangian, we have introduced the gauge fixing
technique and the Faddeev-Popov method in Coulomb gauge. Further, we
have converted to first order formalism. There are two reasons
motivating this choice: firstly, the (unphysical) ghost degrees of
freedom can be formally eliminated and secondly, the system can be
reduced to the ``would-be physical'' degrees of freedom. Formal here
refers to the fact that the resulting equations are nonlocal and hence
very difficult to approach in practical calculations.

Using functional derivation techniques, in Chapter~\ref{chap:ds} we
have explicitly derived the quark field equations of motion. Based on
these equations, we have obtained the Feynman rules and the relevant
tree-level proper two-point and vertex functions. We have also
discussed the general decomposition of the quark (proper and
connected) two-point functions.  Based on the Lagrange transformation,
we have derived exact relations between the corresponding dressing
functions. Given that at one-loop order in perturbation theory the
fourth Dirac structure $\ga^0\ga^i$ of the quark gap equation
vanishes, the set of equations relating these functions is
significantly reduced.  Starting with the quark field equation of
motion, we have derived the \DS equations for the quark proper
two-point function, quark contributions to the gluon proper two-point
functions and quark-gluon vertex functions.  Moreover, in
Chapter~\ref{chap:nGreen} we have presented the explicit derivation of
the four-point Green's function for quark-antiquark and diquark
systems.

Since the \DS equations build an infinite tower of coupled integral
equations, they cannot be solved exactly, and hence approximation
schemes have to be employed. These have to preserve the symmetries of
the theory, which are reflected in the \ST identities. Thus, starting
with the \emph{time-dependent} BRST transform we have explicitly
derived the \ST identities for the quark-gluon vertices and, in the
limit of heavy quark mass, for the four-quark-gluon vertex.

\section{One-loop perturbative  results}

In the second part of this thesis, a one-loop perturbative analysis of
the quark gap equation, quark contribution to the gluonic two-point
functions and quark-gluon vertex functions has been undertaken. The
various propagator and two-point dressing functions have been
evaluated at this order. To this end, the (dimensionally regularized)
results for the required noncovariant massive integrals have been
obtained, using differential equations and integration by part
techniques.

The results for the various two-point functions are rather
illuminating. Since the singularities are absent in both Euclidean and
spacelike Minkowski regions, the analytic continuation between
Euclidian and Minkowski is justified.  The second important physical
results is the renormalization of the quark mass and
propagator. Namely, we have verified that the one-loop renormalized
quark mass agrees explicitly with the calculation performed in linear
covariant gauges.  Also, up to color factors the results for the quark
propagator dressing function agree with the corresponding results
obtained in Quantum Electrodynamics. Finally, the correct one-loop
coefficient of the perturbative $\ba$-function has been obtained.

Turning to the quark-gluon vertex functions, we have explicitly
evaluated their divergent parts and considered them in conjunction
with the corresponding \ST identity. This identity contains some
peculiar objects, the quark-ghost scattering kernels (analogous to
ghost-gluon kernels known from \YM theory). In order to gain some more
information about them, it has been helpful to analyze the \ST
identity at one-loop perturbative order. This has been done in two
different ways: firstly by using the translation invariance of the
loop integrals, without explicitly evaluating them, and secondly by
considering only the divergent parts of the integrals entering the
identity.  In this later investigation, we have employed the results
for the divergent parts of the quark-gluon vertex functions previously
derived and have found that the equation is satisfied.  Also, the
quark-ghost kernels appear not to contain divergences, in agreement
with the calculations performed using the method of split dimensional
regularization.

\section{Heavy quarks}

In the third part of the thesis, we have considered the limit of the
heavy quark mass. After performing a heavy quark transform of the
quark field (adapted for Coulomb gauge), we have expanded the
generating functional of the theory in the mass parameter and retained
only the leading order. In this case, the system has simplified
dramatically: we have found that only the temporal gluon propagator
contributes at leading order, whereas the spatial gluon is suppressed
by the mass expansion.  Further, we have truncated the \YM sector to
include only the non-perturbative (energy-independent) temporal gluon
propagator and neglect all the higher order \YM Green's functions.

In this setting, we have used the full nonperturbative quark equations
of QCD in order to study the confinement properties. The gap equation,
supplemented by the \ST identity, has been solved and we have shown
that the rainbow-ladder approximation to the quark (and antiquark)
propagators is exact in this case. These equations have been used
along with the \BS equation for mesons and Faddeev equation for
baryons, and we have demonstrated that the ladder approximation to the
\BS equation is also exact.

Using the assumption that the temporal gluon propagator is energy
independent, it was then straightforward to find solutions for these
equations.  In the meson case, from the \BS equation we have found
that only color singlet mesons and $SU(2)$ baryons have finite energy
that increases linearly with the distance, i.e. confining solution
(and otherwise the system is physically not allowed). Further, we have
found that there exist a direct connection between the temporal gluon
propagator and the string tension (at least within the approximation
scheme considered here).

Turning to the baryon case, we have considered the Faddeev equation
for three-quark states in a symmetric configuration.  As in the case
of the $\bar qq$ systems, we have found that the bound state energy
between three quarks increases linearly with the separation and we
have provided a direct connection between the temporal gluon
propagator and the physical string tension. Furthermore, we predict
that the string tension for three-quark states is 3/2 times that of
the $\bar qq$ system.

At this point, we make a short comment regarding our approximations.
Firstly, we have assumed that the temporal gluon propagator is energy
independent. As explained in the text, this assumption is supported by
the lattice results and moreover, from formal arguments it can be
inferred that this propagator must have at least a part that is energy
independent, in order to cancel the ghost loops in the \YM expressions
and to solve the energy divergence problem of Coulomb gauge. The
second approximation was to neglect the \YM vertices and, given that
the spatial quark-gluon vertices are suppressed by the mass, we have
found that the temporal vertex remains nonperturbatively bare. Since
the temporal gluon propagator is dressed, we have the situation of a
gluon string connecting two naked color charges, and thus the
inclusion the \YM vertices would represent the dressing of these naked
charges, i.e. the screening mechanism.  Thus, we anticipate that this
would only lower the value of the string tension, and would not alter
the linear behavior of the confining potential. The fact that the
three-gluon vertex is irrelevant for the infrared properties of the
theory has been also shown in the Hamiltonian approach
\cite{Feuchter:2004mk}.

 Finally, in the heavy mass limit and within the same truncation
scheme, we have also considered the 4-point quark Green's functions,
and in particular we have studied the role of singularities of these
functions. Firstly, we have found that both for $\bar qq$ and diquark
systems the poles (physical and unphysical) naturally separate --- as
is well-known, this would not be the case in the usual QCD
calculations.  Also, we have found that the bound state energies
emerge as a pole of the resonant component, thus providing and
explicit connection between the (nonperturbative) 4-point Green's
function and the bound state energy resulting from the \BS equation.

%% file: app.not.tex
\appendix

\chapter{Conventions and useful formulae for the gauge group \texorpdfstring{$SU(N_c)$}{SU(N)}}
\label{chap:app.not}

In this work, natural units are used:
\be
\hbar=c=1.
\ee
Initially we work in Minkowski space, with the following metric:
\bea
g_{\mu\nu}=\textrm{diag}(1, -\vec 1).
\eea
Where perturbative integrals are to be explicitly evaluated, we
analytically continue to Euclidean space, i.e.  $k_0\to\imath k_4$, as
explained in the text.

 The group elements of $SU(N_c)$ are unitary $N_c\times N_c$ matrices
with determinant one and can be written in the form
\be
U(x)=\exp\{-\imath \theta^{a}(x)T^a\}
\label{eq:Umatr}
\ee
where $\theta^{a}(x)$ specify the angle of rotation in color space.

The $N_c^2-1$ (Hermitian) generators of $SU(N_c) $ obey the Lie
algebra
\be
[T^a,T^b]=\imath f^{abc}T^c
\label{eq:com}
\ee 
where the numbers $f^{abc}$ are the completely antisymmetric structure
constants of the group. From the relation
\be
\mbox{det~} U=\exp\left\{\mbox{Tr} \ln U\right\}
=\exp\{-\imath
\theta^{a}(x) \mbox{Tr~}  T^a\}
\ee
it follows that the generators are traceless matrices:
\be
\mbox{Tr~} T^a=0. 
\ee
In addition, we have the normalization condition
\be
\mbox{Tr} (T^aT^b)=\frac{1}{2}\delta^{ab}.
\ee
The sum rules used in the text are:
\bea
T^aT^a&=&C_F \mathds{1}\label{eq:sumrul1}\\
T^bT^cf^{dbc}&=&\frac{\imath}{2}T^d N_c \label{eq:sumrul2}\\
T^aT^bT^a&=&\left ( C_F-\frac{N_c}{2}\right)T^a \label{eq:sumrul3}
\eea
where the Casimir invariant is given by
\be
C_F=\frac{N_c^2-1}{2N_c}.\label{eq:casimir}
\ee
We note also the Fierz identity:
\be
2\left[T^a\right]_{\al\ba}\left[T^a\right]_{\de\ga}=\de_{\al\ga}
\de_{\de\ba}-\frac{1}{N_c}\de_{\al\ba}\de_{\de\ga}.
\label{eq:fierz}
\ee

%% file: app.standard.tex
\chapter{Standard massive integrals }
\label{chap:app2}

In this Appendix, we present the derivation of some standard Coulomb
gauge massive integrals used in Chapter~\ref{chap:g^2}.  Consider the
integral:
\be
J_m(k^2)=\int\frac{\dk{\w}}{[\w^2+m^2]^\mu[(k-\w)^2+m^2]^\nu}.
\label{eq:smi_example}
\ee
In the case $\mu=\nu=1$ this gives the scalar integral associated
with, for example, the fermion loop in quantum electrodynamics
\cite{Muta:1998vi}.

We present here a method to evaluate such integrals for arbitrary
denominator powers (developed originally in Ref.~\cite{Boos:1990rg})
and generalize to the various additional noncovariant integrals.  We
start by writing the Taylor expansion of the massive propagator in
terms of a hypergeometric function in the following way:
\be
\frac{1}{[\w^2+m^2]^{\mu}}=\frac{1}{[\w^2]^{\mu}}
{}_{1}F_{0} \left(\mu;-\frac{m^2}{\w^2}\right).
\label{eq:prop}
\ee
Now, the idea is to use the Mellin-Barnes representation of the
hypergeometric function ${}_{1}F_{0} (\mu;z)$:
\be
{}_{1}F_{0} (\mu;z)=
\frac{1}{\G(\mu)}\frac{1}{2\pi\imath}\int
\limits_{-\imath\infty}^{\imath\infty}d s (-z)^{s}\G(-s)\G(\mu+s),
\label{eq:hyprep}
\ee
where the contour in the complex plane separates the left poles of the
$\G$ functions from the right poles. A first advantage of this
representation is that the ``mass term'' gets separated from the
massless propagator and the remaining integrals can be calculated with
the Cauchy residue theorem, as we shall see below.  Also, in order to
study various momentum regimes, the results can be written as a
function of either $k^2/m^2$, or $m^2/k^2$. This we do by using the
standard formulas of analytic continuation of the hypergeometric
function (for an extended discussion, see \cite{Boos:1990rg}).

Applying \eq{eq:hyprep} to the massive propagator we can rewrite the
integral $J_m(k^2)$ as:
\bea
\hspace{-1,2cm}J_m(k^2)&=&
\frac{1}{(2\pi\imath)^2}\frac{1}{\G(\mu)\G(\nu)}
\int\!\!\!\!\!\int\limits_{-\imath\infty}^{\imath\infty}ds\,dt 
(m^2)^{s+t}\G(-s)\G(-t)\G(\mu+s)\G(\nu+t)\nonumber\\
&&\times\int\frac{\dk{\w}}{(\w^2)^{\mu+s}[(k-\w)^2]^{\nu+t}}.
\eea
Inserting the general result for the massless integral, derived in 
Ref.~\cite{Watson:2007mz}:
\be
\int\frac{\dk{\w}}{[\w^2]^\mu[(k-\w)^2]^\nu}
=\frac{[k^2]^{2-\mu-\nu-\e}}{(4\pi)^{2-\e}}
\frac{\G(\mu+\nu+\e-2)}{\G(\mu)\G(\nu)}
\frac{\G(2-\mu-\e)\G(2-\nu-\e)}{\G(4-\mu-\nu-2\e)},
\label{eq:si_ex}
\ee
we get for the integral $J_m$:
\bea
\hspace{-1,5cm}J_m(k^2)&=&\frac{[k^2]^{2-\nu-\mu-\e}}{(4\pi)^{2-\e}}
\frac{1}{(2\pi\imath)^2}\frac{1}{\G(\mu)\G(\nu)}
\int\!\!\!\!\!\int\limits_{-\imath\infty}^{\imath\infty}
\!\!ds\, dt \left(\frac{m^2}{k^2}\right)^{s+t}
\G(-s)\G(-t)\nonumber\\ 
&&\times \G(2-\e-\mu-s)\G(2-\e-\nu-t)
\frac{\G(\mu+\nu+s+t-2+\e)}{\G(4-2\e-\mu-\nu-s-t)}.
\eea
With the change of variable $t=2-\e -\mu-\nu-u-s$ (for such a
replacement, the left and right poles of the $\G$ function are simply
interchanged and therefore the condition of separating the poles is
not contradicted) we obtain:
\bea
J_m(k^2)&=&\frac{[m^2]^{2-\mu-\nu-\e}}{(4\pi)^{2-\e}}
\frac{1}{(2\pi\imath)}\frac{1}{\G(\mu)\G(\nu)}
\int\limits_{-\imath\infty}^{\imath\infty}du
\left(\frac{m^2}{k^2}\right)^{-u}\!\frac{\G(-u)}{\G(2-\e+u)}
\nonumber\\
&&\!\!\!\times\frac{1}{(2\pi\imath)}
\int\limits_{-\imath\infty}^{\imath\infty}ds 
\G(-s)\G(2-\e-\mu-s)\G(-2+\e+\nu+\mu+u+s)\G(\mu+u+s).
 \nonumber\\
\label{eq:Jintermediate}
\eea
To evaluate the integral over $s$ we use the Barnes Lemma:
\be
\frac{1}{(2\pi\imath)}\int
\limits_{-\imath\infty}^{\imath\infty}ds\textrm{~}
\G(a+s)\G(b+s)\G(c-s)\G(d-s)=
\frac{\G(a+c)\G(a+d)\G(b+c)\G(b+d)}{\G(a+b+c+d)}
\ee
and for the integral \eq{eq:Jintermediate} it follows immediately that
\bea
\hspace{-1,5cm}J_m(k^2)&=&\frac{[m^2]^{2-\mu-\nu-\e}}{(4\pi)^{2-\e}}
\frac{1}{(2\pi\imath)}\frac{1}{\G(\mu)\G(\nu)}\nonumber\\
&&\int\limits_{-\imath\infty}^{\imath\infty} du
\left(\frac{m^2}{k^2}\right)^{-u}
\frac{\G(-u)\G(\mu+u)\G(\nu+u)\G(\mu+\nu-2+\e+u)}{\G(\mu+\nu+2u)}. 
\eea
Closing the integration contour on the right we have:
\bea
\hspace{-1,5cm}J_m(k^2)&=&\frac{[m^2]^{2-\mu-\nu-\e}}{(4\pi)^{2-\e}}
\frac{1}{(2\pi\imath)}\frac{1}{\G(\mu)\G(\nu)}\nonumber\\
&&(2\pi\imath)
\sum\limits_{j=0}^{\infty}
\left(-\frac{m^2}{k^2}\right)^{-j}\frac{1}{j!}
\frac{\G(\mu+j)\G(\nu+j)\G(\mu+\nu-2+\e+j)}{\G(\mu+\nu+2j)}.
\eea
With the help of the duplication formula
\be
\G(2z)=2^{2z-1}\pi^{-1/2}\G(z)\G\left(z+\frac 12\right),
\ee
we can rewrite $J_m$ as:
\bea
\lefteqn{J_m(k^2)=\frac{[m^2]^{2-\mu-\nu-\e}}{(4\pi)^{2-\e}}
\frac{\G(\mu+\nu-2+\e)}{\G(\mu+\nu)}}\nonumber\\
&&\!\!\sum\limits_{j=0}^{\infty}
\left(-\frac{m^2}{k^2}\right)^{-j}\frac{1}{2^{2j}}\frac{1}{j!}
\frac{\G(\mu+j)}{\G(\mu)}\frac{\G(\nu+j)}{\G(\nu)}
\frac{\G(\mu+\nu-2+\e+j)}{\G(\mu+\nu-2+\e)}
\frac{\G(\frac{\mu+\nu}{2})} {\G(\frac{\mu+\nu}{2}+j)}
\frac{\G(\frac{\mu+\nu+1}{2})} {\G(\frac{\mu+\nu+1}{2}+j)}. 
\nonumber\\
\eea
The sum is clearly a representation of the hypergeometric
${}_{3}F_{2}(a,b,c;d,e;z)$ (see, for example, Ref.~\cite{abramowitz})
and we finally obtain:
\bea
\hspace{-2,5cm}
J_m(k^2)&=&\frac{[m^2]^{2-\mu-\nu-\e}}{(4\pi)^{2-\e}}
\frac{\G(\mu+\nu-2+\e)}{\G(\mu+\nu)}\nonumber\\&&
\times{}_{3}F_{2} \left(\mu,\nu,\mu+\nu-2+\e;
\frac{\mu+\nu}{2},\frac{\mu+\nu+1}{2}
;-\frac{k^2}{4m^2}\right). 
\label{eq:smi1}
\eea
A trivial computation shows that the result \eq{eq:smi1} is consistent
with the known results in the limit $m=0$.  All we have to do is to
invert the argument of the hypergeometric according to the formula
(found, for example, in Ref.~\cite{bateman})
\bea
\lefteqn{{}_{3}F_{2}(a_1,a_2,a_3;b_1,b_2;z)=
\frac{\G(b_1)\G(b_2)}{\G(a_1)\G(a_2)\G(a_3)}
}\nonumber\\
&&
\times\bigg\{
\frac{\G(a_1)\G(a_2-a_1)\G(a_3-a_1)}{\G(b_1-a_1)
\G(b_2-a_1)}(-z)^{-a_1}\nonumber\\
&&\times{}_{3}F_{2}
\left(a_1,a_1-b_1+1,a_1-b_2+1;a_1-a_2+1,a_1-a_3+1;\frac 1z\right)
\nonumber\\
&&+\frac{\G(a_2)\G(a_1-a_2)\G(a_3-a_2)}{\G(b_1-a_2)
\G(b_2-a_2)}(-z)^{-a_2}\nonumber\\
&&\times{}_{3}F_{2}
\left(a_2,a_2-b_1+1,a_2-b_2+1;-a_1+a_2+1,a_2-a_3+1;\frac 1z\right)
\nonumber\\
&&+
\frac{\G(a_3)\G(a_1-a_3)\G(a_2-a_3)}{\G(b_1-a_3)
\G(b_2-a_3)}(-z)^{-a_3}\nonumber\\
&&\times{}_{3}F_{2}\left(a_3,a_3-b_1+1,a_3-b_2+1;
-a_1+a_3+1,-a_2+a_3+1;\frac 1z\right)
\bigg\}.\hspace{2cm}
\eea
The same method can be applied to the other noncovariant integrals
with different denominator structures that appear in the text and the
general expressions for these read:
\bea
\int\frac{\dk{\w}}{[\w^2]^\mu[(k-\w)^2+m^2]^\nu}
&=&
\frac{[m^2]^{2-\mu-\nu-\e}}{(4\pi)^{2-\e}}
\frac{\G(2-\mu-\e) \G(\mu+\nu+\e-2)}{\G(\nu)\G(2-\e)}
\nonumber\\
&&\times{}_{2}F_{1} \left(\mu,\mu+\nu+\e-2;2-\e;
-\frac{k^2}{m^2}\right),
\label{eq:smi2}\\
\int\frac{\dk{\w}}{[\vec{\w}^2]^\mu[(k-\w)^2+m^2]^\nu}
&=&
\frac{[m^2]^{2-\mu-\nu-\e}}{(4\pi)^{2-\e}}
\frac{\G(\frac32-\mu-\e) \G(\mu+\nu+\e-2)}
{\G(\nu)\G(3/2-\e)}\nonumber\\
&&\times{}_{2}F_{1} \left(\mu,\mu+\nu+\e-2;3/2-\e;
-\frac{\vec{k}^2}{m^2}\right).
\label{eq:smi3}
\eea
Differentiation with respect to $k_4$ or $k_i$ gives rise to
expressions for integrals with more complicated numerator structure.
For completeness, we list here the first order $\e$ expansion of the
integrals arising into the one-loop perturbative expressions
considered in this work:
\bea
\int\frac{\dk{\w}}{(\w^2+m^2)[(k-\w)^2+m^2]}
&=&\frac{[m^2]^{-\e}}{(4\pi)^{2-\e}}
\left\{
\frac1\e-\ga +2 \right. \nonumber\\
&&\hspace{-0,5cm}\left.-\sqrt{1+\frac{4m^2}{k^2}}
\ln\left(\frac{\sqrt{1+\frac{4m^2}{k^2}}+1}
{\sqrt{1+\frac{4m^2}{k^2}}-1}
\right)
+\cal{O}(\e)\right\},\\
\int\frac{\dk{\w}}{\w^2[(k-\w)^2+m^2]}
&=&\frac{[m^2]^{-\e}}{(4\pi)^{2-\e}}
\left\{
\frac1\e-\ga+2 \right.\nonumber\\
&&\hspace{-0,5cm}\left.-\left(1+\frac{m^2}{k^2}\right)
\ln \left( 1+\frac{k^2}{m^2}
\right)+\cal{O}(\e)\right\},\\
\int\frac{\dk{\w} \w_i}{\w^2[(k-\w)^2+m^2]}
&=&k_i\frac{[m^2]^{-\e}}{(4\pi)^{2-\e}}
\left\{
\frac{1}{2\e}-{\ga}{2}+1 \right.\nonumber\\
&&\hspace{-0,5cm}\left.+\frac12\frac{m^2}{k^2}
-\frac12\!\left(1+\frac{m^2}{k^2}\right)^2\!
\ln \left( 1+\frac{k^2}{m^2}
\right)\!+\cal{O}(\e)
\!\right\}\!\!,\\
\hspace{-1cm}\int\frac{\dk{\w}}{\vec\w^2[(k-\w)^2+m^2]}
&=&\frac{[m^2]^{-\e}}{(4\pi)^{2-\e}}
\left\{
\frac2\e-2\ga+8 \right.\nonumber\\
&&\hspace{-0,5cm}\left.-2\sqrt{1+\frac{m^2}{\vec k^2}}
\ln\left(\frac{\sqrt{1+\frac{m^2}{\vec k^2}}+1}
{\sqrt{1+\frac{m^2}{\vec k^2}}-1}\right)
+\cal{O}(\e)
\right\}.
\eea

%% file: app.check.noncov2pct.tex
\chapter{Checking the nonstandard integrals}
\label{chap:app3}

One way to check analytically the results for the nonstandard
integrals introduced in Chapter~\ref{chap:g^2}, $A_m$ and $A_m^4$,
\eq{eq:asol} and \eq{eq:a4sol}, respectively, is to make an expansion
around $k_4^2=0$ and evaluate the resulting integrals with the help of
the Schwinger parametrization.  Let us first consider the integral
$A_m$, \eq{eq:adef}.  Using Schwinger parameters \cite{pascual} (see
also \cite{Collins:1984xc}), we can rewrite the denominator factors as
exponential functions to give:
\bea
A_m&=&\int_0^\infty\,d\al\, d\ba\, d\ga \int\dk{\w}
\exp\left\{-(\al+\ba)\w_4^2
+2\ba k_4\w_4-\ba k_4^2-(\al+\ba+\ga)\vec{\w}^2 \right.
\nonumber\\
&&\left.+2\ba\s{\vec{k}}{\vec{\w}}
-\ba\vec{k}^2-\ba m^2\right\}.
\eea
Applying similar reasoning as in Ref.~\cite{Watson:2007mz}, we come to
the following parametric form of the integral (recall that $x=k_4^2$,
$y=\vec{k}^2$):
\bea
\lefteqn{A_m=\frac{(x+y+m^2)^{-1-\e}}{(4\pi)^{2-\e}}\G(1+\e)}
\nonumber\\
&&\times \int_0^1 d\ba\int_0^{1-\ba}\,\frac{d\al}{ (\al+\ba)^{1/2}}
\left[\frac{\al\ba}{(\al+\ba)}
\frac{x}{(x+y+m^2)}+\frac{\ba(1-\ba)y+\ba m^2}{x+y+m^2}\right]
^{-1-\e}\!\!\!.
\eea
For general values of $x$, the integral above cannot be solved because
of the highly nontrivial denominator factor. Since there can be no
singularities at $x=0$ (this would invalidate the Wick rotation which,
as discussed in Chapter~\ref{chap:g^2}, does hold in this case), we
can safely make an expansion around this point and then integrate.  To
first order in powers of $x$ we have:
\bea
\lefteqn{
A_m\stackrel{x\rightarrow0}{=}
\frac{(x+y+m^2)^{-1-\e}}{(4\pi)^{2-\e}}
\G(1+\e)\int_0^1 d\ba\int_0^{1-\ba}d\al\,(\al+\ba)^{-1/2}
\left\{\left[\ba \frac{m^2+y(1-\ba)}{m^2+y}\right]^{-1-\e}\right.}
\nonumber\\
&&-\left.\left[\ba \frac{m^2+y(1-\ba)}{m^2+y}\right]^{-2-\e}
\left[ \frac{\al\ba}{(m^2+y)(\al+\ba)}
-\ba \frac{m^2+y(1-\ba)}{(m^2+y)^2}\right](1+\e)x
\frac{}{}+{\cal O}(x^2)\right\}. \nonumber\\
\eea
After performing the integration we get:
\bea
A_m(x,y)&\stackrel{x\rightarrow0}{=}&
\frac{(x+y+m^2)^{-1-\e}}{(4\pi)^{2-\e}}
(-2)\left\{
\frac1\e \G(1+\e){}_{2}F_1\left(-\e,2+\e;1-\e;-\frac{y}{m^2+y}\right)
\right.\nonumber\\
&&\left. -\frac{x}{m^2+y}+\sqrt{1+\frac{m^2}{y}}
\ln\left(\frac{\sqrt{1+\frac{m^2}{y}}+1}
{\sqrt{1+\frac{m^2}{y}}-1}\right)
+{\cal O}(x^2)+{\cal O}(\e)
\right\}.
\label{eq:a4schw}
\eea
In the above formula, we isolated the hypergeometric term in order to
evaluate the $\e$ expansion separately.  For this purpose, we
differentiate the hypergeometric function with respect to the
parameters.  In general, differentiation of ${}_{2}F_1(a,b;c;z)$ with
respect to, e.g. the parameter $b$, gives (similar expressions are
obtained for differentiation with respect to $a,c$):
\bea
{}_{2}F_1^{(0,1,0,0)}\left(a,b;c;z\right)&=&
\sum_{k=0}^{\infty}\frac{(a)_k (b)_k \Psi(b+k)}{(c)_k}\frac {z^k}{k!}
-\Psi(b){}_{2}F_1\left(a,b;c;z\right),
\label{eq:hypexp}
\eea
where $\Psi(k)$ is the digamma function and $(a)_k$ is the so-called
Pochhammer symbol (defined, for instance, in the standard textbook
\cite{bateman}):
\be
(a)_k=\frac{\G(a+k)}{\G(a)}.
\ee
 With the help of  formula \eq{eq:hypexp}, we obtain:
\bea
{}_{2}F_1\left(-\e,2+\e;1-\e;z\right)=1+\e\ln(1-z)+{\cal O}(\e).
\eea
Inserting this back into \eq{eq:a4schw}, we can write down the result
for the integral $A_m$ (to first order in powers of $x$):
\bea
A_m(x,y)&\stackrel{x\rightarrow0}{=}
&\frac{(x+y+m^2)^{-1-\e}}{(4\pi)^{2-\e}}
\left\{-\frac{2}{\e}+2\ga+2
\left[-\ln\left(\frac{m^2}{m^2+y}\right)
+\frac{x}{m^2+y}
\right]\right.\nonumber\\
&&\left.-2\sqrt{1+\frac{m^2}{y}}
\ln\left(\frac{\sqrt{1+\frac{m^2}{y}}+1}
{\sqrt{1+\frac{m^2}{y}}-1}\right)
+{\cal O}(x^2)+\cal O(\e)\right\},
\eea
which agrees explicitly with the corresponding expansion of the result
given in \eq{eq:asol}.

We now turn to the integral $A^4_m$, given by \eq{eq:a4def}. The
parametric form has the expression:
\bea
\lefteqn{A^4_m=k_4\frac{(x+y+m^2)^{-1-\e}\G(1+\e)}{(4\pi)^{2-\e}}}
\nonumber\\
&&\times\int_0^1 d\ba
\int_0^{1-\ba}d\al\,\frac{\ba}{(\al+\ba)^{3/2}}
\left[\frac{\al\ba}{(\al+\ba)}
\frac{x}{(x+y+m^2)}
+\frac{\ba(1-\ba)y+\ba m^2}{x+y+m^2}\right]^{-1-\e}.
\eea
Calculations similar to the integral $A_m$ bring us to the following
result (to first order in $x$):
\bea
\lefteqn{
A^4_m(x,y)\stackrel{x\rightarrow0}{=}
k_4\frac{(x+y+m^2)^{-1-\e}}{(4\pi)^{2-\e}}
\left\{
2\sqrt{1+\frac{m^2}{y}}
\ln\left(\frac{\sqrt{1+\frac{m^2}{y}}+1}
{\sqrt{1+\frac{m^2}{y}}-1}\right)
\right.
}\nonumber\\
&&-\frac23\frac{x}{y}\left[
1+\left(\frac{m^2}{y}-2\right)\ln\frac{m^2}{m^2+y}-
\frac{2}{\sqrt{1+\frac{m^2}{y}}}
\ln\left(\frac{\sqrt{1+\frac{m^2}{y}}+1}
{\sqrt{1+\frac{m^2}{y}}-1}\right)
\right]
\nonumber\\
&&\left.
+2\left(1+\frac{m^2}{y}\right)
\ln\left(\frac{m^2}{m^2+y}\right)
+{\cal O}(x^2)
+{\cal O}(\e)\right\},\hspace{5cm}
\eea
again in agreement with the expansion of the result given in
\eq{eq:a4sol}.

Another useful check comes from the study of the mass differential
equation, i.e., the mass is in this case regarded as a
variable.\footnote {Recall that when deriving the differential
equations presented in Chapter~\ref{chap:g^2}, the mass has been
treated as a parameter, and only the differentiation with respect to
$k_4$ and $k_i$ has been considered.}

With $I^n$ given by \eq{eq:idef}, the derivative with respect to the
mass reads:
\be
m\frac{\pd I^n}{\pd m}=
\int\frac{\dk{\w}\,\w_4^n}{\w^2[(k-\w)^2+m^2]\vec{\w}^2}
\left\{-2\frac{m^2}{(k-\w)^2+m^2}\right\}.
\label{eq:ademass}
\ee
From the relations \eq{eq:ade0}, \eq{eq:adevec} and \eq{eq:ademass} we
get the following relation:
\be
k_4\frac{\pd I^n}{\pd k_4}
+k_k\frac{\pd I^n}{\pd k_k}+m\frac{\pd I^n}{\pd m}=(d+n-5)I^n.
\ee
Using the same procedures as described in Chapter~\ref{chap:g^2}, we
can then derive a differential equation for the integral in terms of
the mass:
\be
m^2\frac{\pd I^n}{\pd m^2}=(d+n-4)\frac{m^2}{k^2+m^2}I^n
-\frac{m^2}{k^2+m^2}
\int\frac{\dk{\w}\,\w_4^n}{[(k-\w)^2+m^2]^2\vec{\w}^2}.
\label{eq:adeint}
\ee

Starting with the case $n=0$ where $I^0\equiv A_m$, we see that by
inserting the solution, \eq{eq:asol}, we have that in the limit
$\e\rightarrow0$
\bea
m^2\frac{\pd A_m}{\pd m^2}+(1+2\e)\frac{m^2}{k^2+m^2}A_m
\!\!\!\!&=&\!\!\!\!-\frac{m^2(x+y+m^2)^{-2-\e}}{(4\pi)^{2-\e}}
\frac{1}{y\sqrt{1+\frac{m^2}{y}}}
\ln\left(\frac{\sqrt{1+\frac{m^2}{y}}+1}
{\sqrt{1+\frac{m^2}{y}}-1}\right)\nonumber\\
&&+{\cal O}(\e).
\label{eq:masszero1}
\eea
In terms of Schwinger parameters, the explicit integral of
\eq{eq:adeint} reads:
\be
\frac{m^2}{k^2+m^2}\int\frac{\dk{\w}
}{[(k-\w)^2+m^2]^2\vec{\w}^2}
=\frac{m^2}{x+y+m^2}\frac{\G(1+\e)}{(4\pi)^{2-\e}}
\int_0^1d\al\,(1-\al)^{-1/2-\e}(m^2+\al y)^{-1-\e}
\label{eq:masszero2}
\ee
and for $m^2\neq0$ indeed
\bea
\frac{m^2}{k^2+m^2}\int\frac{\dk{\w}}{[(k-\w)^2+m^2]^2\vec{\w}^2}
\!\!&=&\!\!\frac{m^2(x+y+m^2)^{-2-\e}}{(4\pi)^{2-\e}}
\frac{1}{y\sqrt{1+\frac{m^2}{y}}}
\ln\left(\frac{\sqrt{1+\frac{m^2}{y}}+1}{\sqrt{1+\frac{m^2}{y}}-1}\right)
\nonumber\\
&&+{\cal O}(\e),
\eea
showing that the mass differential equation is satisfied.  In fact, in
the differential equation \eq{eq:adeint}, there is a potential
ambiguity arising from the ordering of the limits $m^2\rightarrow0$
and $\e\rightarrow0$. Namely, for $m^2=0$, the right-hand side of
\eq{eq:masszero1} vanishes as $m^2\ln m^2$, whereas the parametric
form of the integral in \eq{eq:masszero2} goes like $m^2/\e$.
However, this problem is not manifest because of multiplication with
the overall factor $m^2$ in the differential equation. Since the
solution of the mass differential equation is in principle formally
derived as the integral over $m^2$ and $m^2=0$ is the only the limit
of this integral, the ambiguity encountered may be regarded as an
integrable singularity and presents no problem.

Turning now to the case $n=1$ where $I^1\equiv A_m^4$, we first
extract the overall $k_4$ factor as before by defining
$A^4=k_4\ov{A}_m$ such that the differential equation is
\be
m^2\frac{\pd\ov{A}_m}{\pd m^2}=-2\e\frac{m^2}{k^2+m^2}\ov{A}_m
-\frac{m^2}{k^2+m^2}\int\frac{\dk{\w}}{[(k-\w)^2+m^2]^2\vec{\w}^2}.
\label{eq:masszero3}
\ee
Notice that in the integral term we have used the identities
\be
\int\frac{\dk{\w}\,\w_4}{[(k-\w)^2+m^2]^2\vec{\w}^2}=
\int\frac{\dk{\w}\,(k_4-\w_4)}{[\w^2+m^2]^2
\left(\vec{k}-\vec{\w}\right)^2}
=k_4\int\frac{\dk{\w}}{[(k-\w)^2+m^2]^2\vec{\w}^2}.
\ee
Now, for $m^2\neq0$, the integral term of \eq{eq:masszero3} is finite
as $\e\rightarrow0$; however, the $m^2=0$ limit is again ambiguous but
as above this can be regarded as an integrable singularity.  Also,
when $m^2=0$, $\ov{A}$ is known to be $\e$ finite (it is the massless
integral considered in Ref.~\cite{Watson:2007mz}).  This means that as
$\e\rightarrow0$ we have the simple integral expression
\be
m^2\frac{\pd\ov{A}_m}{\pd m^2}=
-\frac{1}{x+y+m^2}\frac{1}{(4\pi)^{2-\e}}
\frac{1}{y\sqrt{1+\frac{m^2}{y}}}
\ln\left(\frac{\sqrt{1+\frac{m^2}{y}}+1}
{\sqrt{1+\frac{m^2}{y}}-1}\right).
\ee
Knowing the solution, \eq{eq:a4sol}, it is straightforward to show
that when $m^2=0$ the original massless integral from
Ref.~\cite{Watson:2007mz} is reproduced and that the derivative of the
massive solution satisfies the above.

%% file: app.noncov3pct.tex
\chapter{One-loop non-covariant vertex integrals}
\label{app:app4}

In order to extract the divergence of the non-covariant three-point
integrals appearing in the quark-gluon vertex functions presented in
Chapter~\ref{chap:g^2}, we use a method based on the Schwinger
parametrization \cite{pascual}. As an example, consider the divergent
integral:
\be
I=\int\dk{\w}\frac{\w_i\w_j}{\w^2(\vec\w+\vec k_3)^2(\w-k_1)^2}.
\label{eq:vertexloop1}
\ee
In principle, one can also consider massive quark
propagators. i.e. integrals of the form $1/\w^2[(\vec\w+\vec
k_3)^2][(\w-k_1)^2-m^2]$. However, as will shortly become clear, the
mass factor does not contribute to the divergent part and hence in the
following we will set $m=0$.  Using the Schwinger parametrization, the
formula \eq{eq:vertexloop1} can be rewritten as:
\bea
\lefteqn{I=\int_0^\infty\,d\al\, d\ba\, d\ga \int\dk{\w}
\left[
\frac{1}{2}\frac{\de_{ij}}{\al+\ba+\ga}+\frac{(\ba k_{3i}-\ga
  k_{1i})(\ba k_{3j}-\ga k_{1j})}{(\al+\ba+\ga)^2}
\right]}\nonumber\\
&&\times\exp{\left\{-(\al+\ga)\w_4^2-(\al+\ba+\ga)\vec{\w}^2
+2\ga\,\w_4 k_4-2\vec{\w}\cdot(\ba \vec k_3-\ga\vec k_1)
-\ba\vec{k_3}^2-\ga k_1^2\right\}}. \nonumber\\
\label{eq:int_wiwj}
\eea

We start by considering the first term (the factor $\de_{ij}/2$ has
been left aside):
\bea
I_0&=&\int_0^\infty\,d\al\, d\ba\, d\ga \int\dk{\w}
\frac{1}{\al+\ba+\ga}
\exp \left\{-(\al+\ga)\w_4^2-(\al+\ba+\ga)\vec{\w}^2
+2\ga\w_4 k_4\right.\nonumber\\
&&\left.-2\vec{\w}\cdot (\ba \vec k_3-\ga\vec k_1)
-\ba\vec{k_3}^2-\ga k_1^2\right\}
\eea
(in fact, it turn out that the second term in \eq{eq:int_wiwj} is
convergent and therefore not interesting for our purpose). After
making a shift
\be 
\vec\w\rightarrow\vec\w-\frac{\ba\vec
  k_3-\ga\vec k_1}{\al+\ba+\ga},\,\,\,
\w_4\rightarrow\w_4+\frac{\ga k_{14}}{\al+\ga} 
\ee
and performing the momentum integration, we are left with the
parametric integral
\bea
\hspace{-1cm}I_0&=&\frac{1}{(4\pi)^{2-\e}}
\int_0^\infty\,d\al\, d\ba\, d\ga 
\frac{1}{(\al+\ba+\ga)^{5/2-\e}}\frac{1}{(\al+\ga)^{1/2}}
\nonumber\\
&&\hspace{2cm}\times
\exp\left\{\frac{(\ba\vec k_3-\ga\vec
  k_1)^2}{\al+\ba+\ga}+\frac{\ga^2k_{14}^2}{\al+\ga}-\ba\vec
  k_3^2-\ga k_1^2\right\}.
\eea
Following the usual procedure (also used in the
Appendix~\ref{chap:app3} to check the nonstandard two-point
integrals), we insert the identity $1=\int_0^{\infty}
d\la\,\de(\la-\al-\ba-\ga)$, make the rescaling $\al\rightarrow\al\la,
\ba\rightarrow\ba\la, \ga\rightarrow\ga\la$ and perform the integral
over $\la$. After rearranging the terms, we get:
\bea
I_0&=&\frac{\G(\e)}{(4\pi)^{2-\e}}
\int_0^1\,d\al\,  d\ba \, d\ga \de(1-\al-\ba-\ga)
\frac{1}{(\al+\ga)^{1/2}}\nonumber\\
&&
\hspace{2cm}\times\left\{\ba(1-\ba)\vec k_3^2
+\ga(1-\ga)\vec k_1^2
+2\ba\ga \vec  k_1\cdot\vec k_3 
+\frac{\al\ga k_{14}^2}{\al+\ga}\right\}^{-\e}.
\eea
In the above equation, we see that the integrand is convergent in the
limit $\e\rightarrow 0$, and the divergence of $I_0$ is only given by
the overall factor $\G(\e)$ and thus in the exponent we can set
$\e=0$.  It also becomes clear that the mass term (which would enter
the argument of the exponential) cannot not influence this result, and
hence our initial setting with $m=0$ is justified.  The result is:
\be
I_0=\frac{1}{(4\pi)^2}\frac23\frac{1}{\e}+\textrm{finite}.
\label{eq:int_res}
\ee
By a similar calculation, one can show that the second term in
\eq{eq:int_wiwj} is convergent and therefore the overall divergent
structure of the original integral is
\be
I=\frac{1}{(4\pi)^2}\frac{\de_{ij}}{3\e}+\textrm{finite}.
\ee

A straightforward example is the covariant integral
\be
I_1=\int\frac{\dk{\w}\,\w_i\w_l}{\w^2(k_1-\w)^2(k_2+\w)^2}.
\ee
This can be calculated without explicitly writing out the parametric
form, by using the following simple relation:
\bea
\int\frac{\dk{\w}\,\w_i(\w+k_2)^2}{\w^2(k_1-\w)^2(k_2+\w)^2}&=&
k_2^2\int\frac{\dk{\w}\,\w_i}{\w^2(k_1-\w)^2(k_2+\w)^2}+
2k_{\mu}\int\frac{\dk{\w}\,\w_i\w_\mu}{\w^2(k_1-\w)^2(k_2+\w)^2}
\nonumber\\
&&+\int\frac{\dk{\w}\,\w_i}{(k_1-\w)^2(k_2+\w)^2},
\label{eq:cami}
\eea
where the first term and the temporal part of the second term on the
right hand side are convergent.  Using the fact that the result has
the form $\de_{il} I^{*}$ (where $I^{*}$ includes the factor $1/d$),
it is simple to get the $\e$ coefficient by combining the terms on the
right hand side of \eq{eq:cami}. We obtain:
\be
I_1=\frac{\de_{il}}{(4\pi)^2}\frac{1}{4\e}+\textrm{finite}.
\ee

Finally, let us examine the integral 
\be
I_2=\int\frac{\dk{\w}\,\w_i\w_l\w_j\w_k}
{\w^2(k_1-\w)^2(k_2+\w)^2\vec\w^2}=
I^{*}(\de_{ij}\de_{kl}+\de_{il}\de_{jk}+\de_{ik}\de_{jl}).
\ee
As before, we rearrange the terms and obtain
\bea
2k_{2k}I^{*}(\de_{ij}\de_{kl}+\de_{il}\de_{jk}+\de_{ik}\de_{jl})
&=&\int\frac{\dk{\w}\,\w_i\w_j\w_l}{\w^2(k_1-\w)^2\vec \w^2}-
\int\frac{\dk{\w}\,\w_i\w_j\w_l}{\w^2(k_3+\w)^2(\vec k_2-\vec \w)^2}
\nonumber\\
&&\hspace{-0.3cm}
+\int\dk{\w}\frac{\w_i\w_lk_{2j}+\w_i\w_jk_{2l}+\w_l\w_jk_{2l}}
{\w^2(k_3+\w)^2(\vec k_2-\vec\w)^2}+\textrm{finite}\!.
\eea
The last integral on the right hand side has been already evaluated
and hence we only need to consider the integral
\be
J=\int\dk{\w}\frac{\w_i\w_j\w_l}{\w^2(\vec k_3+\vec\w)^2(k_1-\w)^2}.
\label{eq:wiwjwl}
\ee
This we evaluate by differentiating the parametric form of the
integral \eq{eq:int_wiwj} (i.e., the integral with the factor
$\w_i\w_j$ in the numerator) with respect to $k_{1l}$. The result is
\bea
J&=&\int\dk{\w}\left\{[\ga(\de_{ij}k_{1l}+\de_{il}k_{1j}
+\de_{jl}k_{1i})
-\ba(\de_{ij}k_{3l}+\de_{il}k_{3j}+\de_{jl}k_{3i})]\right.
\nonumber\\
&&\hspace{1cm}\left.+\frac{(\ga k_{1j}-\ba k_{3j})(\ga k_{1i}
-\ba k_{3i})(\ga k_{1l}-\ba k_{3l})}{(\al+\ba+\ga)^3}\right\}
\exp\left\{\dots\right\}
\eea
(the argument of the exponential function is identical to the one of
\eq{eq:int_wiwj}).  After making the usual manipulations, we extract
the $\e$ coefficient from the parametric integral and obtain:
\be
I_2=\frac{1}{(4\pi)^2}\frac{1}{20\e}(\de_{ij}\de_{kl}
+\de_{il}\de_{jk}+\de_{ik}\de_{jl})+\textrm{finite}.
\ee

We list here all the \emph{divergent} (covariant and non-covariant)
one-loop vertex functions used in the evaluation the Coulomb gauge
quark-gluon vertices:
\bea
\int\frac{\dk{\w}\,\w_i\w_l}{\w^2(k_1-\w)^2(k_2+\w)^2}&=&
\frac{\de_{il}}{(4\pi)^2}\frac{1}{4\e}
+\textrm{finite}\\
\int\frac{\dk{\w}\,\w_4^2}{\w^2(k_1-\w)^2(k_2+\w)^2}&=&
\frac{1}{(4\pi)^2}\frac{1}{4\e}+\textrm{finite}\\
\int\frac{\dk{\w}\,\w_i\w_l}{(k_1-\w)^2(k_2+\w)^2\vec\w^2}&=&
\frac{\de_{il}}{(4\pi)^2}\frac{1}{3\e}+\textrm{finite}\\
\int\frac{\dk{\w}\,\w_4^2}{(k_1-\w)^2(k_2+\w)^2\vec\w^2}&=&
\frac{1}{(4\pi)^2}\frac{1}{\e}+\textrm{finite}\\
\int\frac{\dk{\w}\,\w_i\w_l\w_j\w_k}
{\w^2(k_1-\w)^2(k_2+\w)^2\vec\w^2}&=&
\frac{1}{(4\pi)^2}\frac{1}{20\e}(\de_{ij}\de_{kl}
+\de_{il}\de_{jk}+\de_{ik}\de_{jl})+\textrm{finite}\\
\int\frac{\dk{\w}\,\w_4^2\w_j\w_k}
{\w^2(k_1-\w)^2(k_2+\w)^2\vec\w^2}&=&
\frac{\de_{jk}}{(4\pi)^2}\frac{1}{12\e}+\textrm{finite}.
\eea

%% file: app.energy.vertex.tex
\chapter{Temporal component of the quark-baryon vertex}
\label{chap:qbvertex}

In this Appendix we present the explicit derivation of the
energy-dependent part of the Faddeev vertex, \eq{eq:gammat}. We start
with \eq{eq:faddeev.simpl} and consider the first of the permutations
of the energy integral:
\be
I=-\int\dk{k_0}\frac{1}
{\left[p_1^0+k_0-m-{\cal I}_r+\imath\e\right]
\left[p_2^0-k_0-m-{\cal I}_r+\imath\e\right]}
\G_{t}(p_1^0+k_0,p_2^0-k_0,p_3^0).
\label{eq:temporal1}
\ee
Using
\be
\frac{1}{\left[z+a+\imath\e\right]
\left[z+b+\imath\e\right]}=\frac{1}{(b-a)}
\left\{\frac{1}{z+a+\imath\e}
-\frac{1}{z+b+\imath\e}\right\}
\label{eq:intid1}
\ee
and shifting the integration variables, we find that the integral $I$,
\eq{eq:temporal1}, depends only on the momentum $p_3^0$ (and
implicitly on the bound state energy of the system $P_0$).
Explicitly, it reads (using the symmetry of $\G_t$):
\bea
I=-\frac{2}{\left[P_0-p_3^0-2(m+{\cal I}_r)\right]}
\int\dk{k_0}\frac{\G_{t}(P_0-p_3^0+k_0,-k_0,p_3^0)}
{\left[k_0+P_0-p_3^0-m-{\cal I}_r+\imath\e\right]}
\eea
Replacing this in the equation \eq{eq:faddeev.simpl}, we find:
\bea
\G_{t}(p_1^0,p_2^0,p_3^0)&=&-2g^2C_B W_{\si\si}(|\vec r|) 
\sum_{i=1,2,3}
\frac{1}{\left[P_0-p_i^0-2(m+{\cal I}_r)\right]}\nonumber\\
&&\times\int \dk{k_0}\frac{\G_{t}(P_0-p_i^0+k_0,-k_0,p_i^0)}
{\left[P_0-p_i^0+k_0-m-{\cal  I}_r+\imath\e\right]}\hspace{2cm}
\label{eq:gammat1}
\eea
The form of the equation \eq{eq:gammat1} suggests that the function
$\G_t$ can be expressed as a symmetric sum
\be
\G_{t}(p_1^0,p_2^0,p_3^0)=f(p_1^0)+f(p_2^0)+f(p_3^0),
\ee
such that the integral equation for $\G_t$ (function of three
variables) is reduced to an integral equation for the function $f$ (of
only one variable). The function $f(p_i^0)$ should be chosen such that
the integral on the right hand side of the equation \eq{eq:gammat1}
generates a factor proportional to $\left[P_0-p_i^0-2(m+{\cal
I}_r)\right]$, to cancel the corresponding factor in the
denominator. To examine this possibility, we impose the following
condition:
\be
\frac{1}{\left[P_0-p_i^0-2(m+{\cal I}_r)\right]}
\int \!\!\dk{k_0}\frac{f(P_0-p_i^0+k_0)+f(-k_0)+f(p_i^0)}
{\left[P_0-p_i^0+k_0-m-{\cal  I}_r+\imath\e\right]}
=-\frac{\al\; \imath}{P_0-3(m+{\cal I}_r)} f(p_i^0)
\ee
where $\al$ is a (dimensionless) positive constant which remains to be
determined. Rearranging the terms to factorize the function $f(k_0)$,
the above equation can be rewritten as
\bea
\lefteqn{
\int \dk{k_0} f(k_0)
\left[
\frac{1}{k_0-m-{\cal  I}_r+\imath\e}+
\frac{1}{P_0-p_i^0-k_0-m-{\cal  I}_r+\imath\e}
\right]
}\nonumber\\
&&=
(-\imath)\frac{(2\al -1)P_0-2\al p_i^0 
+ (3-4\al) (m+{\cal  I}_r)}{ 2 [P_0-3 (m+{\cal  I}_r)]} f(p_i^0).
\hspace{3cm}
\label{eq:gammat2}
\eea
Then the most obvious ansatz for  the function $f$ is
\be
f(k_0)=\frac{1}{(2\al -1)P_0-2\al k_0 + (3-4\al) 
(m+{\cal  I}_r) +\imath\e}
\ee
such that on the right hand side of the equation \eq{eq:gammat2}
the numerator is cancelled by $f(p_i)$. The next step is to complete
the integration on the on the left hand side, which gives (note that
the $\e$ prescription is chosen such that only the first term in the
bracket survives -- the integration must not give rise to any new
terms containing the energy $p_i^0$):
\bea
\int \dk{k_0} f(k_0) \frac{1}{k_0-m-{\cal  I}_r+\imath\e}
&=&-\frac{\imath}{(2\al -1)P_0 + (3-6\al) (m+{\cal  I}_r)}.
\label{eq:lhs}
\eea
It is then straightforward to compare \eq{eq:gammat2} and \eq{eq:lhs}
and find that the equality is satisfied for $\al=3/2$, leading to the
expression for the vertex $\G_t$ used in the text.